\documentclass{article}

\usepackage{arxiv}

\usepackage[utf8]{inputenc} % allow utf-8 input
\usepackage[T1]{fontenc}    % use 8-bit T1 fonts
\usepackage{hyperref}       % hyperlinks
\usepackage{url}            % simple URL typesetting
\usepackage{booktabs}       % professional-quality tables
\usepackage{amsfonts}       % blackboard math symbols
\usepackage{amsmath}       % blackboard math symbols
\usepackage{nicefrac}       % compact symbols for 1/2, etc.
\usepackage{microtype}      % microtypography
\usepackage{lipsum}		% Can be removed after putting your text content
\usepackage{doi}
\usepackage{soul}
\usepackage{dsfont}

\usepackage[numbers]{natbib}
\usepackage{graphicx}
\usepackage{blindtext}

\title{Behavior-based dependency networks between places shape urban economic resilience}

\author{ 
    \href{https://orcid.org/0000-0001-8967-1967}{\includegraphics[scale=0.06]{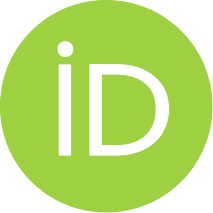}\hspace{1mm}Takahiro Yabe}\thanks{Corresponding authors: Takahiro Yabe (tyabe@mit.edu) and Esteban Moro (emoro@mit.edu)}  \\
    Institute for Data, Systems, and Society \\
	Massachusetts Institute of Technology\\
	Cambridge, MA 02139 \\
	\texttt{tyabe@mit.edu} \\
	\And
    \href{https://orcid.org/0000-0002-8660-7123}{\includegraphics[scale=0.06]{orcid.pdf}\hspace{1mm}Bernardo Garc\'ia Bulle Bueno} \\
    Institute for Data, Systems, and Society \\
	Massachusetts Institute of Technology\\
	Cambridge, MA 02139 \\
	\texttt{bernard0@mit.edu} \\
	\And
    \href{https://orcid.org/0000-0001-9487-9359}{\includegraphics[scale=0.06]{orcid.pdf}\hspace{1mm}Morgan R.~Frank} \\
    Department of Informatics and Networked Systems \\ 
    University of Pittsburgh \\
    Pittsburgh, PA 15260, USA \\
	\texttt{mrfrank@pitt.edu} \\
	\And
    \href{https://orcid.org/0000-0002-8053-9983}{\includegraphics[scale=0.06]{orcid.pdf}\hspace{1mm}Alex Pentland} \\
    Institute for Data, Systems, and Society \\
    Media Lab \\ 
	Massachusetts Institute of Technology\\
	Cambridge, MA 02139 \\
	\texttt{sandy@media.mit.edu} 
 \And
    \href{https://orcid.org/0000-0003-2894-1024}{\includegraphics[scale=0.06]{orcid.pdf}\hspace{1mm}Esteban Moro$^*$} \\
    Institute for Data, Systems, and Society \\
	Massachusetts Institute of Technology\\
	Cambridge, MA 02139 \\
    Departamento de Matemáticas \& GISC, \\ Universidad Carlos III de Madrid \\
    28911 Leganés, Madrid, Spain \\ 
	\texttt{emoro@mit.edu} \\
}

\renewcommand{\shorttitle}{\textit{arXiv} Template}

\hypersetup{
pdftitle={Behavior-based dependency networks between places shape urban economic resilience},
pdfsubject={physics.soc-ph},
pdfauthor={T. Yabe et al.},
pdfkeywords={human mobility, economic resilience, urban networks},
}

\begin{document}
\maketitle

\begin{abstract}
Urban economic resilience is intricately linked to how disruptions caused by pandemics, disasters, and technological shifts ripple through businesses and urban amenities.
Disruptions, such as closures of non-essential businesses during the COVID-19 pandemic, not only affect those places directly but also influence how people live and move, spreading the impact on other businesses and increasing the overall economic shock. However, it is unclear how much businesses depend on each other in these situations.
Leveraging large-scale human mobility data and millions of same-day visits in New York, Boston, Los Angeles, Seattle, and Dallas, we quantify dependencies between points-of-interest (POIs) encompassing businesses, stores, and amenities. 
Compared to places' physical proximity, dependency networks computed from human mobility exhibit significantly higher rates of long-distance connections and biases towards specific pairs of POI categories. 
We show that using behavior-based dependency relationships improves the predictability of business resilience during shocks, such as the COVID-19 pandemic, by around 40\% compared to distance-based models. 
Simulating hypothetical urban shocks reveals that neglecting behavior-based dependencies can lead to a substantial underestimation of the spatial cascades of disruptions on businesses and urban amenities. 
Our findings underscore the importance of measuring the complex relationships woven through behavioral patterns in human mobility to foster urban economic resilience to shocks. 
\end{abstract}

% keywords can be removed
\keywords{human mobility \and economic resilience \and urban networks}

\section*{Introduction}

Cities are central drivers of economic growth, owing to their agglomeration of businesses and amenities connected through dense social interactions, financial transactions, and human activities \cite{balland2020complex}. 
The high level of complexity enables ideas, innovations, and information to spread across organizations and communities \cite{pentland2014social,vespignani2012modelling}. 
At the same time, high urban connectivity allows shocks to cascade across time, space, and system components. 
Examples of urban networks include labor markets \cite{moro2021universal}, global supply chains \cite{inoue2019firm,supplyScience2023}, networks of social encounters \cite{yabe2023behavioral}, and infrastructure networks \cite{ouyang2012time}. 
Understanding the spread of shocks across urban spaces, among businesses and urban amenities, is crucial for resilient urban planning policies, which aim to mitigate disruptions and to improve the recovery speed and quality of businesses and organizations in cities \cite{elmqvist2019sustainability}.

From global~\cite{modica2015spatial,hidalgo2021economic} to regional~\cite{rose2004defining} scale, studies of the economic resilience of industries focus on predicting the inter-industry cascading along connections in the supply chain networks \cite{verschuur2022ports,klimek2019quantifying,colon2021criticality,inoue2019firm,levermann2014climate}.
However, just like supply chains, the demand side also has a network of dependencies, mediated by human mobility, that can curb or amplify shocks caused by changes in consumer behavior.
For example, sustained remote work during the COVID-19 pandemic \cite{barrero2021working,lucchini2021living} has been associated with decreased foot traffic to cafes near central business districts \cite{latte}, and surveys have indicated that patronage to office-district restaurants is likely to decrease \cite{salon2021potential}. Such dependencies between industries could cause shocks to cascade, thus posing a significant threat to the economic resilience of business ecosystems and their corresponding development and design \cite{zhai2022economic}.  

Dependencies among businesses are typically measured by physical proximity, assuming similar patronage to nearby businesses. As a result, several studies have investigated the resilience of business areas based only on the type and diversity of businesses and amenities \cite{hidalgo2020amenity,bahrami2022using,sevtsuk2020street}. However, these studies often do not incorporate the actual patterns of how individuals visit and interact with different businesses and places.
Recently, studies have used large-scale human mobility data (e.g., mobile phone GPS \cite{gonzalez2008understanding,blondel2015survey}) as scalable low-cost proxies for visitation patterns to various places in urban environments to study behavioral segregation \cite{moro2021mobility}, pandemic response \cite{chang2021mobility}, and disaster recovery \cite{yabe2022toward}. 
Moreover, mobile phone location data have been used to estimate the losses of businesses that rely on foot traffic (e.g., restaurants and cafes) during disaster events \cite{yabe2020quantifying,podesta2021quantifying}. 
Analysis of human behavior patterns using mobility and credit card purchasing data have revealed that activity patterns are clustered into a mixture of behavioral lifestyles (e.g., health and exercise, local trips, shopping weekends, etc.) \cite{di2018sequences,yang2022identifying}, suggesting that certain industries or place categories could have a high dependency on visitors arriving from other specific industries or places. 
For example, a gas station on a commuting route to business districts might be affected as well as the cafe in that business district if people change their visitation patterns to offices.
However, studies on economic resilience have so far neglected such interdependent relationships that human behavior patterns may generate between businesses and other amenities. 
 
This study investigates the interdependencies between urban businesses and amenities through a novel approach using human mobility data and presents a novel approach to measuring economic resilience to large-scale behavioral changes.
Using a large and longitudinal dataset of GPS location records in five major metropolitan areas in the US (New York, Boston, Los Angeles, Seattle, and Dallas), we construct and analyze behavior-based dependency networks between businesses and amenities, and further use the empirical networks to analyze and simulate the cascades of urban shocks. 
We find that empirical dependency networks generated via movements between POIs contain a significantly higher rate of long-distance connections between places, and are biased towards specific POI category pairs in comparison to a baseline network based on the gravity model.
This means shocks in one part of the economy have a greater potential to cascade across a city than would be expected.
Analysis reveals that using the behavior-based dependency network improves the predictability of the resilience of businesses during shocks, such as COVID-19, compared to dependency networks based only on physical proximity. 
We further predict the propagation of changes in visits to POIs under hypothetical external shock scenarios via network simulations and demonstrate how neglecting such dependency relationships may underestimate the extent of the shock. 
Behavior-based dependency networks enable better measurements of the effects of urban shocks, including natural hazards, novel technology, and urban development policies mediated by human behavior.

\section*{Results}

Using a large and longitudinal dataset of GPS location records in five major US metropolitan areas, we construct the behavior-based dependency network at the level of points-of-interest (POIs; e.g., businesses and amenities) in cities. Mobility data was provided by Spectus, who supplied anonymous, privacy-enhanced, and high-resolution mobile location pings for more than 1 million devices across five U.S. Census core-based statistical areas (CBSAs). All devices within the study opted-in to anonymous data collection for research purposes under a GDPR and CCPA-compliant framework. Our second data source is a collection of over 1 million verified places across five CBSAs, obtained via Safegraph. 
Within the mobility dataset, we identified stays at places that were detected to be between 10 minutes and 10 hours in duration and we spatially matched those stays with the closest place locations within 100 meters to infer visits to specific POIs. Detailed methods for visit attribution are shown in Supplementary Note 1. We implemented post-stratification to ensure the representativeness of the data across regions and income levels (Supplementary Note 1).

\begin{figure}[!t]
\centering
\includegraphics[width=.85\linewidth]{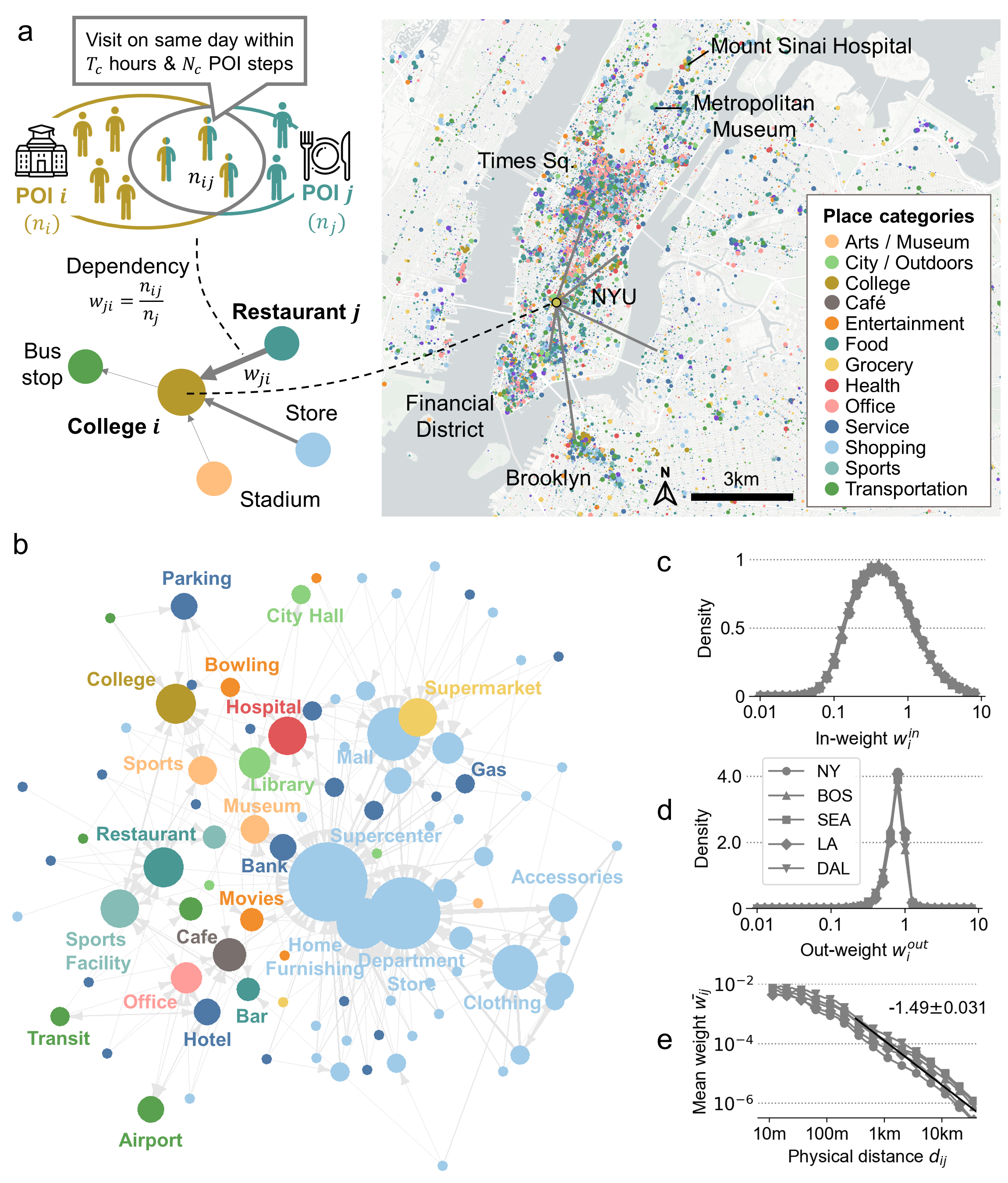}
\caption{\textbf{Behavior-based dependency networks between places in cities.} 
a. The dependency $w_{ji}$ of point-of-interest (POI) $j$ (e.g., a restaurant) on POI $i$ (e.g., a college) is computed as the proportion of the intersection of individuals that visit both the college $i$ and cafe $j$ ($n_{ij}$) on the same day within 6 hours, out of the total count of individuals who visit cafe $j$, $n_j$. Note that dependencies $w_{ij}$ and $w_{ji}$ are asymmetrical. Visualization of total in-weight ($w_i^{in} = \sum_{j} w_{ji}$) for all POIs in the Manhattan area in New York. Colors show the POI category and the node sizes show the total in-weight. 
% Areas such as Times Square, Hudson Yards, and the Financial District, and places such as the Metropolitan Museum, New York University (NYU), and Mount Sinai Hospital are revealed to have a high concentration of in-dependency from other places. 
b. Network diagram showing the average dependencies between POI subcategories in Boston (other cities shown in Supplementary Note 2). Each node is a POI subcategory and the three largest outgoing dependency edges are shown for each node. Node sizes show the in-degree of the constructed network. 
% Many shopping subcategories including supercenters, department stores, malls, and clothing stores, and colleges, cafes, restaurants are depended by many other subcategories.
c,d. Probability density distribution of in-weight $w_i^{in}$ and out-weight ($w_i^{out} = \sum_{j} w_{ij}$) per node in the five cities. The total in-weight $w_i^{in}$ has a substantially larger variance compared to the out-weight $w_i^{out}$, indicating the existence of nodes with a large attraction of dependencies, such as NYU (annotated in panel a). 
e. Average dependency $\overline{w_{ij}}$ of all POIs $i$ and $j$ with a Haversine distance of $d_{ij}$. Average dependency decays with $d_{ij}$ with a slope of $\overline{w_{ij}} \propto d_{ij}^{-1.49}$.
}
\label{fig:fig1}
\end{figure}

In this study, the dependence of a POI $i$ on another POI $j$ is defined as $w_{ij} = n_{ij} / n_i$, where $n_i$ denotes the number of individuals who visit POI $i$ and $n_{ij}$ denotes the number of occurrences that both POIs $i$ and $j$ were visited on the same day, within a 6 hour period and visited directly before or after, without any intermediate visits to other POIs (see Figure \ref{fig:fig1}). 
Because the denominator is based on the number of users who visit the target POI, $w_{ij} \neq w_{ji}$. This simple but intuitive measure considers the asymmetric nature of dependencies between POIs, for example, a cafe could have the majority of the customers coming from a nearby college, but the opposite is rarely the case. 
%The common visitors between POIs $i$ and $j$ are computed using the mobility data, by taking the set of users who visit both POIs $i$ and $j$ on the same day within a 6-hour time period and visited directly before or after, without any intermediate visits to other POIs. 
By computing the dependency weights $w_{ij}~ \forall i,j$, we obtain the adjacency matrix of the behavior-based dependency network $W \in \mathbb{R}^{N \times N}$ where $N$ is the total number of POIs present in the CBSA. The network is weighted and directed with directions indicating dependence (e.g., a link from POI $i$ to POI $j$ indicates that $i$ depends on $j$).
Bootstrap method was used to compute the standard errors for each $w_{ij}$, and to remove the edges with statistically insignificant weights from the dependency network (Supplementary Note 2).
We demonstrate the robustness of the dependency network against the choice of the visit attribution parameters, co-visit detection threshold parameters, and the choice of the POI dataset in Supplementary Note 3.

\subsection*{Behavior-based dependency networks}

The behavior-based dependency networks embed the complex spatial and functional dependencies between businesses and amenities. 
For each place $i$, we compute the total in-weight $w_i^{in} = \sum_{j} w_{ji}$ and out-weight $w_i^{out} = \sum_{j} w_{ij}$, across the five cities. 
Figure \ref{fig:fig1}a visualizes the total in-weight of all nodes in the behavior-based dependency network in the Manhattan area of New York, showing the total in-weight $w^{in}_{i}$ with node sizes, colored by place categories. Areas such as Times Square, Hudson Yards, and the Financial District, and places such as the Met Museum, New York University, and Mount Sinai Hospital have a high concentration of in-dependency from other places.
To obtain a more qualitative understanding of the dependency network, the network diagram in Figure \ref{fig:fig1}b shows the average dependencies between POI subcategories in Boston (other cities shown in Supplementary Note 2). Each node represents a POI subcategory (there are 96 of them in the dataset) and the three largest outgoing dependency edges are shown for each node. Node sizes show the in-degree of the constructed network (i.e., how many other POI categories depend on that node). Many shopping subcategories including supercenters, department stores, malls, and clothing stores, and colleges, cafes, restaurants are depended by many other subcategories. 

Figure \ref{fig:fig1}c and d show the probability density distribution of $w_i^{in}$ and $w_i^{out}$, respectively. 
Despite the geographical, sociodemographic, and economic differences across the five metropolitan areas, we observe striking similarities in the in- and out-weight distributions across cities.  
The total in-weight $w_i^{in}$ has a substantially larger variance, ranging from $0.01$ to $10$, compared to the out-weight $w_i^{out}$, which mostly ranges between $0.1$ and $1$.  
This is mainly due to the functional form of $w_{ij}$, which allows an arbitrary number of places to depend on a specific place, but limits how much one place can depend on others. 
The long tail of $P(w_i^{in})$ indicates the existence of nodes with a large attraction of dependencies, such as the Metropolitan Museum in New York City (annotated in panel a), as well as major retail stores such as Walmart, Market Basket, and Home Depot (Supplementary Note 2.2 shows a list of places with high $w_i^{in}$ and $w_i^{out}$).

\begin{figure}[!t]
\centering
\includegraphics[width=\linewidth]{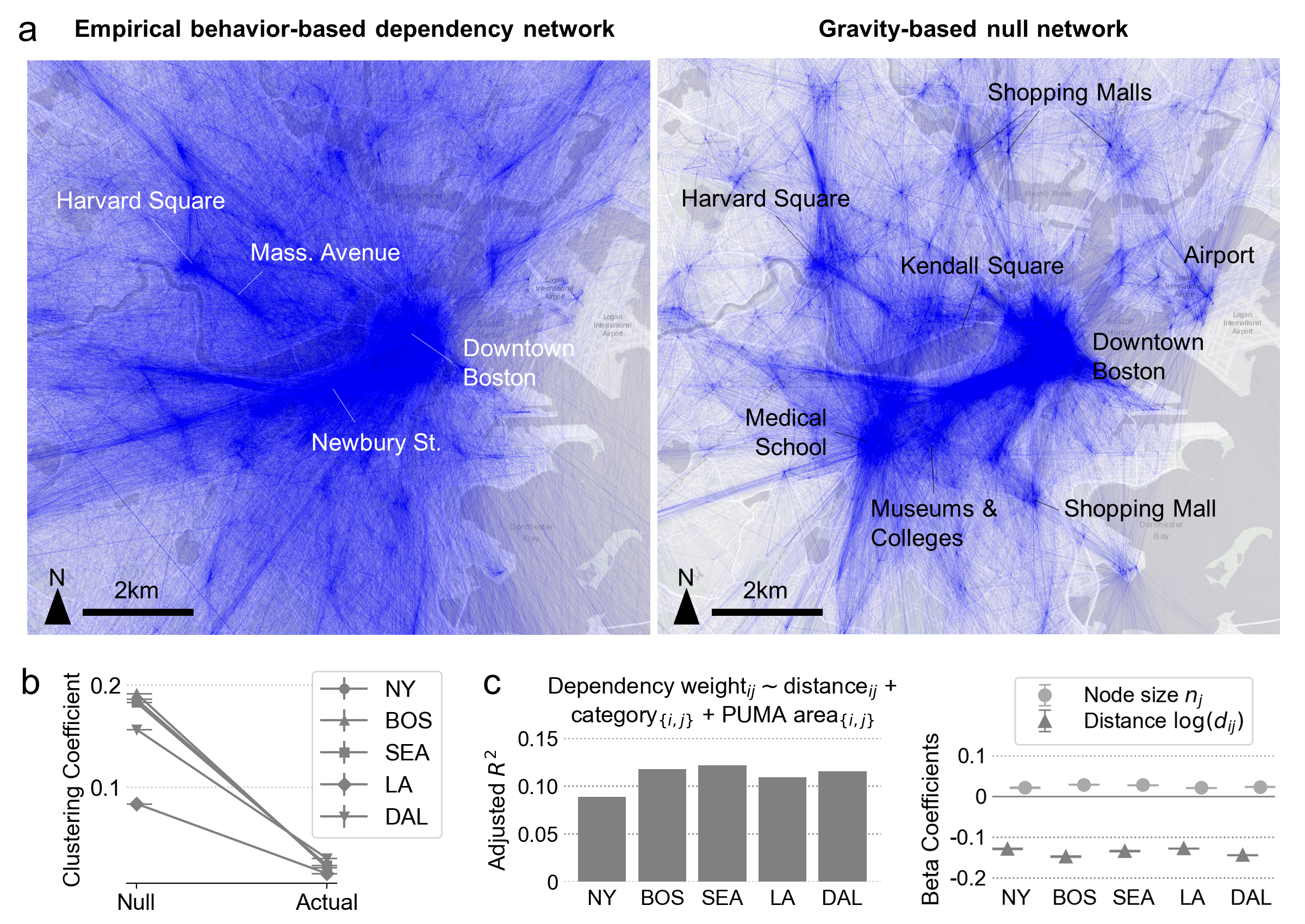}
\caption{\textbf{Behavior-based dependency networks are different from co-location networks.} 
a. A visual comparison of the empirical behavior-based dependency network (left) and the simulated null network (right) that stochastically generates network weights between places based on the fitted gravity law, while controlling for the in-degree, in-weight of the nodes. Although the number of links and the total weight of each node are consistent, the empirical network is more spatially dispersed compared to the null, indicating the existence of long-distance dependencies between places. On the other hand, the null network exhibits clustered local connections around large hubs including university campuses and shopping malls. 
b. The null network has a higher average clustering coefficient compared to the empirical network. c. Adjusted R squared of the OLS model that regresses logged dependency weight $\log{w_{ij}}$ by physical factors, including the logged distance between POIs $i$ and $j$, the total number of visits, and POI subcategories of POIs $i$ and $j$. The low R squared between 0.1 to 0.15 indicates that the dependency weights have distinct characteristics other than physical factors. Beta coefficients of the size of the node $n_j$ and the physical distance $d_{ij}$ between nodes $i$ and $j$ in the OLS model. 
Dependency weights are larger when the node has more visitors and when the distance is shorter.
}
\label{fig:fig2}
\end{figure}

\subsection*{Behavior-based dependency networks are different from co-location networks}

``Everything is related to everything else, but near things are more related than distant things'', according to Tobler's First Law of Geography \cite{tobler1970computer}.
The behavior-based dependency network is no exception, due to the spatial limits in human mobility \cite{gonzalez2008understanding}. 
Figure \ref{fig:fig1}e plots the average dependency $\overline{w_{ij}}$ of all POIs $i$ and $j$ separated by a Haversine distance of $d_{ij}$ (in kilometers). In all five metropolitan areas, the average dependency $\overline{w_{ij}}$ of all POIs $i$ and $j$ with a Haversine distance of $d_{ij}$ is relatively constant until around 100 meters, but decays with $d_{ij}$ in a power law trend with a slope of $\overline{w_{ij}} \propto d_{ij}^{-1.49}$, estimated by fitting the domain $d_{ij} \in (0.5km,100km)$. The slow decay shows that dependency between businesses extends far beyond their local area. However, empirical dependency cannot be described by distance only. To test that idea we compare empirical dependency networks to null networks generated from principles of physical co-location. 

To generate null networks that have similar density and physical characteristics, the total in-weight and in-degree of each node were conserved from the empirical network. Links of the null networks were generated stochastically according to probabilities proportional to the gravity law $g_{ij} = n_i n_j / (d_0 + d_{ij})^\gamma$, where $n_i$ and $n_j$ are the total number of visits to POIs $i$ and $j$, $d_{ij}$ is the physical distance between POIs $i$ and $j$, $d_0$ is the distance cutoff parameter, and $\gamma$ is the exponent parameter of the gravity model (see Methods). 
Parameters $d_0$ and $\gamma$ were fitted empirically to maximize the correlation between $g_{ij}$ and $n_{ij}$, which is the total number of common visitors between POIs $i$ and $j$, and were estimated as $d_0=0.2$ and $\gamma=1.5$  (Supplementary Note 4.1). 
% The weights of the generated edges in the null network were estimated by distributing the total in-weight of each node $w_i^{in} = \sum_{j \in N(i)} w_{ji}$ proportionally to $g_{ij}$. Thus, the generated edge weight of the null network is computed as $\hat{w_{ij}} = w_i^{in} * (g_{ij} / \sum_j g_{ij})$.
Figure \ref{fig:fig2}a compares the empirical behavior-based dependency network (left) and the simulated null network (right). Despite controlling for the in-degree and total in-weight of each node, the empirical network is more spatially dispersed compared to the null, indicating the existence of longer-distance dependencies between places. On the other hand, the null network exhibits clustered local connections around large hubs including university campuses and shopping malls as shown quantitatively using the average clustering coefficients in Figure \ref{fig:fig2}b, even though all nodes have the same in-degree and in-weight in both networks.

Furthermore, to disentangle the physical and behavioral factors that generate the observed dependency networks, we tested a simple regression model with the specification: $\log{w_{ij}} \sim \log{d_{ij}} + n_j + \eta_i + \eta_j + \theta_i + \theta_j$, where $n_j$ is the total number of visits to place $j$, and $\eta_i$ and $\theta_i$ denote the place category and Public Use Microdata Area (PUMA) of place $i$, respectively. 
Figures \ref{fig:fig2}c show the summary of the regression results. 
The regression results reveal that distance, POI type, and neighborhood only explain around 9\% to 12\% of the variance observed in the dependency weights, suggesting that the empirical behavior-based dependency network contains much more nuanced and specific information about the relationships between places, which cannot be fully captured using physical factors of the places. 
% and that the dependency weights are larger when the dependee has more visitors $n_j$ and when the distance $d_{ij}$ is shorter.
% Figure \ref{fig:fig2}a demonstrates, using an example of MIT, how physical distance does not necessarily strongly correlate with dependency. 
Full regression results, as well as robustness tests against the choice of model parameters and selected time period are shown in Supplementary Note 4.

\begin{figure}
\centering
\includegraphics[width=\linewidth]{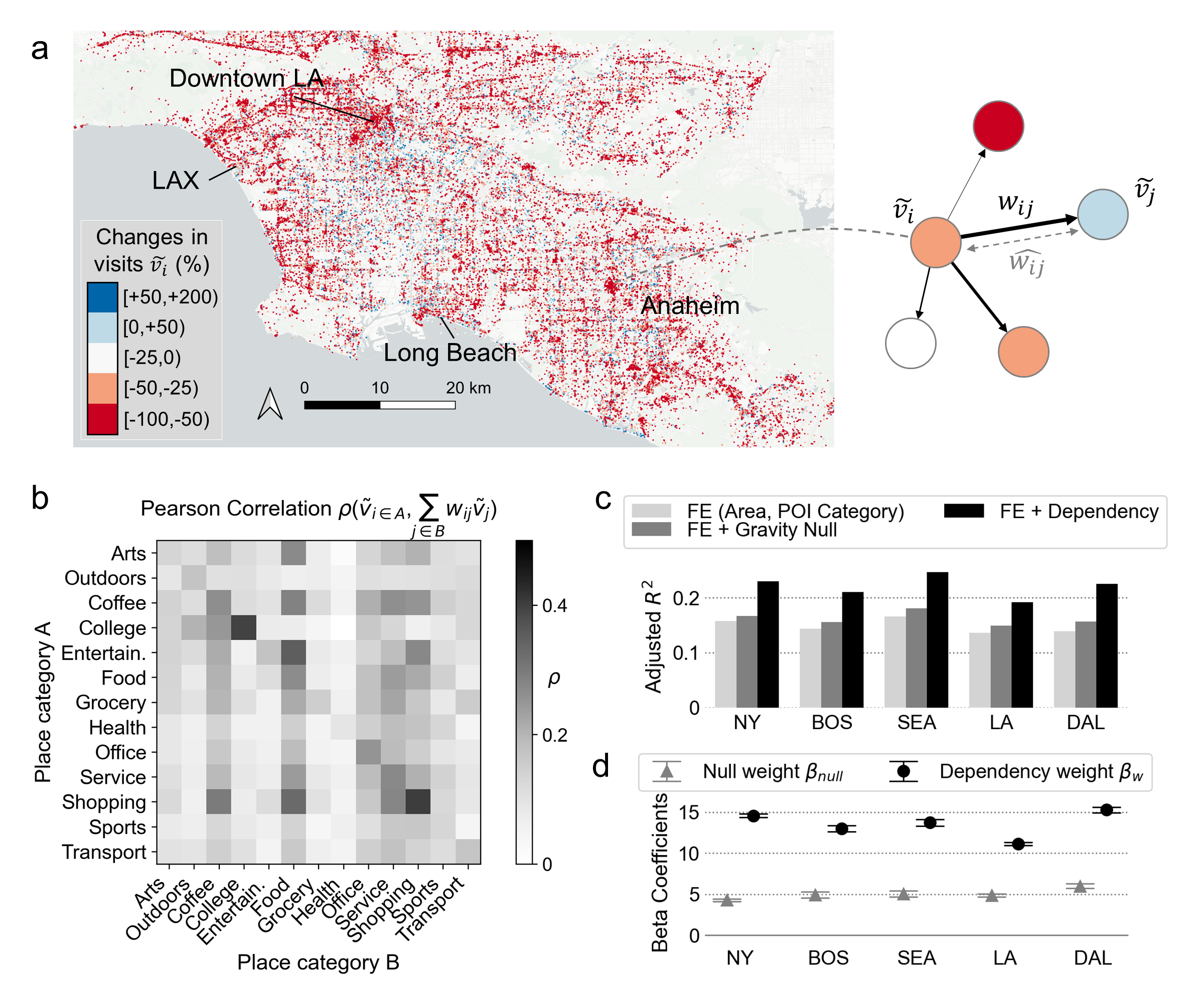}
\caption{\textbf{Behavior-based dependency networks shape economic resilience.}
a. Map visualizing the change in visitation patterns to places $\tilde{v_i} = (v_i^{COVID}/v_i^{pre} - 1)*100\%$, during the pandemic (March - May 2020) compared to the pre-pandemic period (September - December 2019) in Los Angeles. The schematic on the left shows the specification of the model, where the normalized visits at place $i$ are regressed using the sum of dependency weights weighted by the normalized visits of the neighboring nodes, $\sum_{j} w_{ij} \tilde{v_j}$, and PUMA and place category fixed effects for POI $i$. 
b. Correlation between the change in visits to the ego $\tilde{v_{i\in A}}$, which belongs to category A, and the weighted sum of the neighbors that belong to category B, $\sum_{j in B} w_{ij} \tilde{v_j}$. Indeed, Figure \ref{fig:fig3}b shows that most coefficients are significant and positive indicating that loss in visits in the neighbors is shared with the ego.
% (only estimates with $p<0.01$ are shown). 
% Loss of visits to service and shopping stores has the most substantial cascading effect through the behavior-based dependency network.
c. Adjusted $R^2$ of the normalized visits using i) area and category fixed effects, ii) fixed effects and gravity model-based dependency weights $\hat{w_{ij}}$, and iii) fixed effects with behavior-based dependency weights $w_{ij}$. Using the behavior-based dependency network significantly improves the $R^2$ by 40\% compared to using the distance-based null network. 
d. Regression coefficients for the physical distance-based dependency ($\beta_{null}$) and behavior-based dependency ($\beta_w$) for the five cities and the pooled model. All variables were centered and standardized. The effects of the behavior-based dependency are 2-3 times in magnitude compared to the physical distance-based dependency. 
}
\label{fig:fig3}
\end{figure}

\subsection*{Predictability of economic resilience via behavior-based dependency}

The results so far show that the behavior-based dependency networks encode place-specific relationships between businesses and urban amenities that cannot be fully captured by physical characteristics alone. 
Here, we empirically investigate whether using the dependency network could improve the predictability of how shocks cascade across places in cities, using the COVID-19 pandemic as an example of an extreme empirical shock. 
The observed change in visits to different places is computed by $\tilde{v_i} = (v_i^{COVID}/v_i^{pre} - 1)*100\%$, where $v_i^{COVID}$ and $v_i^{pre}$ denote the number of visits to place $i$ during the pandemic (March - June 2020) and the pre-pandemic period (September - December 2019), respectively. 
Figure \ref{fig:fig3}a plots the change in visits $\tilde v_i$ in the Los Angeles metropolitan statistical area, which indicates that most, but not all, places experienced a negative effect. The probability density distributions for $\tilde v_i$ for each metropolitan area, and for different periods during the pandemic can be found in Supplementary Note 5.1. 

Given the dependency between places, we model the change in visits to $i$, $\tilde v_i$, as the sum of the direct loss of visits that place $i$ experienced due to the pandemic and the network effects from its neighbors. The network effects component, which is the weighted sum of the change in visits that its depended neighbors $j$ experienced, is described as $\tilde v_i = \sum_{j} w_{ij} \tilde v_j $, as depicted in the schematic on the right panel of Figure \ref{fig:fig3}a. 
The dependency weights $w_{ij}$ are computed using mobility co-visit data from prior to the pandemic (September - December 2019). 
We test whether there is a significant correlation between the change in visits to the ego $\tilde v_i$ and the weighted sum of the neighbors $\sum_{j} w_{ij} \tilde v_j$ for different place category pairs. Indeed, Figure \ref{fig:fig3}b shows that most coefficients are significant and positive indicating that loss in visits in the neighbor nodes is shared with the ego node. Loss of visits to service and shopping stores has the most substantial correlation with other nodes through the behavior-based dependency network.

To demonstrate that these correlations between POI types are not the result of the blanket impact of COVID on all POIs, we consider several alternative null models to predict $\tilde v_i$ and find that the empirical dependency network is most predictive.
Combining the network effects with the fixed effects of the place $i$'s subcategory $\eta_i$ and the Public Use Microdata Area (PUMA) $\theta_i$, we model the change in visits of $i$, $\tilde v_i$  as:
\begin{equation}\label{modelspec}
    \tilde v_i  = \beta_0 + \underbrace{\beta_W \sum_{j} w_{ij} \tilde v_j} _\textrm{network effects} + \underbrace{\eta_i + \theta_i }_\textrm{ego fixed effects} + \epsilon
\end{equation}
As a baseline for comparison, we also prepare an alternative model that models the dependency between places based on the gravity-based null network weights $\hat w_{ij} $ instead of the empirical dependency weight $w_{ij}$. %thus $\tilde v_i = \beta_0  + \beta_{null} \sum_{j} \hat w_{ij} \tilde v_j + \gamma_i + \theta_i$. 
% The functional form of $g_{ij} = n_i n_j /(0.2+d_{ij})^1.5$ was used to generate null weights $\hat{w_{ij}}$, similar to the results in Figure \ref{fig:fig3}, which assigns uniform weights for places located within approximately 200 meters but quickly decays beyond 200 meters, mimicking the patterns shown in Figure 1c. 

Figure \ref{fig:fig3}c shows the adjusted $R^2$ of the normalized visits using i) area and category fixed effects, ii) fixed effects and gravity-based null model $\hat w_{ij}$, and iii) fixed effects with behavior-based dependency $w_{ij}$. Using the behavior-based dependency network significantly improves the $R^2$ by at least 40\% compared to using the physical proximity-based network model. 
Figure \ref{fig:fig3}d shows the estimated regression coefficients for the physical distance-based dependency ($\hat \beta_{null}$) and behavior-based dependency ($\hat \beta_w$) for the five cities and the pooled model. All variables were standardized by subtracting the mean and dividing by the standard deviation for the coefficients to be comparable. Full regression tables are shown in Supplementary Note 5.3.
The effects of the behavior-based dependency are 2-3 times in magnitude compared to the physical distance-based dependency. 
Supplementary Note 5.4 shows the estimation results for the different time periods, which all showed better predictability using the behavior-based dependency network. 
Furthermore, similar results were obtained when using summer breaks as exogenous shocks. The dependency network was able to predict the changes in visits to places that have high dependency on college campuses (see Supplementary Note 5.7).

\begin{figure}[!t]
\centering
\includegraphics[width=\linewidth]{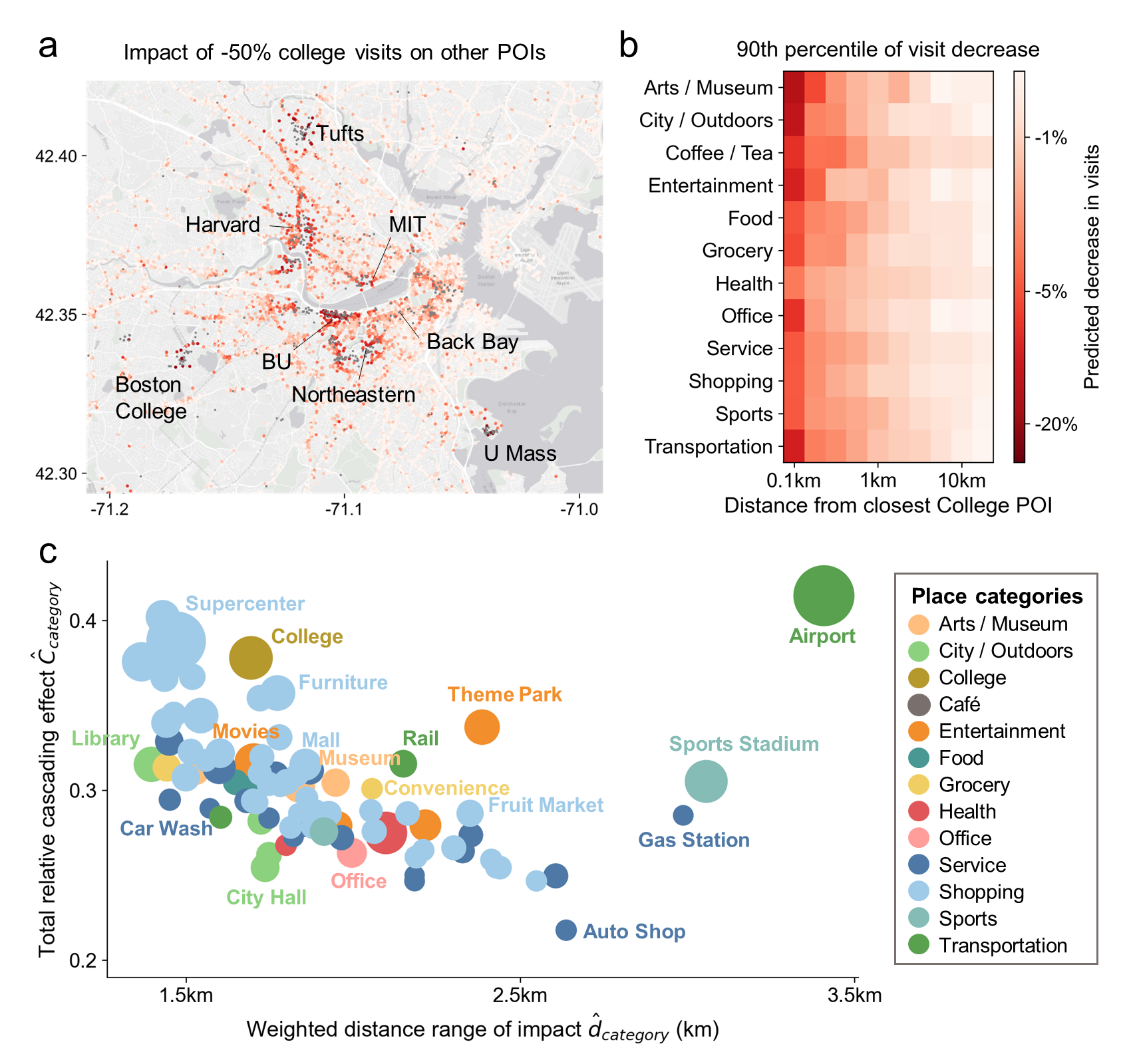}
\caption{\textbf{Cascading impacts of hypothetical urban shocks.} 
a. Simulated effects of a 50\% reduction in visits to college POIs (gray points) on nearby non-college POIs (red points; darker red indicates larger negative impacts), using the fitted Leontief Open Model. Impacted POIs are not limited to those in proximity to college POIs, but also in locations that are popular with college students.
% , for example, Massachusetts Avenue which connects MIT and Harvard University. 
Neglecting the behavior-based dependencies results in a substantial underestimation of the effects on POIs that are located further away from colleges. 
b. Impacts of the 50\% visit reduction to colleges on places by category and distance (90th percentile decrease in visits are shown, log scaled). 
While most significant impacts occur within 0.5 km, places such as arts and museums, food, and service locations experience substantial long-distance impacts. 
c. Total cascading impact of closing places on other locations, relative to its own size (x-axis) and the weighted distance range of the impact (y-axis) for different POI subcategories. Supercenters and colleges have high cascading effects but are focused locally ($\sim$ 1.5km around the POI). On the other hand, the impacts of airports, stadiums, theme parks, and gas stations are both large and far-reaching (around 2.5km to 3.5km). 
% Understanding the magnitude and spatial extent of the cascading effects could be applied to design emergency management policies to effectively close places while minimizing economic losses. 
}
\label{fig:fig4}
\end{figure}

\subsection*{Cascading impacts of hypothetical urban shocks}

Besides the COVID-19 pandemic, what can the behavior-based dependency network tell us about other types of future shocks, such as the increase in online education, remote health services, and fewer visits to gas stations due to higher adoption of electrical vehicles?
To explore these questions, we apply the network effects model (equation \ref{modelspec}) to simulate the spatial cascades of such shocks in different cities. More specifically, re-writing and reorganizing equation \ref{modelspec} in matrix form, we obtain $\Vec{v} = W \Vec{v} + \Vec{f}$, where $\Vec{v}$ is a vector of $\tilde v_i$ for all $N$ places, $W$ is an $N \times N$ matrix where each element is $\tilde w_{ij} = \hat \beta_W  w_{ij}$, and vector $\Vec{f}$ is an aggregation of all fixed effects $\beta_0$, $\eta_i$, and $\theta_i$. This model specification is known as the Leontief Open Model, which is a simplified and linear economic model for an economy in which input equals output \cite{leontief1986input}.
To predict the propagation of shocks throughout places in the city, the shocks are modeled in the fixed effect vector $\Vec{f}$ (e.g., all colleges experience an external shock where visits are reduced by $50\%$ due to uptake of online education), and the production vector $\Vec{v}$ is computed by solving the linear system $\hat{\Vec{v}} = (I-W)^{-1} \Vec{f}$.% via the generalized minimal residual iteration method.

The shift to online education which occurred during the pandemic is reported to have a continuing effect, with roughly 20 percent of school systems planning to or have already started online school programs \cite{RR-A956-1}. 
Previous studies \cite{bonner1968economic} as well as analysis in Figure \ref{fig:fig2} have pointed out that college campuses have a substantial impact on the local economy. 
If online learning and remote education were permanent and increased with the help of advanced technology (e.g., augmented reality), what impacts would it have on other businesses and amenities?
Figure \ref{fig:fig4}a shows the simulated effects of a 50\% reduction in visits to college POIs (gray points) on nearby non-college POIs (red points; darker red indicates larger negative impacts). Impacted POIs are not limited to those in proximity to college POIs, but also in locations that are popular with college students, for example, Massachusetts Avenue which connects MIT and Harvard University. 
For comparison, we simulated the shocks to non-college POIs using the physical distance network $\hat{W}$, where $\hat w_{ij} $ is used as the matrix elements instead of behavior-based dependency $w_{ij}$ (Supplementary Note 6.1). 
Comparing the simulation results using the dependency network and the null network shows that neglecting the behavior-based dependencies results in a substantial underestimation of the effects on POIs that are located further away from colleges. 

The effects of online education were heterogeneous for different place categories located at different distances from colleges. Figure \ref{fig:fig4}b shows the 90th percentile of impacts on POIs by category and distance (log scaled). 
While most significant impacts occur within 0.5 km, places such as arts and museums, food, and service places experience substantial long-distance impacts. 
% To quantitatively compare the decay of impact with respect to the distance from the closest college between the two models, Figure \ref{fig:fig4}d plots the normalized (divided by the average impact to POIs within 100 meters) visit decrease at different distance intervals. The figure highlights the long-distance cascade of shocks from colleges mediated by the complex interdependencies created by human mobility behavior. 
Simulations assuming different levels of visit decrease to colleges (e.g., -100\%, -25\%), and other case studies of hypothetical shocks, including increased rate of remote health services and decreased visits to gas stations due to the adoption of electric vehicles, show a similar long-distance cascade of shocks (Supplementary Note 6.1).
These persistent spatial cascades emphasize the importance of considering behavior-based dependency relationships between places to grasp the holistic impact of such urban shocks for resilient urban planning.

Further leveraging the network model, we are able to simulate the impacts of POI closure scenarios and identify the seed nodes (POIs) that have the largest cascading effects on other POIs if inflicted by other urban shocks. 
For each node, we simulate the cascading impacts of a 100\% visit change to a single POI $i$, by computing $\hat{\vec{\nu}}^{(i)} = (I-W)^{-1} \vec e^{(i)}$, where $\vec e^{(i)}$ is a one-hot encoding vector of the initial shock that assigns a change in visits of $+1$ to node $i$ and $0$ otherwise, and $\hat{\vec{\nu}}^{(i)}$ is the resulting vector of the cascading impacts, where each element measures the impacts of the initial shock to all nodes.
The total impacts of changes in the number of visits to all nodes can be computed by multiplying $\hat{\vec{\nu}}^{(i)} = (\hat \nu_1^{(i)},\ldots,\hat \nu_N^{(i)})$ with the vector of total visits to each POI, $\vec n = (n_1, \cdots, n_N)$. Thus, the total impacts of the initial shock to node $i$ can be computed by $C_i = \sum_{j;j\neq i} \hat{\nu}^{(i)}_j  n_j$. 
By further scaling the impact to its own size $n_i$, we obtain the total relative cascading effect as $\hat{C_i} = C_i/n_i$. 
For example, $\hat{C_i}=0.3$ would indicate that increasing the number of visits to node $i$ by 100\% ($=n_i$) results in a total of 30\%$ \times n_i$ increase in visits across all other nodes. 
The mean relative cascading impacts of each POI category, $\hat{C}_{category}$ are shown in the y-axis of Figure \ref{fig:fig4}c. 
POI categories such as airports, supercenters, colleges, furniture stores, theme parks, railway stations, and sports stadiums have a high impact on other POIs in urban areas propagated through behavior-based dependency networks. 

When implementing policies to close down certain POIs for emergency response (e.g., lockdowns during pandemics), it is important to understand the spatial extent of the cascade. 
To quantify this, we defined the distance range of the cascade by computing the average distance to impacted nodes, weighted by the magnitude of the impacts. More specifically, we compute the weighted distance range of POI $i$ by $\hat{d}_{i} = \sum_{j; j\neq i} \hat{\nu}^{(i)}_j d_{ij} / \sum_{j; j\neq i} \hat{\nu}^{(i)}_j$.
The weighted distance range of impact for each POI category, $\hat{d}_{category}$ are shown in the x-axis of Figure \ref{fig:fig4}c. 
Supercenters and colleges have high cascading effects but are focused locally ($\sim$ 1.5km around the POI). On the other hand, the impacts of airports, stadiums, theme parks, and gas stations are both large and far-reaching (around 2.5km to 3.5km). 
Estimation results for all cities are shown in Supplementary Note 6.2.
Understanding the magnitude and spatial extent of the cascading effects could be applied to design emergency management policies to effectively close places while minimizing economic losses. 
The large magnitude of the spatial cascades that occur due to behavior-based dependency networks calls for new urban policy-making approaches that balance the benefits of mobility restriction measures (e.g., preventing the spread of diseases) while minimizing the total cascading economic impacts to urban places and amenities.

\section*{Discussion}

Fostering the resilience of urban systems to shocks is an urgent challenge for cities and communities, with increasing risks of climate change-induced disasters, long-lasting effects of the COVID-19 pandemic, and unprecedented technological shifts in how we move (electric vehicles, autonomous vehicles), work\cite{hansen2023remote}, shop, and learn\cite{schwartz}. 
Such urban shocks could induce significant shifts in human behavior and urban activity patterns by changing the various incentive and cost mechanisms that motivate activities in cities.
A plethora of research has focused on modeling the spillover effects of disruptions on supply chains across industries (e.g., \cite{inoue2019firm}), however, there has been limited investigation into the spillover effects that could be mediated through the movement of people in cities.
The broader socioeconomic impacts that such behavioral changes could have on cities, for example on the social fabric of communities \cite{yabe2023behavioral}, economic networks and local businesses, and urban infrastructure systems, are not well understood. 
Motivated by these critical challenges, we used empirical data from mobile phone devices collected from five major metropolitan areas in the US to quantitatively measure and analyze the urban economic networks mediated by human behavior, and their resilience to potential urban shocks. 

In this context, our paper contains three important contributions toward understanding the economic network dependencies in cities. 
First, our approach measures and reveals that the dependent relationships that exist between businesses and places are highly complex products of human behavioral preferences and decisions rather than a measure determined solely by the urban form, including the physical distance between places, and their physical locations, popularity, and categories (which only explain around 10\% of the variance). We observe the existence of places that are highly dependent on hundreds of other places (e.g., gas stations, gyms) and others that are depended by many other locations (e.g., major shopping centers, universities, major hospitals). 
Our results show that urban economic networks are determined by the behavior generated by individual activity patterns, connecting distant businesses and amenities due to the combination of work, leisure, and shopping activities on the same day. However, businesses that are next door to each other are not necessarily dependent on each other since they target and attract people with different interests and lifestyles.
Second, using different periods of the COVID-19 pandemic as external shocks, we showed that using the pre-pandemic behavior-based dependency network with a Leontief input-output formulation improves the predictability of the spillover shocks to different businesses, compared to the distance-based co-location network. 
Third, simulations of hypothetical urban shocks showed that the dependencies generated by human behavior significantly amplify the shocks to places that are located further away from the origins of the shocks. 
Our results show that for example, while supercenters affect mostly local POIs, airports, sports stadiums, and gas stations have a significant wide-range effect on POIs across the city. 
Policies to contain the spread of pandemics or future urban shocks need to incorporate the spatial extent of the impacts mediated by the dependency network.    
This points to the importance of shifting from a place-based approach to a network-based approach in designing urban interventions and, in general, in understanding the socioeconomic impacts of urban changes. 

Our study has several limitations.  
Dependency weights between places were computed using all mobile phone users that were observed to visit both places (`co-visit'), however, we were not able to differentiate between visitations of different natures. We could further classify co-visits into different types such as routine and exploration behavior. Moreover, following recent studies focusing on the substantial differences in mobility behavior across sociodemographic groups (e.g., \cite{gauvin2020gender,wang2018urban}), we could decompose the dependencies into different income ranges, to better understand which sociodemographic segments are contributing to different types of dependency relationships. Therefore, the dependency metric computed in our study should be interpreted as an aggregated measure of all co-visits that occur in cities, and further decomposition and contextualization of the dependency metric could be conducted when applying this approach to analyze specific urban shocks and policies. 
Another challenge lies in understanding how the dependency networks between places reorganize due to various urban shocks. The assumption in this study did not consider such dynamic reorganization because of the sudden and short-term nature of the shocks we analyzed (e.g., COVID-19, closures of colleges). 
Modeling the dynamics of the behavior-based dependency networks using human behavior data observed across a longer time frame could be an interesting topic for future research. 

Our findings have implications for our understanding of the resilience of urban systems to shocks. While spillover effects of urban shocks (e.g., disasters, power outages) have been often studied with a focus on supply chains that connect firms and industries, our results show that human behavior and mobility patterns also contribute to the cascade of shocks across businesses and amenities in different industries. To better understand how the effects of urban shocks or technological shifts would manifest in cities, a spatial understanding of risk (which in itself underestimates the broader impacts of shocks) should be complemented by how the flow of individuals connects different firms and places. This framework could be applied to assess the impacts of both negative and positive shocks on cities. Examples of negative shocks include natural hazards, and pandemics, while urban transportation (e.g., fare-free bus programs) and land use policies (e.g., pedestrian-only streets) could have positive shocks on businesses in the neighborhoods. Urban planners could leverage the observed dependency relationships to target such policies, mitigate potential disruptions better, and amplify the positive impacts on local businesses and the well-being of communities. It also calls for a more holistic understanding of shopping, innovation, or shopping districts, since the vibrancy of those places and their impact on other areas might depend on business and amenities across the city.

\section*{Data Availability}
The data that support the findings of this study are available from Spectus through their Social Impact program, but restrictions apply to the availability of these data, which were used under the license for the current study and are therefore not publicly available. Information about how to request access to the data and its conditions and limitations can be found in \url{https://spectus.ai/social-impact/}.  
Data access requests should be submitted through Spectus' Social Impact customer page \url{https://spectus.ai/lp/book-a-demo/}, where the Sales team at Spectus may be contacted in a timely manner. Other data including the American Community Survey is available for download at \url{https://data.census.gov/}, and Tiger shapefiles can be downloaded from the US Census Bureau \url{https://www.census.gov/programs-surveys/geography/guidance/tiger-data-products-guide.html}.

\section*{Code Availability}
The analysis was conducted using Python. Code to reproduce the main results in the figures from the aggregated data is publicly available on GitHub \url{https://github/takayabe0505/dependencynetwork}.

\section*{Methods}

\subsection*{Mobility and point-of-interest data}
We utilize an anonymized location dataset of mobile phones and smartphone devices provided by Spectus Inc., a location data intelligence company that collects anonymous, privacy-compliant location data of mobile devices using their software development kit (SDK) technology in mobile applications and ironclad privacy framework. Spectus processes data collected from mobile devices whose owners have actively opted in to share their location, and require all application partners to disclose their relationship with Spectus, directly or by category, in the privacy policy. With this commitment to privacy, the dataset contains location data for roughly 15 million daily active users in the United States. Through Spectus’ Social Impact program, Spectus provides mobility insights for academic research and humanitarian initiatives. All data analyzed in this study are aggregated to preserve privacy. 
To measure the visitation patterns of individuals in urban environments, we attribute the stops of individual users to specific places in the city. To study the stops at different places, we use stops that are longer than 10 minutes but shorter than ten hours. In our study, we use location data of places collected by Safegraph. To protect the users' privacy, we have removed various privacy-sensitive places from our places database, including health-related places, places where the vulnerable population are located, military-related, religious facilities, places that are related to sexual-orientation, and adult-oriented places. As a result, we have a total of over 1 million places in the five cities.
The home locations of individual users are estimated at the CBG level using different variables including the number of days spent in a given location in the last month, the daily average number of hours spent in that location, and the time of the day spent in the location during nighttime. See Supplementary Note 1.1 for more details. 
The representativeness of this data has been tested and corrected in Supplementary Notes 1.3 and 1.4 using post-stratification techniques.
Since the data used was anonymized and spatially aggregated at places, categories, or census areas, we were granted an Exemption by the MIT Committee on the Use of Humans as Experimental Subjects (COUHES protocol \#1812635935) and its extension \#E-2962.

\subsection*{Behavior-based dependency networks}
We define the dependence of a POI $i$ on another POI $j$ as $w_{ij} = \frac{n_{ij}}{n_i}$, where $n_i$ denotes the number of visits to POI $i$ and $n_{ij}$ denotes the number of `co-visits' between POIs $i$ and $j$. A co-visit is defined as an instance in which POIs $i$ and $j$ were visited by the same individual 1) on the same day, 2) within $T_c$ hours from exiting POI $i$ ($j$) to entering POI $j$ ($i$), and 3)  within $T_s$ intermediate POIs. 
Because the denominator is based on the number of visits to the target POI, $w_{ij} \neq w_{ji}$. This simple but intuitive measure considers the asymmetric nature of dependencies between POIs. By computing the dependency weights $w_{ij}~ \forall i,j$, we obtain the behavior-based dependency matrix $W \in \mathbb{R}^{N \times N}$ where $N$ is the total number of POIs present in the CBSA. 
As a baseline parameter setting, we use $T_c = 6$ hours and $T_s = 1$ POIs. The sensitivity and statistical robustness of the dependency network when using different co-visit detection threshold parameters $T_c$ and $T_s$, only short non-work visits, and data from different time periods, are tested and discussed in Supplementary Notes 2 and 3. 

\subsection*{Distance-based null networks}

To generate null networks that preserve basic structural properties 1) the weight $w_{ij}$ decays with physical distance and 2) the in-weight $w_{ij}$ is larger for nodes with larger visitation $n_i$, we generated edges based on the generalized gravity law: $g_{ij} = n_i n_j / (d_0 + d_{ij})^\gamma$, where $n_i$ and $n_j$ are the total number of visits to POIs $i$ and $j$, $d_{ij}$ is the physical distance between POIs $i$ and $j$, $d_0$ is the distance cutoff parameter, and $\gamma$ is the exponent parameter of the gravity model. Parameters $d_0=0.2$ and $\gamma=1.5$ were fitted empirically to maximize the correlation between $g_{ij}$ and $n_{ij}$, which is the total number of common visitors between POIs $i$ and $j$ (Supplementary note 4.1). 
For each edge in the actual dependency network connecting $i$ and $j$ with a dependency weight $w_{ij}$, we compute its gravity component using the empirical fit between $g_{ij}$ and $w_{ij}$ and select an alternative node with the same level of corresponding gravity weight from its 10,000 closest nodes, and is assigned the same weight $w_{ij}$. 
This algorithm enables us to construct a null network where we 1) maintain the linear relationship between $w_{ij}$ and $g_{ij}$, 2) the same number of in-edges are selected for each node, and 3) the total in-weight for each node is kept consistent. Details about the null network generation procedure, as well as statistical analysis of the null network and their differences from the actual dependency network can be found in Supplementary Note 4.

\subsection*{Modeling impacts of COVID-19 using dependency networks}
To investigate the utility of the dependency network for predicting the resilience of businesses, we construct regression models that predict the change in visitation patterns to a POI using information about the change in visitation patterns to its alters and the dependency network. 
The observed change in visits to different places is computed by $\tilde v_i = v_i^{after}/v_i^{before} - 1  * 100 (\%)$, where $v_i^{before}$ and $v_i^{after}$ denote the number of visits to place $i$ before the pandemic (September - December 2019) and during different periods of the pandemic period (March - November 2020), respectively.  
We build a simple linear regression model of the form:
\begin{equation}
    \tilde v_i \sim \sum_j w_{ij} \tilde v_j + \sum_j \hat w_{ij} \tilde v_j + \eta_i + \theta_i 
\end{equation}
where $\tilde v_i$ denotes the change in visitations to POI $i$ during the different stages of the pandemic in 2020, $\sum_j w_{ij} \tilde v_j$ is the sum of the neighbors' (POIs $j$) change in visitations ($\tilde v_j$) weighted by the dependency network weights $w_{ij}$, $\sum_j \hat w_{ij} \tilde v_j$ is the sum of the neighbors' (POIs $j$) change in visitations ($\tilde v_j$) weighted by the distance based null network weights $\hat w_{ij}$, $\eta_i$ is the fixed effect (FE) for POI $i$'s subcategory, and $\theta_i$ is the fixed effect for POI $i$'s located Public Use Microdata Area (PUMA). 
Robustness of regression results to the choice of the time period used to generate the dependency network, different model parameters, and similar results obtained using the case study of college summer breaks are shown in Supplementary Note 5.

\subsection*{Simulating cascades of hypothetical urban shocks}
To simulate the spatial cascades of shocks in different cities, we use the model specification of the Leontief Open Model. Re-writing and reorganizing the regression model in matrix form, we obtain
$\Vec{v} = W \Vec{v} + \Vec{f}$, where $\Vec{v}$ is a vector of $\tilde v_i$ for all $N$ places, $W$ is an $N \times N$ matrix where each element is $\tilde w_{ij} = \hat \beta_W w_{ij}$, and vector $\Vec{f}$ is an aggregation of all fixed effects $\beta_0$, $\eta_i$, and $\theta_i$. 
To predict the propagation of shocks throughout places in the city, the shocks are modeled in the fixed effect vector $f$ (e.g., all colleges experience an external shock of $-50\%$ visits reduction due to uptake of online education), and the production vector $\Vec{v}$ is computed by solving the linear system $\hat{\Vec{v}} = (I-W)^{-1} \Vec{f}$ via the generalized minimal residual iteration method. Details and parameter sensitivity analysis can be found in Supplementary Note 6.1.

Furthermore, we simulate the impacts of POI closure scenarios and identify the seed nodes (POIs) that have the largest cascading effects on other POIs if inflicted by other urban shocks. 
For each node, we simulate the cascading impacts of a 100\% visit change to a single node $i$, by computing $\hat{\vec{\nu}}^{(i)} = (I-W)^{-1} \vec e^{(i)}$, where $\vec e^{(i)}$ is a one-hot encoding vector of the initial shock that assigns a change in visits of $+1$ to node $i$ and $0$ otherwise, and $\hat{\vec{\nu}}^{(i)}$ is the resulting vector of the cascading impacts, where each element measures the impacts of the initial shock to all nodes.
The total impacts of changes in the number of visits to all nodes can be computed by multiplying $\hat{\vec{\nu}}^{(i)} = (\hat \nu_1^{(i)},\ldots,\hat \nu_N^{(i)})$ with the vector of total visits to each POI, $\vec n = (n_1, \cdots, n_N)$. Thus, the total impacts of the initial shock to node $i$ can be computed by $C_i = \sum_{j;j\neq i} \hat{\nu}^{(i)}_j  n_j$. 
By further scaling the impact to its own size $n_i$, we obtain the total relative cascading effect as $\hat{C_i} = C_i/n_i$. 
The distance range of the cascade is computed as the average distance to impacted nodes, weighted by the magnitude of the impacts. More specifically, we compute the weighted distance range of POI $i$ by $\hat{d}_{i} = \sum_{j; j\neq i} \hat{\nu}^{(i)}_j d_{ij} / \sum_{j; j\neq i} \hat{\nu}^{(i)}_j$. Figure 4c plots the relative cascading impacts and distance ranges of each POI category, $\hat{C}_{category}$ and $\hat{d}_{category}$. More details and results for all cities can be found in Supplementary Note 6.2.

% \bibliographystyle{unsrt}
% \bibliography{ms}  %%% Uncomment this line and comment out the ``thebibliography'' section below to use the external .bib file (using bibtex) .

\section*{Acknowledgements}
We would like to thank Spectus who kindly provided us with the mobility data set for this research through their Data for Good program. E.M. acknowledges support by Ministerio de Ciencia e Innovación/Agencia Española de Investigación\\  (MCIN/AEI/10.13039/501100011033) through grant PID2019-106811GB-C32 and the National Science Foundation under Grant No.~2218748.

\section*{Author contributions statement}
T.Y. designed the algorithms, performed the analysis, and developed models and simulations. B.G.B.B. and E.M. performed part of the analysis, and partially developed models and simulations. M.F., A.P., and E.M. supervised the research. All authors wrote the paper. Company data were processed by T.Y., B.G.B.B., and E. M. All authors had access to aggregated (non-individual) processed data. All authors reviewed the manuscript. 

\section*{Competing Interests}
The authors declare no competing interests.

\end{document}

\maketitle

\tableofcontents
% \addtocontents{toc}{\protect\thispagestyle{empty}}
% \pagenumbering{gobble}
\newpage

\listoffigures

\listoftables

\newpage

\setcounter{figure}{0}
\setcounter{table}{0}

% https://docs.google.com/spreadsheets/d/1xt6MUXLXT1qcxUUfqUIuAjZ7cCSXanxFEFKlJ0jv_x8/edit?usp=sharing

\section{Mobility data analytics and representativeness}

\subsection{Home estimation and stop detection}
In this study, we utilize an anonymized location dataset of mobile phones and smartphone devices provided by Spectus Inc., a location data intelligence company which collects anonymous, privacy-compliant location data of mobile devices using their software development kit (SDK) technology in mobile applications and ironclad privacy framework. Spectus processes data collected from mobile devices whose owners have actively opted in to share their location and requires all application partners to disclose their relationship with Spectus, directly or by category, in the privacy policy. With this commitment to privacy, the data set contains location data for roughly 15 million daily active users in the United States. Through Spectus’ Data for Good program, Spectus provides mobility insights for academic research and humanitarian initiatives. All data analyzed in this study are aggregated to preserve privacy\footnote{\url{https://spectus.ai/privacy/privacy-policy/}}. Each entry in the data table comprises anonymized device ID, location coordinates, start time, and dwell time of the stop for the device.

To define the type of location (Home or Work), different variables are used, including the number of days spent in a given location in the last month, the daily average number of hours spent in that location, and the time of the day spent in the location (nighttime/daytime). To estimate the home position of a user, the algorithm combines the three variables and creates a score that represents the probability that the position points to the home. The more days and the average number of hours spent in the position, the higher the score is. Higher scores will also be assigned to the most common places during the night. The location that maximizes this score is defined as the home of the device.

Once the location of the home location is identified, the algorithm looks for the work position. Note that the algorithm requires the work location to be located at least 100 meters apart from the home location. The same variables used for the detection of the home location are used, but a higher score is given to daytime locations for the work location rather than nighttime locations.
Spectus runs the algorithm every week in order to confirm or update the inferred home and work locations as we observe new data. We will only consider devices that have been present in Spectus’ dataset for at least 15 days. 
Spectus tightly restricts access to the inferred precise home and work locations of devices. Furthermore, it is used as input into various downstream processes to create more privacy-protected versions of Spectus datasets.
For example, we only expose home and work datasets in Spectus Workbench associated with standard Census Block Groups, created by the U.S. Census Bureau, rather than the precise locations. This offers a good balance between utility and privacy: according to the U.S. Census Bureau, there are between 600 and 3000 people living in each block group. Each block group is an aggregate of contiguous U.S. blocks sharing similar socio-demographic characteristics.
The representativeness of this data has been tested and corrected in Section 1.3 in the Supplementary Material.
The stops, which are location clusters where individual users stay for a given duration, are estimated using the Sequence Oriented Clustering approach \cite{xiang2016extracting}. 

In this study, we selected New York, Boston, Seattle, Los Angeles, and Dallas as the five CBSAs given constraints on data collection. The five cities were selected with respect to the diversity of characteristics in terms of geographical locations (2 Northeast, 1 Northwest, 1 Southwest, and 1 Southern), sociodemographic details (population ranges from 3.5M in Seattle to 19.8M in New York), political inclinations, and weather characteristics.
% To quantitatively show the diversity of the four cities’ characteristics, we plot various sociodemographic, geographic, COVID-19 related, and climate related variables of the four cities among the top 15 metropolitan statistical areas in the US. More specifically, Figure \ref{fig:msa_chara} shows the the population (log), population \% change from 2020 to 2021, median household income, poverty rates, \% of workers who use private vehicles to work, \% population with education level with Bachelor degree (all using the American Community Survey), the COVID-19 stringency index in 2022 January provided by the University of Oxford, and annual average temperature of the four cities compared against the largest 15 metropolitan statistical areas in the US. The figure shows that our collection of four cities covers a wide range of values for each of the sociodemographic, political (COVID-19 stringency), and climate characteristics. 

% \textcolor{red}{
% \begin{itemize}
%     \item Figure: plot showing the social, economic, geographical diversity of the five cities [can compute locally] 
% \end{itemize}
% }

\subsection{Safegraph POI data and visit attribution}
To measure the visitation patterns of individuals in urban environments, we attribute the stops of individual users to specific places in the city. To study the stops at different places, we use stops that are longer than 10 minutes but shorter than ten hours. In our study, we use location data of places collected by Safegraph \footnote{\url{https://www.safegraph.com/}}. To protect the users' privacy, we have removed various privacy-sensitive places from our places database. Sensitive places include health-related places, places where the vulnerable population are located, military-related, religious facilities, places that are related to sexual-orientation, and adult-oriented places \footnote{\url{https://spectus.ai/privacy/spoi-policy/}}. As a result, we have a total of 403,669 places in New York, 107,081 places in Boston, 81,140 places in Seattle, 326,332 places in Los Angeles, and 136,686 places in Dallas. The breakdown of the number of places by the place category is shown in Table \ref{table:pois}.
To attribute a stop to a place, we simply attribute each stop the closest place in our dataset. To avoid attributing a stop to place far away, we attribute the stop to a place within $d_{max}=100$ meters from the observed location of the stop. If the stop is further away than 100 meters from any place in the dataset, the stop is discarded from our dataset and not used for computing the diversity of encounters. 

\begin{table}
\centering
\caption{Number of places in the Safegraph dataset in the five core-based statistical areas (CBSAs) analyzed in this study.}
\begin{tabular}{lrrrrr}
\toprule
& \multicolumn{5}{c}{CBSAs} \\
\cmidrule{2-6}
Place category & New York & Boston & Seattle & Los Angeles & Dallas  \\
\midrule
Arts and Museums  & 4,580  & 1,183  & 799 & 3,697  & 1,073 \\
City and Outdoors & 7,201  & 5,311  & 2,792 & 4,817 & 2,174 \\
Coffee and Tea    & 14,799 & 3,481  & 3,392  & 11,395 & 4,360 \\
College           & 3,737  & 1,266  & 766 & 2,661 & 942 \\
Entertainment     & 2,296  &  657  & 584 & 2,101 & 745 \\
Food              & 66,348 & 12,687 & 9,924 & 39,907 & 16,795 \\
Grocery           & 14,763  & 2,942  & 1,964  & 7,636  & 3,792 \\
Health            & 20,654   & 4,799   & 3,913   & 14,212  & 5,806 \\
Office            & 7,840   & 3,256  & 2,628  & 10,936 & 6,128 \\
Service           & 114,581  & 33,525 & 23,920 & 104,253 & 45,520 \\
Shopping          & 92,211  & 20,302  & 14,089 & 79,024 & 29,069 \\
Sports            & 12,383  & 5,892  & 2,464 & 10,509 & 3,782 \\
Transportation    & 41,729  & 11,114  & 13,245 & 32,141 & 14,056 \\
\midrule
All places        & 403,669 & 107,081 & 81,140 & 326,332 & 136,686 \\
\bottomrule
\end{tabular}
\label{table:pois}
\end{table}

\begin{figure}[ht]
\centering
\includegraphics[width=\linewidth]{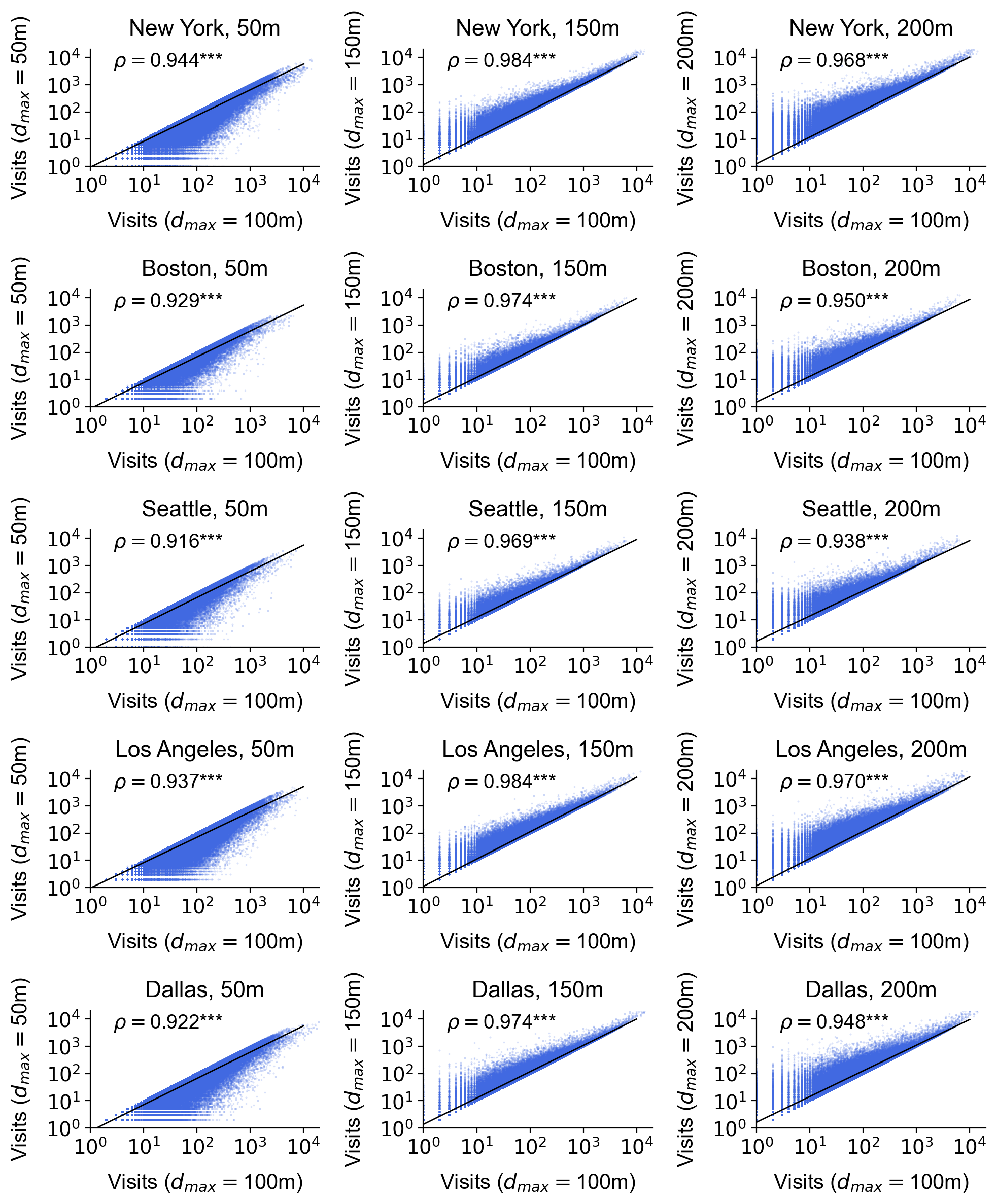}
\caption[Estimates of number of visits to POIs under different maximum spatial threshold parameters]{\textbf{Estimates of number of visits to POIs under different maximum spatial threshold parameters.} Pearson correlation of estimated number of visits to different POIs between different spatial threshold parameters are extremely high across all cities.}
\label{fig:s1visitattri}
\end{figure}

% \subsubsection{Robustness to a selection of spatial threshold}
The robustness of our estimation of the number of visits to different locations has been tested using different spatial thresholds of $d_{max}$. Different levels of $d_{max}$ could change how individuals' stays are attributed to the places and thus could affect our estimates of the income diversity of physical encounters. 
Figure \ref{fig:s1visitattri} compares the estimated number of visits to places over a 1 month period in the five cities when we use different levels of $d_{max}$ (y-axis) with our default parameter $d_{max}=100m$ (x-axis). For all values $d_{max}=\{50,150,200\}$, the Pearson correlation of the number of visits are extremely high, in most cases above $\rho > 0.95$. This robustness check shows that the estimated visitation patterns do not depend on the choice of the spatial threshold parameter for visit attribution.

\subsection{Data representativeness}\label{repre}

The location data used in our study is collected from smartphones via various apps and services. Although a significant portion (85\% according to 2021 data\footnote{\url{https://www.statista.com/topics/2711/us-smartphone-market/#topicHeader__wrapper}}) of the US population owns a smartphone, one could question the representativeness of the 1.16 million user samples across geographical regions and income quantiles. Studies have reported the digital divide and smartphone usage gaps across sociodemographic groups in the US \cite{tsetsi2017smartphone}. In this section, we test whether our group of users in the mobility data is representative of the total population, and further employ post-stratification techniques to correct for any potential biases in the sampling rates across places and socioeconomic status and to test whether the results on income diversity dynamics are robust to such uncertainties concerning data representativeness.

\begin{figure}[!t]
\centering
\subfloat[New York]{\includegraphics[width=0.5\linewidth]{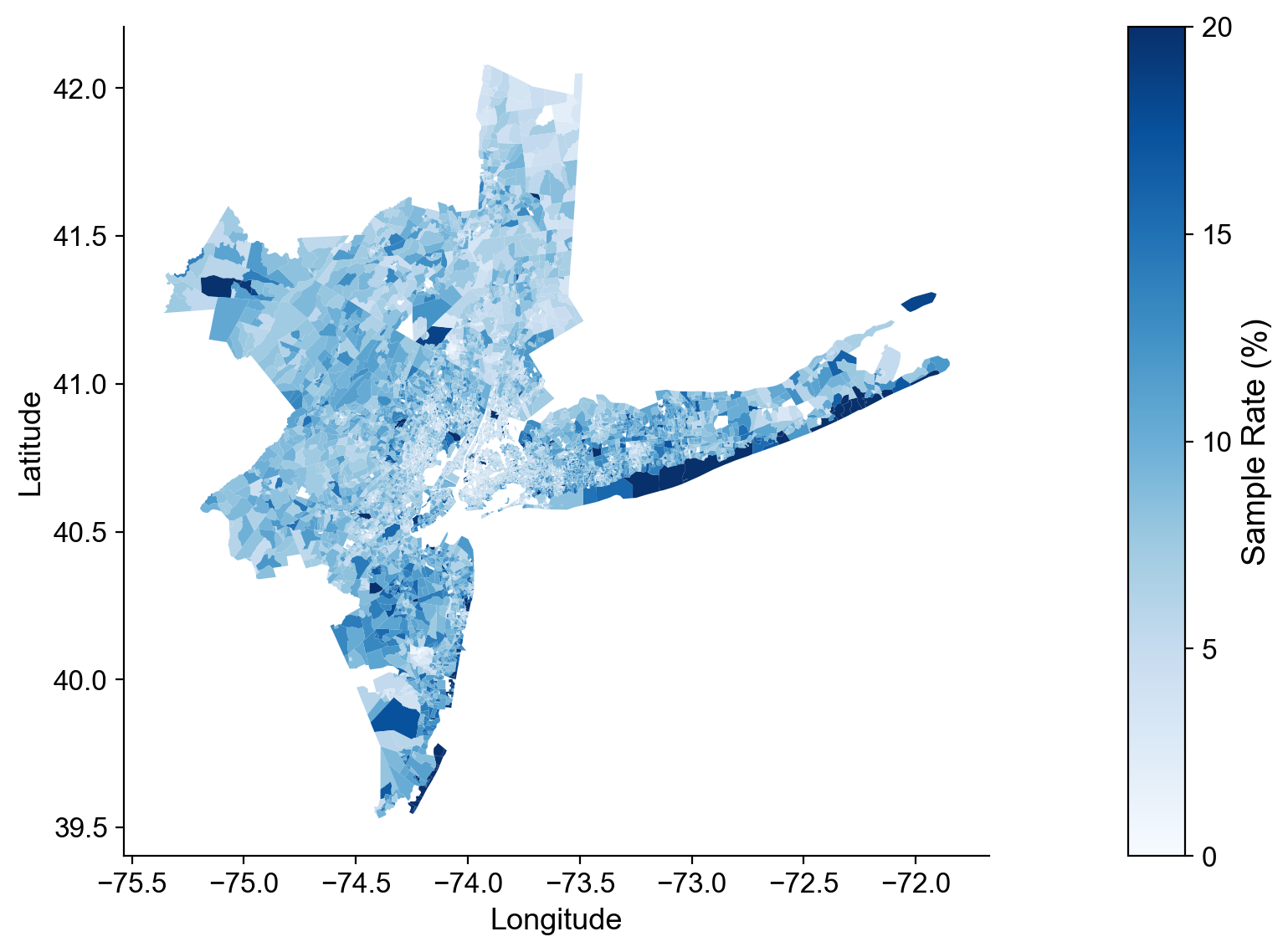}} 
\hspace{0.1\textwidth}
\subfloat[Boston]{\includegraphics[width=0.3\linewidth]{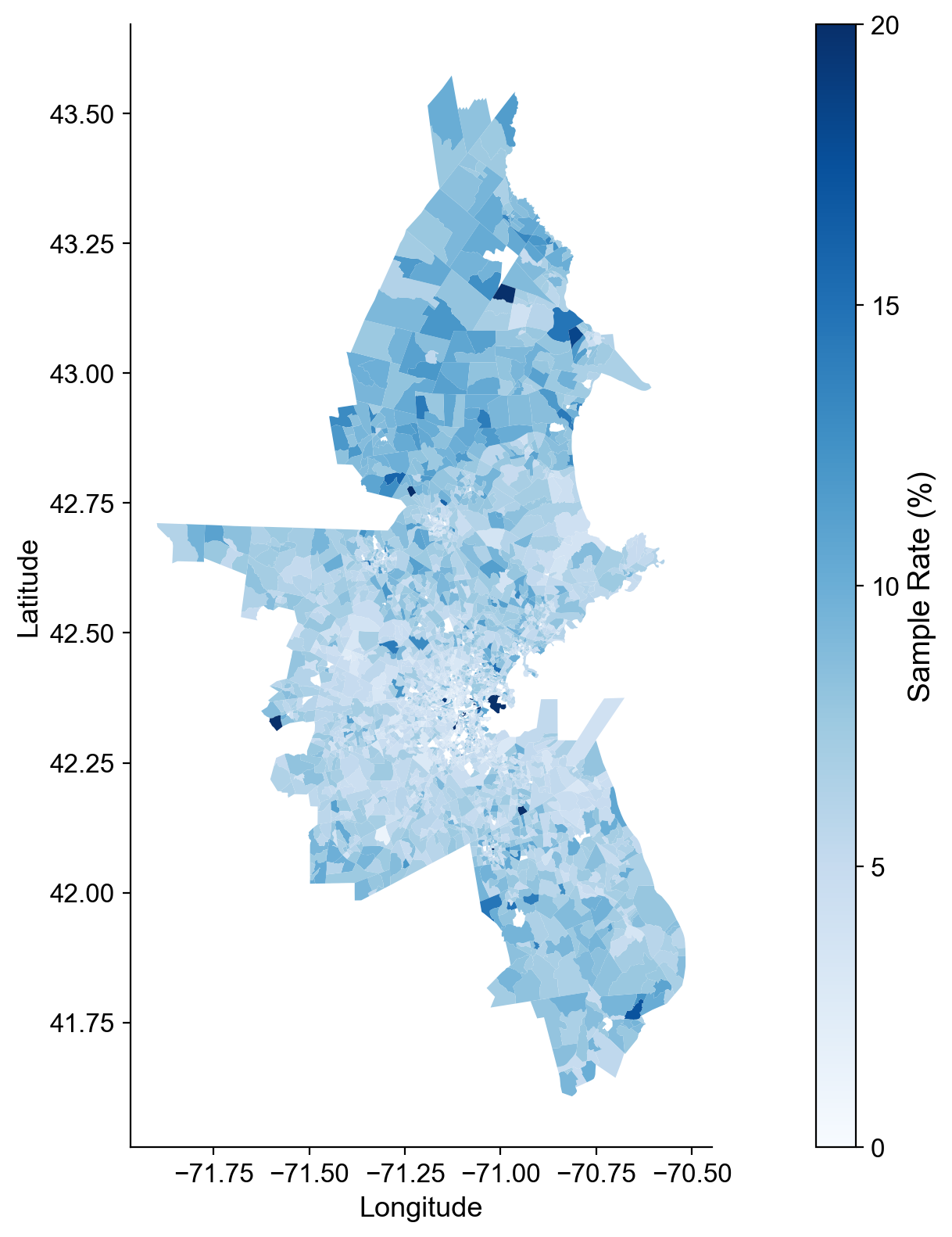}} \\
\subfloat[Seattle]{\includegraphics[width=0.42\linewidth]{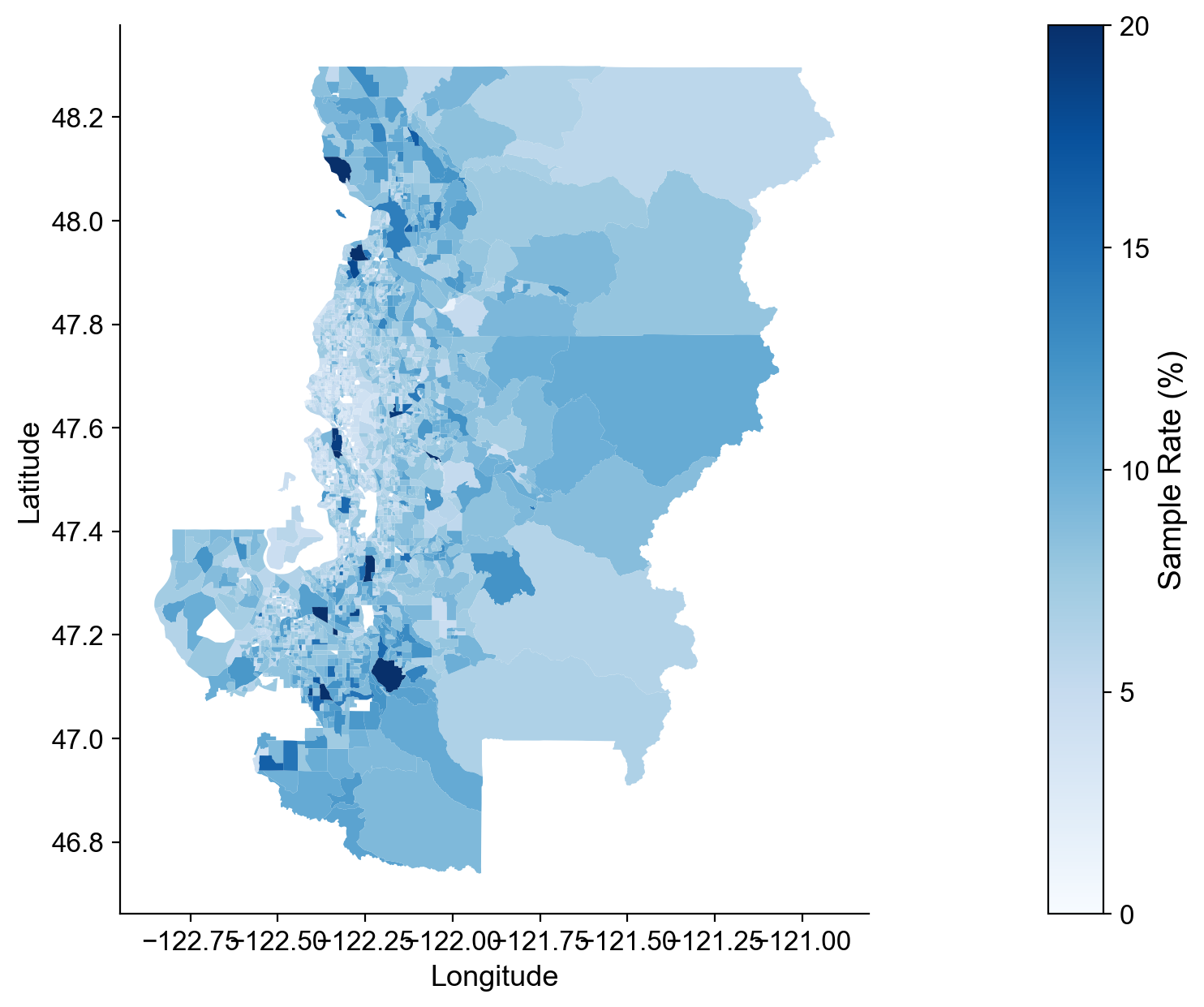}}
\hspace{0.1\textwidth}
\subfloat[Los Angeles]{\includegraphics[width=0.4\linewidth]{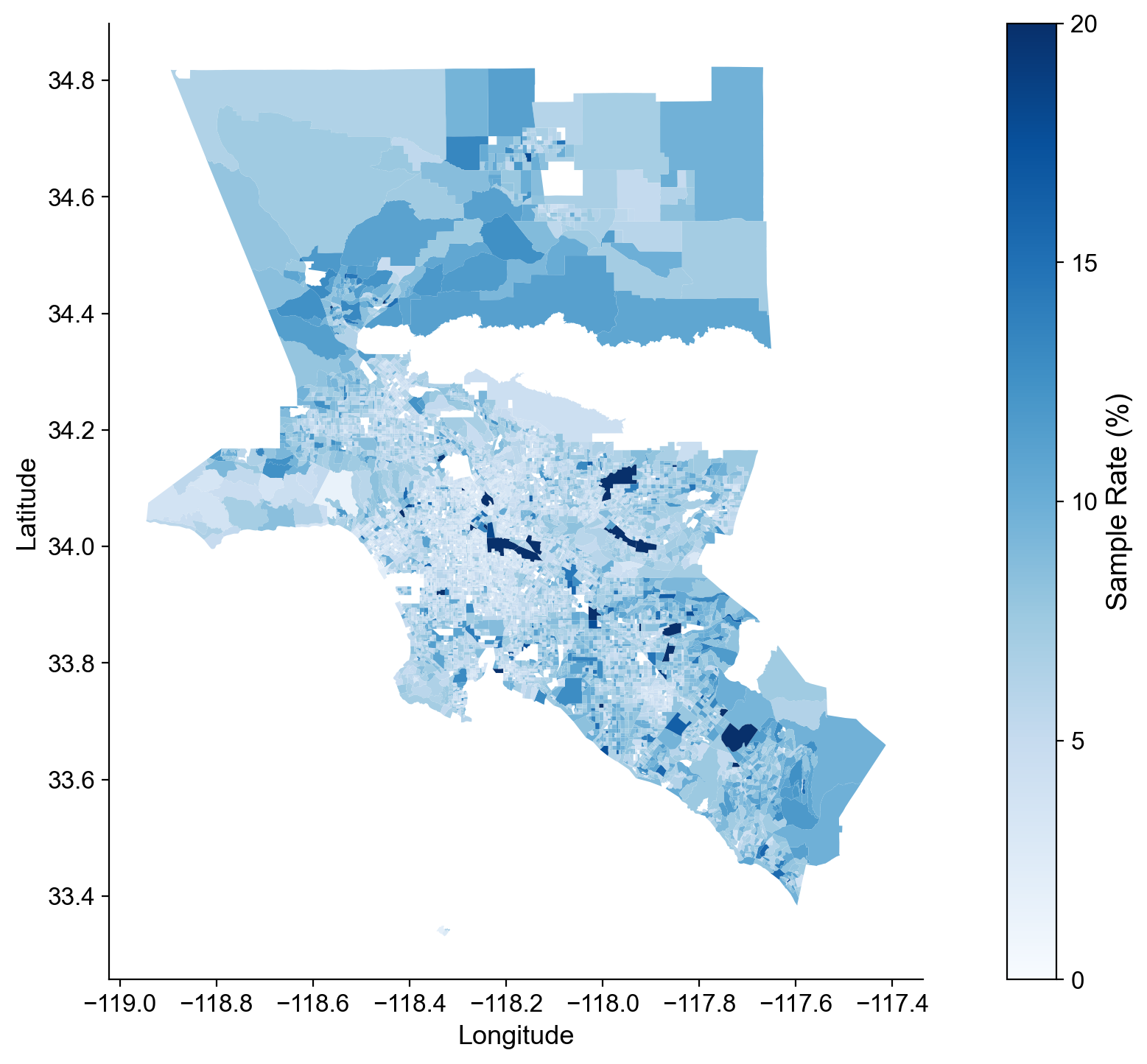}} \\
\subfloat[Dallas]{\includegraphics[width=0.53\linewidth]{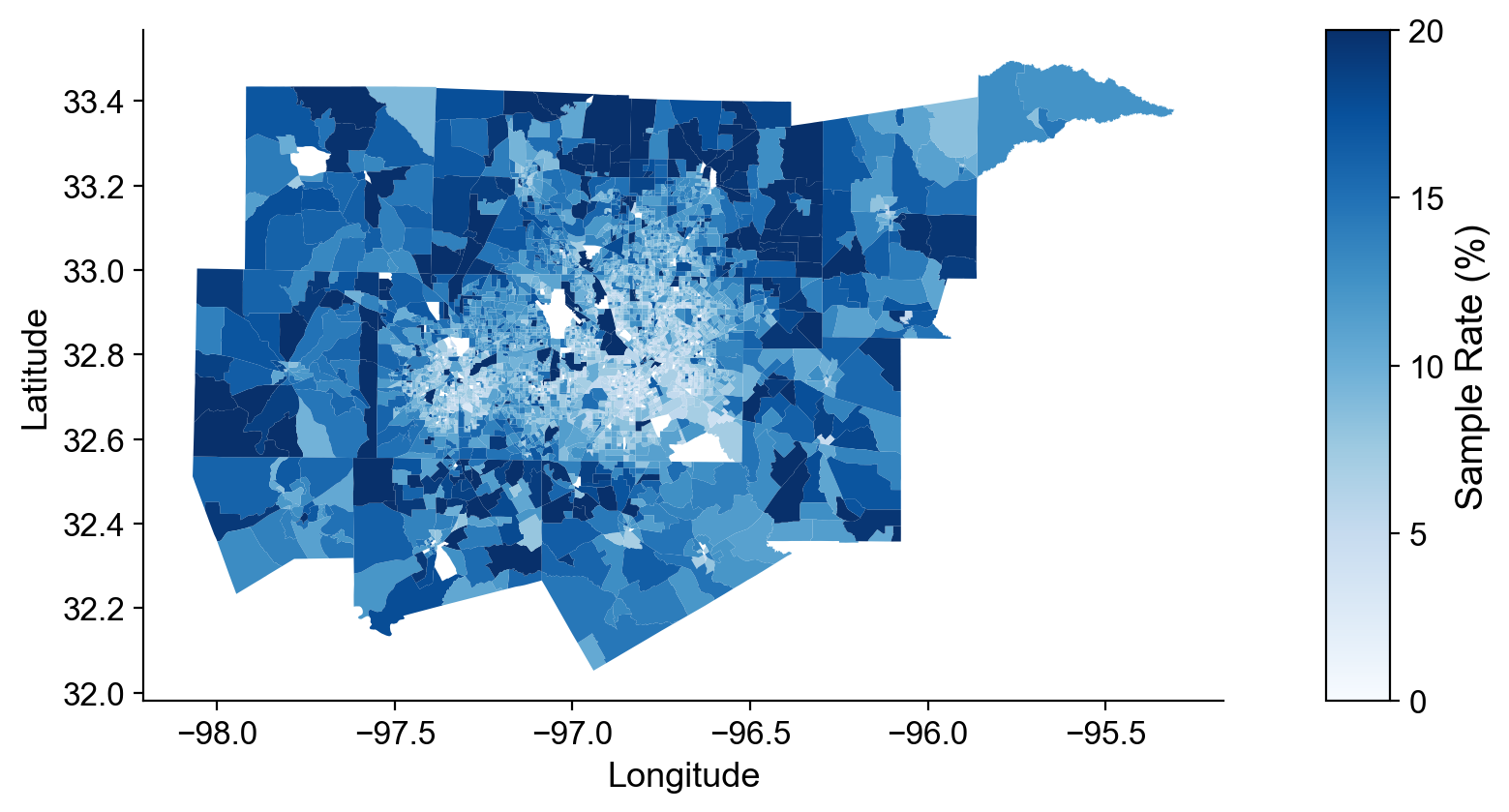}}
\caption[Sample rates of mobile phone location data across the five metropolitan areas.]{\textbf{Sample rates of mobile phone location data across the five metropolitan areas.} Sample rates are calculated by dividing the total number of users who were observed across a three month period in the dataset by the census block group population obtained from the American Community Survey \cite{acs}. Maps were produced in Python using the TIGER shapefiles from the U.S. Census Bureau \cite{tiger}}
\label{fig:s1samplerate}
\end{figure}

\begin{figure}[!t]
\centering
\includegraphics[width=\linewidth]{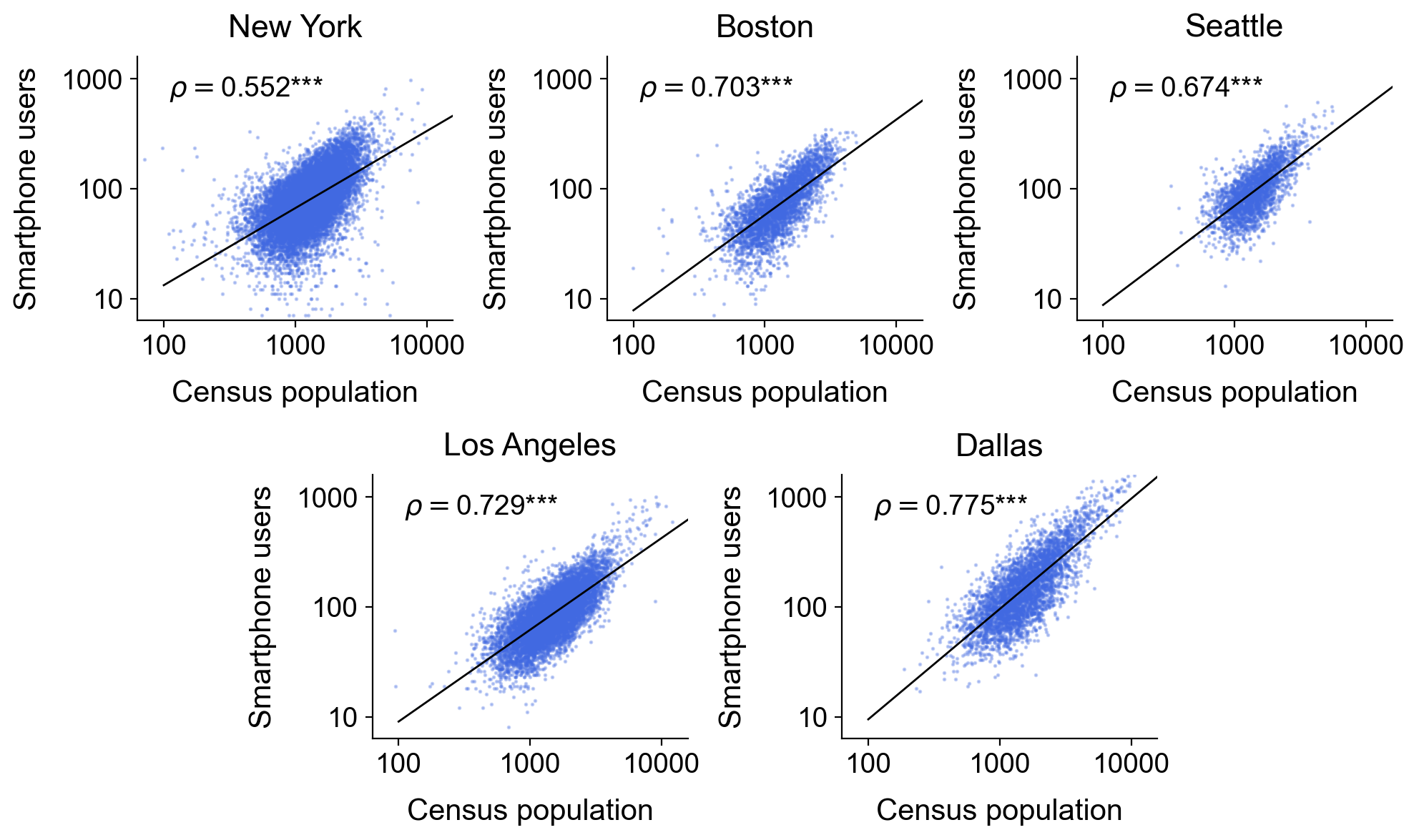}
\caption[Comparison of census population and smartphone users for census block groups in the five metropolitan areas]{\textbf{Comparison of census population and smartphone users for census block groups in the five metropolitan areas.} The correlation is moderately high, between around $\rho=0.55$ and $\rho=0.78$, showing that despite the use of such small census areas and potential bias in the smartphone usage patterns, we are able to obtain a good representation of the population. This bias is corrected using post-stratification techniques.}
\label{fig:s1representativeness}
\end{figure}

% \subsubsection{Population and income representativeness}
The sampling percentage of the mobility data (100\% $\times$ number of observed mobile phone users divided by the total population from the census data) is around 5\% to 10\% across all census block groups (CBGs) in the metropolitan regions, as shown in Figure \ref{fig:s1samplerate}. Sample rates are calculated by dividing the total number of users who were observed across a three-month period in the dataset by the census block group population obtained from the American Community Survey \cite{acs}. 
There is a large variation in the sample rates, within and across metropolitan areas.  
To test whether the users in the location data are representative of the entire population, first, we compare the population detection in our mobility data and the 2019 ACS data for each of the CBGs in the cities. The panels in Figure \ref{fig:s1representativeness} show the comparison between the census population (x-axis) and the number of observed smartphone users (y-axis) on the CBG scale in the months of September 2019 to January 2020 in the five metropolitan areas. The correlation is moderately high, between around $\rho=0.55$ and $\rho=0.78$, showing that despite the use of such small census areas and potential bias in the smartphone usage patterns, we are able to obtain a good representation of the population. 
To overcome this bias, in Section 1.4, we use post-stratification techniques to correct for such differences in the sample percentages across CBGs. Later, in Section 3.3, we assess whether our estimates of visitation patterns are affected by the representativeness of the data.  

\begin{figure}[!t]
\centering
\includegraphics[width=\linewidth]{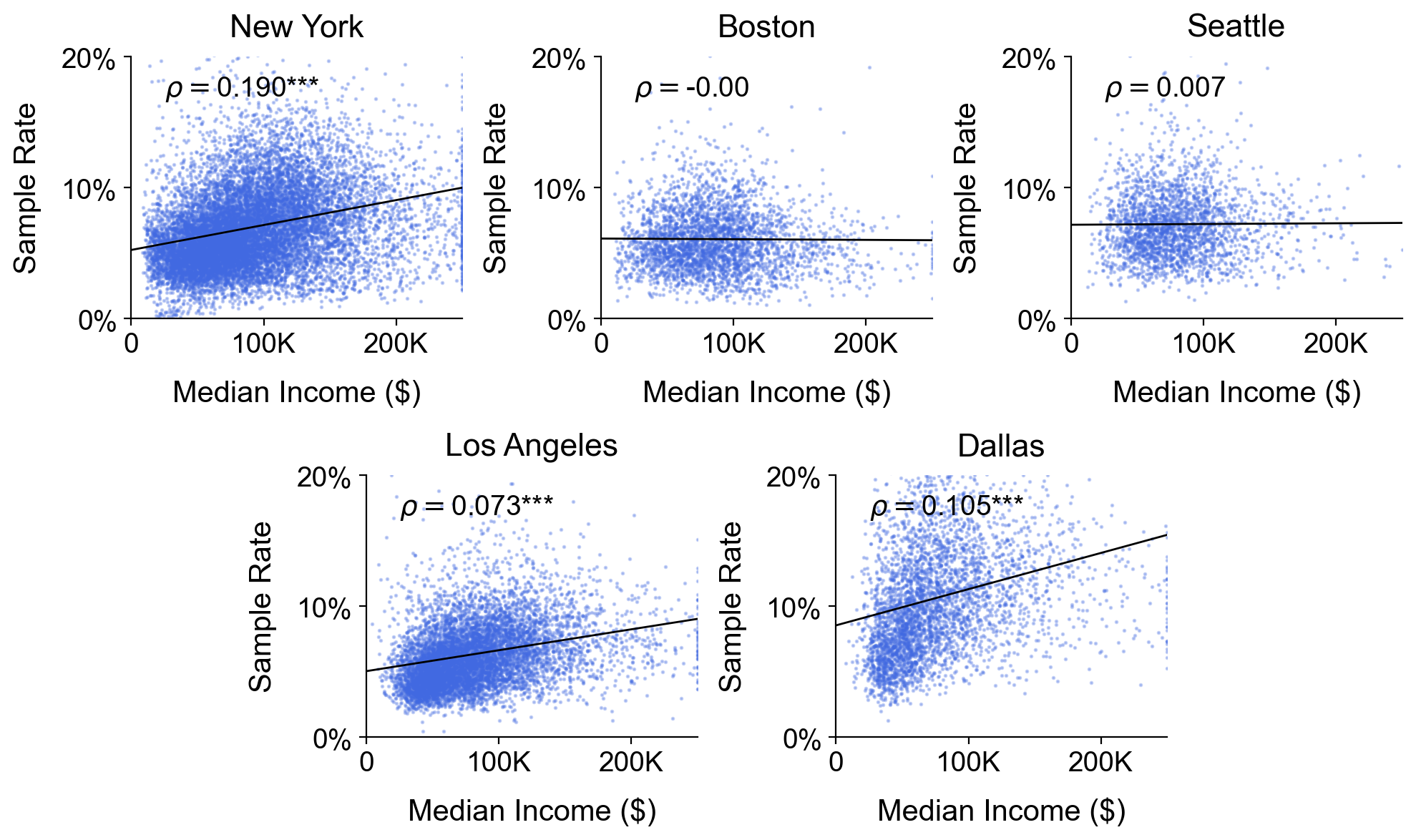}
\caption[Comparison of census block group median income and the sample rate of mobile phone smartphone users in the five metropolitan areas]{\textbf{Comparison of census block group median income and the sample rate of mobile phone smartphone users in the five metropolitan areas.} The correlation is small (below $\rho=0.2$) but significantly positive in New York, Los Angeles, and Dallas. This indicates that in these metropolitan areas, higher income groups are over-represented in the dataset. This bias is corrected using post-stratification techniques.}
\label{fig:s1incomebias}
\end{figure}

% \subsection{Income representativeness}
In addition to the differences in sampling rates across CBGs, differences in representativeness across income groups are important to ensure the findings of the study are not biased toward specific sociodemographic groups. 
To measure the representativeness across income quantiles, we plot the correlation between the median income of the census block group and the sample rate (shown in Figure \ref{fig:s1samplerate}) for the five metropolitan areas, as shown in Figure \ref{fig:s1incomebias}. A balanced dataset would have a nonsignificant correlation between the two metrics. However, we observe a small (below $\rho=0.2$) but significant positive correlation between income and sample rates for New York, Los Angeles, and Dallas. This indicates that in these metropolitan areas, higher income groups are over-represented in the dataset.
In Section 1.4, we use post-stratification techniques to correct for such differences in the sample percentages across CBGs and income groups. 

% \textcolor{red}{
% \begin{itemize}
%     \item Figure: scatter plot between census population and mobile phone users, for each city, in September 2019 -Jan 2020
%     \item Figure: scatter plot between income ranges vs sample rate, for each city, in September 2019 -Jan 2020
%     \item Figure: Spatial plots of sampling rates, for each city, in September 2019 -Jan 2020 ... [can compute locally]
% \end{itemize}
% }

\subsection{Post-stratification of mobility data} \label{poststrat}
To correct for, and to understand the effects of the varying sampling rates across CBGs and income groups on our estimation of visitation patterns to places, we apply a post-stratification technique, which is used in previous studies \cite{moro2021mobility,yabe2023behavioral}. 
Post-stratification is a well know sampling tool \cite{salganik2019bit} and is typically used to study the impact of sampling biases in mobile phone location data \cite{jiang2016timegeo} or (geolocated) social media data \cite{wang2018urban} on various downstream tasks and analyses. 
Following the methods employed in Moro et al. \cite{moro2021mobility}, we denote $w_g$ the expansion factor, which is the ratio of the population of census block group $g$ to the population detected in our mobility data. 
We then weight the visits completed by people from census block group $g$ spends at place $\alpha$ by 
\begin{equation*}
    \hat{n}_{gi} = w_g n_{gi}
\end{equation*}
where the assumption is that $n_{g\alpha}$ is proportional to the number of people visiting the place. Using this method, we could increase (decrease) the visits to places by people coming from census block groups that are under-estimated (over-estimated). 

% Recomputing the income diversity of urban encounters using the corrected duration of stays $\hat{\tau}_{g\alpha}$, as shown in Figure \ref{fig:representativeness}b we observe that the dynamics of the income diversity decrease between the raw mobility data and the post-stratified  data are very similar. These results show the robustness of the insights on income diversity, and that even though the representativeness of mobile phone users are not perfect, the effect on our estimations are very limited. 

\begin{table}
\centering
\caption{Description of the four core-based statistical areas (CBSAs) analyzed in this study.}
\resizebox{\textwidth}{!}{\begin{tabular}{lrrrr}
\toprule
CBSA & Population & \# users (monthly) & \# stays (monthly) & \# places  \\
\midrule
New York-Newark-Jersey City & 19.95M & 1,456K & 36.67M & 403,669  \\
Boston-Cambridge-Newton & 4.64M & 144K & 2.34M & 107,081  \\
Seattle-Tacoma-Bellevue & 3.55M & 141K & 2.23M & 81,140  \\
Los Angeles-Long Beach-Anaheim & 13.05M & 452K & 10.00M & 326,332  \\
Dallas-Fort Worth-Arlington & 6.70M & 425K & 8.83M & 136,686  \\
\midrule
Total & 47.89M & 2.62M & 60.07M & 1,054,908 \\
\bottomrule
\end{tabular}}
\label{table:stats}
\end{table}

\clearpage
\section{Behavior-based dependency networks}

\subsection{Measuring behavior-based dependency between places}
Using the large and longitudinal dataset of GPS location records in five major US metropolitan areas introduced in Supplementary Note 1, we construct the behavior-based dependency network at the level of points-of-interest (POIs; e.g., businesses and amenities) in cities. 
Within the mobility dataset, we identified stays at places that were detected to be between 10 minutes and 10 hours and we spatially matched those stays with the closest place locations within 100 meters to infer visits to specific POIs, as described in Supplementary Note 1. The representativeness of the data across regions and income levels was ensured via post-stratification techniques.

\begin{figure}[ht]
\centering
\includegraphics[width=\linewidth]{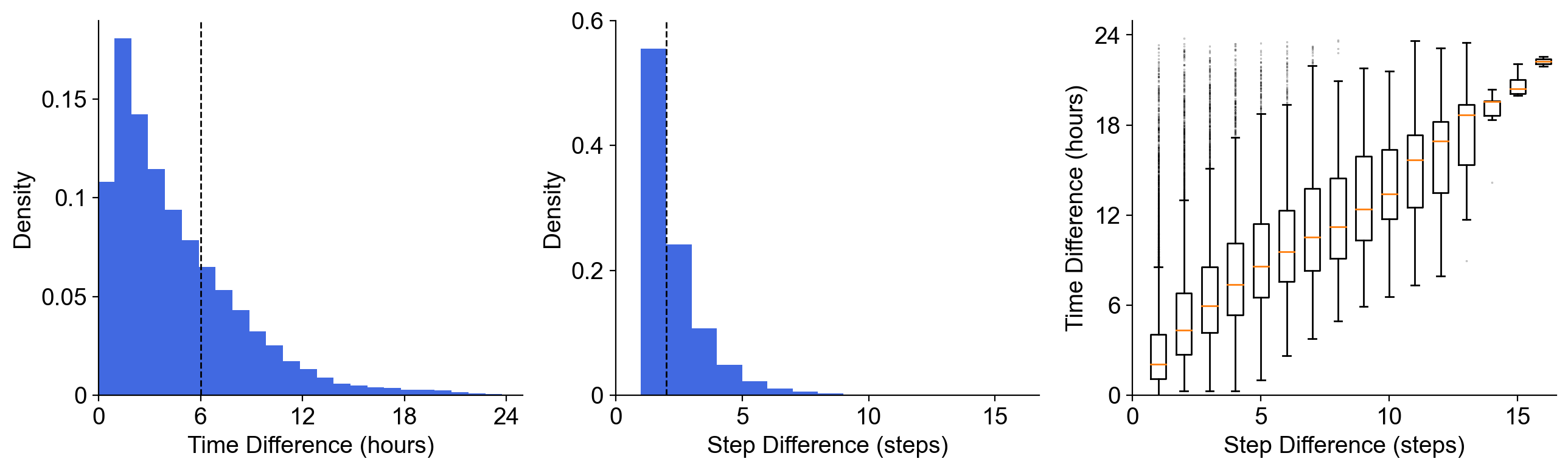}
\caption[Distribution of time and step differences between two POIs visited by the same individual on the same day]{\textbf{Distribution of time and step differences between two POIs visited by the same individual on the same day.} The majority of the co-visits can be captured by using a threshold of 6 hours time difference and 1 step difference. The dependency networks were computed using co-visits that occurred within 6 hours and 1 step. 
}
\label{fig:s2stepdist}
\end{figure}

In this study, the dependence of a POI $i$ on another POI $j$ is defined as 
\begin{equation}
    w_{ij} = \frac{n_{ij}}{n_i}
\end{equation}
where $n_i$ denotes the number of visits to POI $i$ and $n_{ij}$ denotes the number of `co-visits' between POIs $i$ and $j$. A co-visit is defined as an instance in which POIs $i$ and $j$ were visited by the same individual:
\begin{itemize}
    \item on the same day,
    \item within $T_c$ hours from exiting POI $i$ ($j$) to entering POI $j$ ($i$), and
    \item within $T_s$ intermediate POIs 
\end{itemize}
Because the denominator is based on the number of visits to the target POI, $w_{ij} \neq w_{ji}$. This simple but intuitive measure considers the asymmetric nature of dependencies between POIs. 
As shown in Figure \ref{fig:s2stepdist}, the distributions of the time difference or number of intermediate steps between any given two POIs within a movement trajectory follow a long-tailed distribution, where the majority of the pairs of POIs can be included within 6 hours and 1 step difference. 
Therefore, as a baseline parameter setting, we use $T_c = 6$ hours and $T_s = 1$ POIs (which indicates direct visitation from POI $i$ to POI $j$, or vice-versa). 
The characteristics of the dependency network when using different co-visit detection threshold parameters $T_c$ and $T_s$ are tested and discussed in the following sections such as Supplementary Note 2.3 and 4.2. 
By computing the dependency weights $w_{ij}~ \forall i,j$, we obtain the behavior-based dependency matrix $W \in \mathbb{R}^{N \times N}$ where $N$ is the total number of POIs present in the CBSA. 

\subsection{Qualitative characteristics of networks}

\begin{figure}[!t]
\centering
\includegraphics[width=\linewidth]{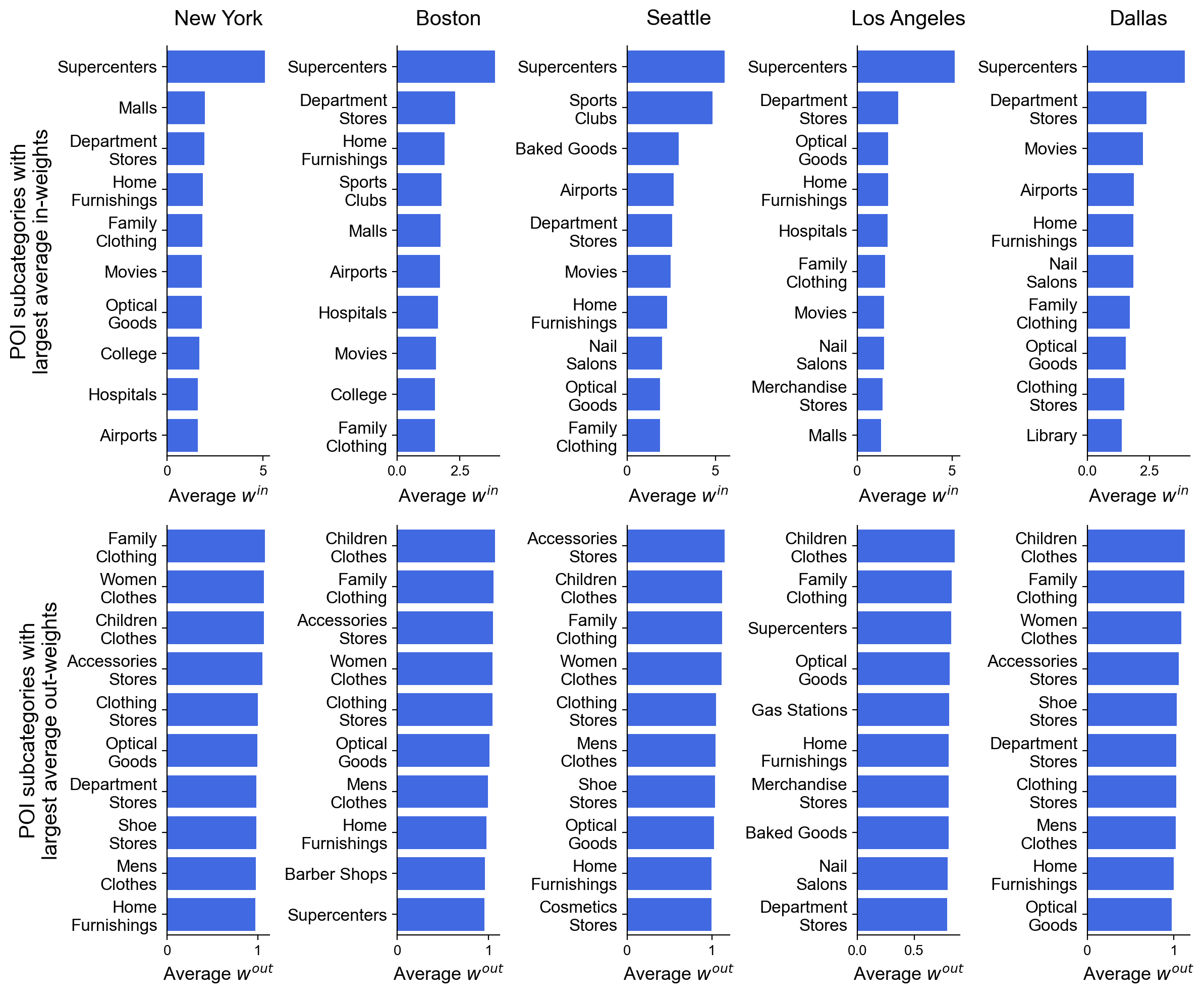}
\caption[POI subcategories with the largest average in- and out-weights]{\textbf{POI subcategories with the largest average in- and out-weights} Supercenters, malls, department malls, and airports had the largest in-weights, while places such as art dealers and various types of retail stores had the largest average out-weight. }
\label{s2categoryrank}
\end{figure}

To obtain a better understanding of the generated behavior-based dependency network, the POI subcategories with the largest average in- and out- weights were plotted in Figure \ref{s2categoryrank}.
Supercenters, malls, department malls, and airports had the largest in-weights, indicating that these places were depended by many other POIs. On the other hand, places such as art dealers and various types of retail stores had the largest average out-weight, indicating that such places had high dependency on other places. The subcategories that appeared in both rankings were consistent across the five metropolitan areas. 
As shown in Figures \ref{s2categoryrank1} and \ref{s2categoryrank3}, the POI subcategories that appeared in each ranking were similar when we changed the co-visit detection parameter to $T_c = 1$ and $T_c = 3$, respectively. 

\begin{figure}[!t]
\centering
\subfloat[New York]{\includegraphics[width=0.43\linewidth]{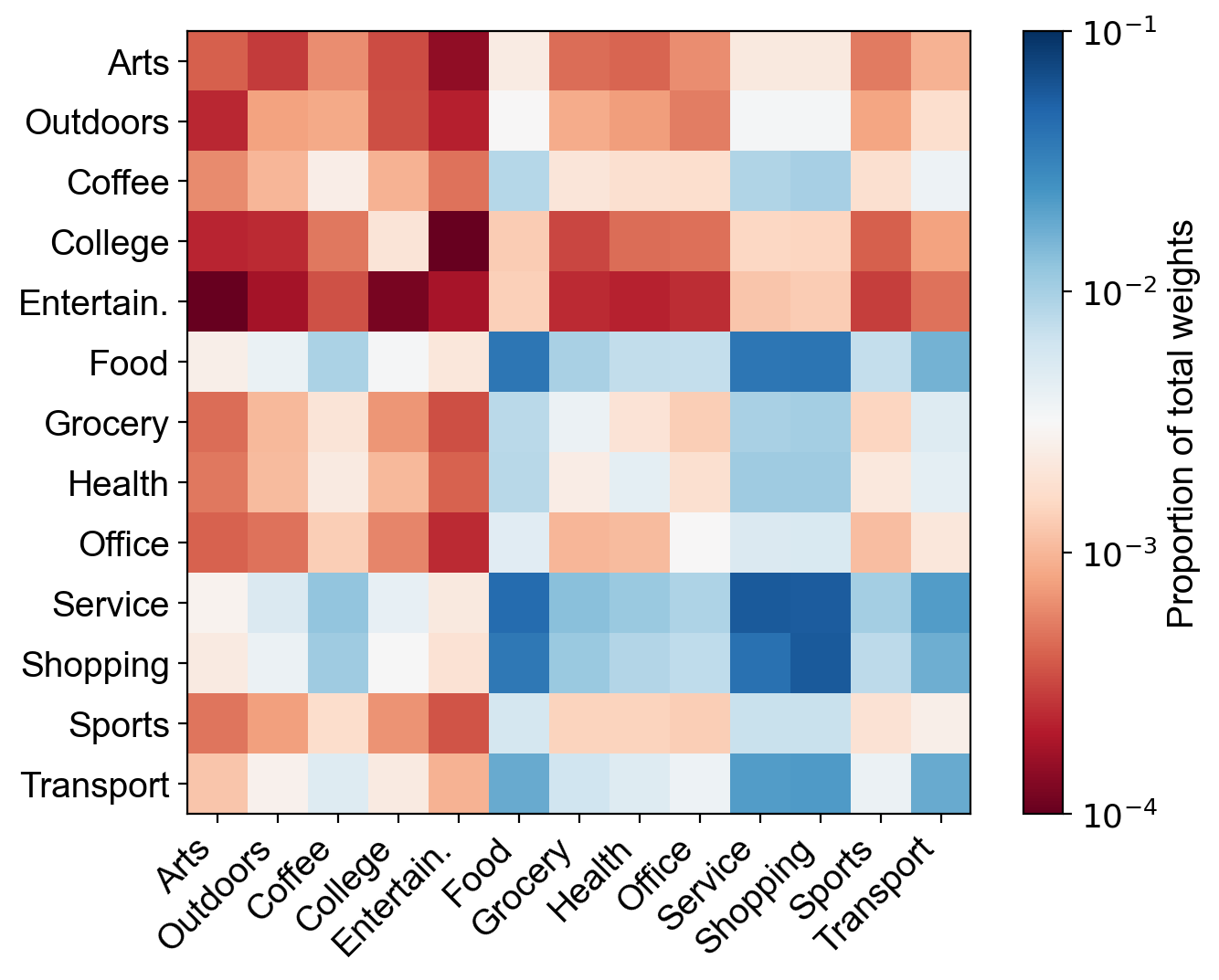}} 
\hspace{0.1\textwidth}
\subfloat[Boston]{\includegraphics[width=0.43\linewidth]{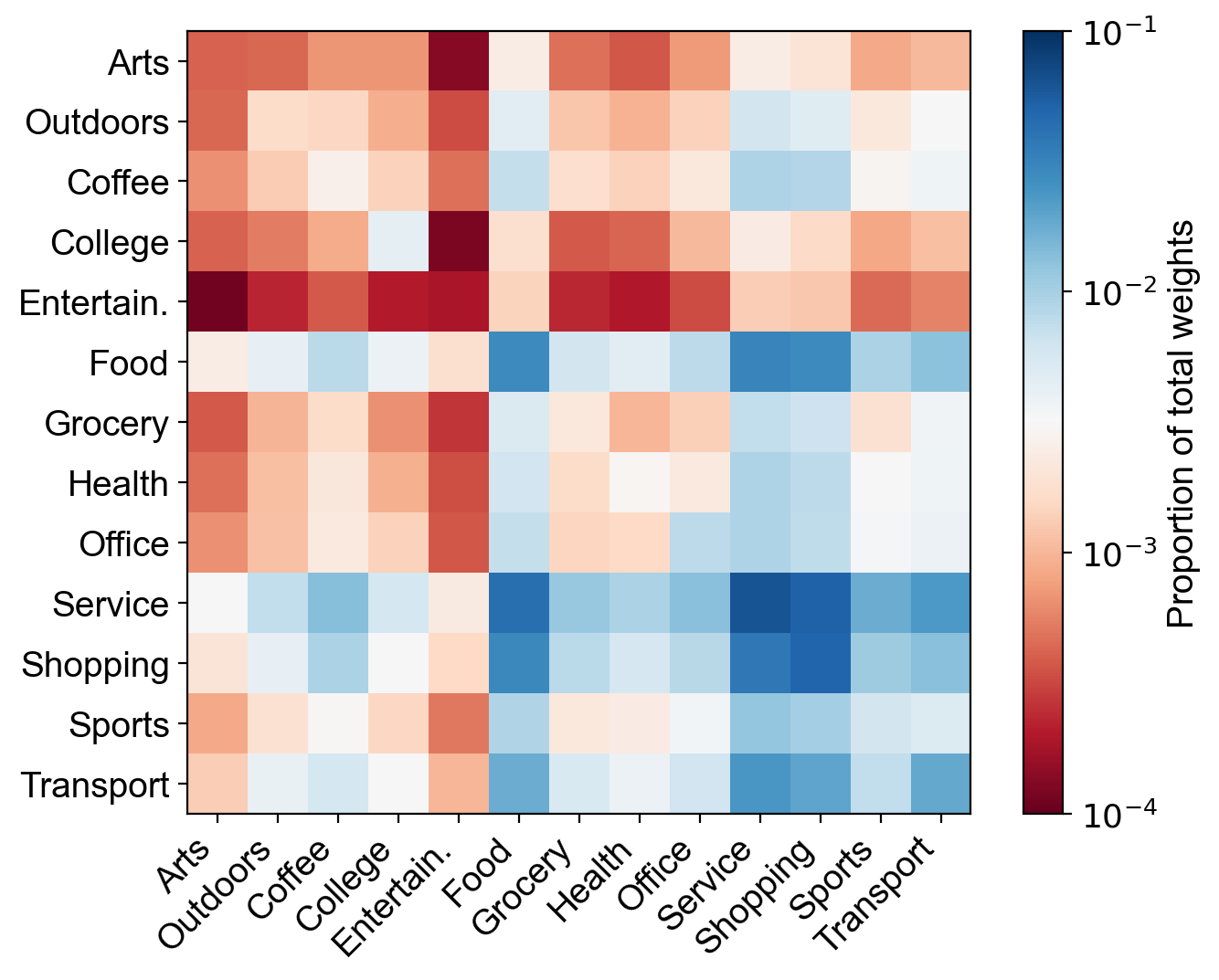}} \\
\subfloat[Seattle]{\includegraphics[width=0.43\linewidth]{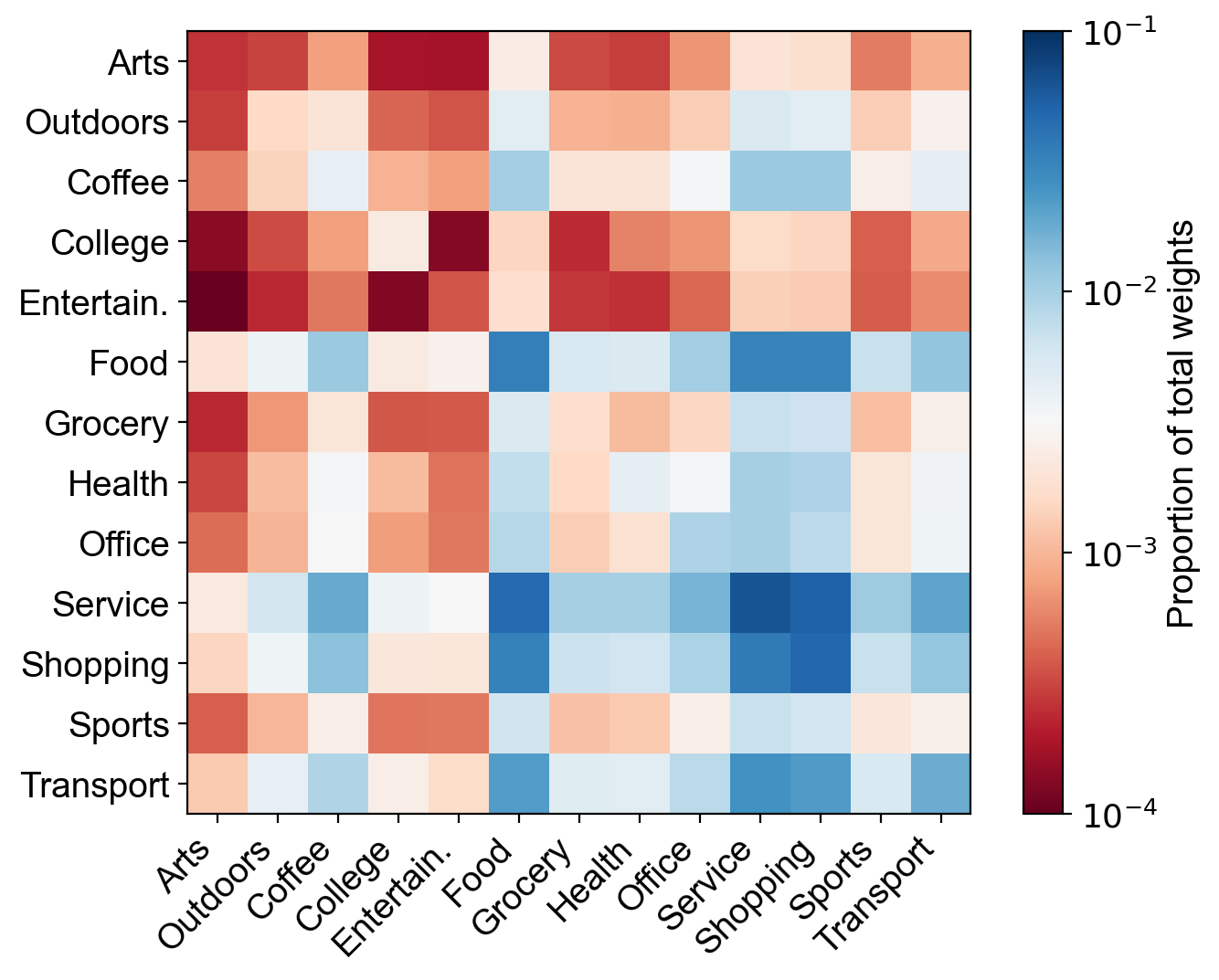}}
\hspace{0.1\textwidth}
\subfloat[Los Angeles]{\includegraphics[width=0.43\linewidth]{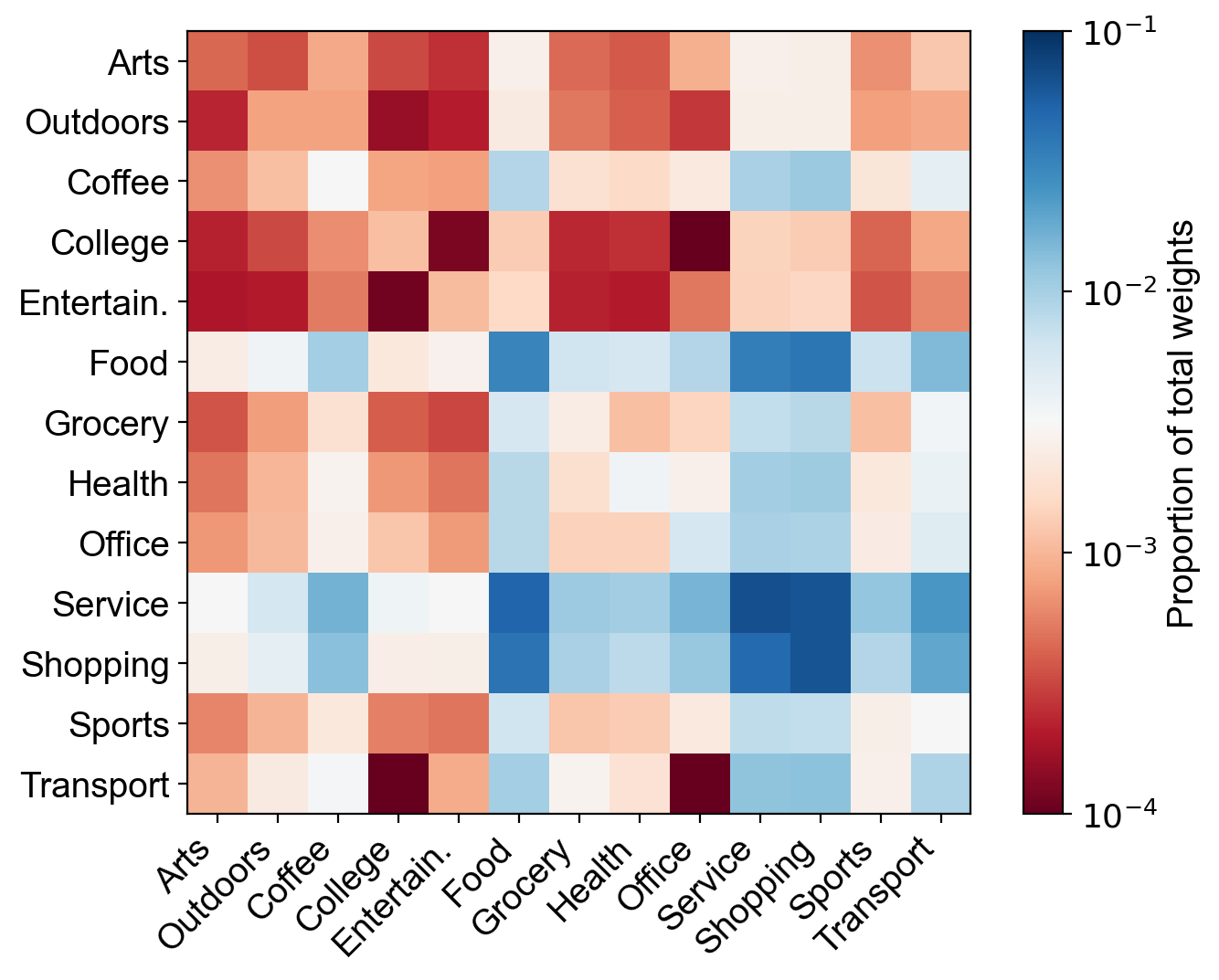}} \\
\subfloat[Dallas]{\includegraphics[width=0.43\linewidth]{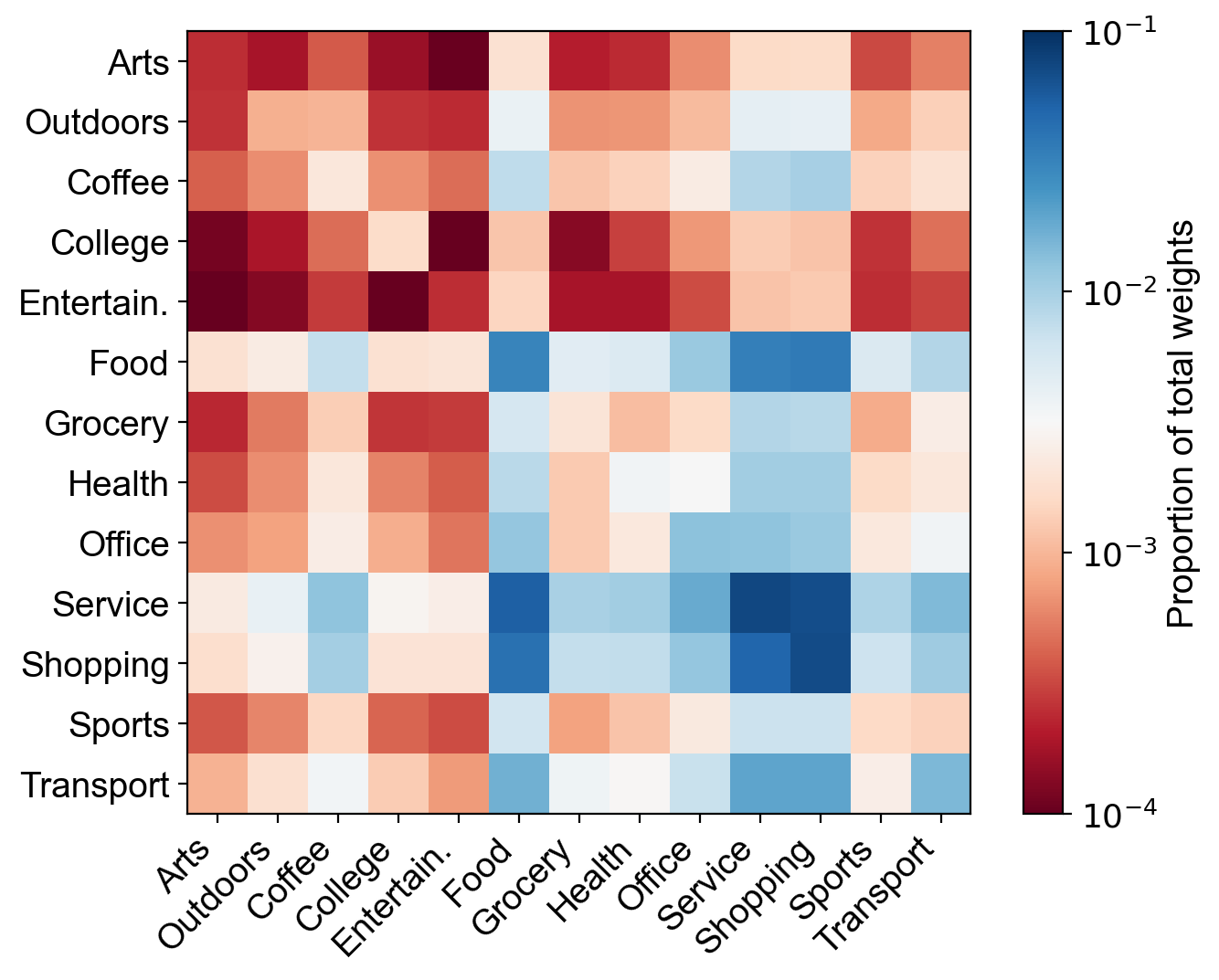}}
\caption[Proportion of edge weights among POI category pairs]{\textbf{Proportion of edge weights among POI category pairs.} The proportion of edge weights was computed by taking the sum of weights that connect the vertical and horizontal categories and dividing that by the total weights that exist in the network. The patterns of dependency are consistent across cities, with Food, Service, Shopping, and Transport playing a big role in both depending on and being depended by others.}
\label{fig:s2wproportion}
\end{figure}

To further observe the dependency relationships between categories, we computed the proportion of edge weights among POI category pairs, shown in Figure \ref{fig:s2wproportion}. 
The proportion of edge weights was computed by taking the sum of weights that connect the vertical and horizontal categories and dividing that by the total weights that exist in the network, more formally, by $\sum_{i\in A, j\in B} w_{ij} / \sum_{i,j} w_{ij}$. 
As the matrices in Figure \ref{fig:s2wproportion} show, the patterns of dependency are consistent across cities, with Food, Service, Shopping, and Transport playing a big role in both depending on and being depended by others.

\subsection{Comparison of networks under different co-visit detection parameters}

\begin{figure}[!t]
\centering
\includegraphics[width=\linewidth]{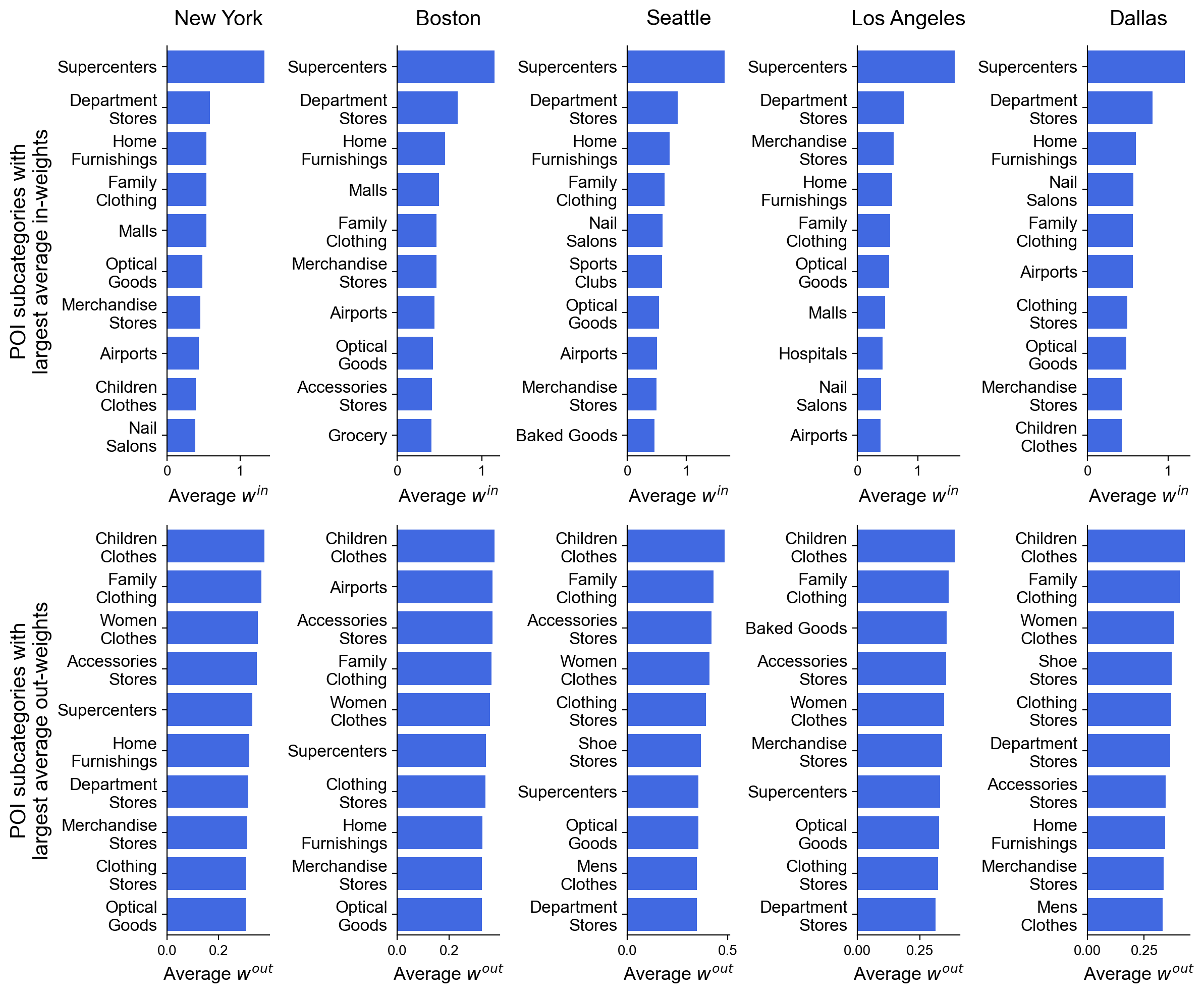}
\caption[POI subcategories with the largest average in- and out-weights, when $T_c = 1$ hour]{\textbf{POI subcategories with the largest average in- and out-weights, when $T_c = 1$ hour} Supercenters, malls, department malls, and airports had the largest in-weights, while places such as art dealers and various types of retail stores had the largest average out-weight. Results were similar to the baseline parameter when $T_c = 6$ hours.}
\label{s2categoryrank1}
\end{figure}

\begin{figure}[!t]
\centering
\includegraphics[width=\linewidth]{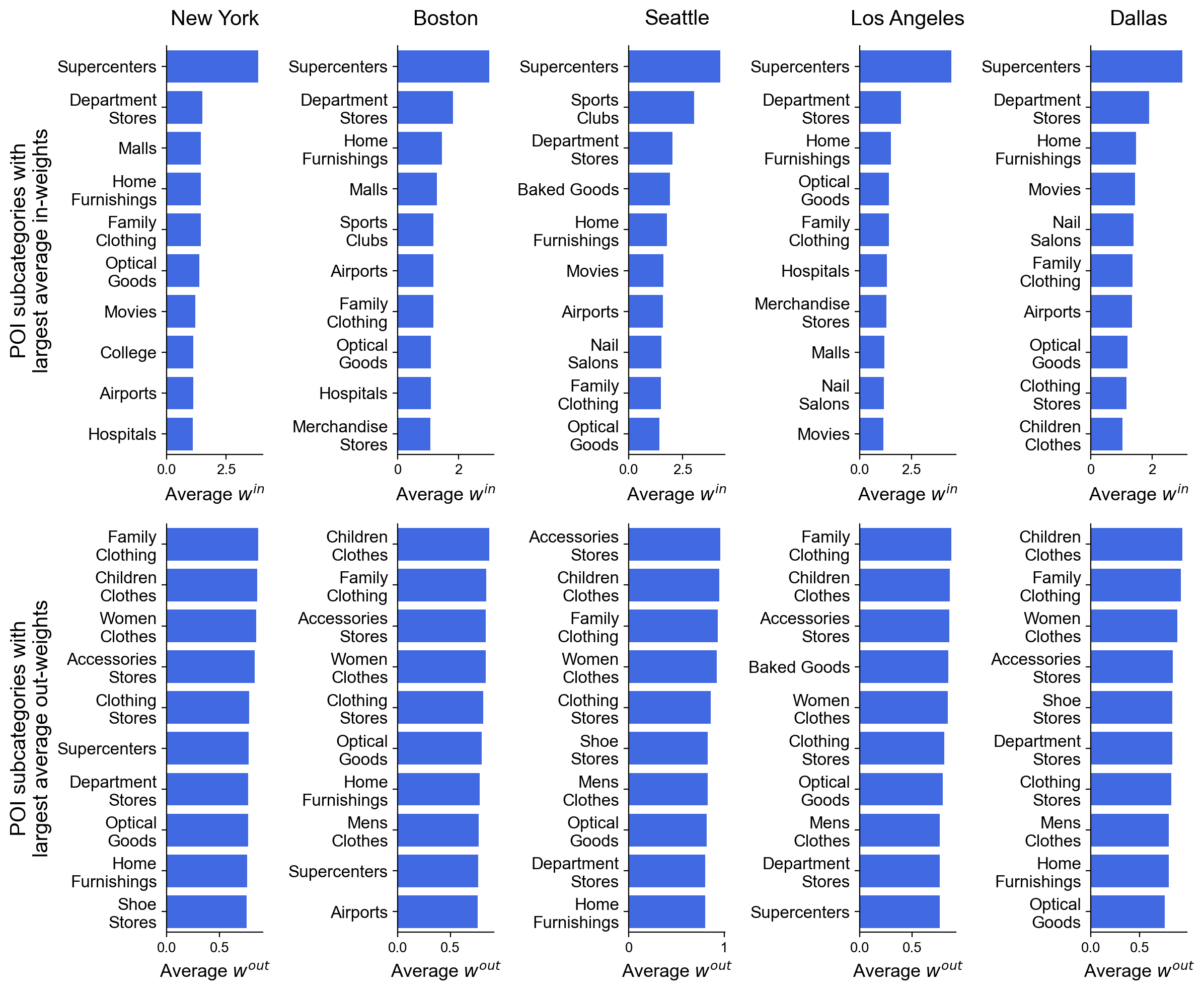}
\caption[POI subcategories with the largest average in- and out-weights, when $T_c = 3$ hour]{\textbf{POI subcategories with the largest average in- and out-weights, when $T_c = 3$ hour} Supercenters, malls, department malls, and airports had the largest in-weights, while places such as art dealers and various types of retail stores had the largest average out-weight. Results were similar to the baseline parameter when $T_c = 6$ hours.}
\label{s2categoryrank3}
\end{figure}

\begin{figure}[!t]
\centering
\subfloat[Network characteristics under different maximum step difference $T_s$]{\includegraphics[width=.9\linewidth]{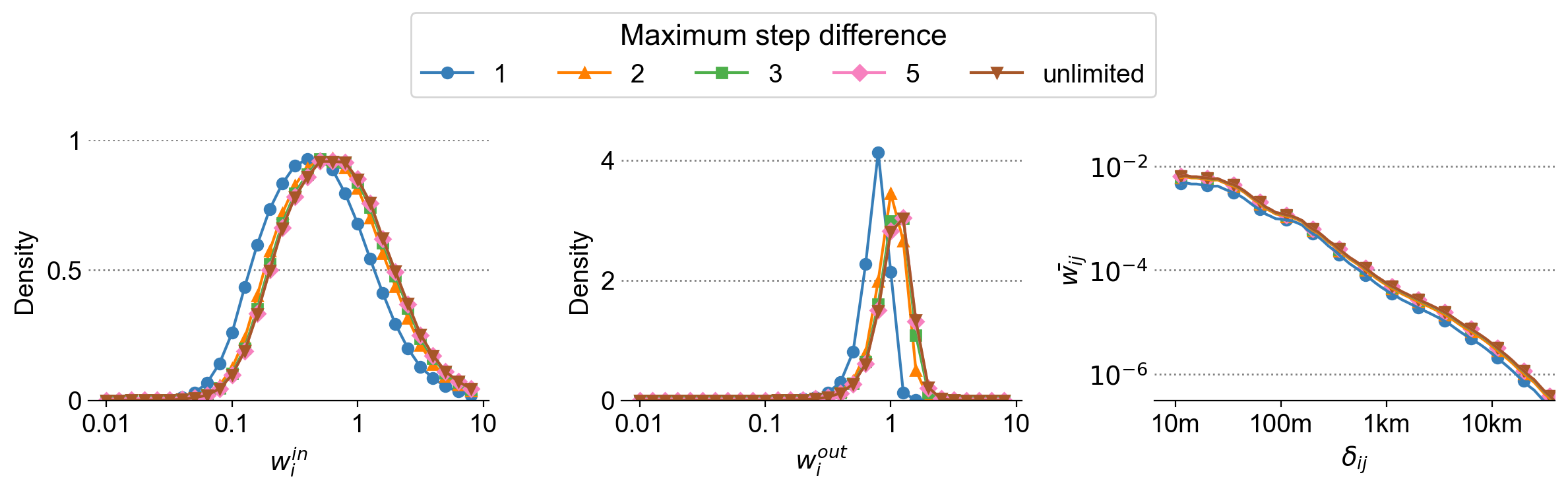}} \\
\subfloat[Network characteristics under different maximum time difference $T_c$]{\includegraphics[width=.9\linewidth]{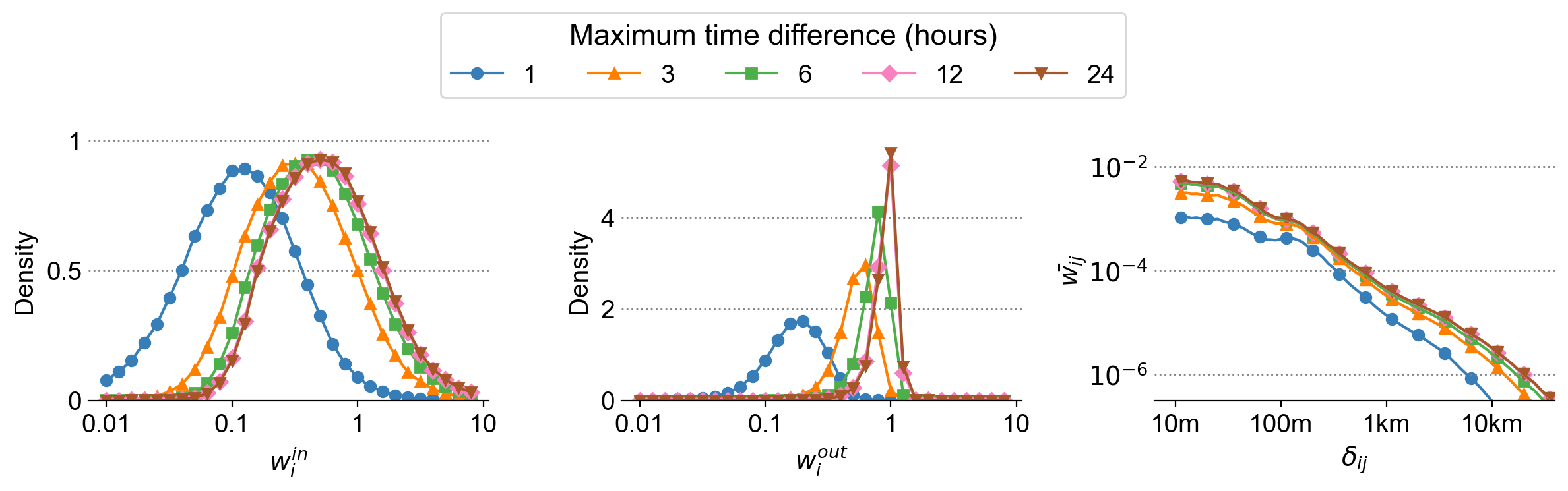}} \\
\subfloat[Difference in category pairwise dependency weights compared to baseline network]{\includegraphics[width=\linewidth]{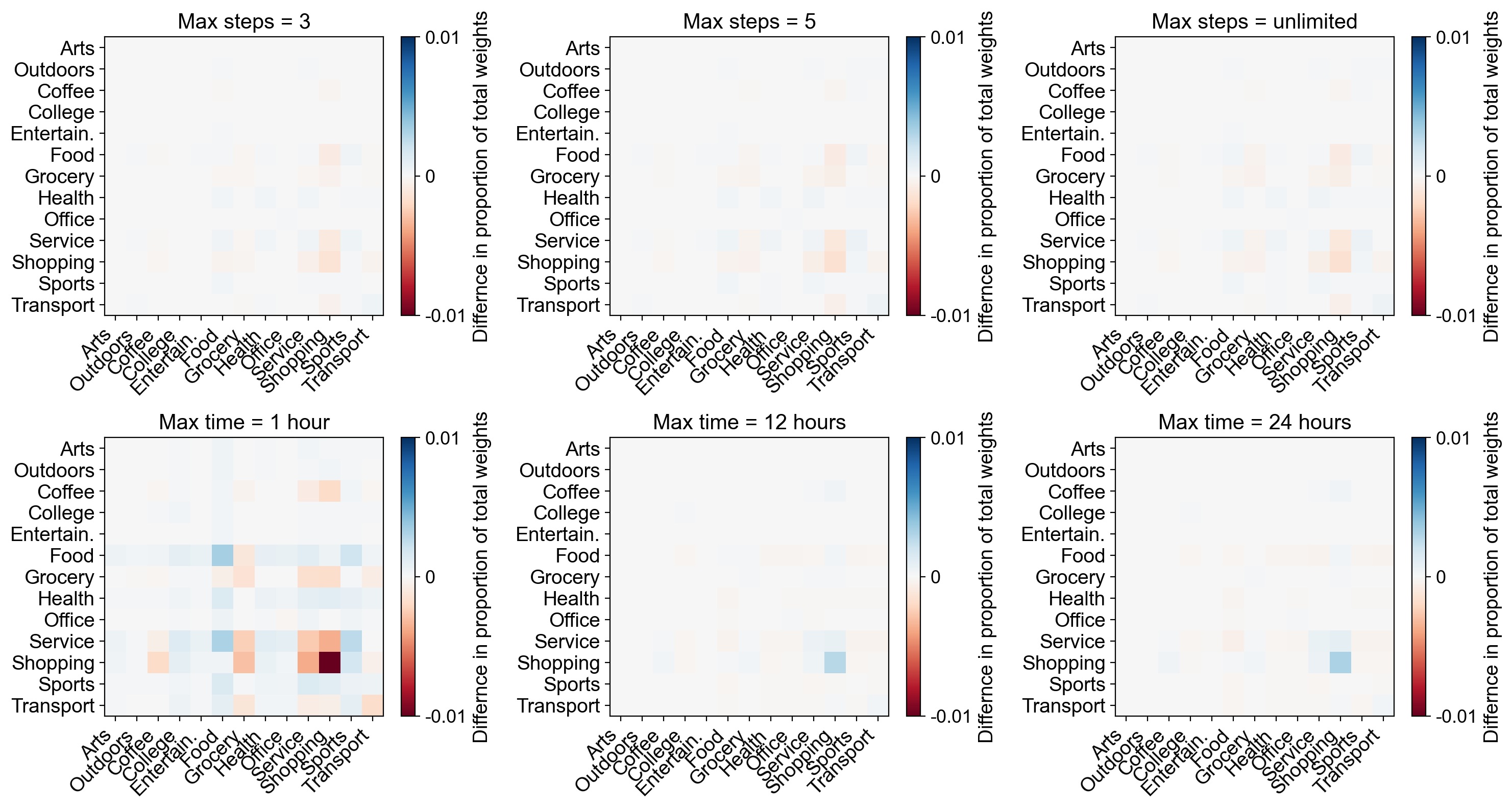}}
\caption[Dependency network in New York under different co-visit detection parameters]{\textbf{Dependency network in New York under different co-visit detection parameters.}}
\label{fig:s2nyc}
\end{figure}

\begin{figure}[!t]
\centering
\subfloat[Network characteristics under different maximum step difference $T_s$]{\includegraphics[width=.9\linewidth]{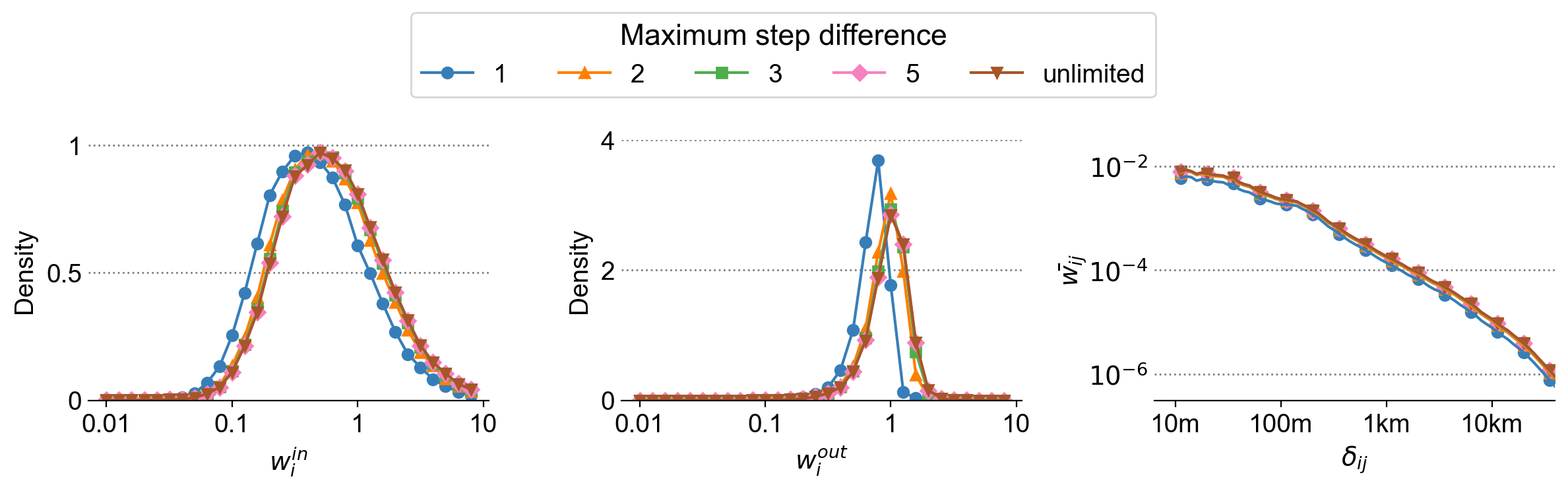}} \\
\subfloat[Network characteristics under different maximum time difference $T_c$]{\includegraphics[width=.9\linewidth]{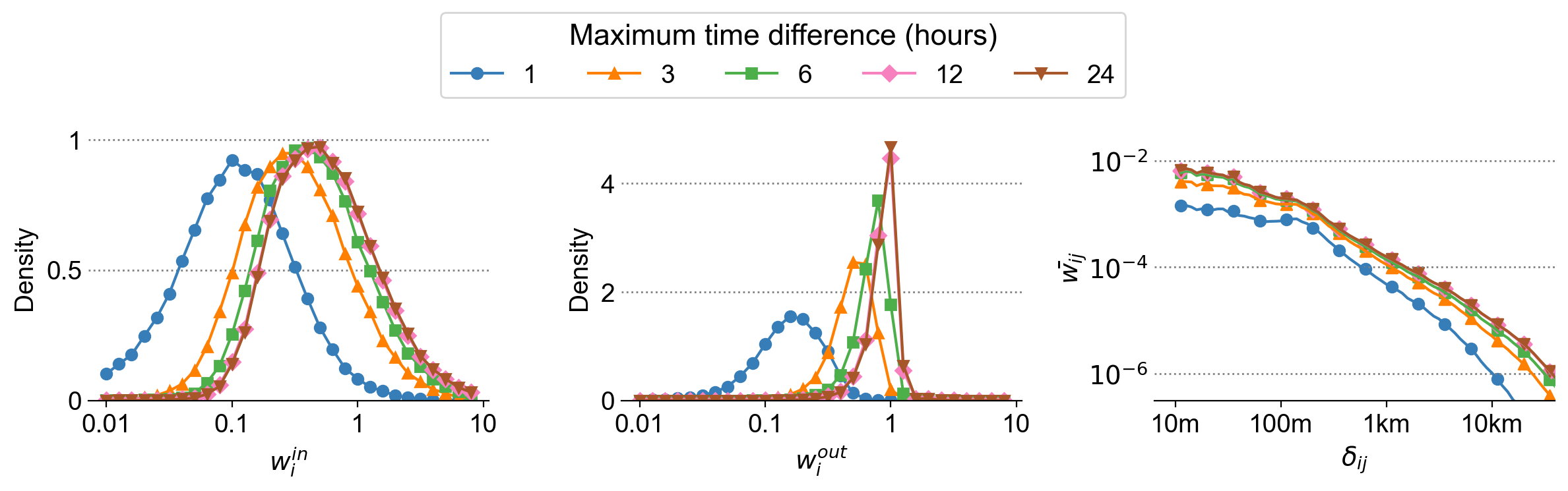}} \\
\subfloat[Difference in category pairwise dependency weights compared to baseline network]{\includegraphics[width=\linewidth]{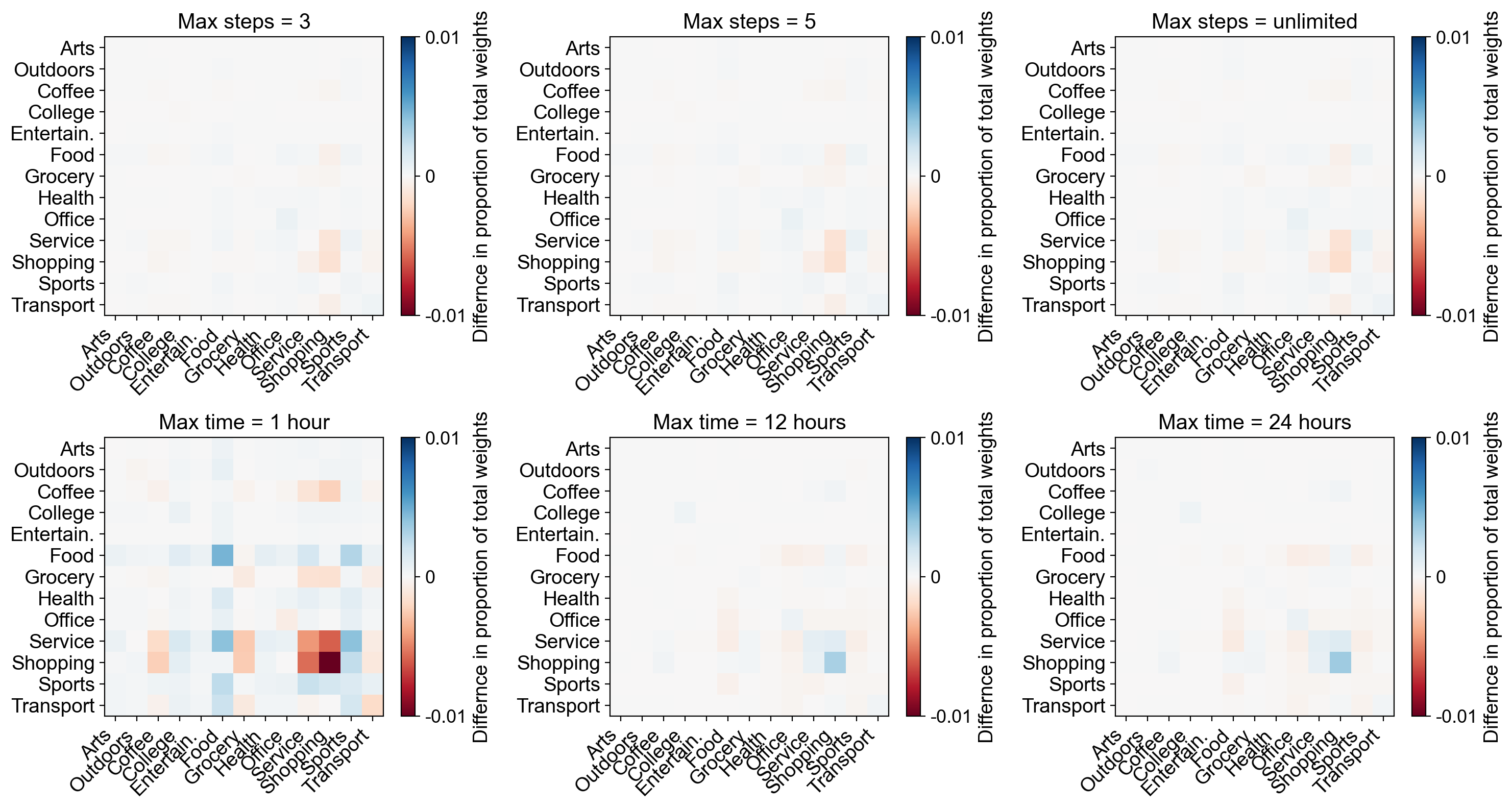}}
\caption[Dependency network in Boston under different co-visit detection parameters]{\textbf{Dependency network in Boston under different co-visit detection parameters.}}
\label{fig:s2boston}
\end{figure}

\begin{figure}[!t]
\centering
\subfloat[Network characteristics under different maximum step difference $T_s$]{\includegraphics[width=.9\linewidth]{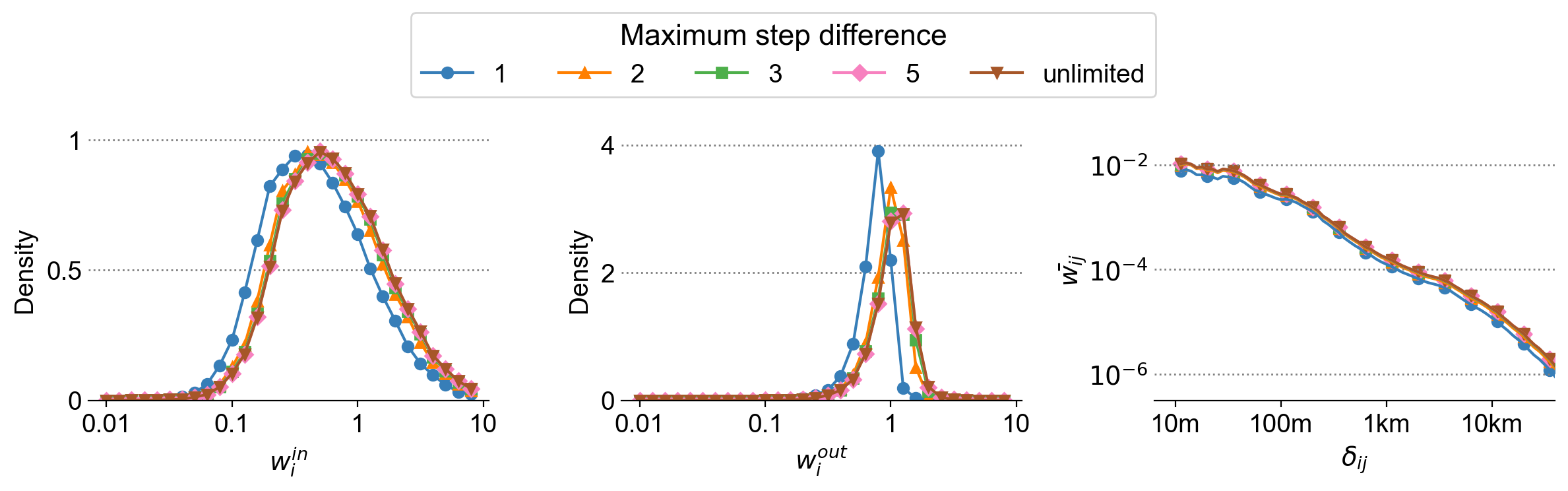}} \\
\subfloat[Network characteristics under different maximum time difference $T_c$]{\includegraphics[width=.9\linewidth]{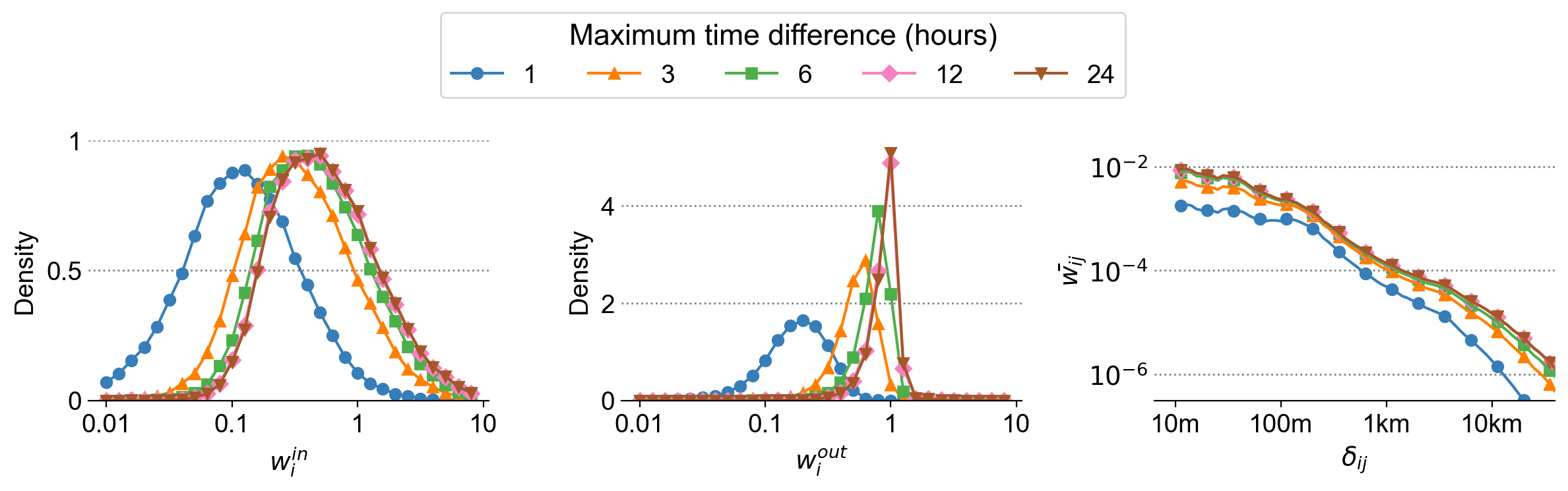}} \\
\subfloat[Difference in category pairwise dependency weights compared to baseline network]{\includegraphics[width=\linewidth]{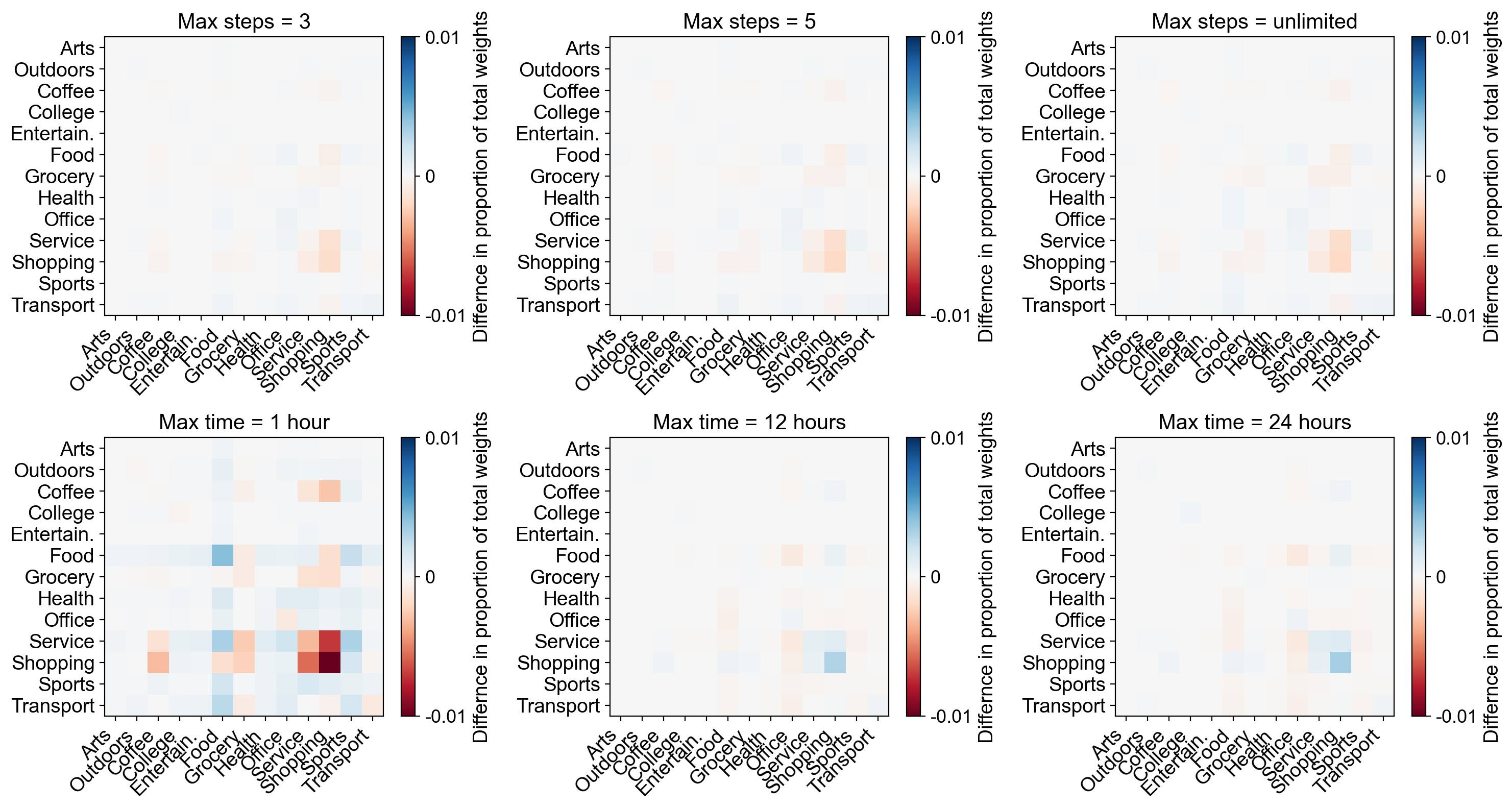}}
\caption[Dependency network in Seattle under different co-visit detection parameters]{\textbf{Dependency network in Seattle under different co-visit detection parameters.}}
\label{fig:s2seattle}
\end{figure}

\begin{figure}[!t]
\centering
\subfloat[Network characteristics under different maximum step difference $T_s$]{\includegraphics[width=.9\linewidth]{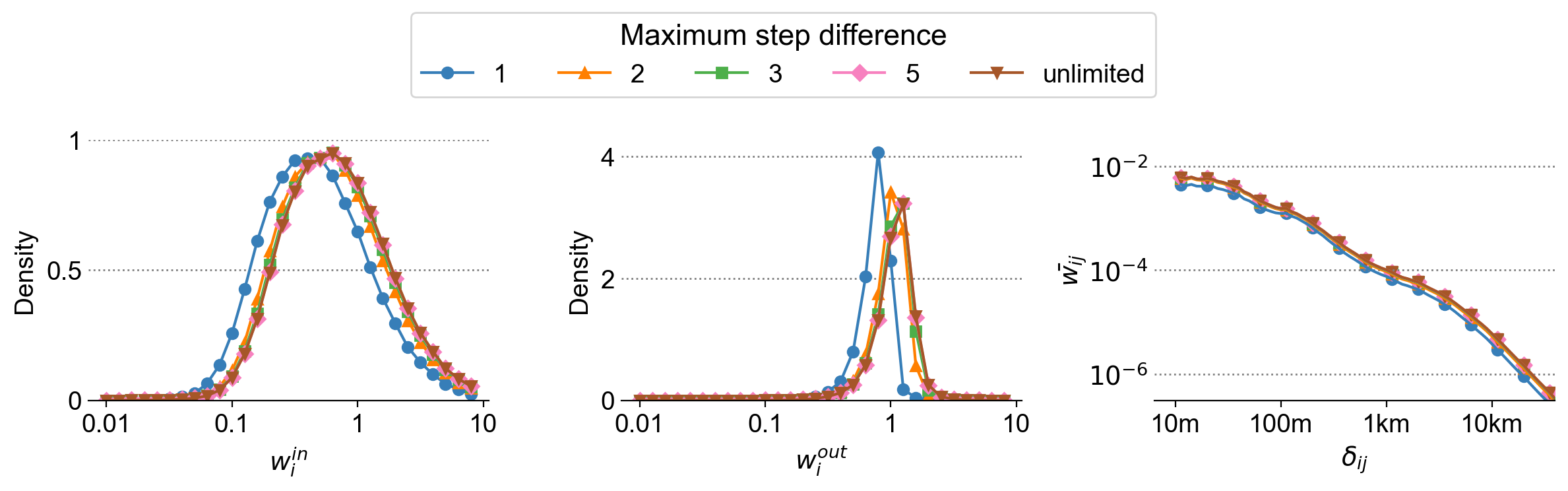}} \\
\subfloat[Network characteristics under different maximum time difference $T_c$]{\includegraphics[width=.9\linewidth]{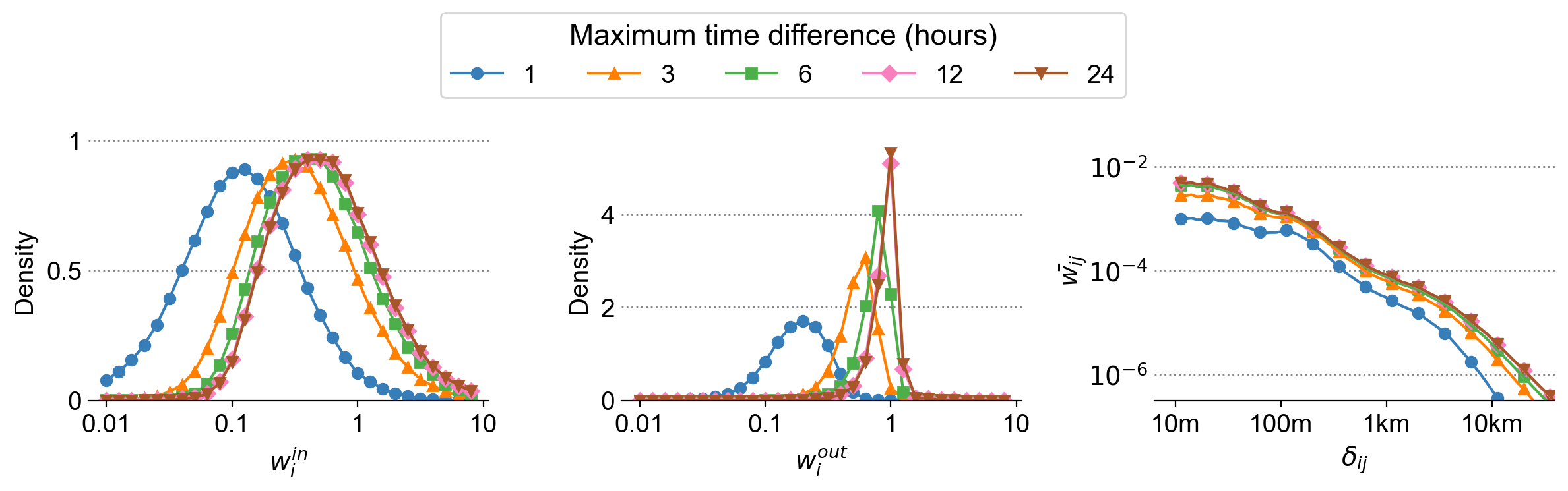}} \\
\subfloat[Difference in category pairwise dependency weights compared to baseline network]{\includegraphics[width=\linewidth]{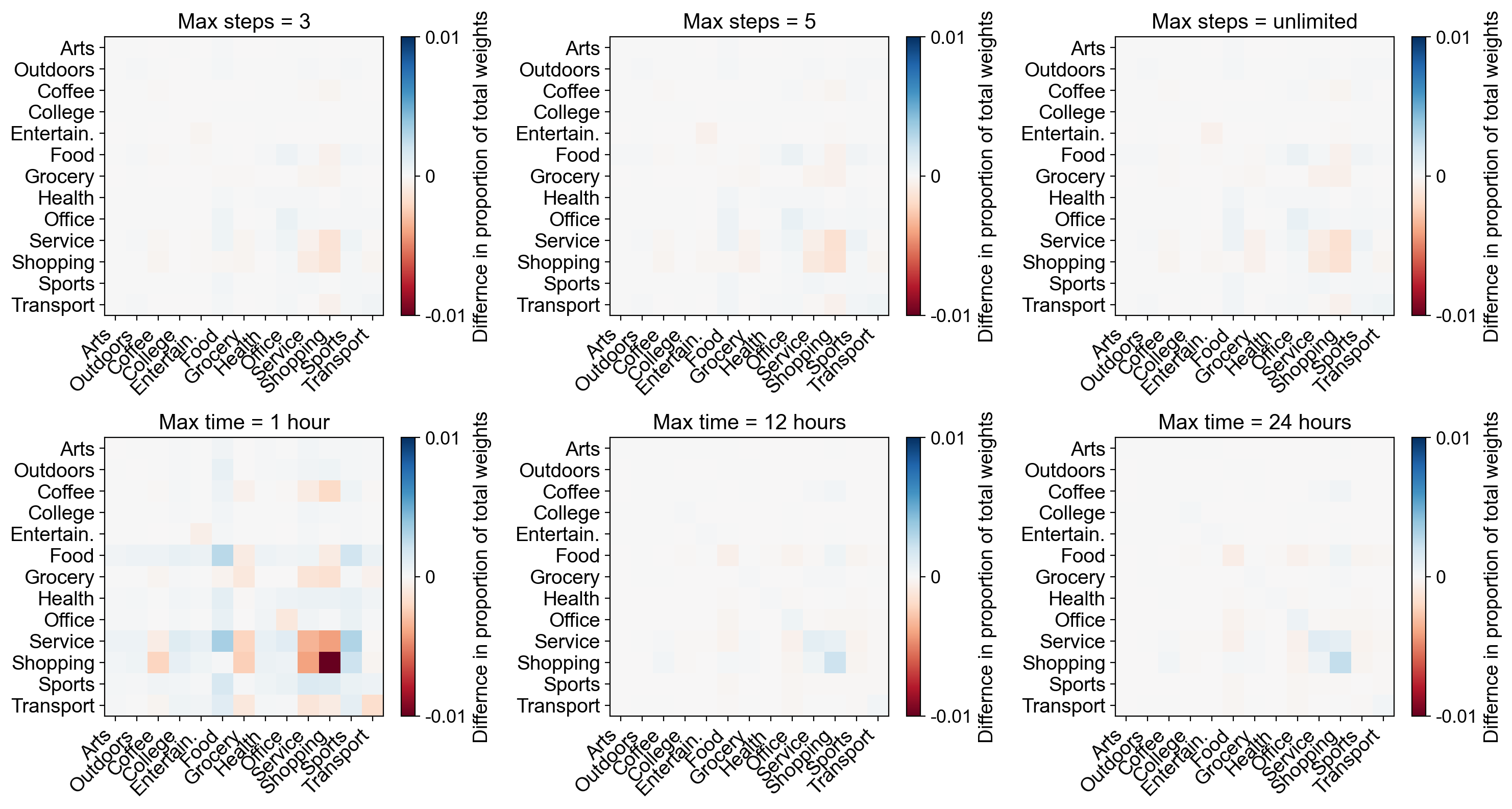}}
\caption[Dependency network in Los Angeles under different co-visit detection parameters]{\textbf{Dependency network in Los Angeles under different co-visit detection parameters.}}
\label{fig:s2la}
\end{figure}

\begin{figure}[!t]
\centering
\subfloat[Network characteristics under different maximum step difference $T_s$]{\includegraphics[width=.9\linewidth]{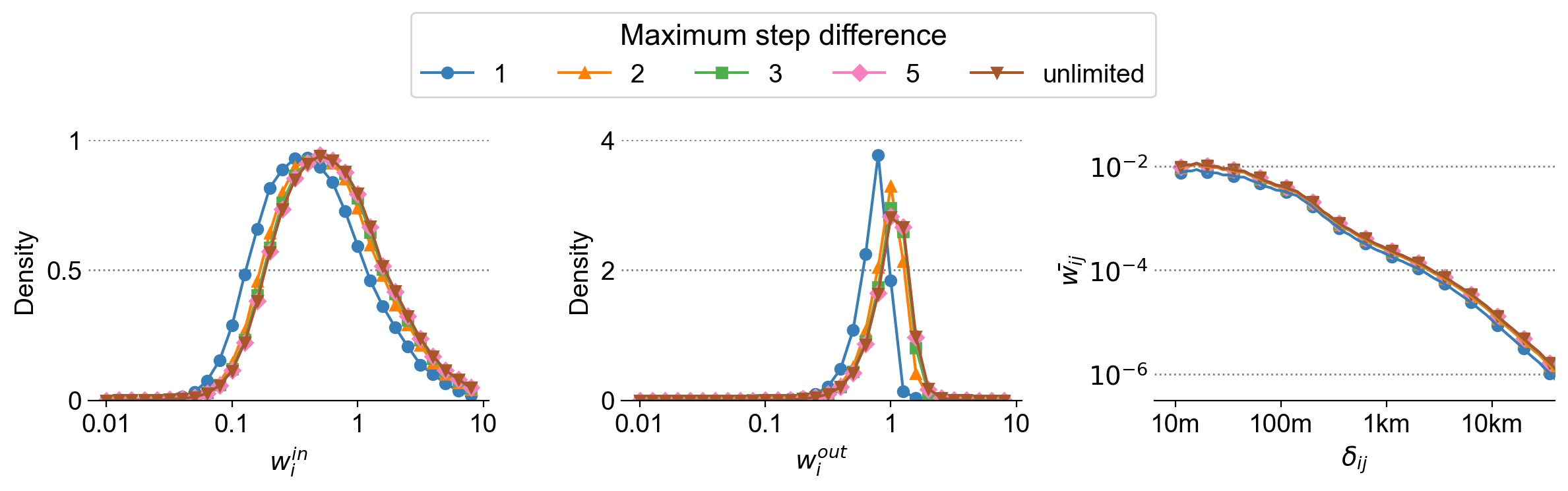}} \\
\subfloat[Network characteristics under different maximum time difference $T_c$]{\includegraphics[width=.9\linewidth]{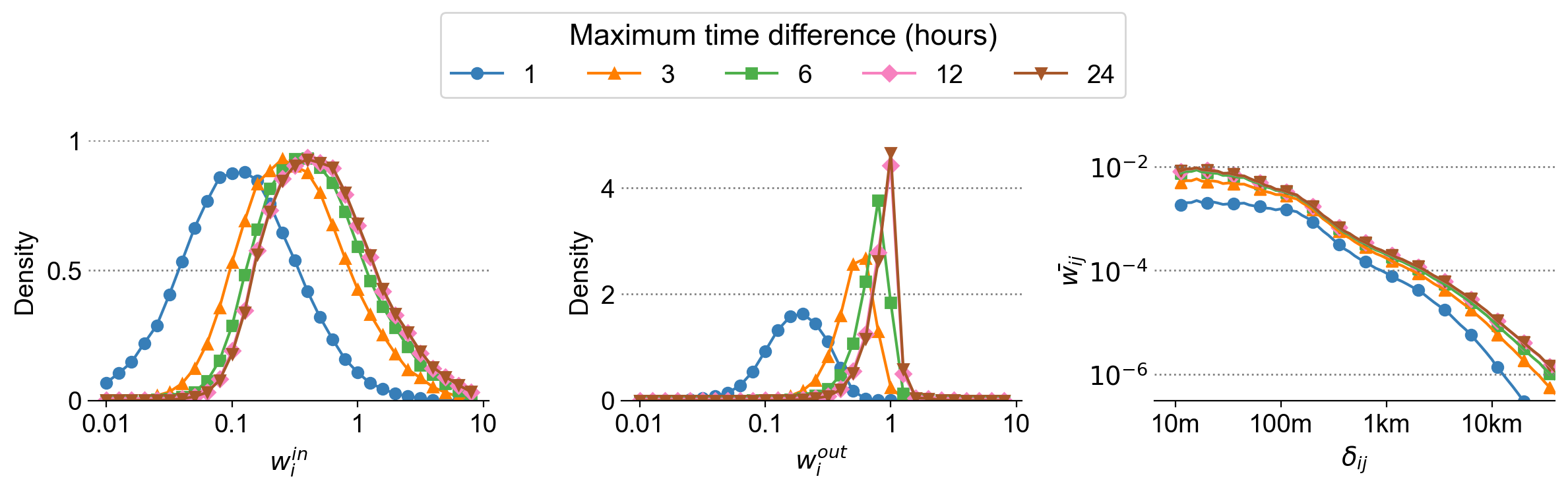}} \\
\subfloat[Difference in category pairwise dependency weights compared to baseline network]{\includegraphics[width=\linewidth]{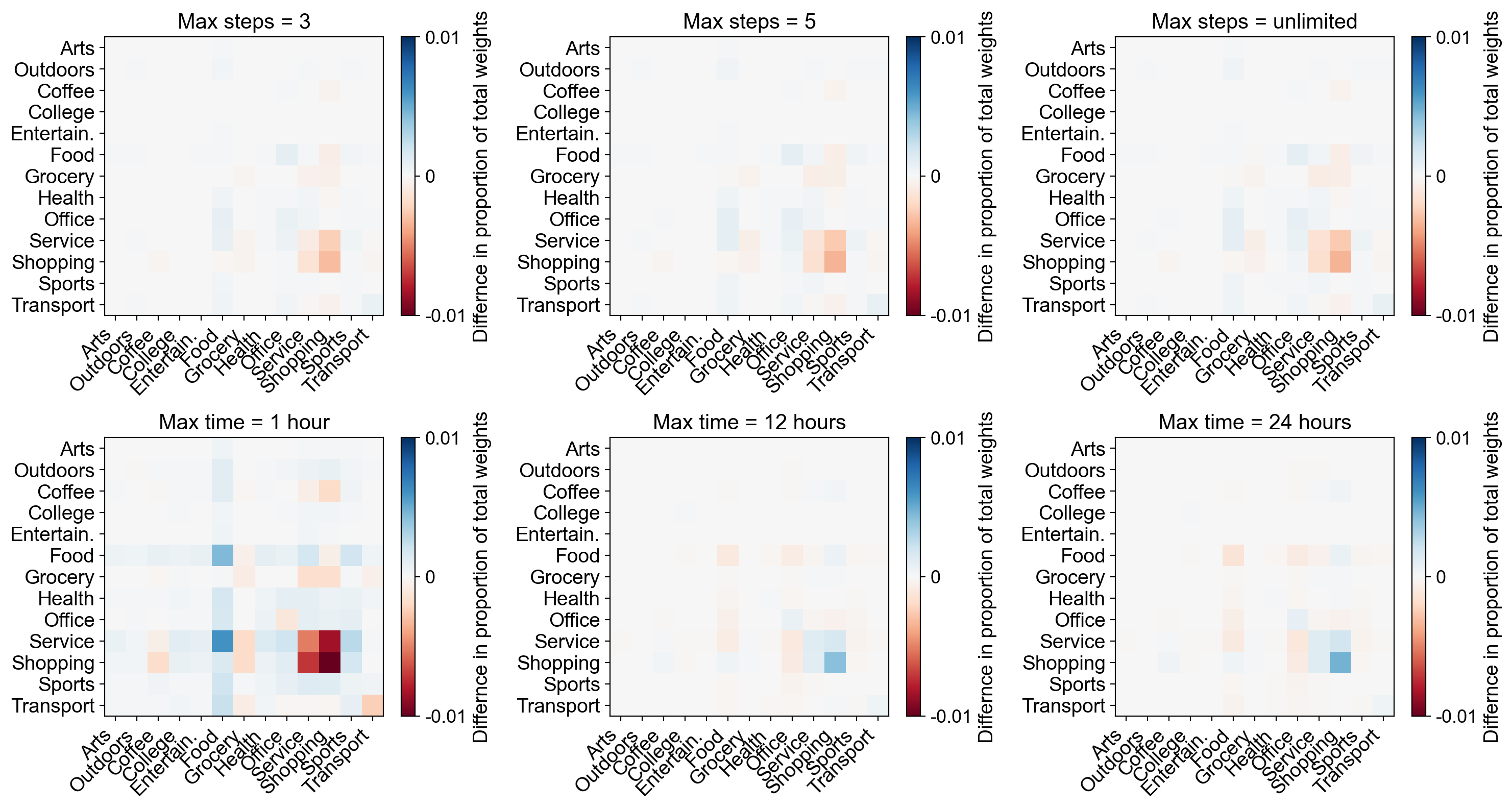}}
\caption[Dependency network in Dallas under different co-visit detection parameters]{\textbf{Dependency network in Dallas under different co-visit detection parameters.}}
\label{fig:s2dallas}
\end{figure}

How much do the co-visit detection parameters ($T_s$,$T_c$) affect the structural properties of the dependency networks? To investigate this, we generated the dependency network using 25 pairs of parameters ($T_s = [1,2,3,5,unlimited]$ and $T_c = [1,3,6,12,24]$ hours). 
The following analysis compares the following characteristics of the dependency networks across different co-vist detection parameters:
\begin{itemize}
    \item POI subcategories with largest in- and out-weights (Figures \ref{s2categoryrank1} and \ref{s2categoryrank3})
    \item Distributions of in-weights and out-weights per POI  (panels (a) and (b) in Figures \ref{fig:s2nyc} to \ref{fig:s2dallas}) 
    \item Relationship between average weight and physical distance between the POIs  (panels (a) and (b) in Figures \ref{fig:s2nyc} to \ref{fig:s2dallas}) 
    \item Difference in the proportion of weights for each category pair (positive means baseline network has more weights between the category pairs) (panel (c) in Figures \ref{fig:s2nyc} to \ref{fig:s2dallas}) 
\end{itemize}

As shown in Figures \ref{s2categoryrank1} and \ref{s2categoryrank3}, the POI subcategories that appeared in each ranking were similar when we changed the co-visit detection parameter to $T_c = 1$ and $T_c = 3$, respectively. Under all parameters, supercenters, malls, department malls, and airports had the largest in-weights, while places such as art dealers and various types of retail stores had the largest average out-weight.

Panels (a) in Figures \ref{fig:s2nyc} to \ref{fig:s2dallas}) show that although the distribution obviously shifts to the right (more steps would include more links, thus more in- and out- weights) the maximum step difference parameter does not make a big difference on the distributions of the in-weights, out-weights, or the decay of average weights with the physical distance. 

On the other hand, Panels (b) in Figures \ref{fig:s2nyc} to \ref{fig:s2dallas}) show that the distributions are significantly affected by the choice of time difference parameter. Obviously, the longer the temporal threshold we use, the distributions shift to the right (more time difference would include more links, thus more in- and out- weights). Especially, the dependency networks with the 1 hour threshold present a significantly sparser network with lower in-weights, out-weights, and faster decay of distance with respect to physical distance. The $T_c =6$ threshold that we chose as the baseline parameter generates network characteristics that are closely in between $T_c = 3$, and $T_c = 12$ and $T_c = 24$.

Panels (c) in Figures \ref{fig:s2nyc} to \ref{fig:s2dallas}) show that despite the changes in co-visit detection parameters, the proportion of weights among category pairs are not affected significantly. Note that positive differences (blue color) mean that the baseline network ($T_s = 1$,$T_c = 6$) contains more weight in the specific category pair. The only exception is $T_c = 1$ (similar to the in-weights, out-weights, and distance decay curves), where we observe much fewer weights among shopping POIs in the baseline network. This indicates that when we limit the time difference to 1 hour, we capture significantly more of the shopping-to-shopping weights compared to other category pairs. 

\begin{figure}[!t]
\centering
\subfloat[New York]{\includegraphics[width=.6\linewidth]{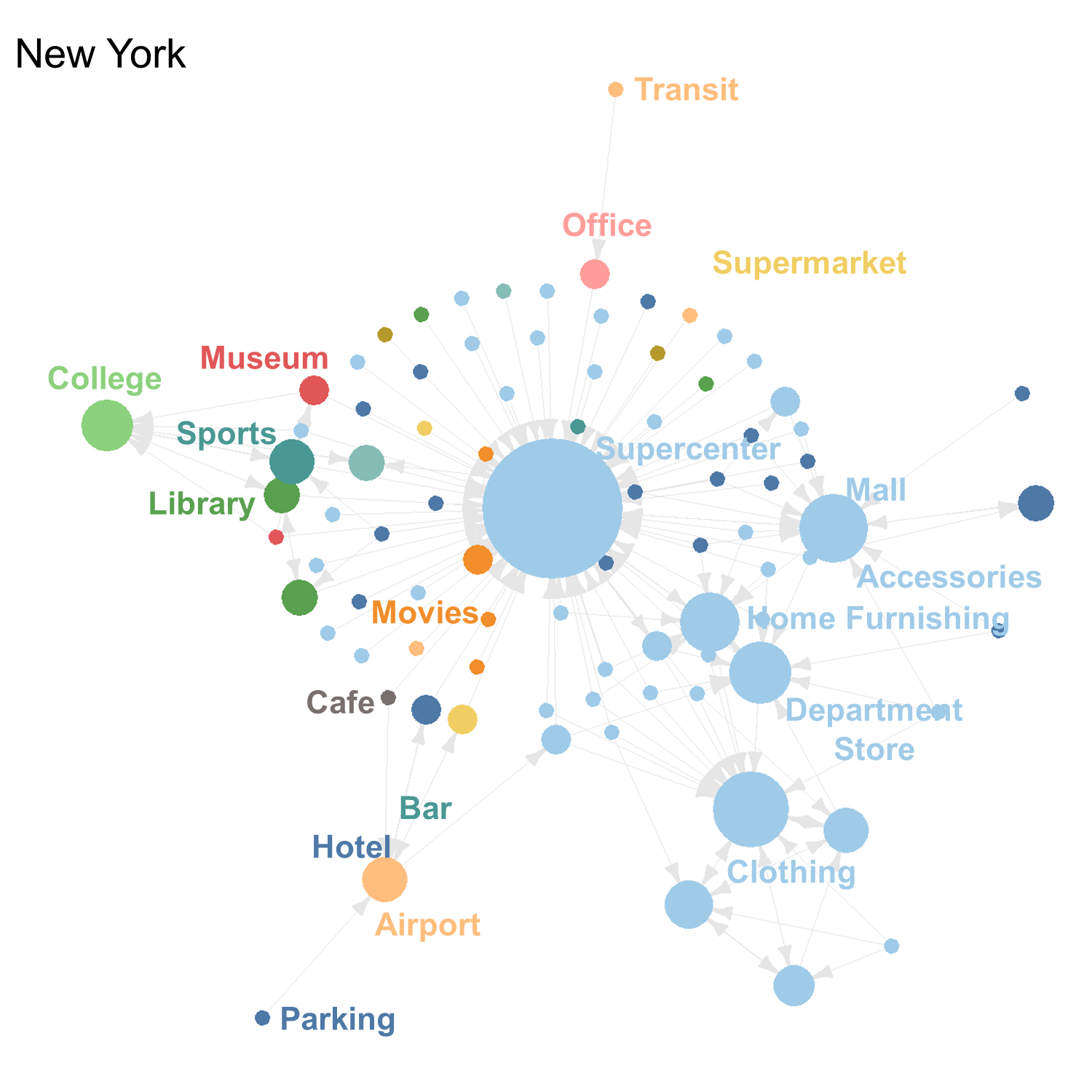}} \\
\subfloat[Boston]{\includegraphics[width=.6\linewidth]{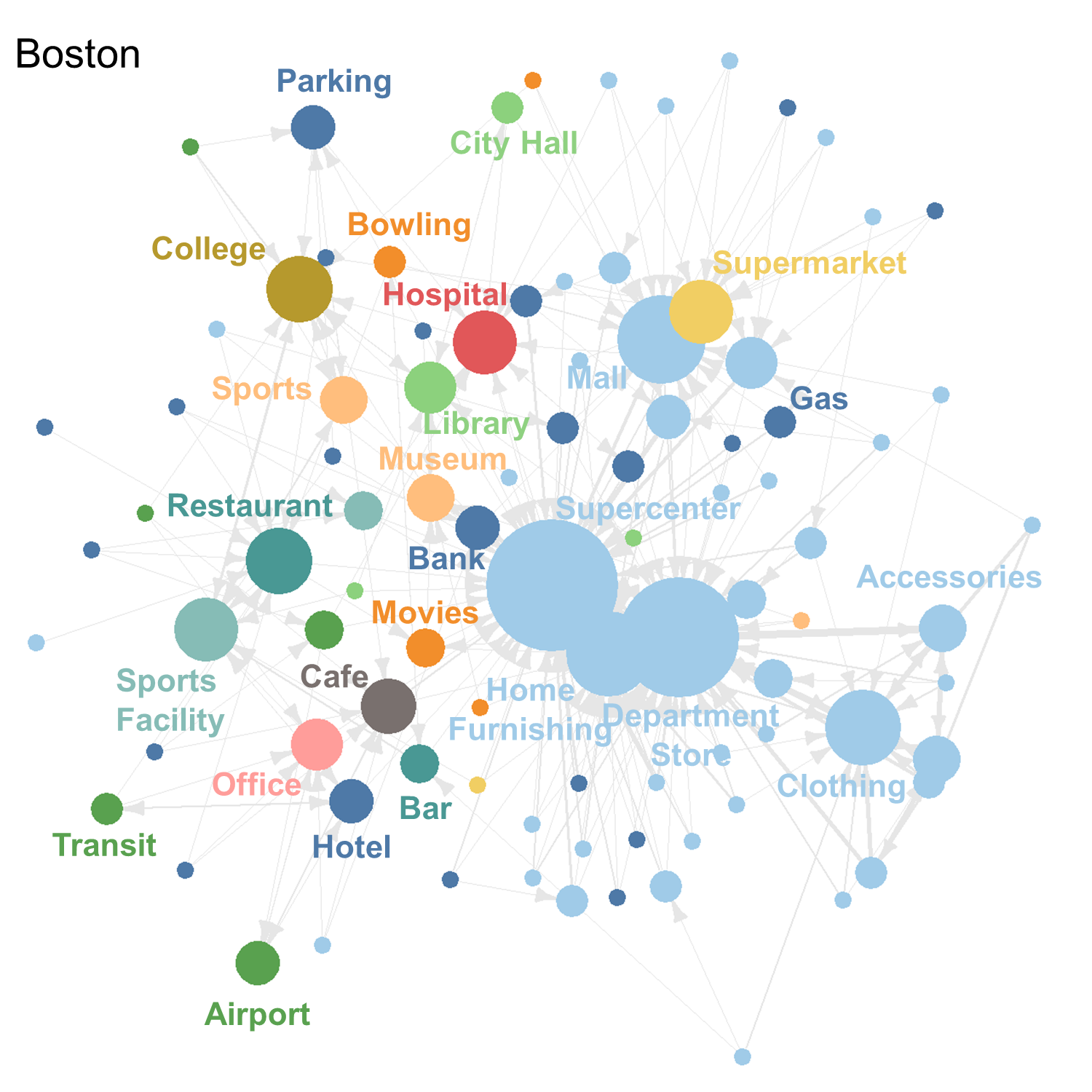}} 
\caption[Network diagram showing the average dependencies between POI subcategories in New York and Boston]{\textbf{Network diagram showing the average dependencies between POI subcategories in New York and Boston.} Each node is a POI subcategory and the three largest outgoing dependency edges are shown for each node. Node sizes show the in-degree of the constructed network.}
\label{fig:catnetwork1}
\end{figure}

\begin{figure}[!t]
\centering
\subfloat[Seattle]{\includegraphics[width=.6\linewidth]{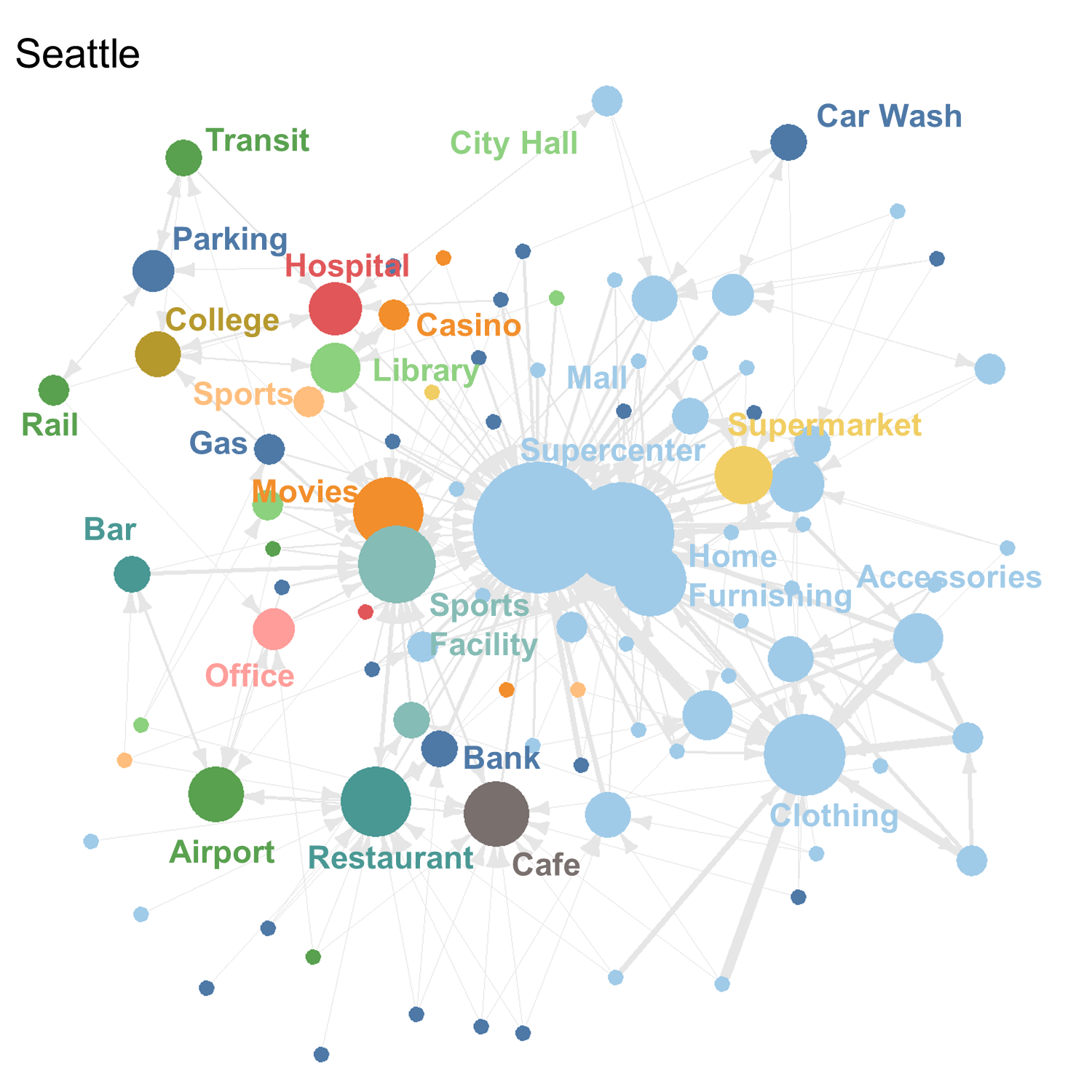}} \\
\subfloat[Los Angeles]{\includegraphics[width=.6\linewidth]{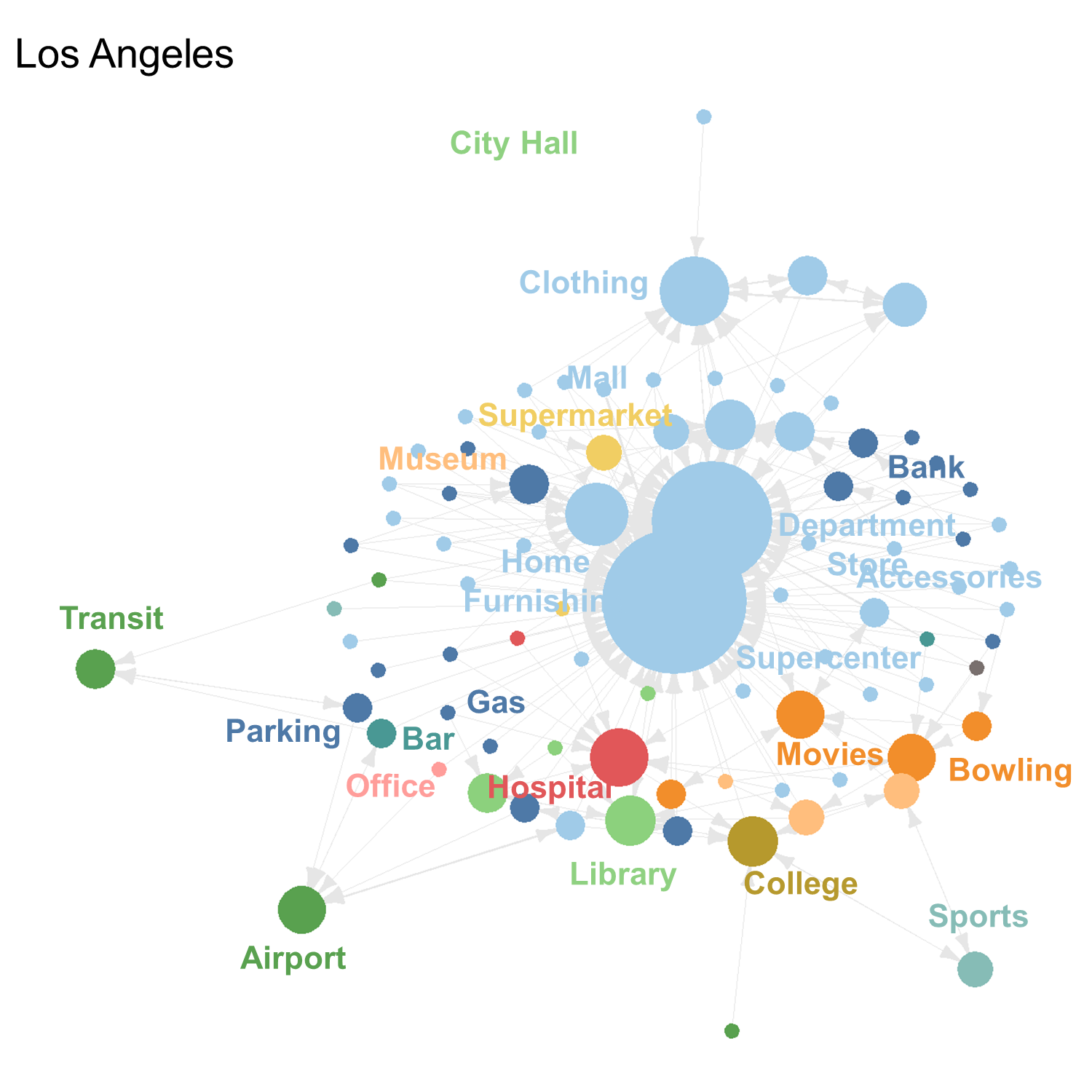}} 
\caption[Network diagram showing the average dependencies between POI subcategories in Seattle and Los Angeles]{\textbf{Network diagram showing the average dependencies between POI subcategories in Seattle and Los Angeles.} Each node is a POI subcategory and the three largest outgoing dependency edges are shown for each node. Node sizes show the in-degree of the constructed network.}
\label{fig:catnetwork2}
\end{figure}

\begin{figure}[!t]
\centering
\subfloat[Dallas]{\includegraphics[width=.6\linewidth]{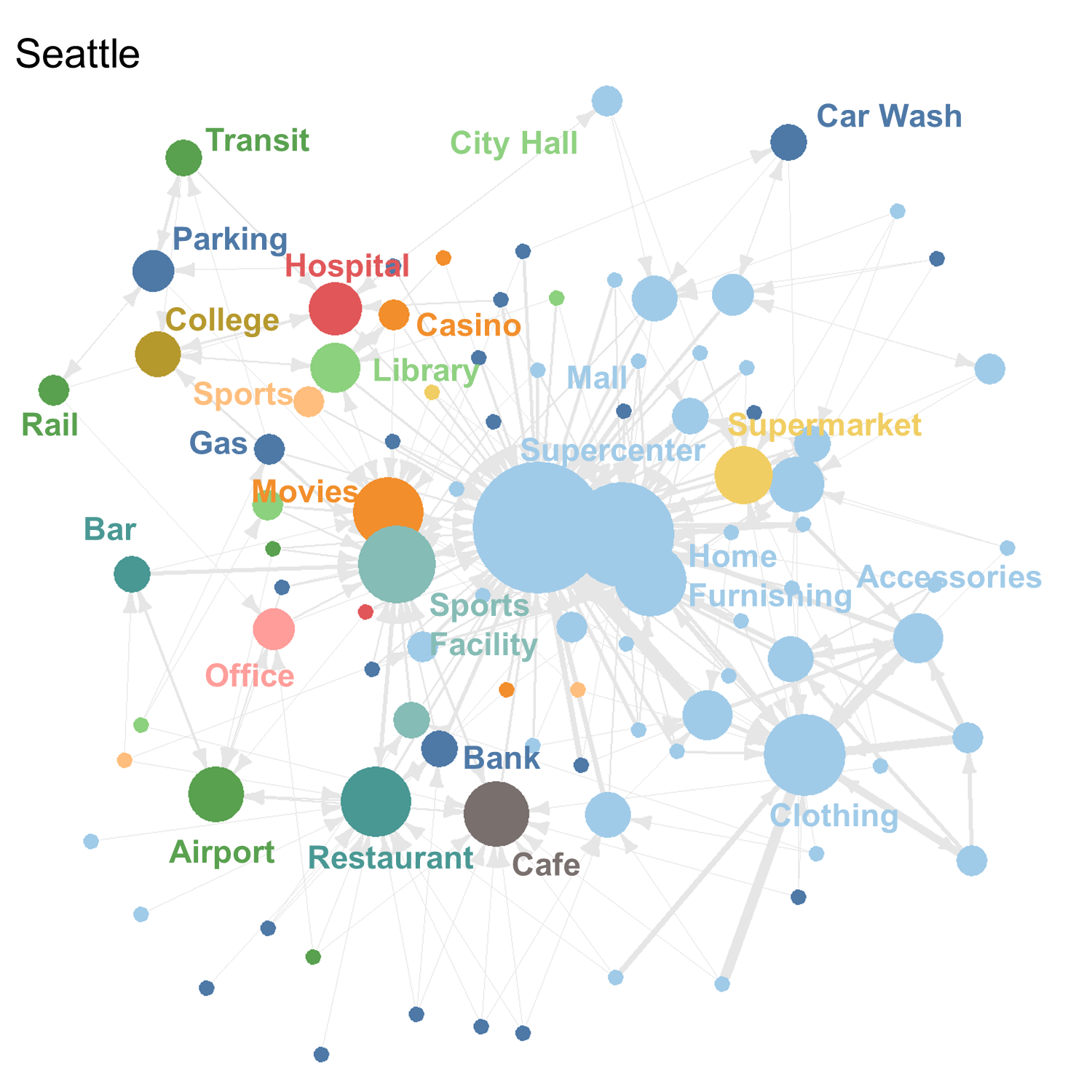}} 
\caption[Network diagram showing the average dependencies between POI subcategories in Dallas]{\textbf{Network diagram showing the average dependencies between POI subcategories in Dallas.} Each node is a POI subcategory and the three largest outgoing dependency edges are shown for each node. Node sizes show the in-degree of the constructed network.}
\label{fig:catnetwork3}
\end{figure}

Network diagrams in Figures \ref{fig:catnetwork1} to \ref{fig:catnetwork3} show the average dependencies between POI subcategories in the five cities. Each node represents a POI subcategory (there are 96 of them in the dataset) and the three largest outgoing dependency edges are shown for each node. Node sizes show the in-degree of the constructed network (i.e., how many other POI categories depend on that node). Many shopping subcategories including supercenters, department stores, malls, and clothing stores, and colleges, cafes, restaurants are depended by many other subcategories. 

\clearpage
\section{Statistical robustness of behavior-based dependency networks}

Apart from co-visit detection parameters $T_c$ and $T_s$, many factors and decision parameters could affect the characteristics of the dependency network and its weights.
In this section, we test the statistical robustness of the edge weights and network characteristics by comparing the dependency network with several variants.
More specifically, we will make comparisons with the following networks:

\begin{itemize}
    \item Dependency network computed using data from a different time period, for example, January -- April 2019, instead of September -- December 2019. 
    \item Dependency network where edges with small weights are removed. This is done by computing the 25th percentile of the weights $w_{ij}$ using bootstrap method. 
    \item Dependency network calculated without the post-stratification procedure. 
\end{itemize}

The following subsections show the details of how each type of network is calculated and comparisons of network statistics with the empirical dependency network.
In this section, comparisons of edge-based and category-aggregated edge weights are compared between the empirical and alternative dependency networks. Further robustness checks on the modeling of network weights, the predictability of shocks during COVID-19, and resilience to future hypothetical shocks are performed in Supplementary Notes 4, 5, and 6, respectively.

\subsection{Robustness against choice of time period}
\begin{figure}[!t]
\centering
\subfloat[New York, in-weight]{\includegraphics[width=.48\linewidth]{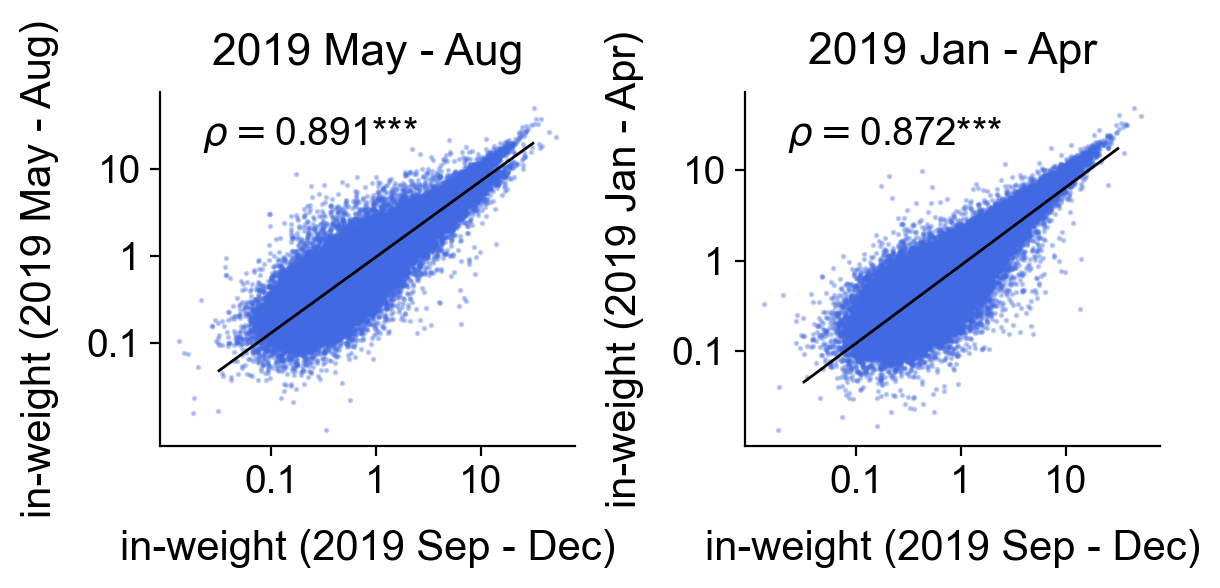}} 
\subfloat[New York, out-weight]{\includegraphics[width=.48\linewidth]{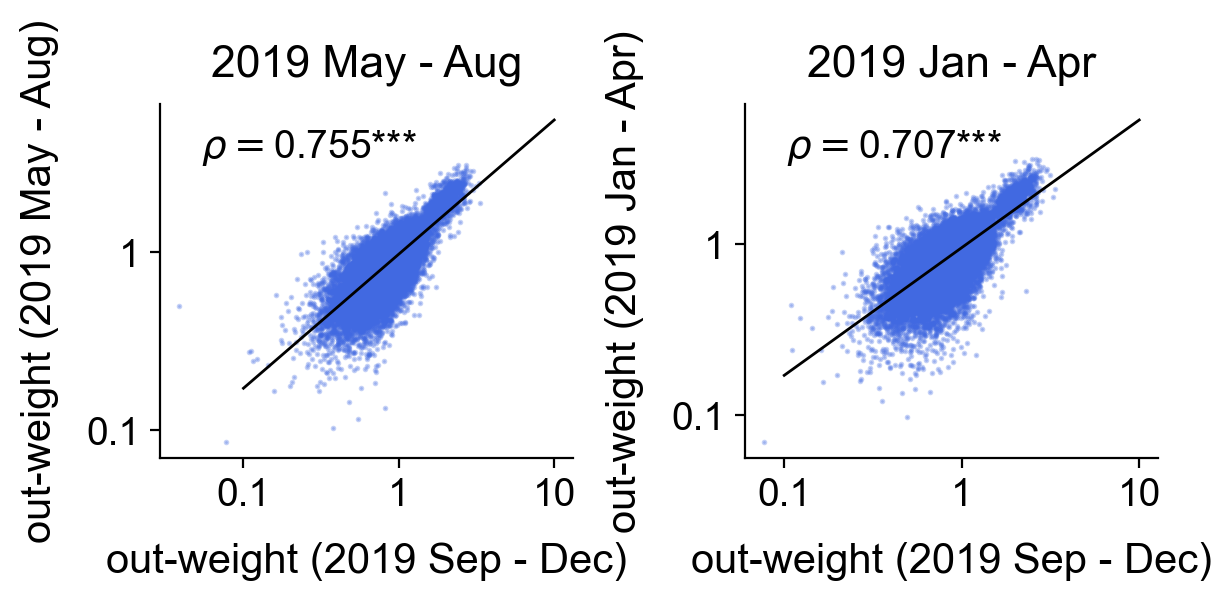}} \\
\subfloat[Boston, in-weight]{\includegraphics[width=.48\linewidth]{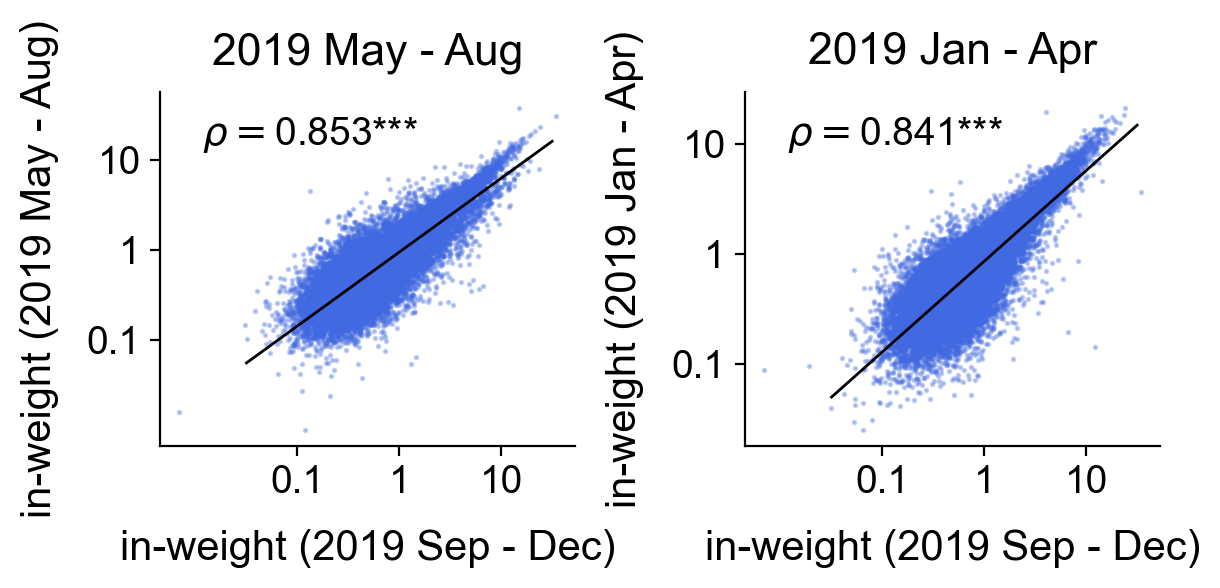}} 
\subfloat[Boston, out-weight]{\includegraphics[width=.48\linewidth]{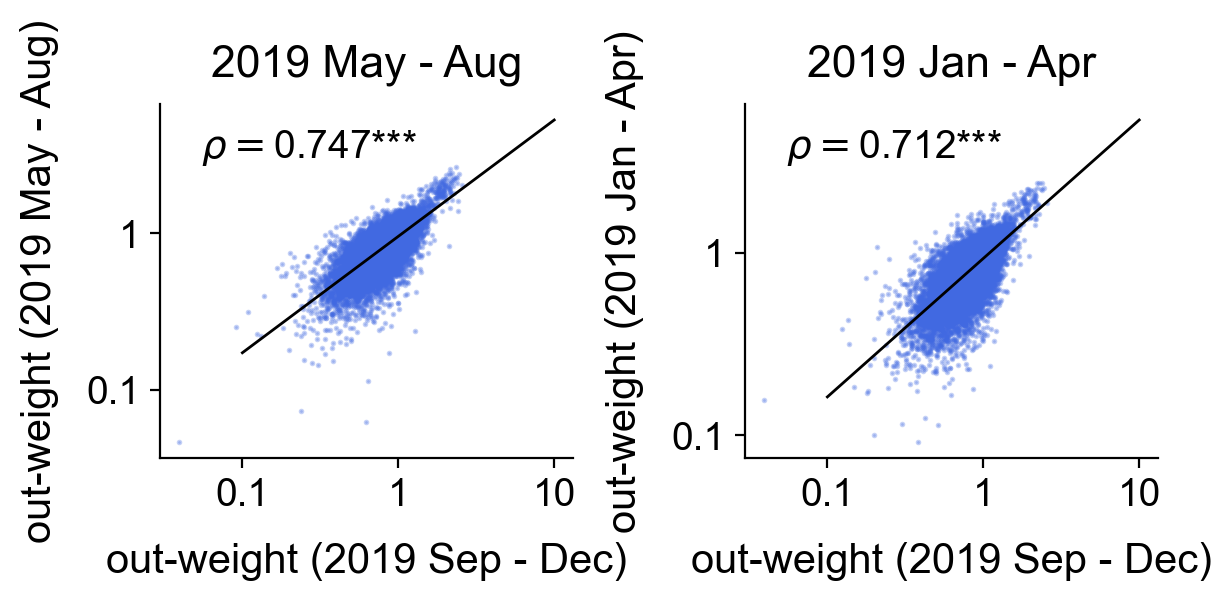}} \\
\subfloat[Seattle, in-weight]{\includegraphics[width=.48\linewidth]{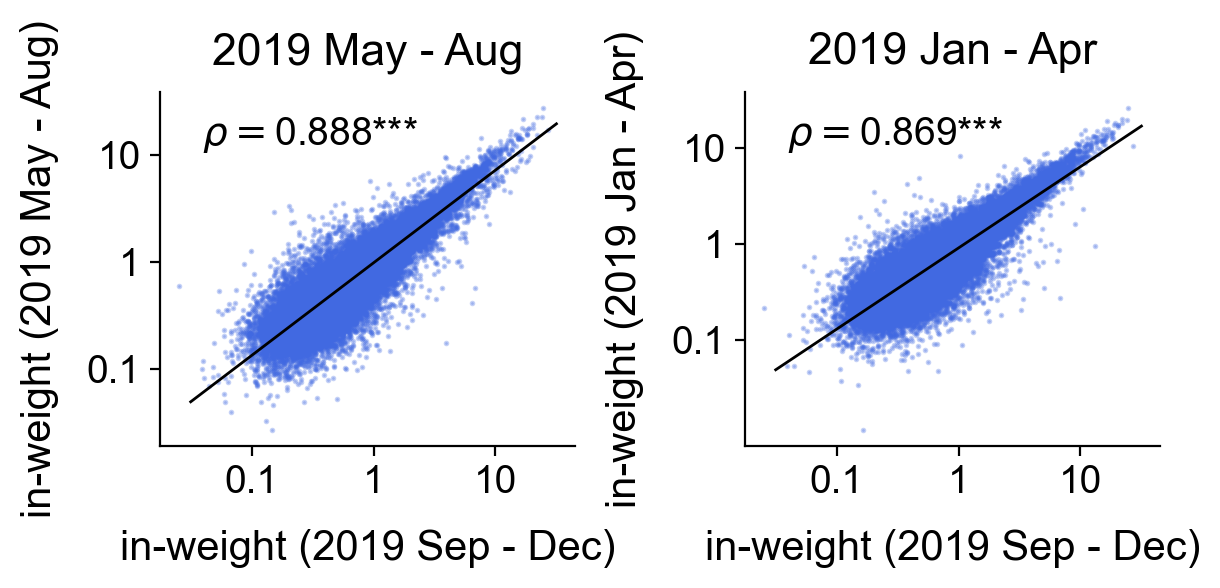}} 
\subfloat[Seattle, out-weight]{\includegraphics[width=.48\linewidth]{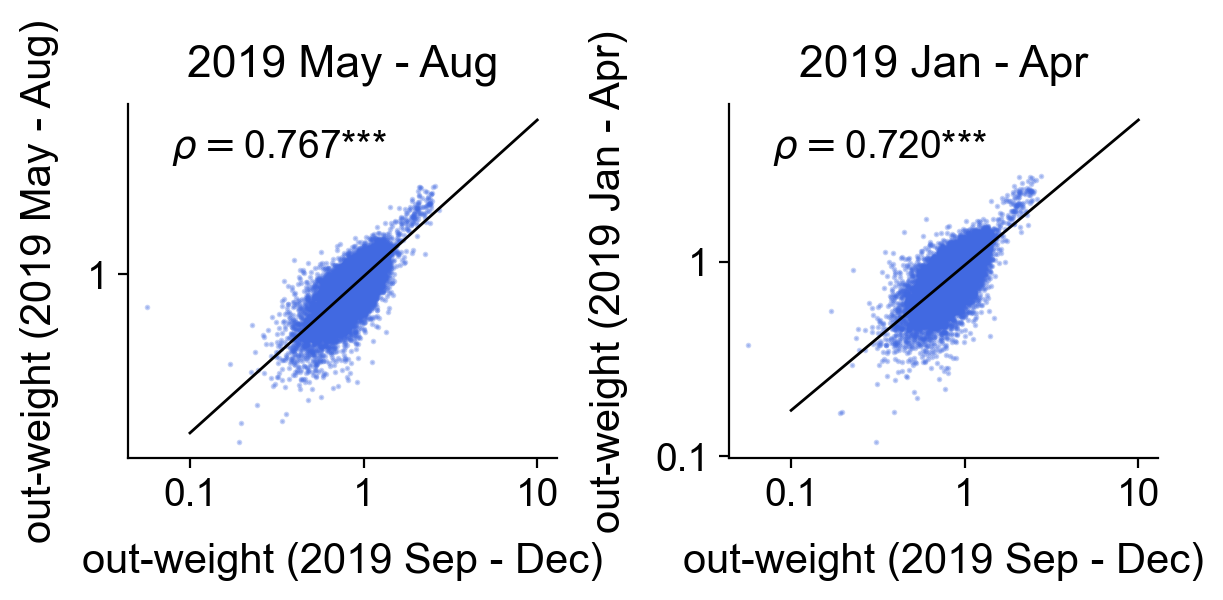}} 
\caption[Comparison of in- and out-weights of each POI across different data collection periods]{\textbf{Comparison of in- and out-weights of each POI across different data collection periods for New York, Boston, and Seattle.} In- and out-weights across different time periods (2019 May -- August and 2019 January -- April) are highly correlated ($\rho>0.7$) with the baseline time period (2019 September -- December).}
\label{fig:s3period1}
\end{figure}

\begin{figure}[!t]
\centering
\subfloat[Los Angeles, in-weight]{\includegraphics[width=.48\linewidth]{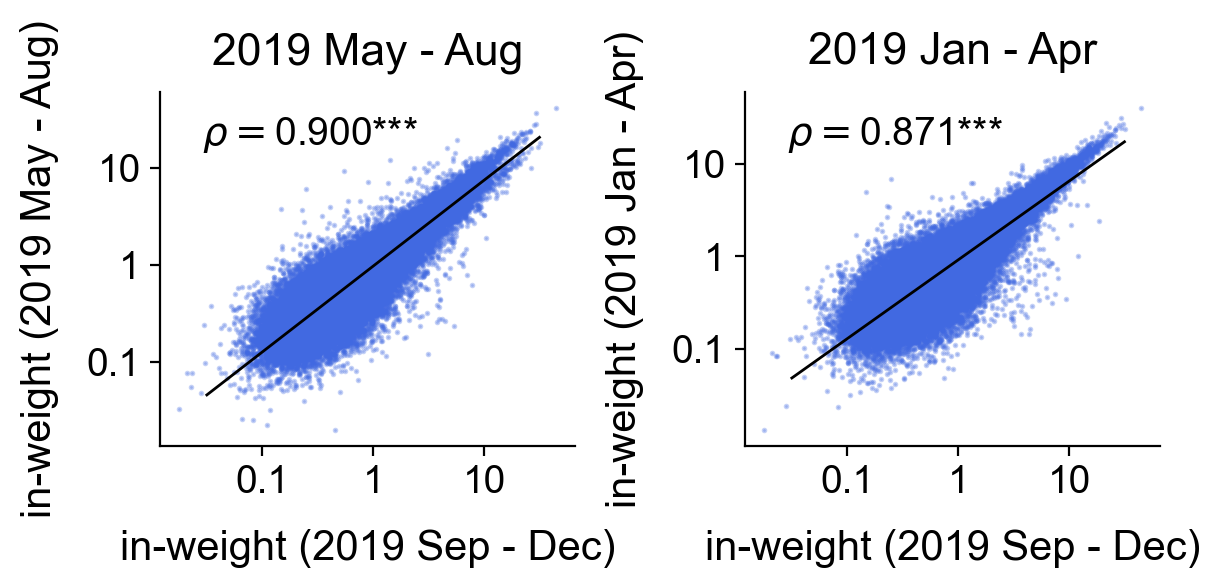}} 
\subfloat[Los Angeles, out-weight]{\includegraphics[width=.48\linewidth]{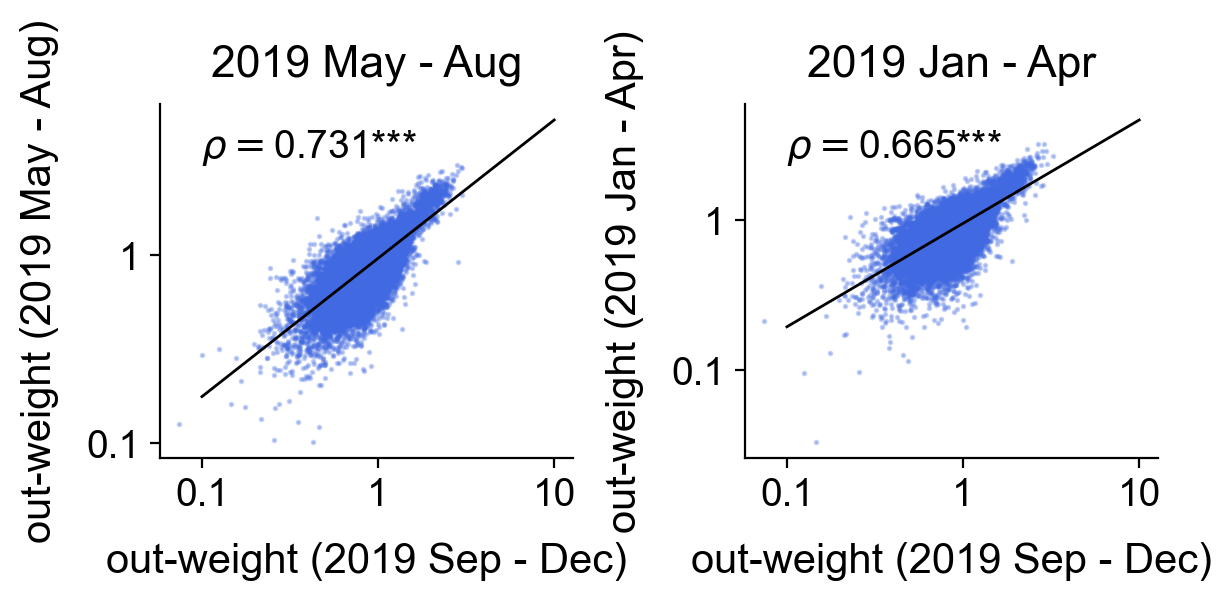}} \\
\subfloat[Dallas, in-weight]{\includegraphics[width=.48\linewidth]{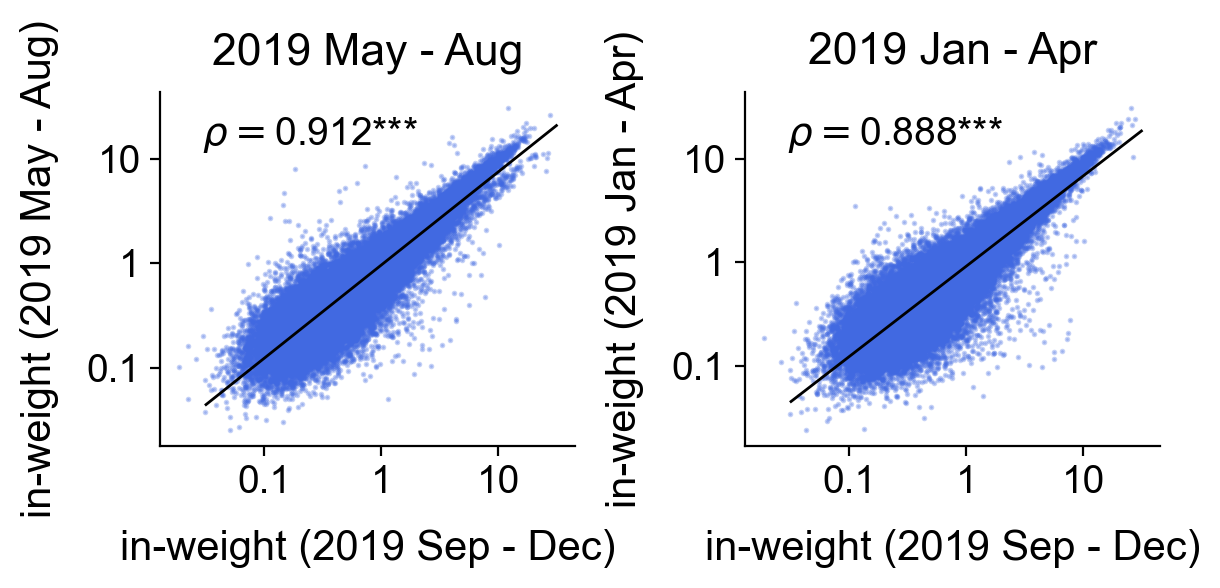}} 
\subfloat[Dallas, out-weight]{\includegraphics[width=.48\linewidth]{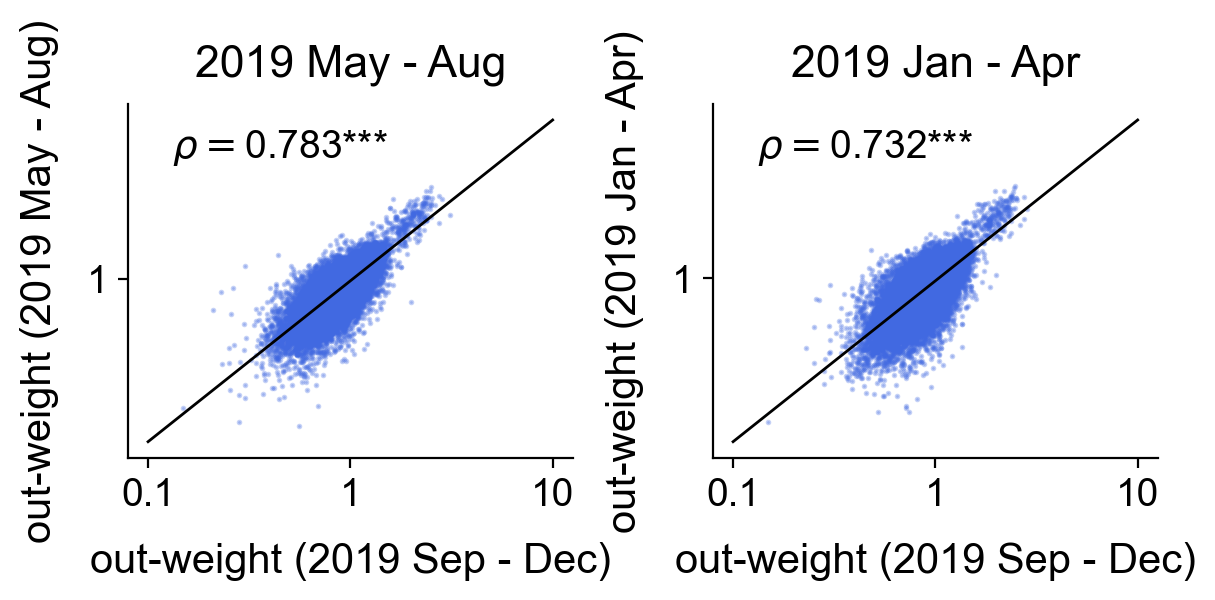}} 
\caption[Comparison of in- and out-weights of each POI across different data collection periods, contd.]{\textbf{Comparison of in- and out-weights of each POI across different data collection periods for Los Angeles and Dallas.} In- and out-weights across different time periods (2019 May -- August and 2019 January -- April) are highly correlated ($\rho>0.7$) with the baseline time period (2019 September -- December).}
\label{fig:s3period2}
\end{figure}

\begin{figure}[!t]
\centering
\subfloat[New York]{\includegraphics[width=\linewidth]{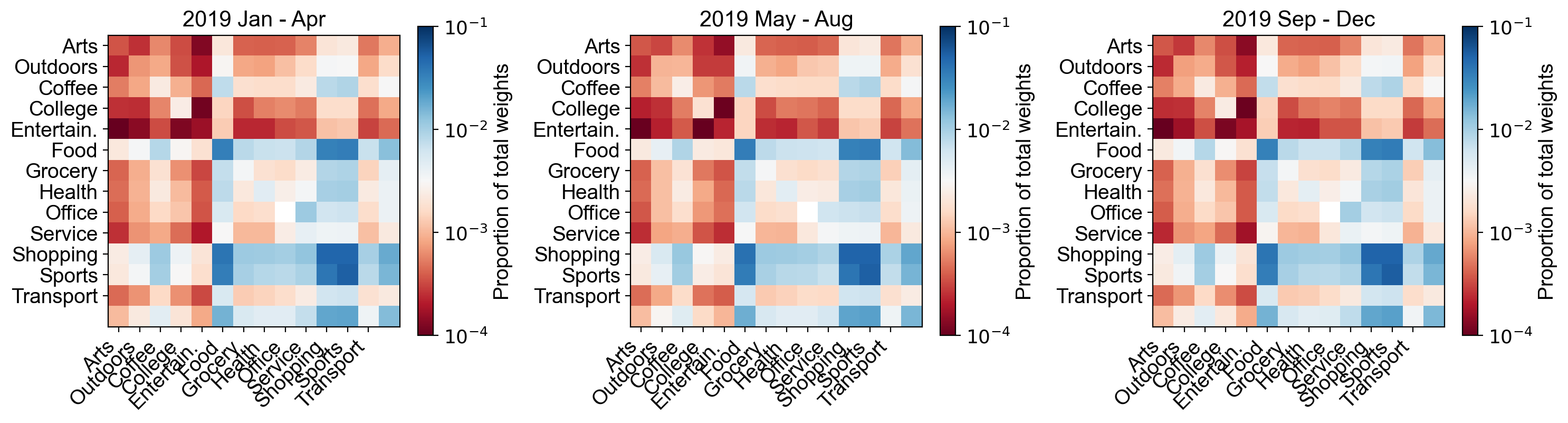}} \\
\subfloat[Boston]{\includegraphics[width=\linewidth]{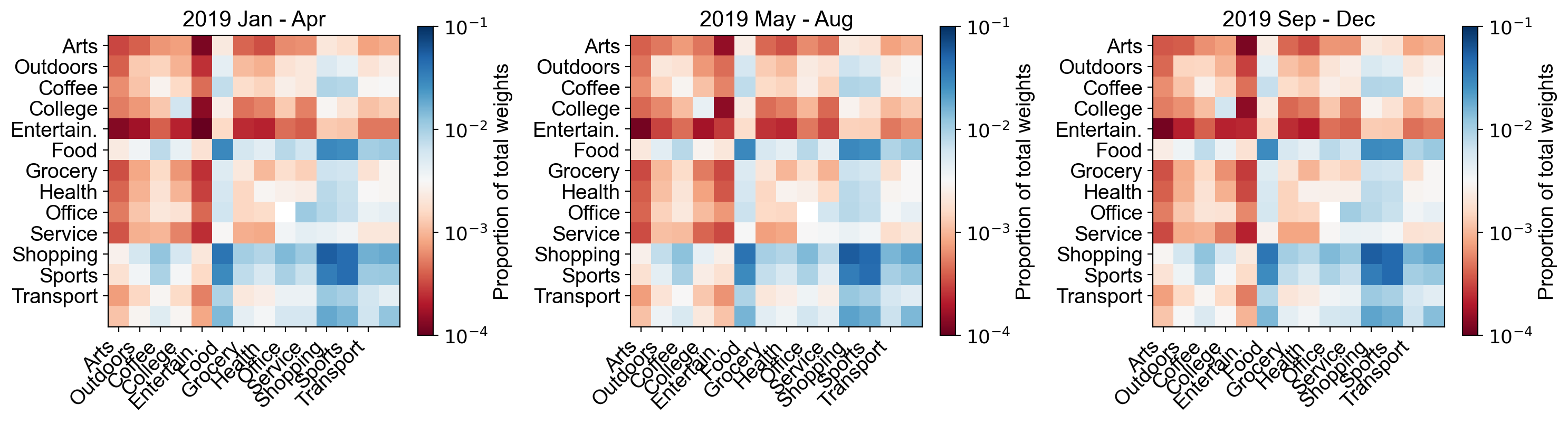}} \\
\subfloat[Seattle]{\includegraphics[width=\linewidth]{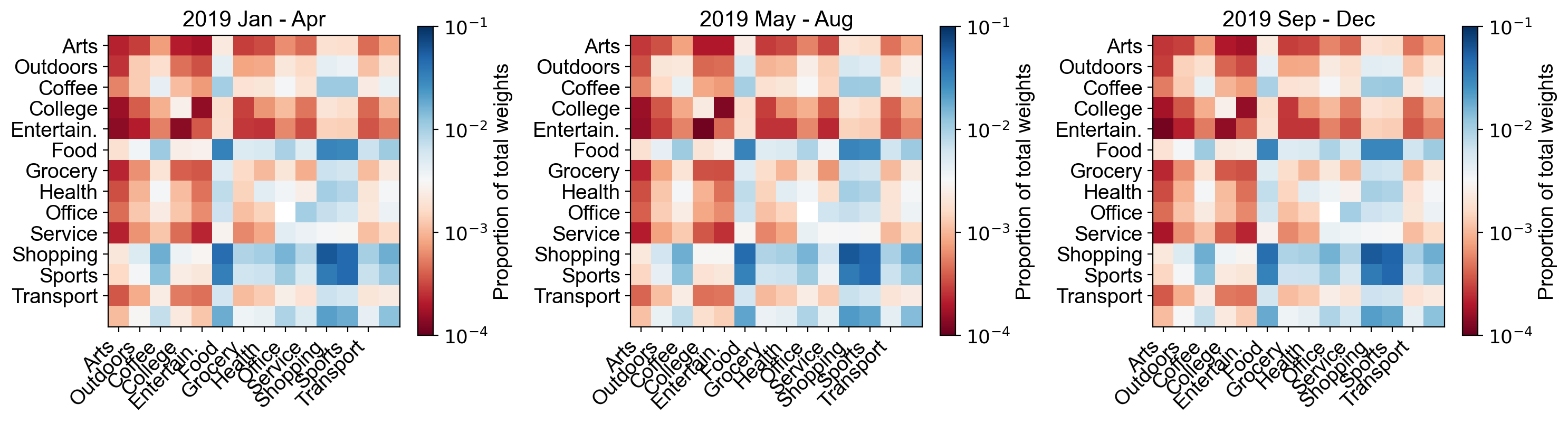}} 
\caption[Comparison of category pairwise weight proportion across different data collection periods]{\textbf{Comparison of category pairwise weight proportion across different data collection periods, for New York, Boston, and Seattle.} The category pairs that are highly dependent on eachother are consistent across different time periods (2019 May -- August and 2019 January -- April) with the baseline time period (2019 September -- December).}
\label{fig:s3matrix1}
\end{figure}

\begin{figure}[!t]
\centering
\subfloat[Boston]{\includegraphics[width=\linewidth]{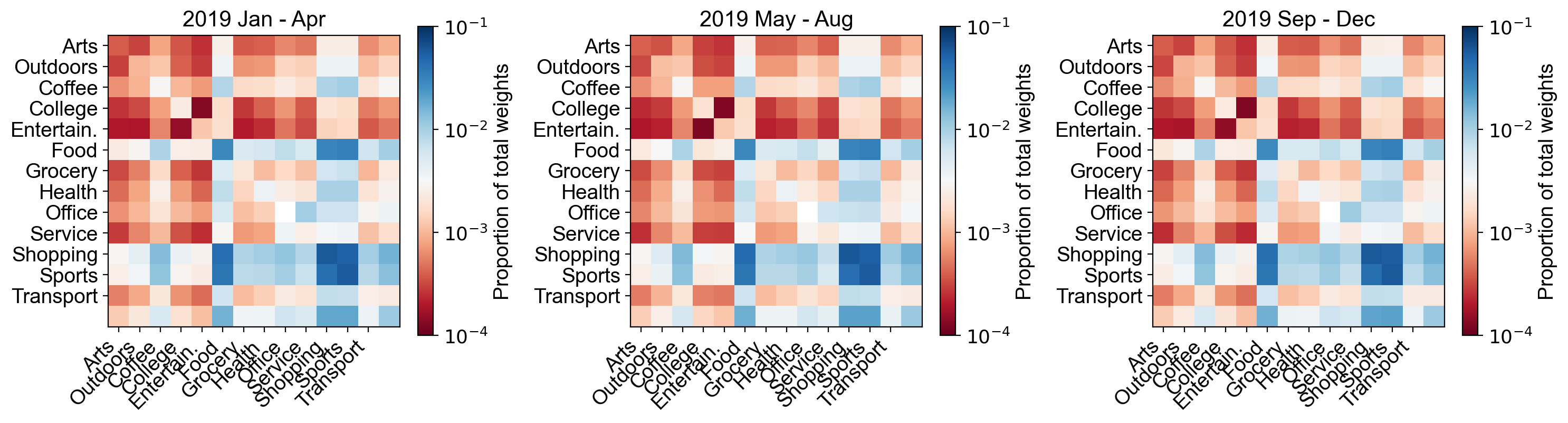}} \\
\subfloat[Boston]{\includegraphics[width=\linewidth]{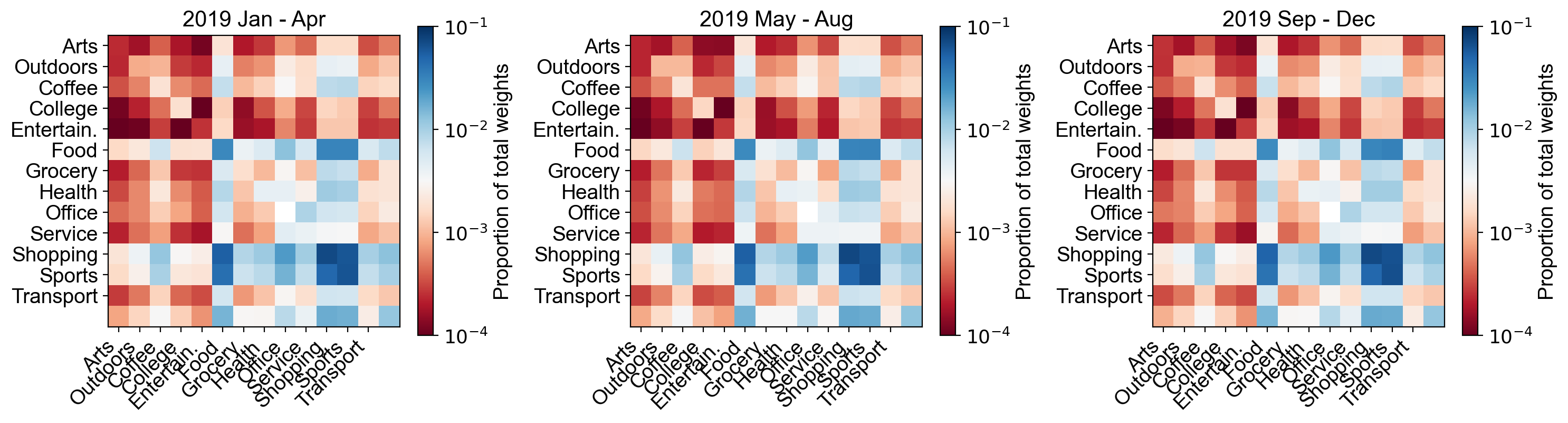}}  
\caption[Comparison of category pairwise weight proportion across different data collection periods]{\textbf{Comparison of category pairwise weight proportion across different data collection periods, for Los Angeles and Dallas.} The category pairs that are highly dependent on eachother are consistent across different time periods (2019 May -- August and 2019 January -- April) with the baseline time period (2019 September -- December).}
\label{fig:s3matrix2}
\end{figure}

The baseline dependency network was generated using data collected during the period of September to December, 2019. To assess whether the dependency networks are dynamic across time, we compared the in- and out-weights of each POI across different data collection periods. Figures \ref{fig:s3period1} and \ref{fig:s3period2} shows that the in- and out-weights across different time periods (2019 May -- August and 2019 January -- April) are highly correlated ($\rho>0.7$) with the baseline time period (2019 September -- December). Moreover, the category pairwise weight proportions were compared across different data collection periods, as shown in Figures \ref{fig:s3matrix1} and \ref{fig:s3matrix2}. The category pairs that are highly dependent on eachother are consistent across different time periods (2019 May -- August and 2019 January -- April) with the baseline time period (2019 September -- December).

\subsection{Computing quartiles of $w_{ij}$ via bootstrap method}

As shown in Figure 1 of the main manuscript and other figures in the Supplementary Material (e.g., Figures \ref{fig:s2nyc} - \ref{fig:s2dallas}), edge weights $w_{ij}$ are distributed across 4 orders of magnitude, and there exist edges with very small weights (i.e., $w_{ij}<0.001$). 
To test the robustness of the results presented in this paper to such extremely small edge weights, we estimate the confidence interval of each edge weight $w_{ij}$ using bootstrap resampling. 

To obtain bootstrapped samples of the dependency weights $w_{ij}$, the co-visits were sampled randomly with replacement from the original co-visit dataset. For each bootstrap sample the bootstrap replication of the dependency weights $w^b_{ij}$ where $b=1,2,3,...,30$ were computed. 
The 25th percentile of the distribution of the weights, denoted by $w^{p25}_{ij}$ were computed from the sample distribution, and only edges with $w^{p25}_{ij}>0$ were used in the 'bootstrap network'.

\begin{figure}[!t]
\centering
\subfloat[Out-weights of POIs]{\includegraphics[width=\linewidth]{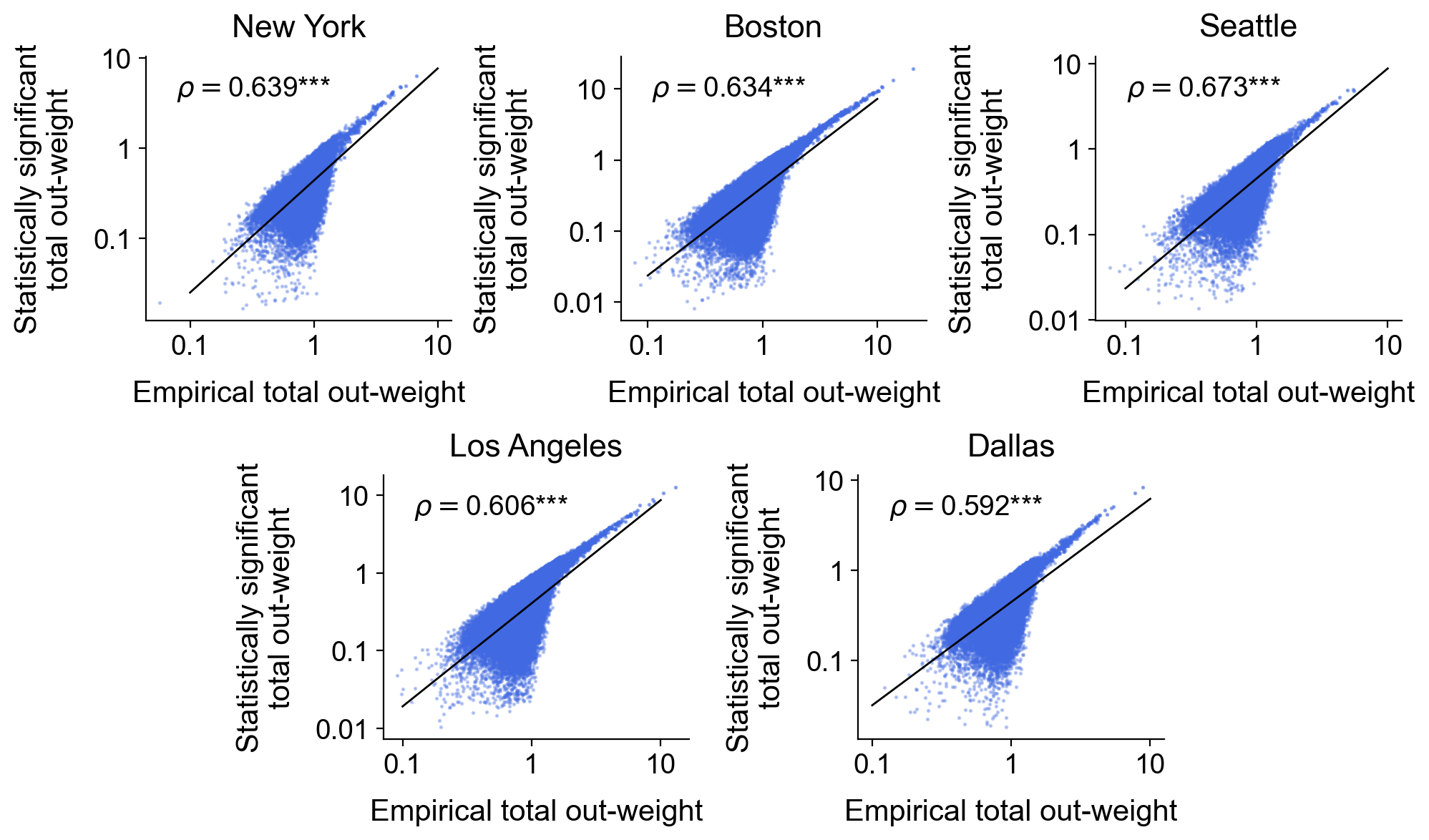}} \\  
\subfloat[In-weights of POIs]{\includegraphics[width=\linewidth]{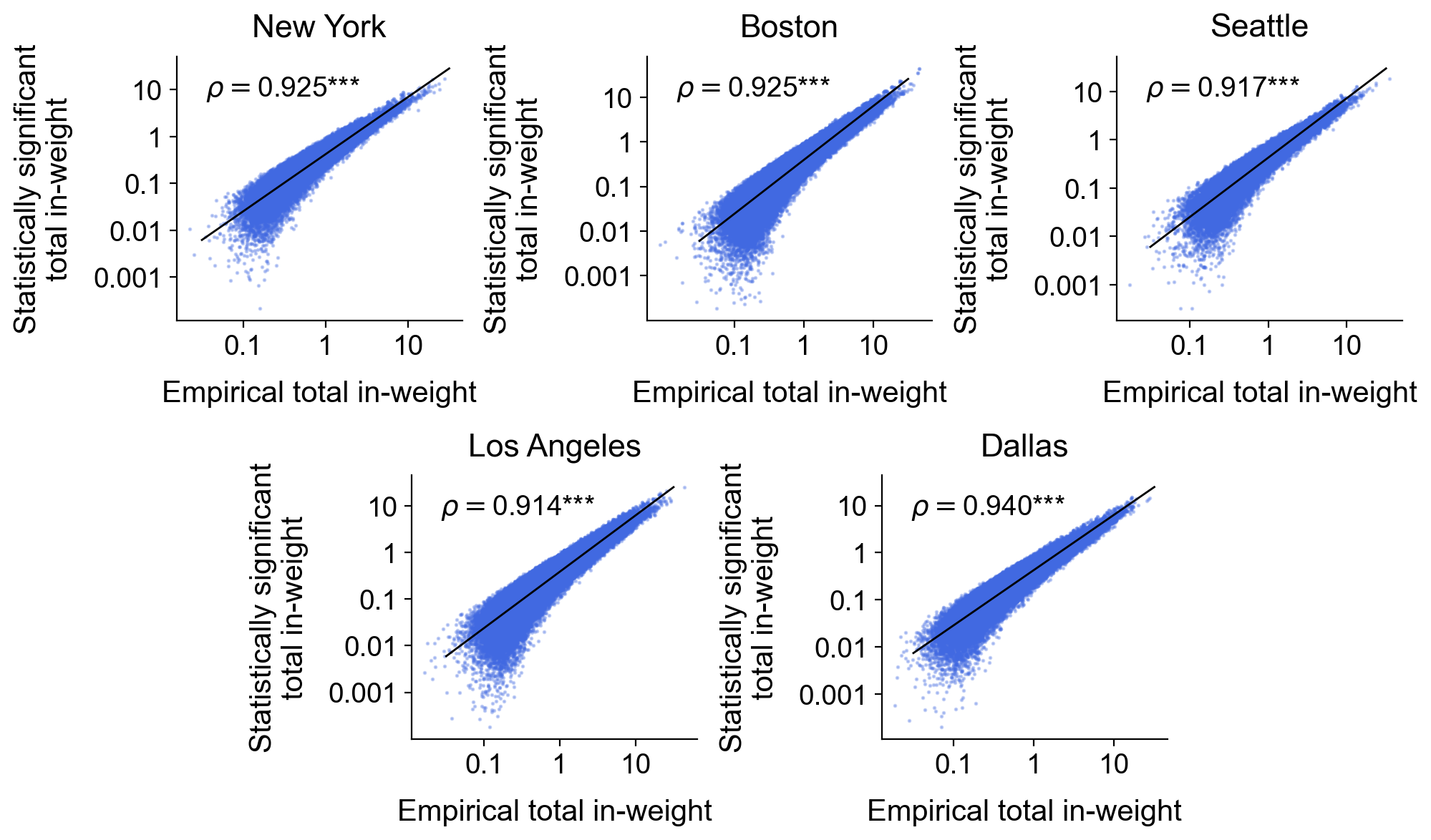}}
\caption[Out- and in-weights of POIs when applying bootstrap method to remove smaller weights]{\textbf{Out- and in-weights of POIs when applying bootstrap method to remove smaller weights.} The correlation is generally moderate ($\rho = 0.6$) for out-weight and high ($\rho = 0.9$) for in-weights.}
\label{fig:s3bstweight}
\end{figure}

\begin{figure}[!t]
\centering
\subfloat[Out-weights of categories]{\includegraphics[width=\linewidth]{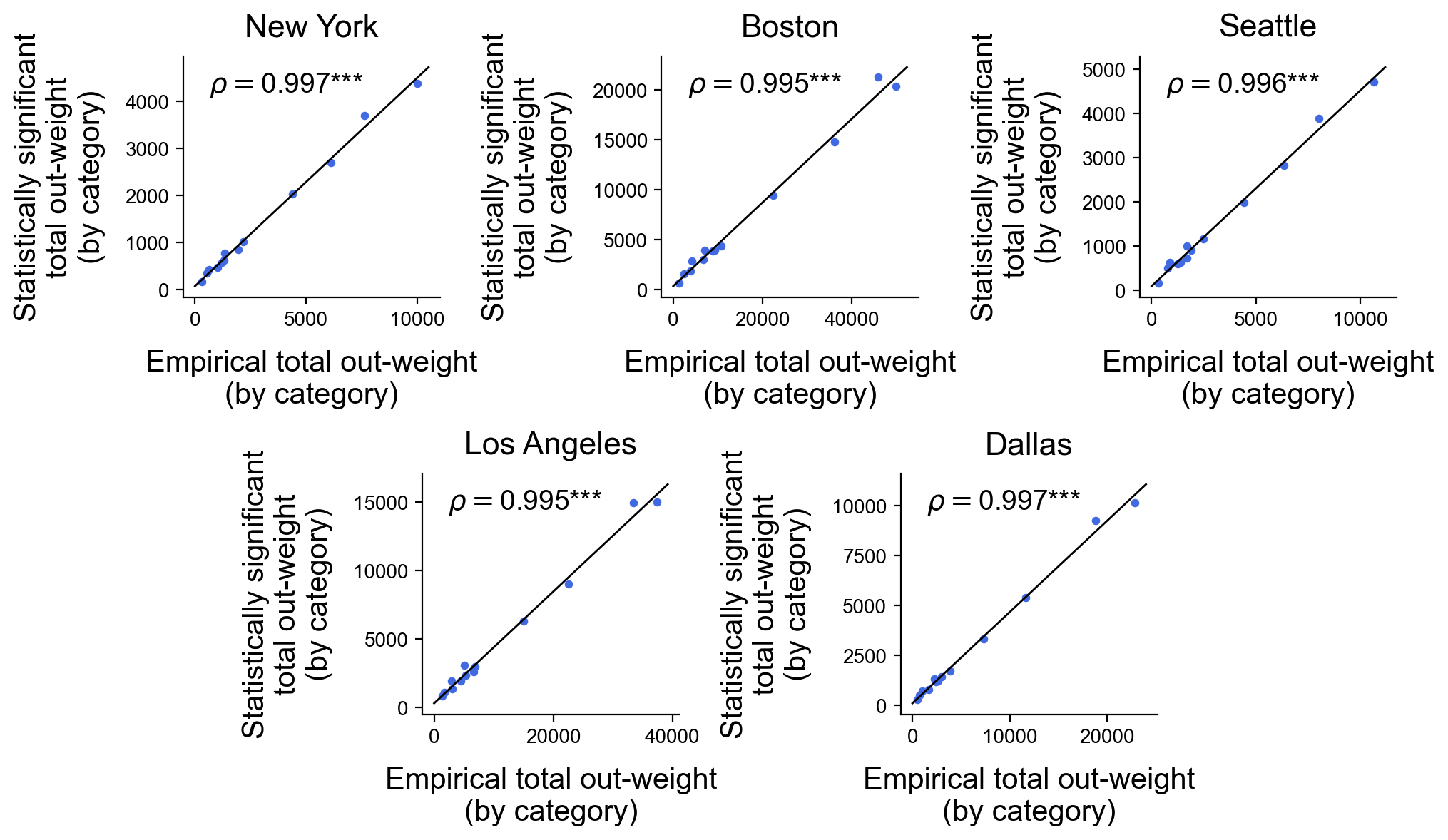}} \\  
\subfloat[In-weights of categories]{\includegraphics[width=\linewidth]{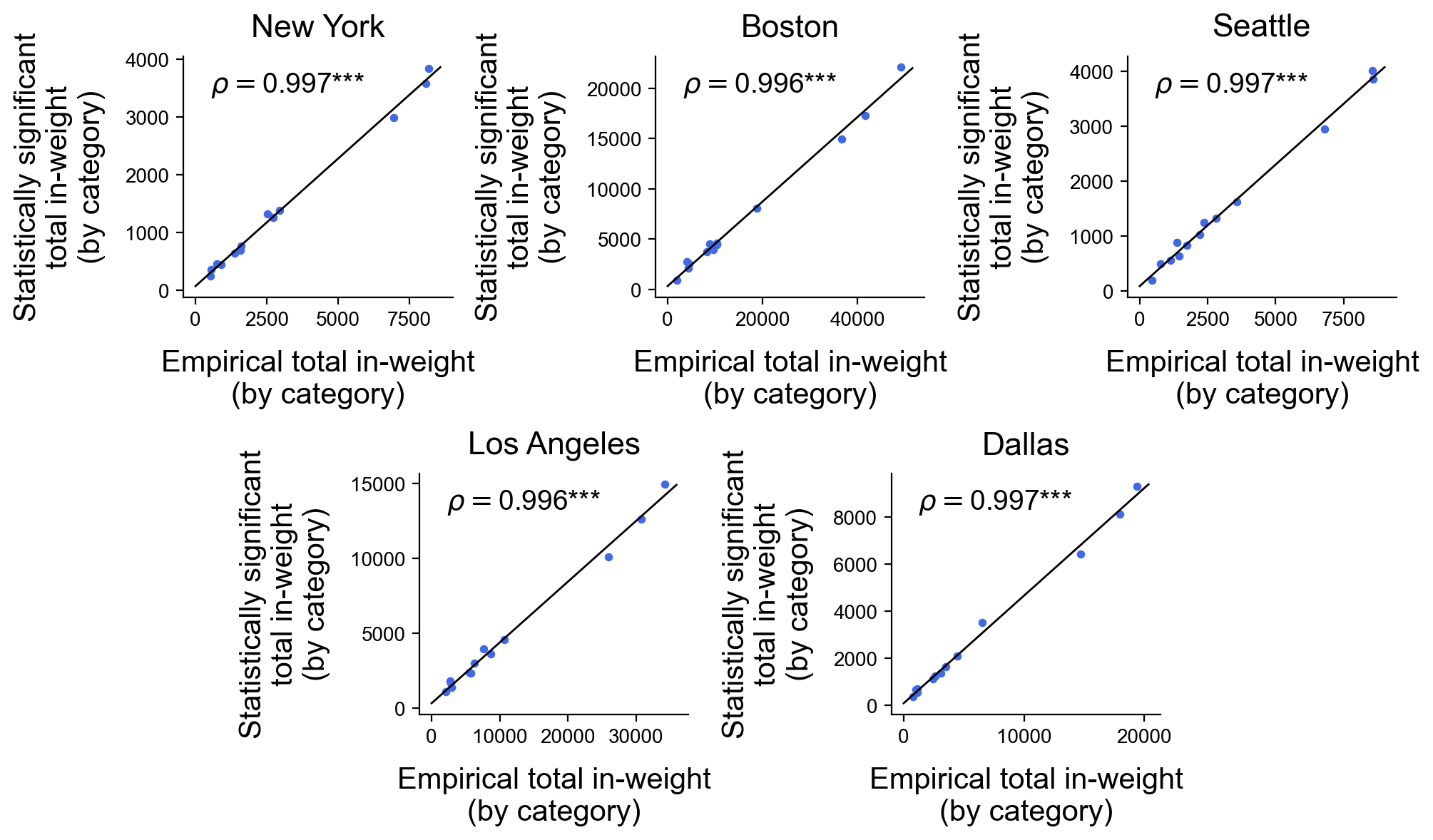}}
\caption[Out- and in-weights of categories when applying bootstrap method to remove insignificant weights]{\textbf{Out- and in-weights of categories when applying bootstrap method to remove insignificant weights.} The correlation is extremely high ($\rho > 0.99$) for both out- and in-weights when aggregated to POI categories.}
\label{fig:s3bstcat}
\end{figure}

Figure \ref{fig:s3bstweight} shows the in- and out- weights of POIs before and after applying the bootstrap method to select statistically significant edges. The correlation is generally moderate ($\rho = 0.6$) for out-weight and high ($\rho = 0.9$) for in-weights. The variance becomes larger especially for the nodes with out-weights below 1. 
Figure \ref{fig:s3bstcat} shows the out- and in-weights of categories when applying bootstrap method to remove insignificant weights. The correlation is extremely high ($\rho > 0.99$) for both out- and in-weights when aggregated to POI categories, indicating that bootstrapping does not have a significant effect when the analysis is aggregated to the category level.

\subsection{Robustness against post-stratification processing}

\begin{figure}[!t]
\centering
\subfloat[Out-weights of POIs]{\includegraphics[width=\linewidth]{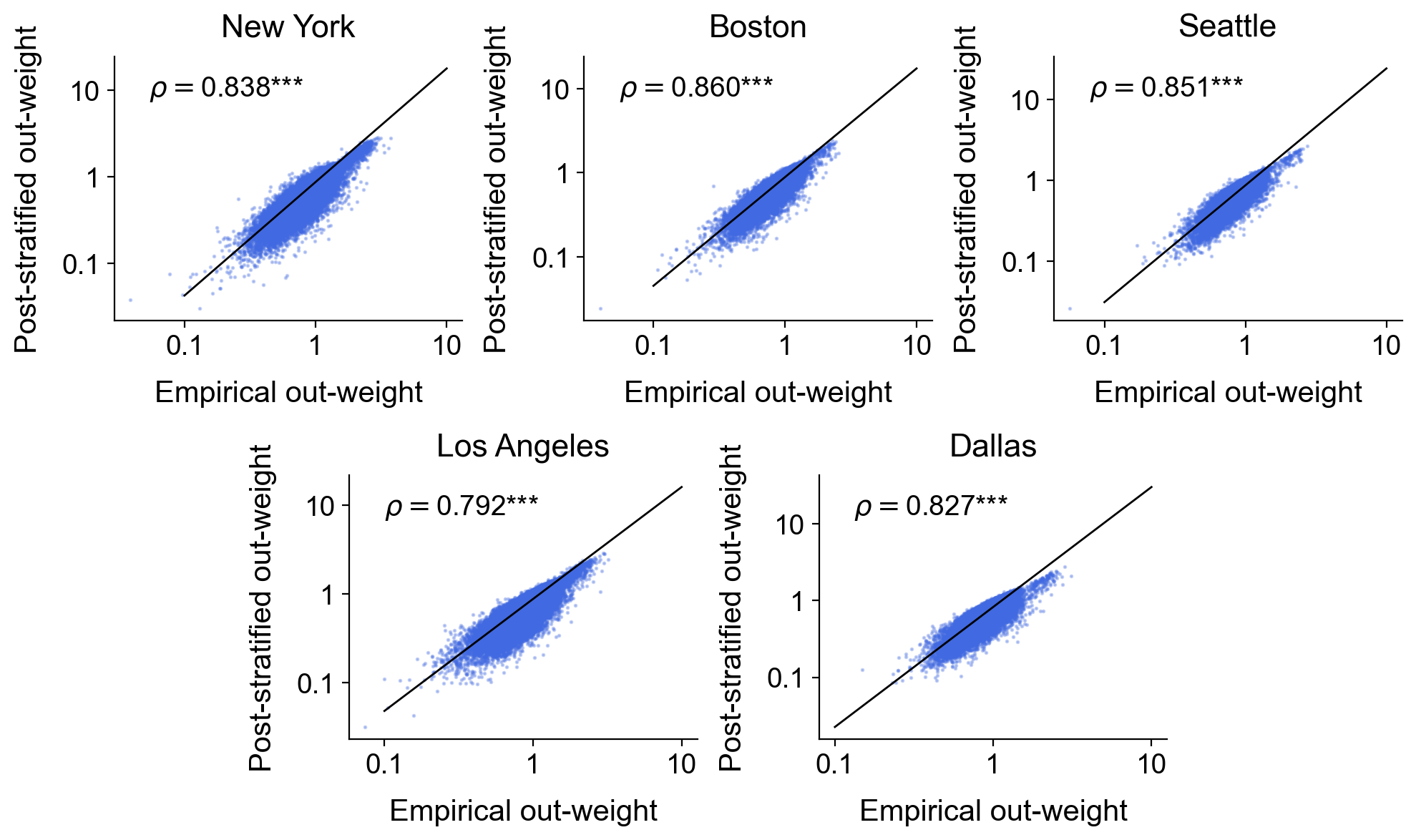}} \\
\subfloat[In-weights of POIs]{\includegraphics[width=\linewidth]{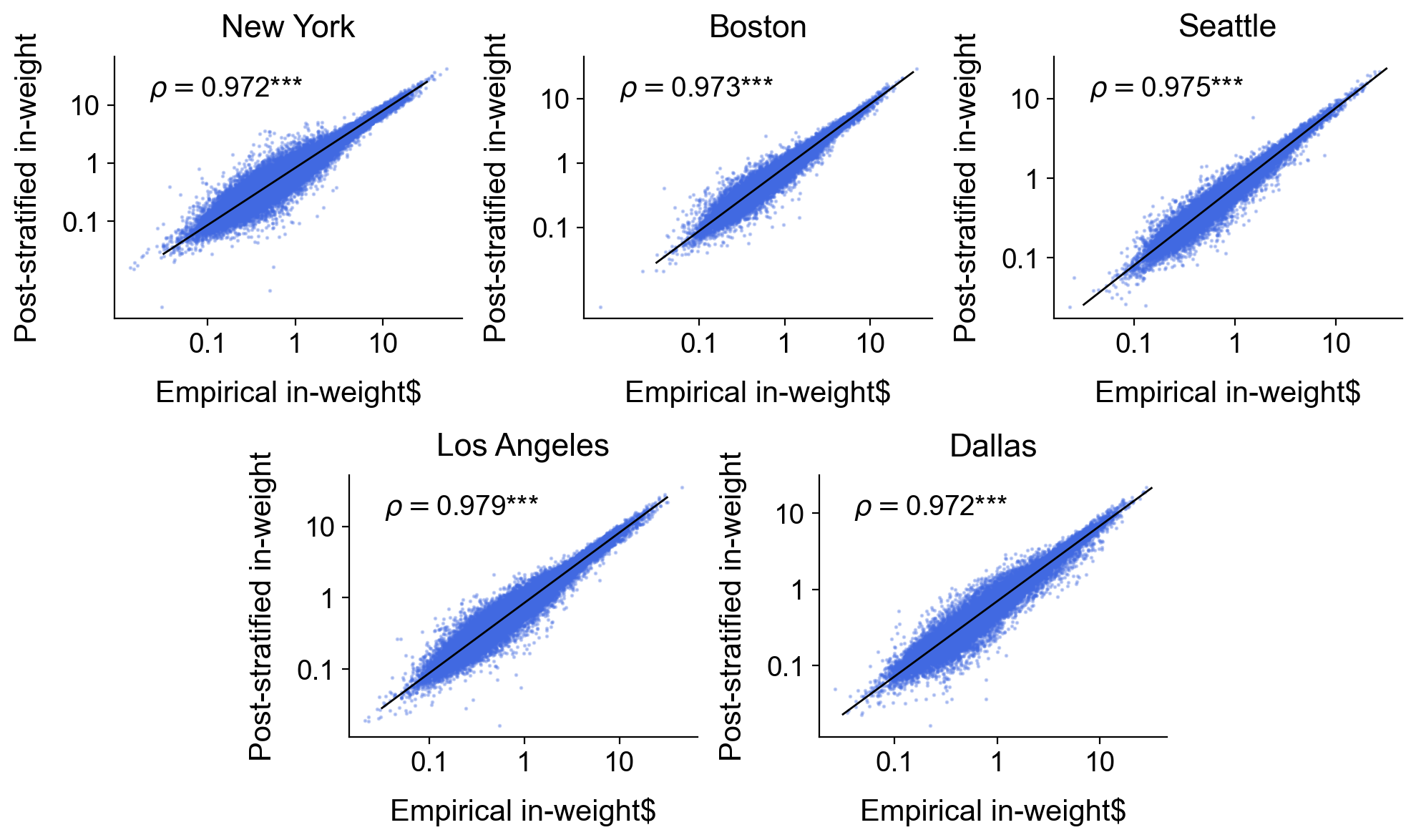}} \\
\caption[Out- and in-weights of POIs when applying post-stratification to correct for biases in mobile phone data]{\textbf{Out- and in-weights of POIs when applying post-stratification to correct for biases in mobile phone data.} The correlation is high ($\rho > 0.8$) for both out- and in-weights even though there was significant bias in sample rates across CBGs and income groups.}
\label{fig:s3poststrat}
\end{figure}

In Supplementary Note 1, analyses showed significant biases in sampling rates of mobile phone users across census block groups and income ranges. To correct for such bias, we conducted post stratification method (Supplementary Note 1.4) and used the corrected data to estimate and analyze the dependency networks. 
Here, we measure how different the estimates of out- and in-weights of POIs would be if we did not apply the post-stratification technique to correct for biases in mobile phone data. In Figure \ref{fig:s3poststrat}, we observe that the correlation is high ($\rho > 0.8$) for both out- and in-weights even though there was significant bias in sample rates across CBGs and income groups.

\clearpage
\section{Modeling dependency network weights}

\subsection{Gravity-based null networks}
\begin{figure}[ht]
\centering
\includegraphics[width=\linewidth]{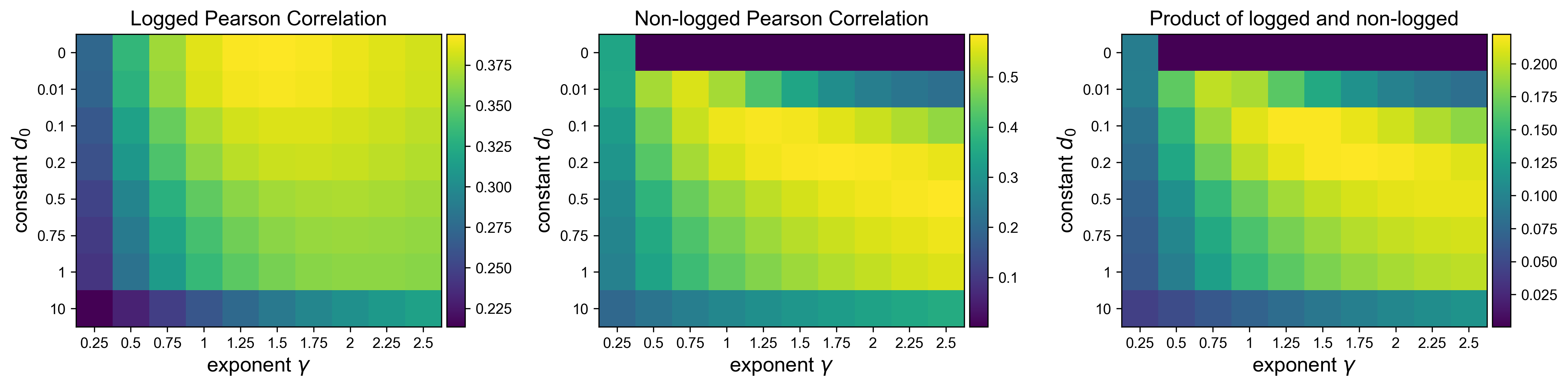}
\caption[Correlation between the observed $n_{ij}$ values and the gravity component $g_{ij}$ under different model parameters]{\textbf{Correlation between the observed $n_{ij}$ values and the gravity component $g_{ij}$ under different model parameters} }
\label{fig:s4wijgijcorr}
\end{figure}

To further understand the structural properties of the dependency networks, we generated null networks that satisfy the following properties:
\begin{itemize}
    \item the weight $w_{ij}$ decays with physical distance
    \item the in-weight $w_{ij}$ is larger for nodes with larger visitation $n_i$
\end{itemize}
To generate such null networks that preserve such qualities, we utilized the generalized gravity law as the theoretical starting point, where we can model the number of common visitors as the gravity component $g_{ij}$, which is the product of the total number of visitors to places $i$ and $j$, $n_i$ and $n_j$, divided by a function of the distance between $i$ and $j$, $d_{ij}$, which takes the form $f(d_{ij}) = (d_0 + d_{ij})^\gamma$:  
\begin{equation}
    g_{ij} = n_i n_j / (d_0 + d_{ij})^\gamma    
\end{equation}
where $n_i$ and $n_j$ are the total number of visits to POIs $i$ and $j$, $d_{ij}$ is the physical distance between POIs $i$ and $j$, $d_0$ is the distance cutoff parameter, and $\gamma$ is the exponent parameter of the gravity model. Parameters $d_0$ and $\gamma$ were fitted empirically to maximize the correlation between $g_{ij}$ and $n_{ij}$, which is the total number of common visitors between POIs $i$ and $j$, and as shown in Figure \ref{fig:s4wijgijcorr}, $d_0=0.2$ and $\gamma=1.5$ showed the highest correlation when the values were logged and not logged. 

%%% figures of correlation between gij and nij under different parameters 

Using the fitted gravity model $w_{ij} = n_{ij}/n_i \sim n_j / (d_0 + d_{ij})^\gamma$, we now construct null networks that mimic the physical properties of the actual dependency networks. 
To enable a fair comparison between the actual dependency network and the null network, we constructed the null networks while fulfilling the following two properties:
\begin{enumerate}
    \item the in-degree of each node are kept consistent in the null network
    \item the in-weight of each node are kept consistent in the null network 
\end{enumerate}
To achieve this, first we construct a linear relationship between the gravity component $g_{ij} = n_j / (d_0 + d_{ij})^\gamma$ and $w_{ij}$, $g_{ij} = \phi(w_{ij})$. 
Then, for each edge in the actual dependency network connecting $i$ and $j$ with a dependency weight $w_{ij}$, we compute its theoretical gravity component using the linear relationship $\phi$ and select an alternative node with the same level of corresponding gravity component from its 10,000 closest nodes, and is assigned the same weight $w_{ij}$. 
This algorithm enables us to construct a null network where we 1) maintain the linear relationship between $w_{ij}$ and $g_{ij}$, 2) the same number of in-edges are selected for each node, and 3) the total in-weight for each node is kept consistent. 

\begin{figure}
\centering
\subfloat[New York]{\includegraphics[width=\linewidth]{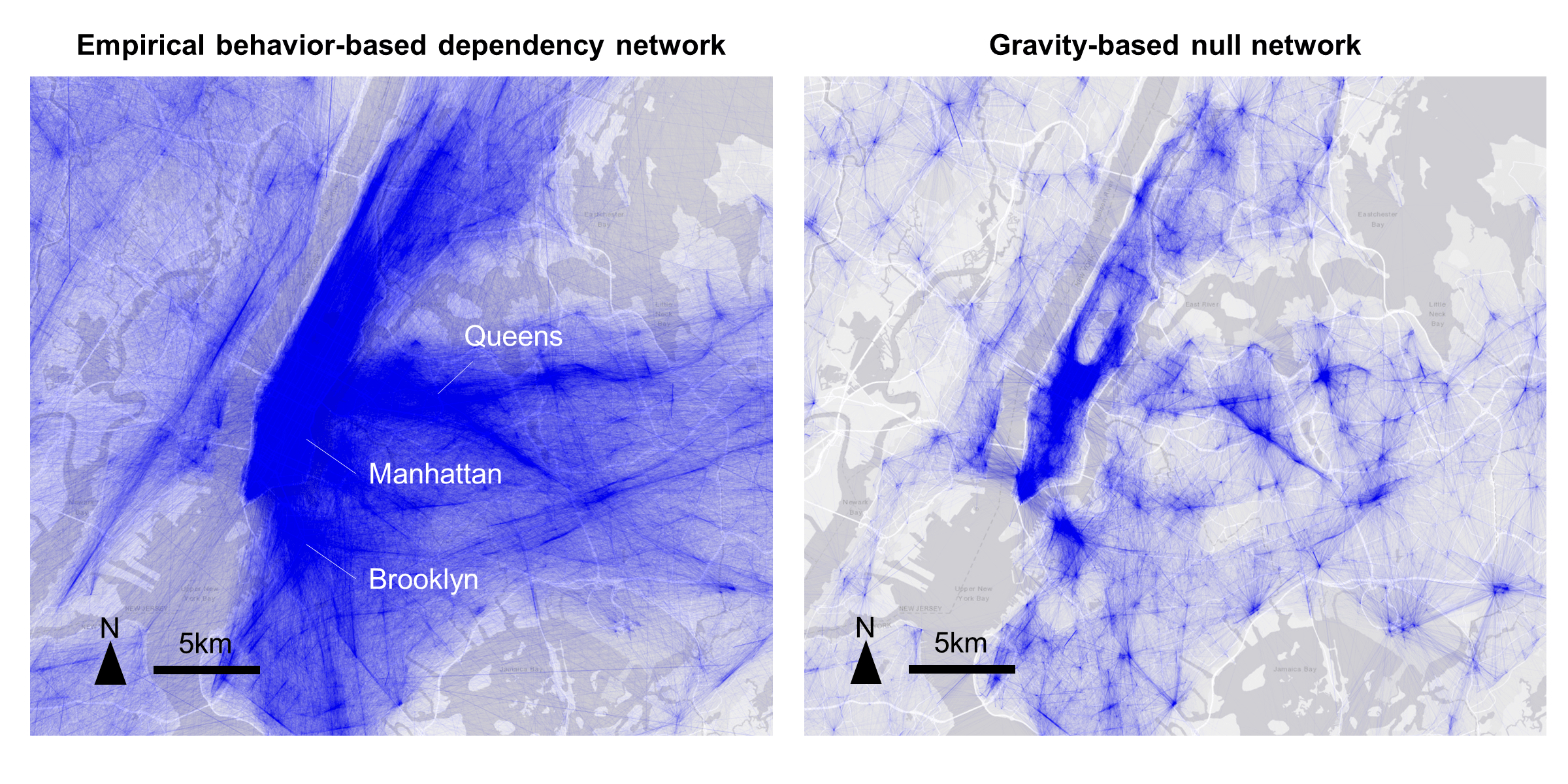}} \\
\subfloat[Boston]{\includegraphics[width=\linewidth]{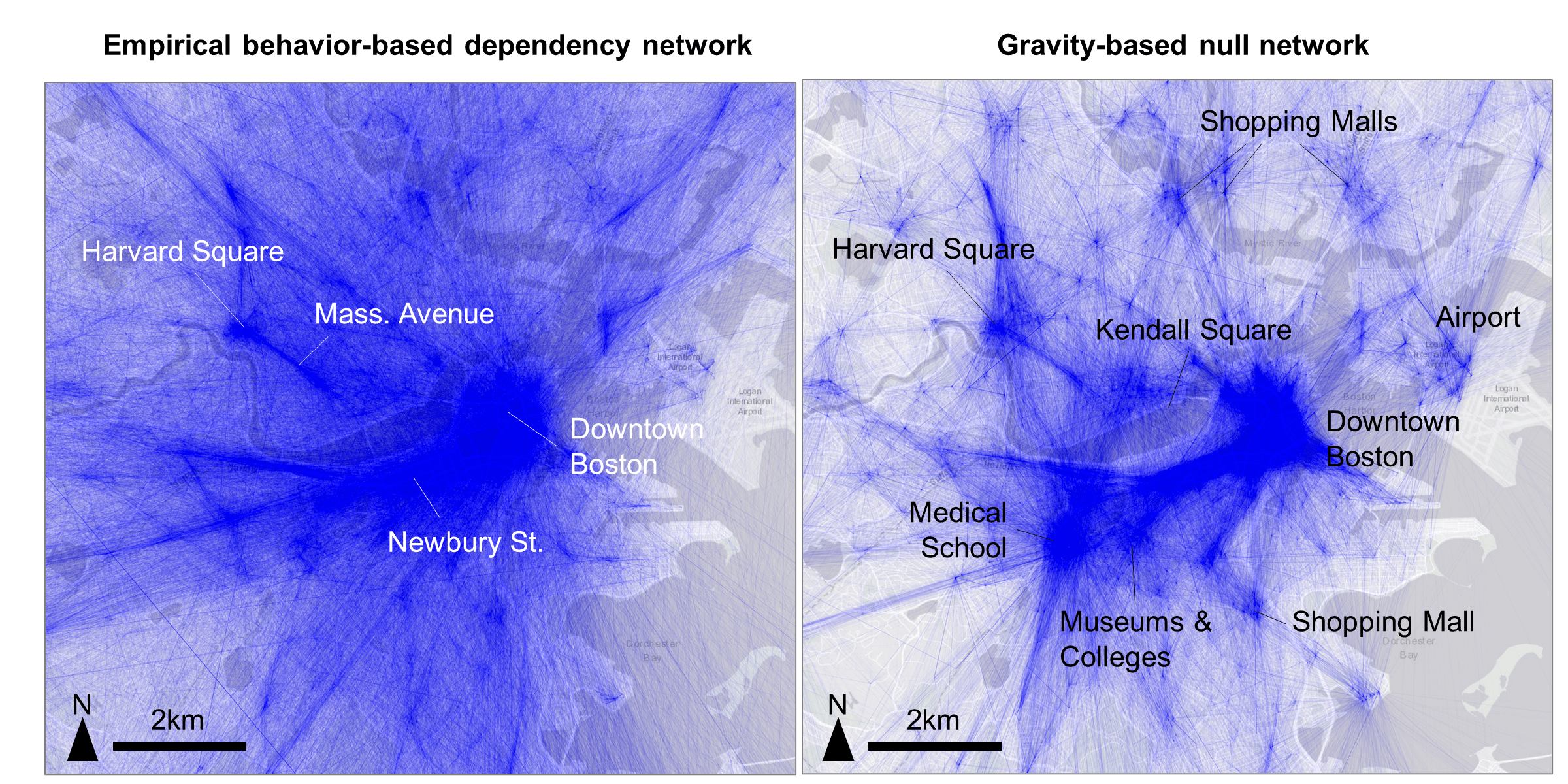}} \\
\caption[Comparison of actual and null networks in New York and Boston]{\textbf{Comparison of actual and null networks in New York and Boston.} For visual purposes, around 1 million links are shown for both the actual and null networks.}
\label{fig:s4nullmaps1}
\end{figure}

\begin{figure}
\centering
\subfloat[New York]{\includegraphics[width=\linewidth]{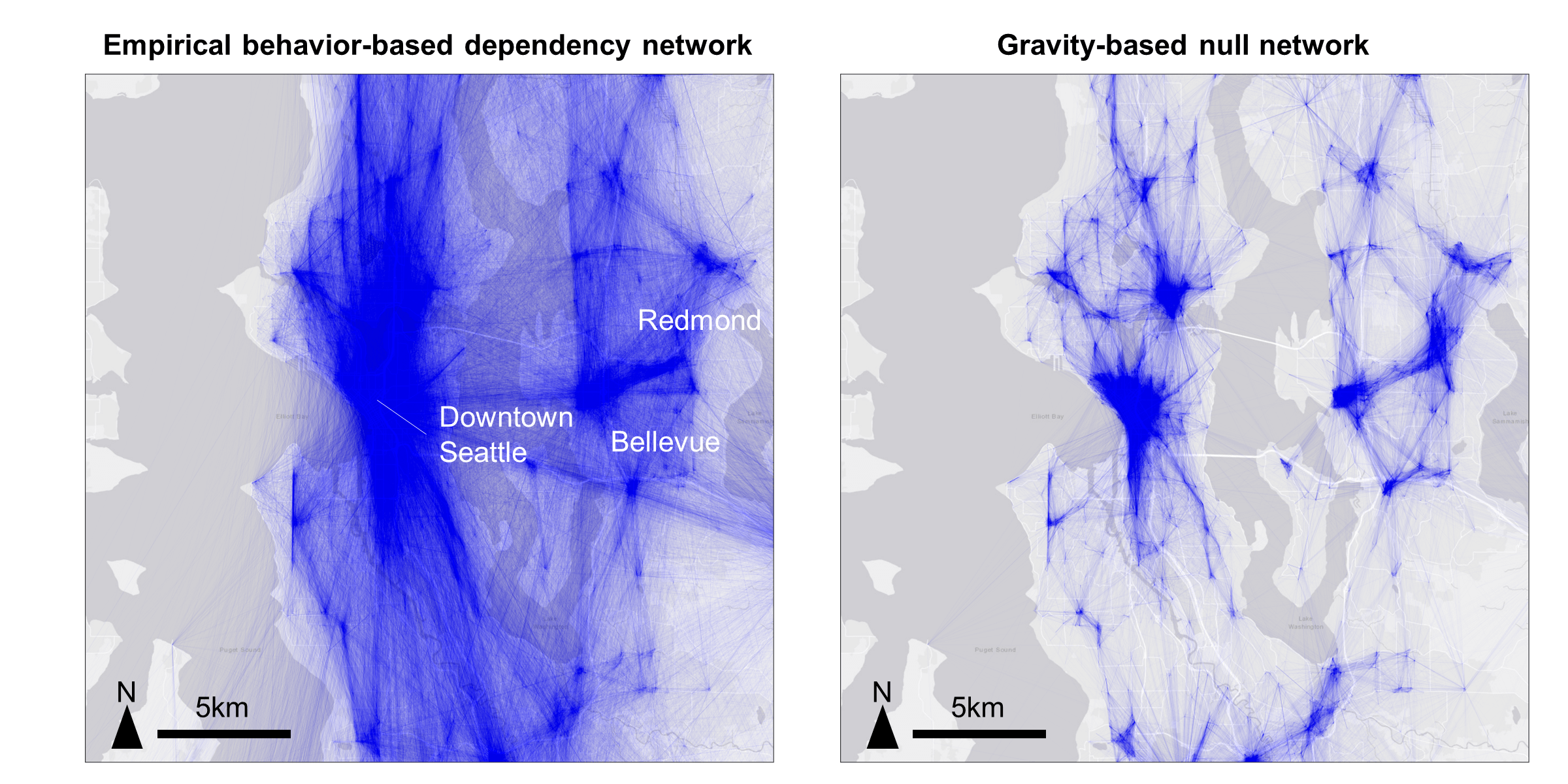}} \\
\subfloat[Boston]{\includegraphics[width=\linewidth]{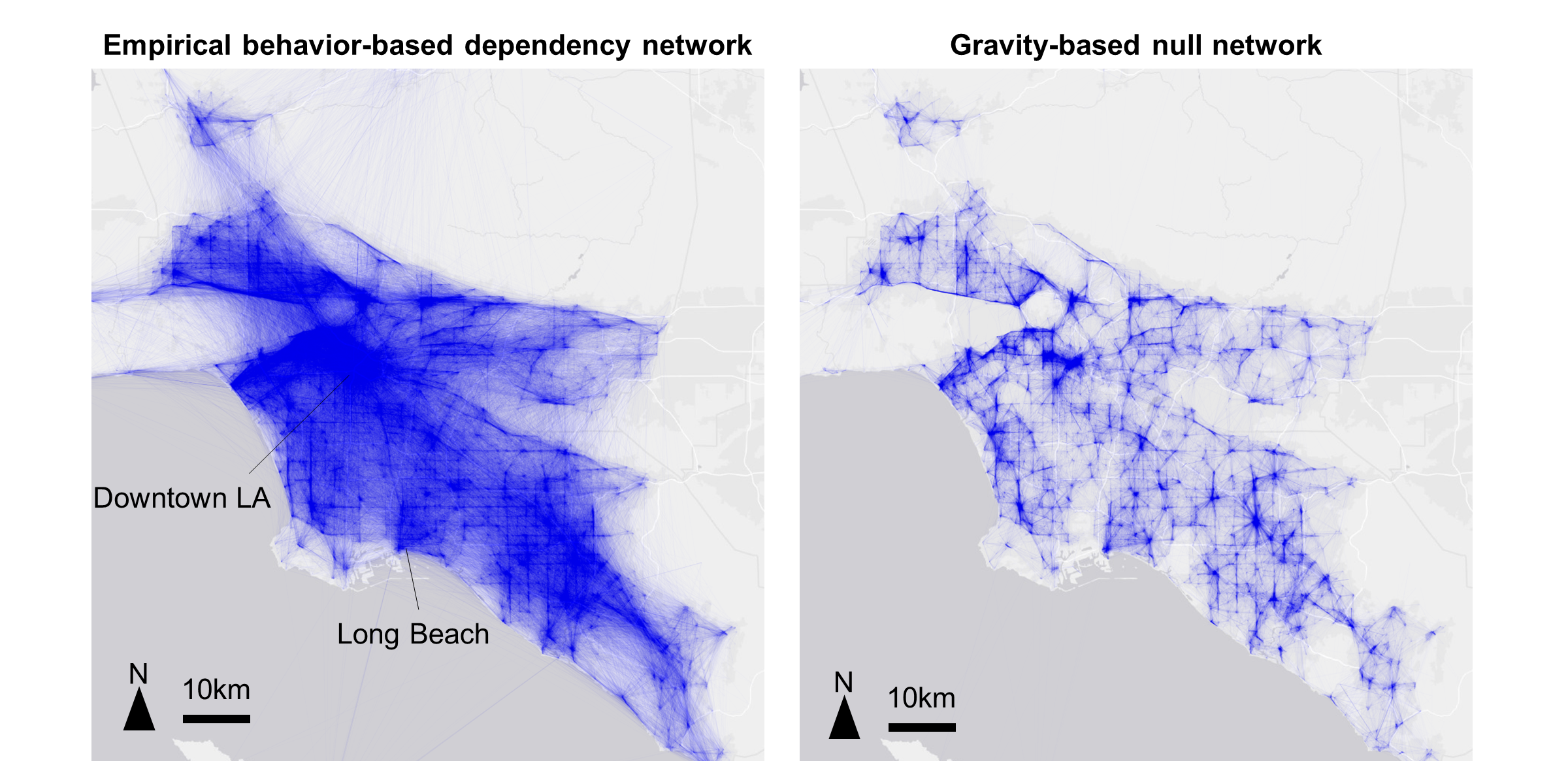}} \\
\caption[Comparison of actual and null networks in Seattle and Los Angeles]{\textbf{Comparison of actual and null networks in Seattle and Los Angeles.} For visual purposes, around 1 million links are shown for both the actual and null networks.}
\label{fig:s4nullmaps2}
\end{figure}

\begin{figure}
\centering
\subfloat[Dallas]{\includegraphics[width=\linewidth]{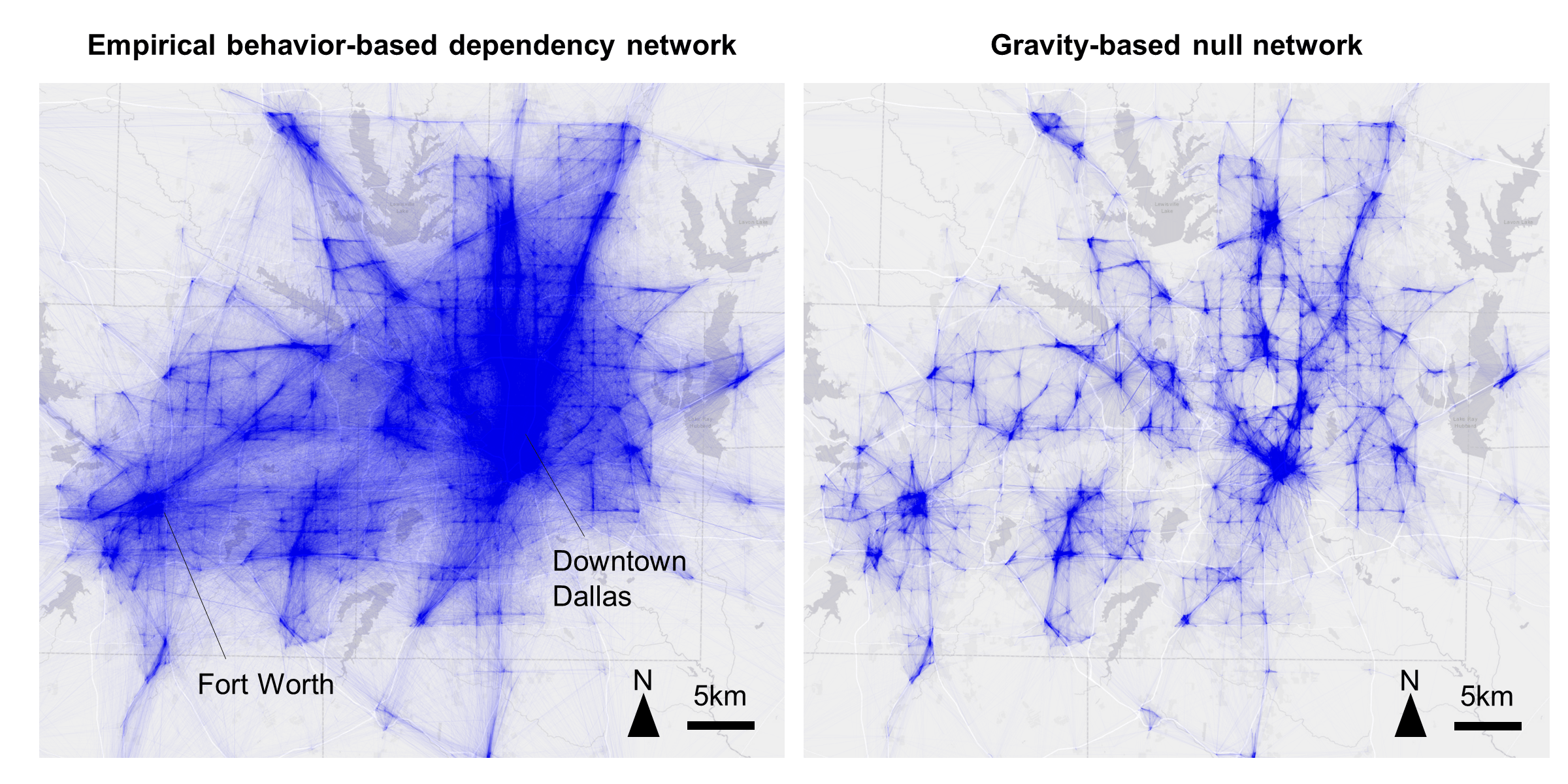}} 
\caption[Comparison of actual and null networks in Dallas]{\textbf{Comparison of actual and null networks in Dallas.} For visual purposes, around 1 million links are shown for both the actual and null networks.}
\label{fig:s4nullmaps3}
\end{figure}

\begin{figure}
\centering
\subfloat[New York]{\includegraphics[width=0.3\linewidth]{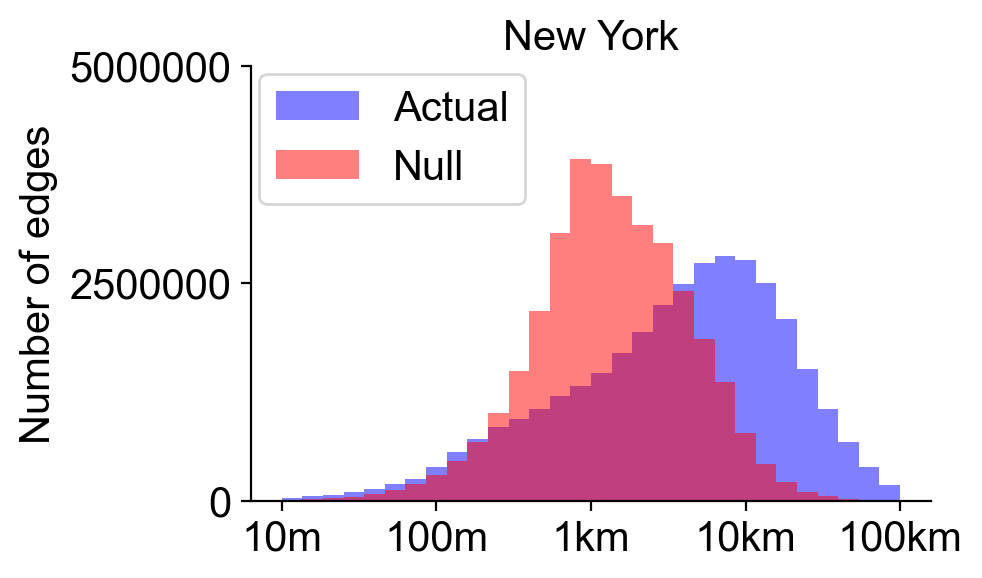}} 
\hspace{0.03\textwidth}
\subfloat[Boston]{\includegraphics[width=0.3\linewidth]{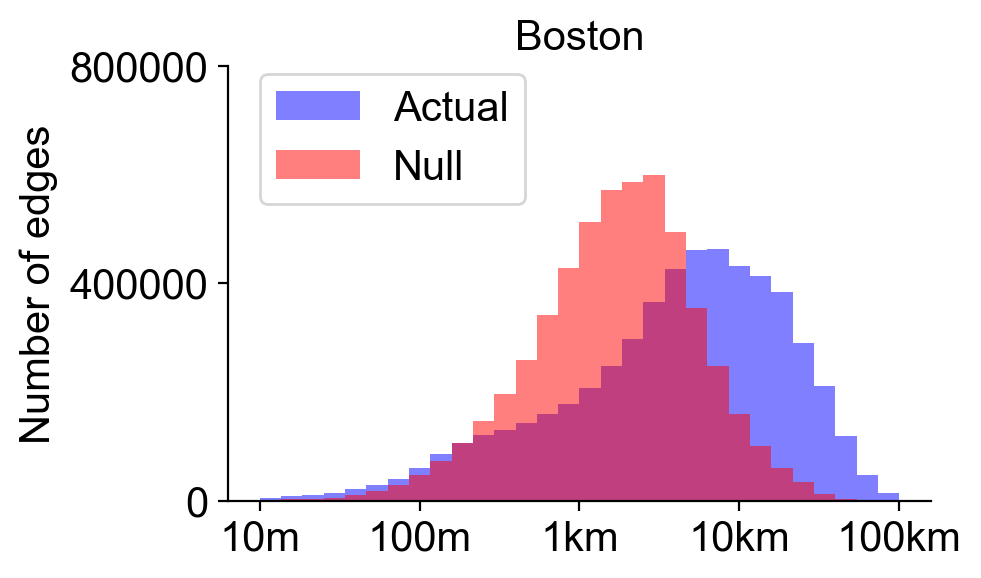}} 
\hspace{0.03\textwidth} 
\subfloat[Seattle]{\includegraphics[width=0.3\linewidth]{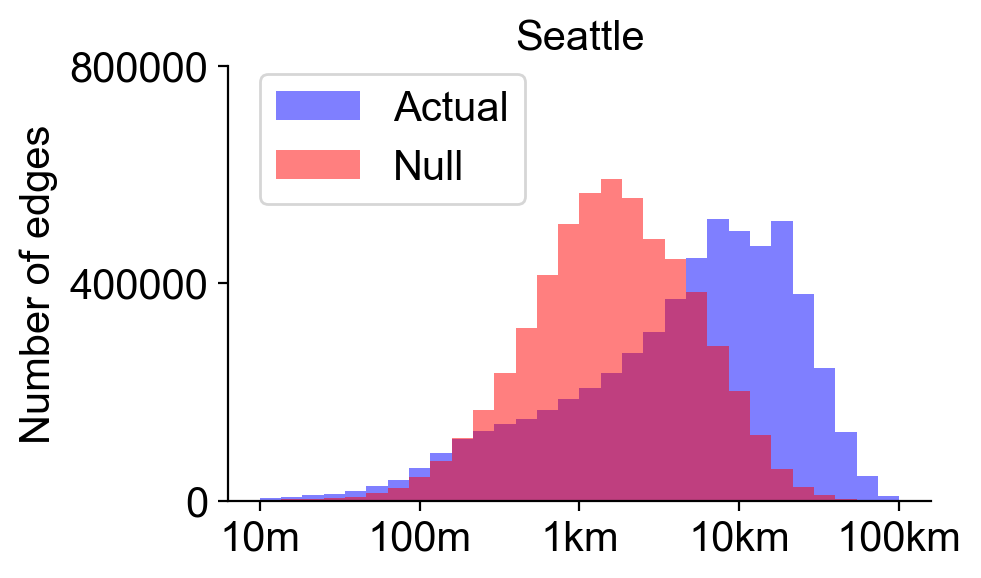}} \\ 
\subfloat[Los Angeles]{\includegraphics[width=0.3\linewidth]{nydistance_distribution_null.png}} \hspace{0.03\textwidth} 
\subfloat[Dallas]{\includegraphics[width=0.3\linewidth]{nydistance_distribution_null.png}}
\caption[Comparison between real and null network on the proportion of edge weights among POI category pairs]{\textbf{Comparison between real and null network on the proportion of edge weights among POI category pairs.} The proportion of edge weights was computed by taking the sum of weights that connect the vertical and horizontal categories and dividing that by the total weights that exist in the network. The patterns of dependency are consistent across cities, with Service and Transport node depending more on other nodes in the real network compared to the null network.}
\label{fig:s4nulldistancedist}
\end{figure}

\begin{figure}
\centering
\subfloat[New York]{\includegraphics[width=0.4\linewidth]{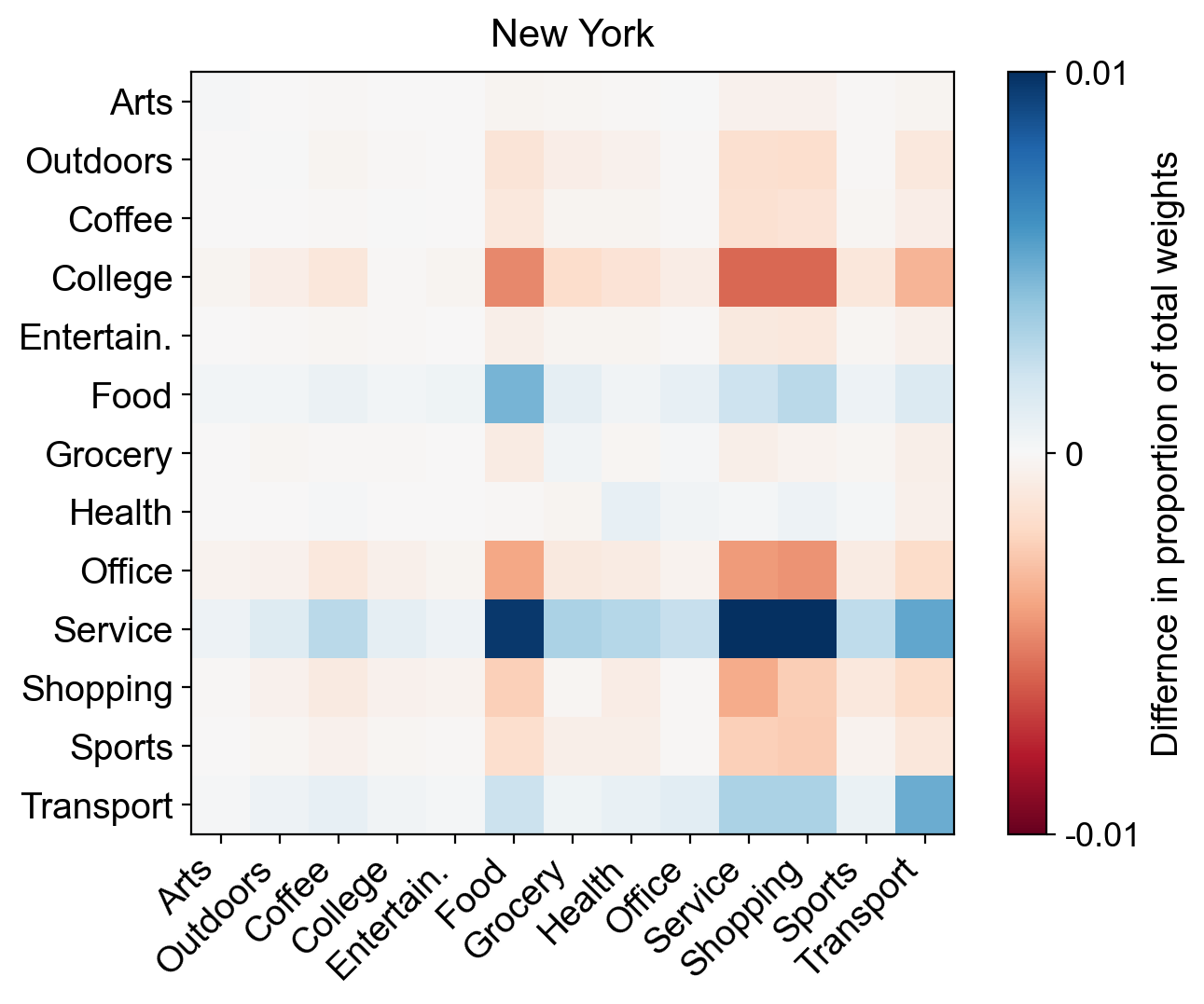}} 
\hspace{0.1\textwidth}
\subfloat[Boston]{\includegraphics[width=0.4\linewidth]{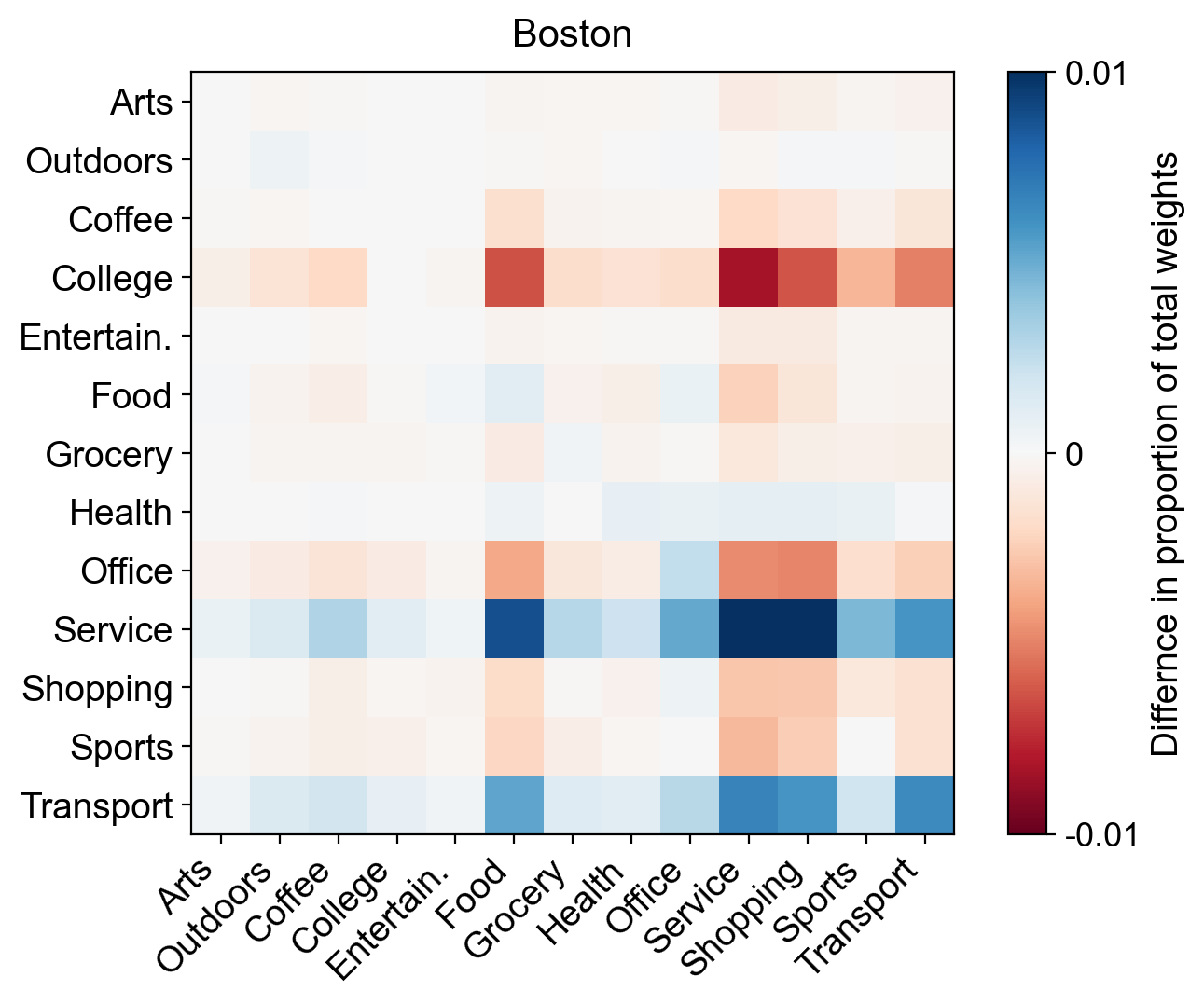}} \\
\subfloat[Seattle]{\includegraphics[width=0.4\linewidth]{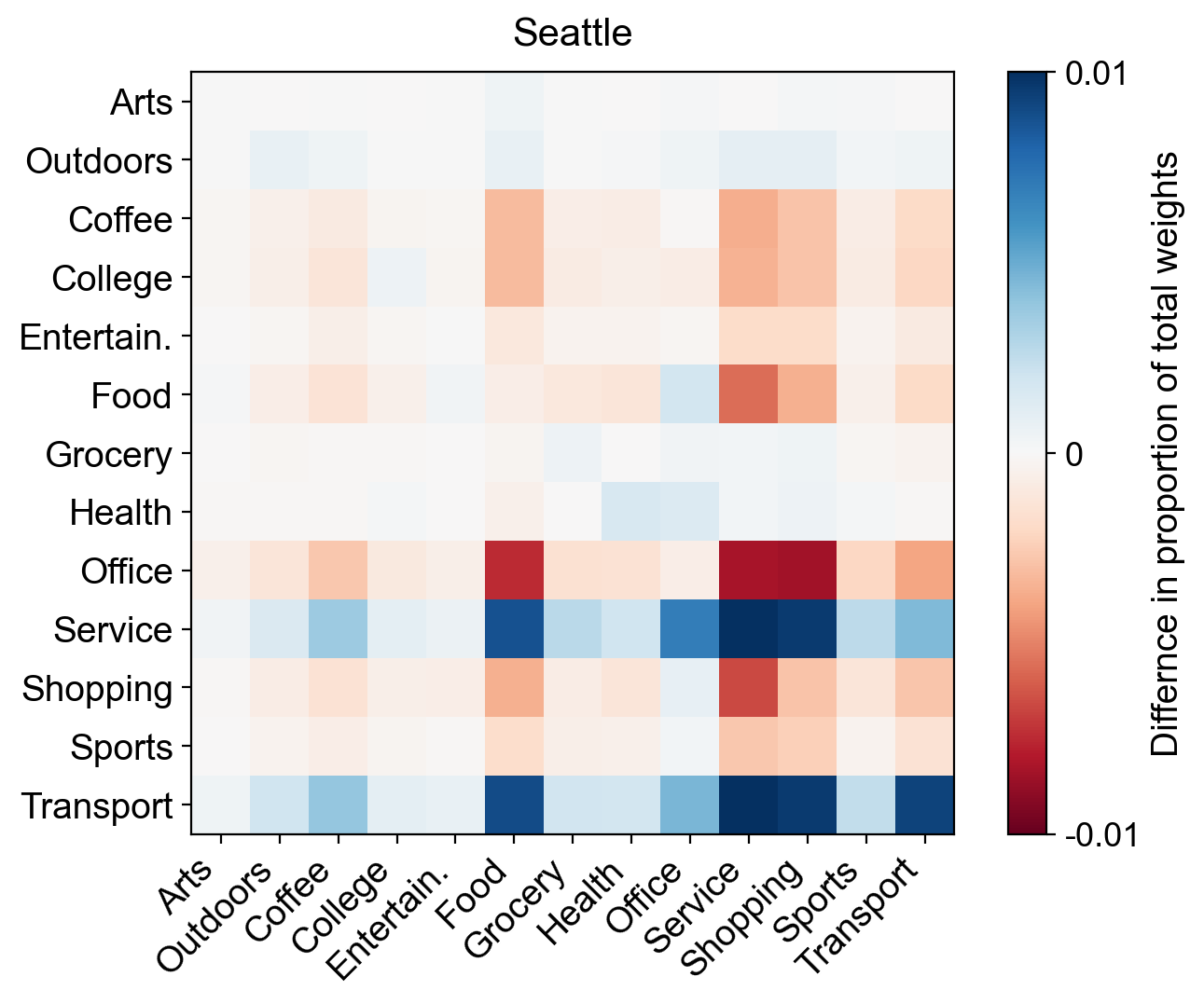}}
\hspace{0.1\textwidth}
\subfloat[Los Angeles]{\includegraphics[width=0.4\linewidth]{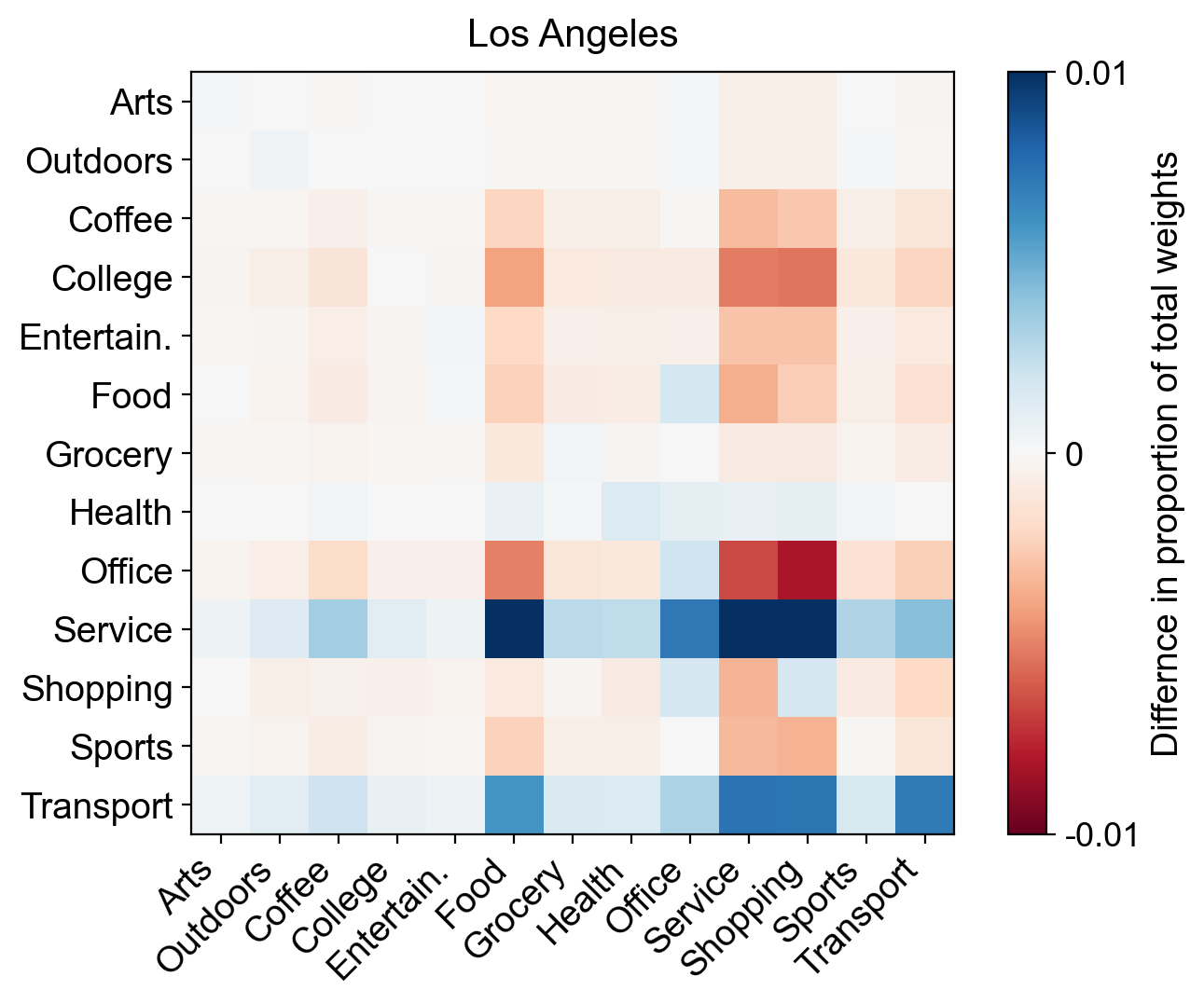}} \\
\subfloat[Dallas]{\includegraphics[width=0.4\linewidth]{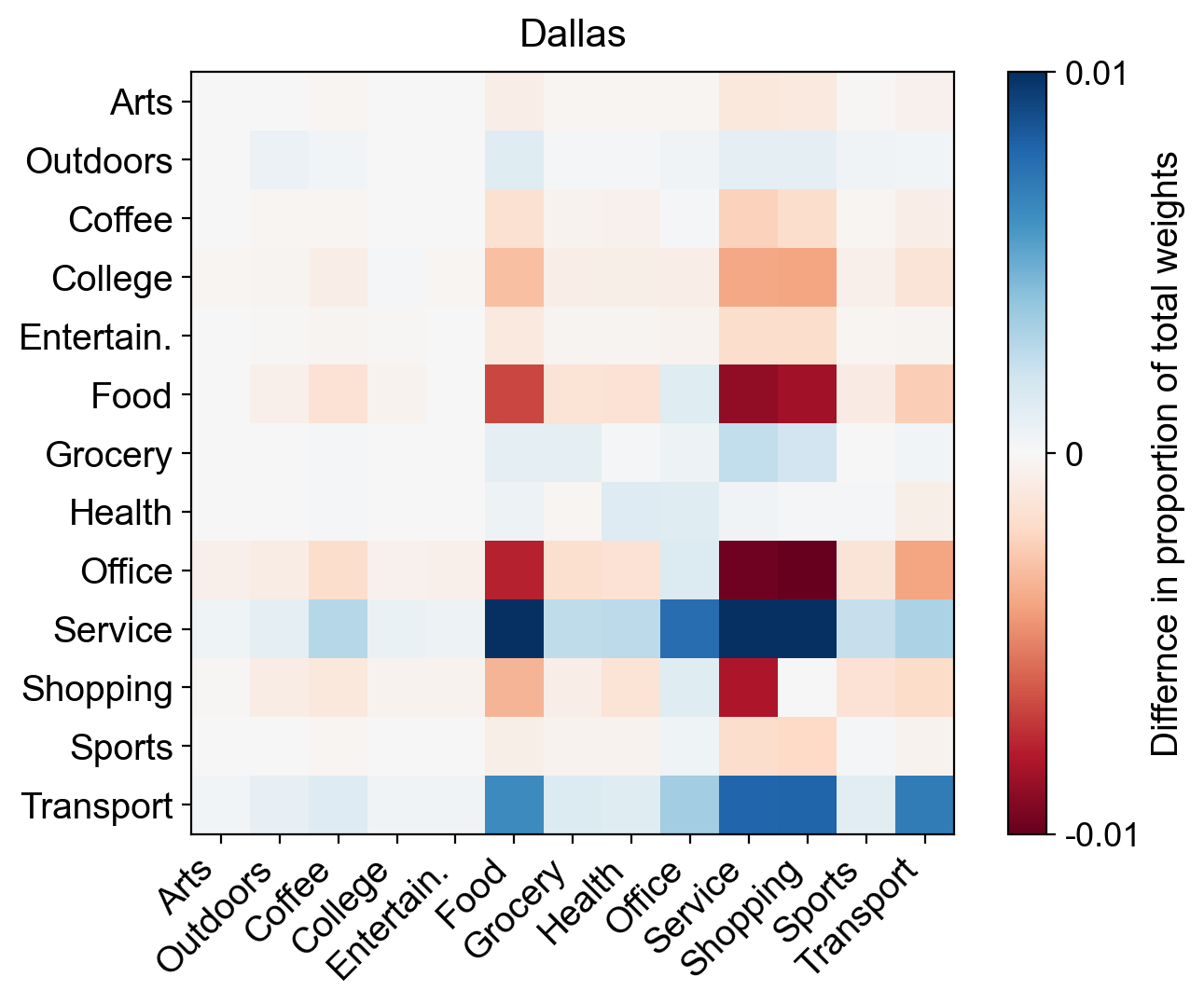}}
\caption[Histograms of edge distances between actual and null networks.]{\textbf{Histograms of edge distances between actual and null networks.} The actual networks contain more long distance edges than the null networks.}
\label{fig:s4nullcatprop}
\end{figure}

Figures \ref{fig:s4nullmaps1} to \ref{fig:s4nullmaps3} show the map visualizations of the actual (left) and null (right) networks for the five cities. 
In all cities, we can observe the null network being more locally clustered around large POIs, while the actual network is more dispersed and contains more long-distance connections. Note that the in-degrees and in-weights of all nodes are equivalent across the two networks. 
Figure \ref{fig:s4nulldistancedist} shows the histogram of physical distances of all edges in the actual and null networks. 
Figure \ref{fig:s4nullcatprop} compares the proportion of edge weights among POI category pairs across the actual and null networks. The proportion of edge weights was computed by taking the sum of weights that connect the vertical and horizontal categories and dividing that by the total weights that exist in the network. The patterns of dependency are consistent across cities, with Service and Transport node depending more on other nodes in the real network compared to the null network.

%% plot map figures of networks 

%% plot distance distributions 

%% plot category wise connections 

\subsection{Regression model of $w_{ij}$}

Results comparing the relationship between dependency weights and physical distance suggest that the dependency weights could be dictated by physical factors. 
Here, we model dependency weights $w_{ij}$ with various physical factors to investigate how much of the variance we observe in dependency weights can be explained by such factors. 
We build simple linear regression models of the form:
\begin{equation}
    \log_{10} w_{ij} \sim \log_{10} d_{ij} + n_j + \eta_i + \eta_j + \theta_i + \theta_j
\end{equation}
where $\log_{10} w_{ij}$ denotes the logged dependency weight of POI $i$ on POI $j$, and:
\begin{itemize}
    \item $\log_{10} d_{ij}$ is the logged Haversine distance between POIs $i$ and $j$. 
    \item $n_j$ is the number of visitors that POI $j$ receives. The premise is that popular POIs are depended more by other POIs. This variable was standardized ($ (x - \mu) / \sigma$) prior to the analysis for comparison across cities. 
    \item $\eta_i$ is the fixed effect (FE) for POI $i$'s subcategory. There are 97 subcategories in the POI dataset, and examples are shown in Figure \ref{s2categoryrank}. 
    \item $\theta_i$ is the fixed effect for POI $i$'s located Public Use Microdata Area (PUMA). 
\end{itemize}

The regression results are shown in Table \ref{table:wijreg1}. To save computation time, 1 million edges were randomly sampled to include in the regression. The resulting $R^2$ or the coefficients were robust against different sampling runs. 
Overall, the adjusted $R^2$ was relatively low (between $0.10$ and $0.16$ for all cities), suggesting that physical factors determine a small \% of the variance of dependency weights. 
The coefficients are similar across cities, with the logged Haversine distance showing a slower decay compared to the decay observed in Figure \ref{fig:s2nyc}, which is due to the samples in the regression being limited to ones that have non-zero weights. 

\begin{table}[ht]
\centering
\caption[Linear regression models predicting the logged dependency weights between POIs $i$ and $j$ ($log_{10} w_{ij}$) using physical factors]{Linear regression models predicting the logged dependency weights between POIs $i$ and $j$ ($log_{10} w_{ij}$) in US cities using the logged Haversine distance ($\log_{10} d_{ij}$), the popularity in visits to place $j$ ($v_j$), and the subcategory fixed effects and PUMA fixed effects of POIs $i$ and $j$. Variables were centered and standardized prior to analysis.}
\begin{tabular}{lrrrrr}
\toprule
& \multicolumn{5}{c}{Dependent Variable: $\log_{10}w_{ij}$} \\
\cmidrule{2-6}
Variable & New York & Boston & Seattle & Los Angeles & Dallas  \\
\midrule
Constant             & -2.268 $^{***}$  & -2.291 $^{***}$ & -2.315 $^{***}$ & -2.312 $^{***}$ & -2.408 $^{***}$  \\
$\log_{10} d_{ij}$   & -0.129 $^{***}$ & -0.147 $^{***}$ & -0.134 $^{***}$ & -0.125 $^{***}$ & -0.143 $^{***}$  \\
$v_j$ (standardized) & 0.022 $^{***}$  & 0.026 $^{***}$  & 0.027 $^{***}$  & 0.020 $^{***}$  & 0.022 $^{***}$  \\
\midrule
Subcategory FE for $i$ and $j$ & Y & Y & Y & Y & Y \\ 
PUMA FE for $i$ and $j$ & Y & Y & Y & Y & Y \\ 
\midrule
Observations (sampled) & 1,000,000 & 1,000,000 & 1,000,000 & 1,000,000 & 1,000,000 \\
Adj. $R^2$ & 0.089 & 0.117 & 0.122 & 0.108 & 0.115 \\
\bottomrule
\end{tabular}
\label{table:wijreg1}
\end{table}

\subsection{Robustness against different network instances}

Table \ref{table:wijregbst} shows the same regression results using the selected network edges using the bootstrap method introduced in Supplementary Note 3.2. 
The bootstrap method removed edges with weights that were almost zero (if the 25th percentile of the bootstrap sample distribution was below zero), and as a result, as shown in Figure \ref{fig:s3bstweight}(a), the POIs with smaller weights (especially out-weights) experienced a significant decrease in edge weights. 
The regression results show a much higher $R^2$ compared to the original results using all of the edges, explaining around 20\% to 30\% of the variance in the dependency weights. 
The results are once again consistent across cities, with regression coefficients in the same range. 

Table \ref{table:wijregshort} shows the same regression results using the dependency network computed using only short stays ($<$ 4 hours). The results of the regression (adjusted R squared and the model coefficients) are consistent with the results from the full network shown in Table \ref{table:wijreg1}. 

\begin{table}[ht]
\centering
\caption[Linear regression models predicting the logged dependency weights between POIs $i$ and $j$ ($log_{10} w_{ij}$) selected using the bootstrap method with physical factors]{Linear regression models predicting the logged dependency weights between POIs $i$ and $j$ ($log_{10} w_{ij}$) in US cities which were selected using the bootstrap method, using the logged Haversine distance ($\log_{10} d_{ij}$), the popularity in visits to place $j$ ($v_j$), and the subcategory fixed effects and PUMA fixed effects of POIs $i$ and $j$. Variables were centered and standardized prior to analysis.}
\begin{tabular}{lrrrrr}
\toprule
& \multicolumn{5}{c}{Dependent Variable: $\log_{10}w_{ij}$} \\
\cmidrule{2-6}
Variable & New York & Boston & Seattle & Los Angeles & Dallas  \\
\midrule
Constant             & -2.067 $^{***}$ & -1.978 $^{***}$ & -2.100 $^{***}$ & -2.249 $^{***}$ & -2.005 $^{***}$  \\
$\log_{10} d_{ij}$   & -0.163 $^{***}$ & -0.199 $^{***}$ & -0.186 $^{***}$ & -0.188 $^{***}$ & -0.232 $^{***}$  \\
$v_j$ (standardized) & 0.025 $^{***}$  & 0.033 $^{***}$  & 0.032 $^{***}$  & 0.030 $^{***}$  & 0.032 $^{***}$  \\
\midrule
Subcategory FE for $i$ and $j$ & Y & Y & Y & Y & Y \\ 
PUMA FE for $i$ and $j$ & Y & Y & Y & Y & Y \\ 
\midrule
Observations (sampled) & 1,000,000 & 1,000,000 & 1,000,000 & 1,000,000 & 1,000,000 \\
Adj. $R^2$ & 0.136 & 0.192 & 0.172 & 0.215 & 0.244 \\
\bottomrule
\end{tabular}
\label{table:wijregbst}
\end{table}

\begin{table}[ht]
\centering
\caption[Linear regression models predicting the logged dependency weights between POIs $i$ and $j$ ($log_{10} w_{ij}$) selected using only short stays ($<$ 4 hrs) with physical factors]{Linear regression models predicting the logged dependency weights between POIs $i$ and $j$ ($log_{10} w_{ij}$) in US cities which were selected using only short stays (< 4 hrs), using the logged Haversine distance ($\log_{10} d_{ij}$), the popularity in visits to place $j$ ($v_j$), and the subcategory fixed effects and PUMA fixed effects of POIs $i$ and $j$. Variables were centered and standardized prior to analysis.}
\begin{tabular}{lrrrrr}
\toprule
& \multicolumn{5}{c}{Dependent Variable: $\log_{10}w_{ij}$} \\
\cmidrule{2-6}
Variable & New York & Boston & Seattle & Los Angeles & Dallas  \\
\midrule
Constant             & -2.182 $^{***}$ & -2.220 $^{***}$ & -2.202 $^{***}$ & -2.280 $^{***}$ & -2.259 $^{***}$  \\
$\log_{10} d_{ij}$   & -0.131 $^{***}$ & -0.149 $^{***}$ & -0.134 $^{***}$ & -0.126 $^{***}$ & -0.145 $^{***}$  \\
$v_j$ (standardized) & 0.022 $^{***}$  & 0.029 $^{***}$  & 0.028 $^{***}$  & 0.021 $^{***}$  & 0.024 $^{***}$  \\
\midrule
Subcategory FE for $i$ and $j$ & Y & Y & Y & Y & Y \\ 
PUMA FE for $i$ and $j$ & Y & Y & Y & Y & Y \\ 
\midrule
Observations (sampled) & 1,000,000 & 1,000,000 & 1,000,000 & 1,000,000 & 1,000,000 \\
Adj. $R^2$ & 0.096 & 0.125 & 0.131 & 0.117 & 0.121 \\
\bottomrule
\end{tabular}
\label{table:wijregshort}
\end{table}

\subsection{Robustness against co-visit detection parameters}

\begin{figure}
\centering
\subfloat[Step difference parameter]{\includegraphics[width=\linewidth]{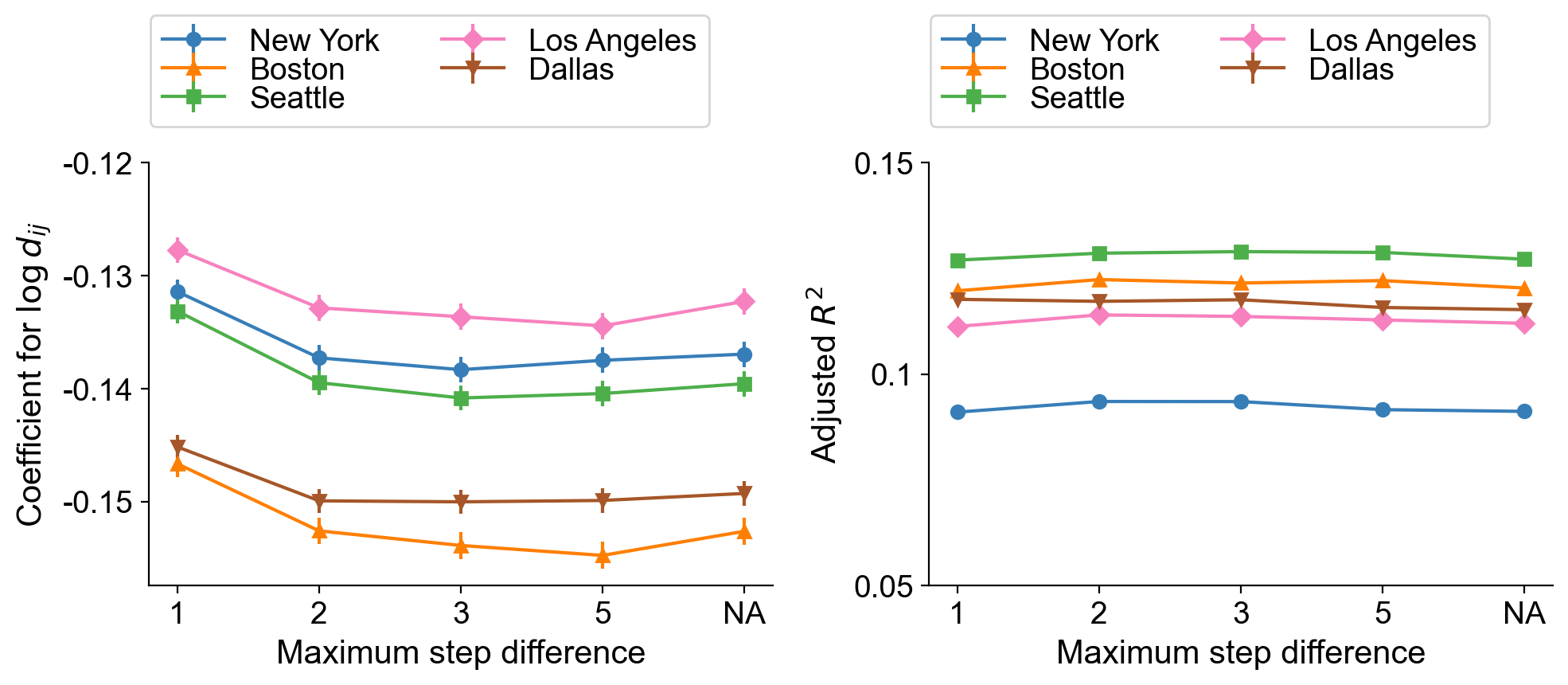}} \\
\subfloat[Time difference parameter]{\includegraphics[width=\linewidth]{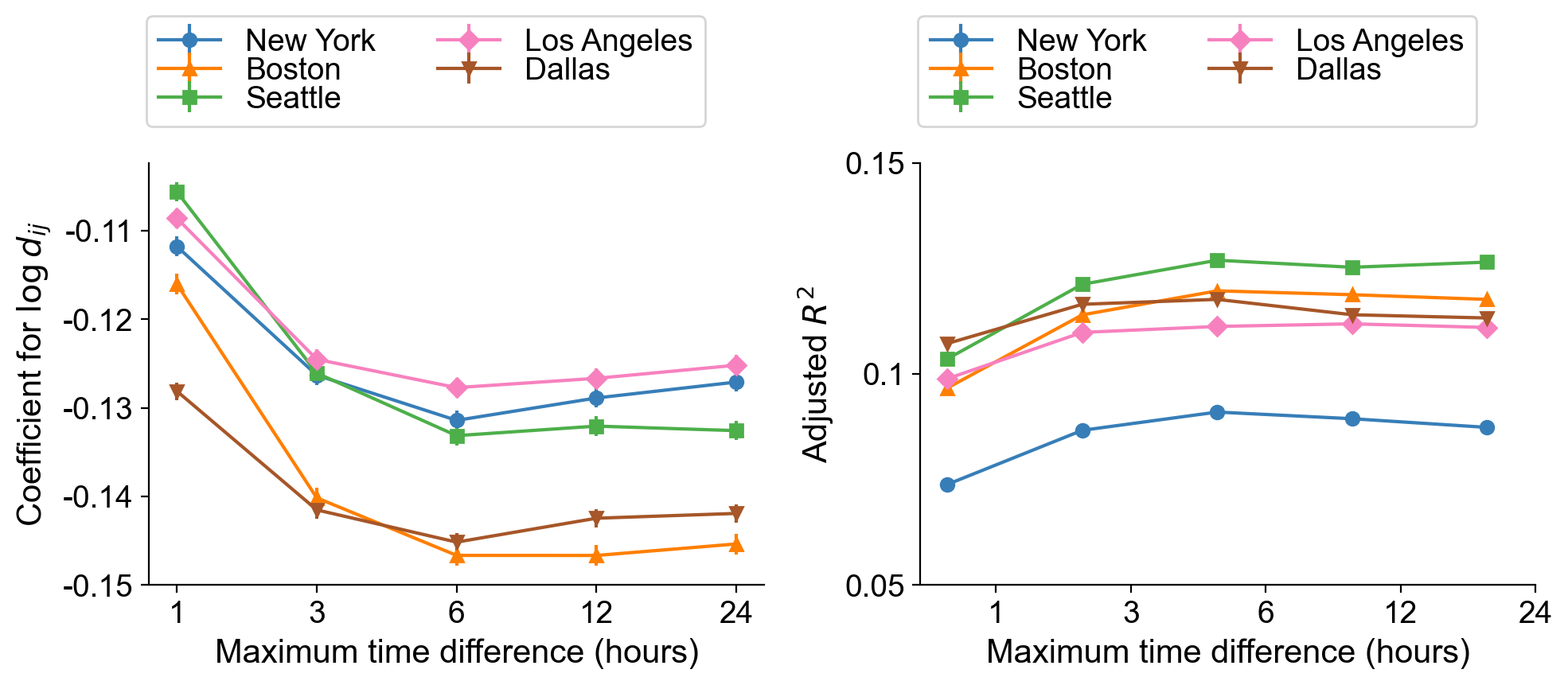}} \\
\caption[Regression results for dependency weights when using different co-visit detection parameters, contd.]{\textbf{Regression results for dependency weights when using different co-visit detection parameters, continued for Los Angeles and Dallas.} Except for the networks generated using maximum time difference of 1 hour, the results (adjusted $R^2$ and coefficient for Haversine distance) are relatively similar as the results in Table \ref{table:wijreg1}.}
\label{fig:s4wijregcoeff2}
\end{figure}

The regression analysis showed two main findings: 1) the physical factors can only explain a fraction of the variance in dependency weights, and 2) the coefficient of the logged Haversine distance is moderate compared to distance exponents used in traditional models such as the Gravity Model (i.e., $-2$). 
To assess the robustness of these findings of the regression model, the regression was performed using dependency networks generated using different co-visit detection parameters (also used in Supplementary Note 2.3). 
The 10 pairs of parameters ($T_s = [1,2,3,5,unlimited]$ and $T_c = [1,3,6,12,24]$ hours) were used to generate the dependency networks for each city, and the regression results (adjusted $R^2$ and coefficient for Haversine distance) are shown in Figure \ref{fig:s4wijregcoeff2}. 
The matrix plots across the five metropolitan areas show similar patterns, both the adjusted $R^2$ and coefficient for Haversine distance are relatively different when the maximum time difference is 1 hour. 
In the rest of the parameter pairs, we obtain a similar result compared to Table \ref{table:wijreg1}.

% \begin{figure}[!t]
% \centering
% \subfloat[New York]{\includegraphics[width=.9\linewidth]{ny_r2_beta_params.png}} \\
% \subfloat[Seattle]{\includegraphics[width=.9\linewidth]{boston_r2_beta_params.png}} \\
% \subfloat[Los Angeles]{\includegraphics[width=.9\linewidth]{seattle_r2_beta_params.png}} 
% \caption[Regression results for dependency weights when using different co-visit detection parameters]{\textbf{Regression results for dependency weights when using different co-visit detection parameters.} Except for the networks generated using maximum time difference of 1 hour, the results (adjusted $R^2$ and coefficient for Haversine distance) are relatively similar as the results in Table \ref{table:wijreg1}.}
% \label{fig:s4wijregcoeff1}
% \end{figure}

% \begin{figure}[!t]
% \centering
% \subfloat[Los Angeles]{\includegraphics[width=\linewidth]{la_r2_beta_params.png}} \\
% \subfloat[Dallas]{\includegraphics[width=\linewidth]{dallas_r2_beta_params.png}} \\
% \caption[Regression results for dependency weights when using different co-visit detection parameters, contd.]{\textbf{Regression results for dependency weights when using different co-visit detection parameters, continued for Los Angeles and Dallas.} Except for the networks generated using maximum time difference of 1 hour, the results (adjusted $R^2$ and coefficient for Haversine distance) are relatively similar as the results in Table \ref{table:wijreg1}.}
% \label{fig:s4wijregcoeff2}
% \end{figure}

\subsection{Robustness against different time periods}

To check the robustness of the results with respect to the time periods used to generate the dependency network, the same regression experiments were performed using 2019 January -- April data and 2019 May -- August data, as shown in Tables \ref{table:wijregJAN} and \ref{table:wijregMAY}.
We can observe that the $R^2$ values and the regression coefficients are similar with those obtained in Table \ref{table:wijreg1}.

\begin{table}[t]
\centering
\caption[Linear regression models predicting the logged dependency weights between POIs $i$ and $j$ ($log_{10} w_{ij}$) using network from 2019 January - April data]{Linear regression models predicting the logged dependency weights between POIs $i$ and $j$ ($log_{10} w_{ij}$) in US cities using network from 2019 January - April data.}
\begin{tabular}{lrrrrr}
\toprule
& \multicolumn{5}{c}{Dependent Variable: $\log_{10}w_{ij}$ (2019 January - April data)} \\
\cmidrule{2-6}
Variable & New York & Boston & Seattle & Los Angeles & Dallas  \\
\midrule
Constant             & -2.157 $^{***}$ & -2.242 $^{***}$ & -2.273 $^{***}$ & -2.264 $^{***}$ & -2.268 $^{***}$  \\
$\log_{10} d_{ij}$   & -0.128 $^{***}$ & -0.146 $^{***}$ & -0.132 $^{***}$ & -0.127 $^{***}$ & -0.144 $^{***}$  \\
$v_j$ (standardized) & 0.019 $^{***}$  & 0.029 $^{***}$  & 0.026 $^{***}$  & 0.019 $^{***}$  & 0.021 $^{***}$  \\
\midrule
Subcategory FE for $i$ and $j$ & Y & Y & Y & Y & Y \\ 
PUMA FE for $i$ and $j$ & Y & Y & Y & Y & Y \\ 
\midrule
Observations (sampled) & 1,000,000 & 1,000,000 & 1,000,000 & 1,000,000 & 1,000,000 \\
Adj. $R^2$ & 0.092 & 0.120 & 0.121 & 0.112 & 0.117 \\
\bottomrule
\end{tabular}
\label{table:wijregJAN}
\end{table}

\begin{table}[t]
\centering
\caption[Linear regression models predicting the logged dependency weights between POIs $i$ and $j$ ($log_{10} w_{ij}$) selected using network from 2019 May - August data]{Linear regression models predicting the logged dependency weights between POIs $i$ and $j$ ($log_{10} w_{ij}$) in US cities using network from 2019 May - August data.}
\begin{tabular}{lrrrrr}
\toprule
& \multicolumn{5}{c}{Dependent Variable: $\log_{10}w_{ij}$ (2019 May - August data)} \\
\cmidrule{2-6}
Variable & New York & Boston & Seattle & Los Angeles & Dallas  \\
\midrule
Constant             & -2.312 $^{***}$ & -2.292 $^{***}$ & -2.320 $^{***}$ & -2.333 $^{***}$ & -2.421 $^{***}$  \\
$\log_{10} d_{ij}$   & -0.130 $^{***}$ & -0.146 $^{***}$ & -0.134 $^{***}$ & -0.127 $^{***}$ & -0.143 $^{***}$  \\
$v_j$ (standardized) & 0.020 $^{***}$  & 0.023 $^{***}$  & 0.024 $^{***}$  & 0.019 $^{***}$  & 0.021 $^{***}$  \\
\midrule
Subcategory FE for $i$ and $j$ & Y & Y & Y & Y & Y \\ 
PUMA FE for $i$ and $j$ & Y & Y & Y & Y & Y \\ 
\midrule
Observations (sampled) & 1,000,000 & 1,000,000 & 1,000,000 & 1,000,000 & 1,000,000 \\
Adj. $R^2$ & 0.090 & 0.114 & 0.118 & 0.108 & 0.114 \\
\bottomrule
\end{tabular}
\label{table:wijregMAY}
\end{table}

\clearpage
\section{Impacts of behavior-based dependency during the COVID-19 pandemic}

\subsection{Analysis of visitation losses in cities}

\begin{figure}
\centering
\subfloat{\includegraphics[width=0.3\linewidth]{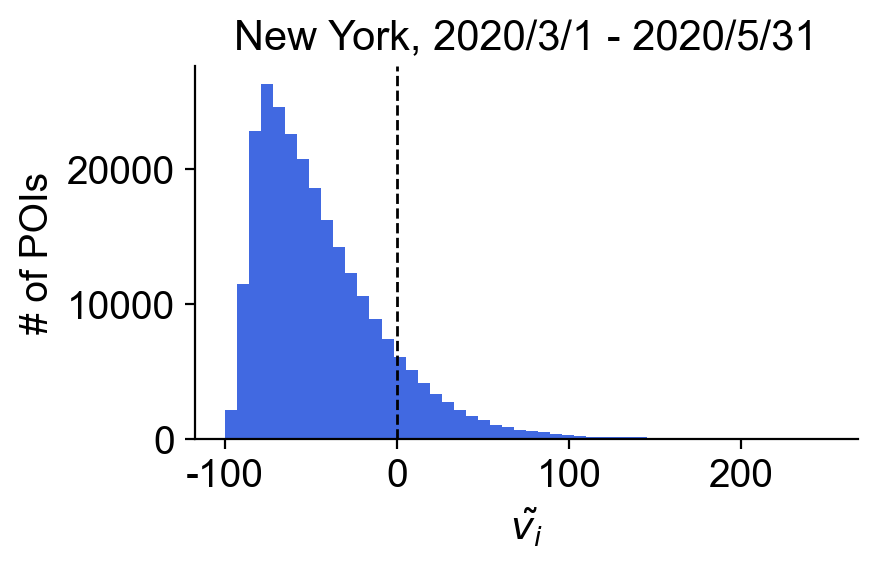}} 
\subfloat{\includegraphics[width=0.4\linewidth]{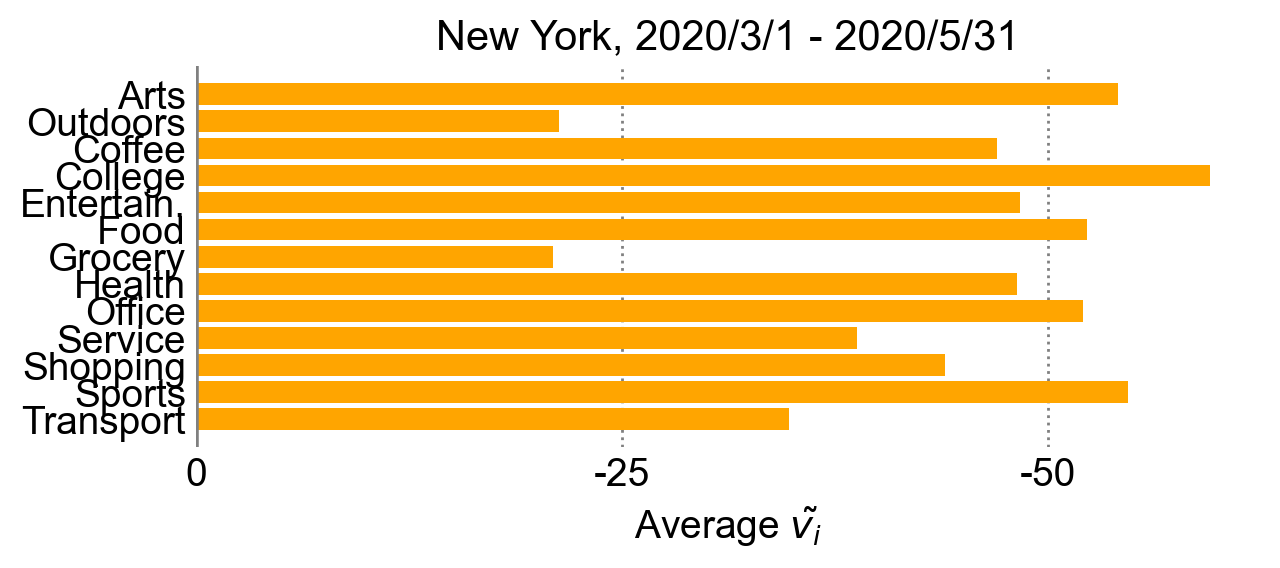}} \\
\subfloat{\includegraphics[width=0.3\linewidth]{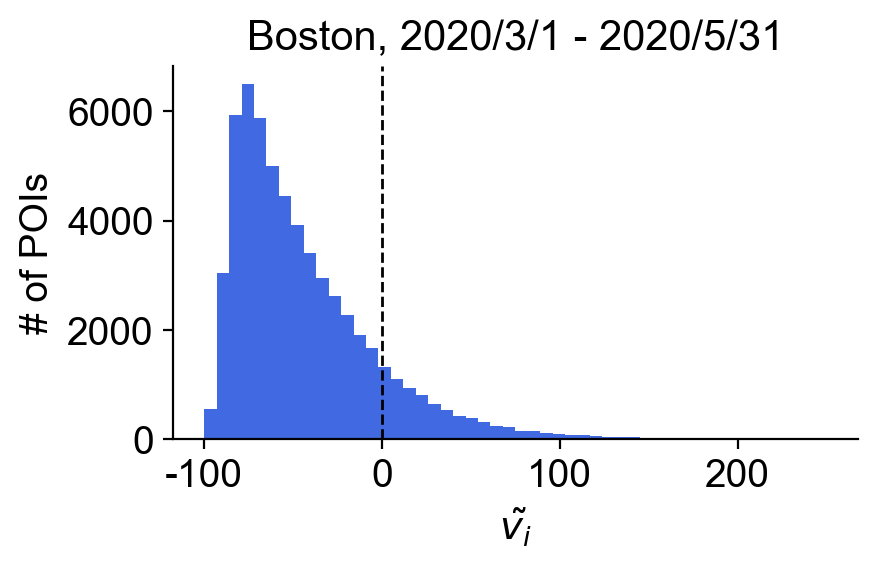}} 
\subfloat{\includegraphics[width=0.4\linewidth]{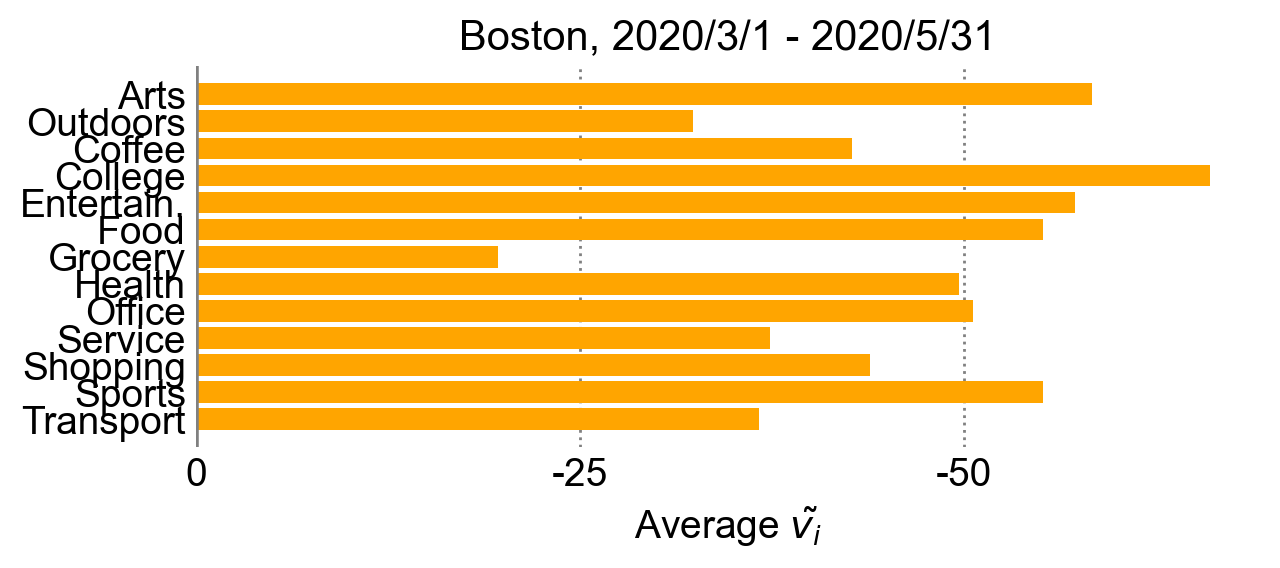}} \\
\subfloat{\includegraphics[width=0.3\linewidth]{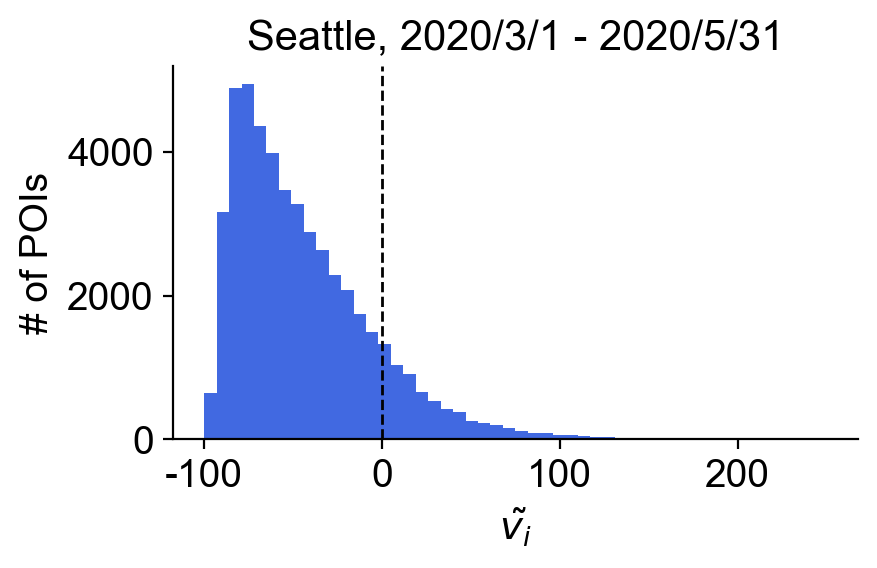}} 
\subfloat{\includegraphics[width=0.4\linewidth]{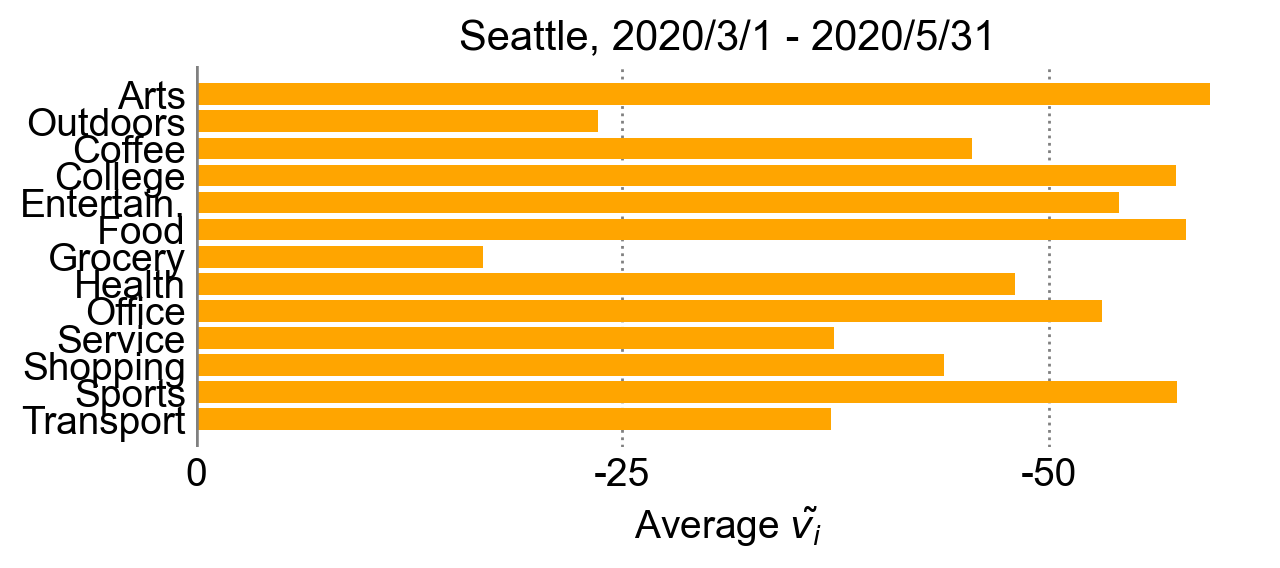}} \\
\subfloat{\includegraphics[width=0.3\linewidth]{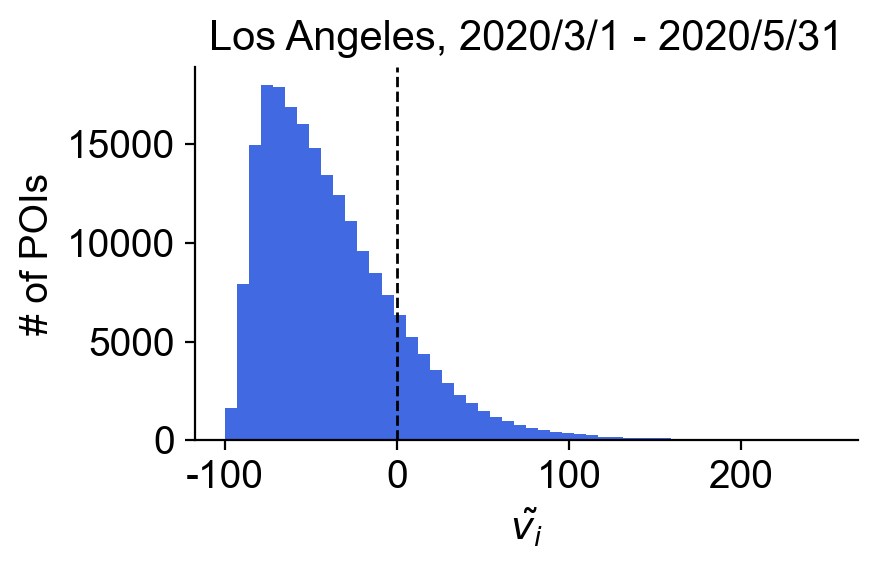}} 
\subfloat{\includegraphics[width=0.4\linewidth]{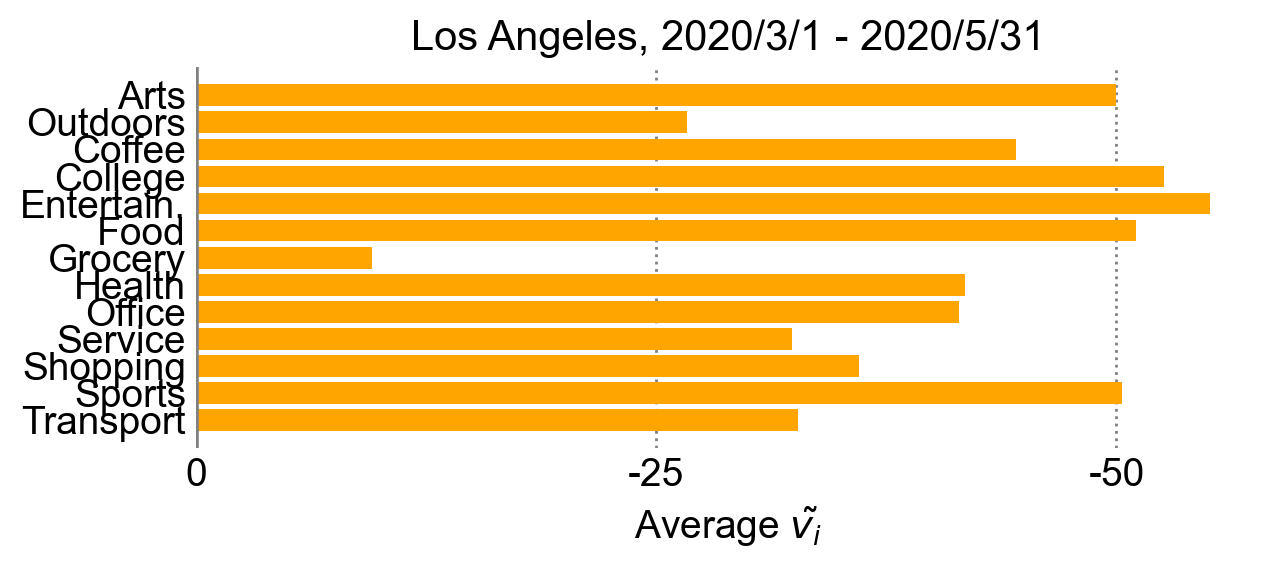}} \\
\subfloat{\includegraphics[width=0.3\linewidth]{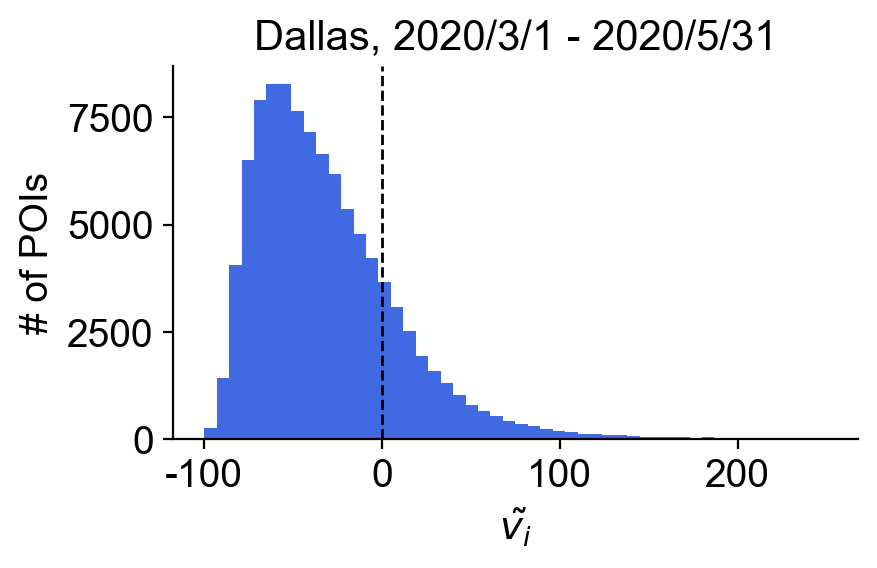}} 
\subfloat{\includegraphics[width=0.4\linewidth]{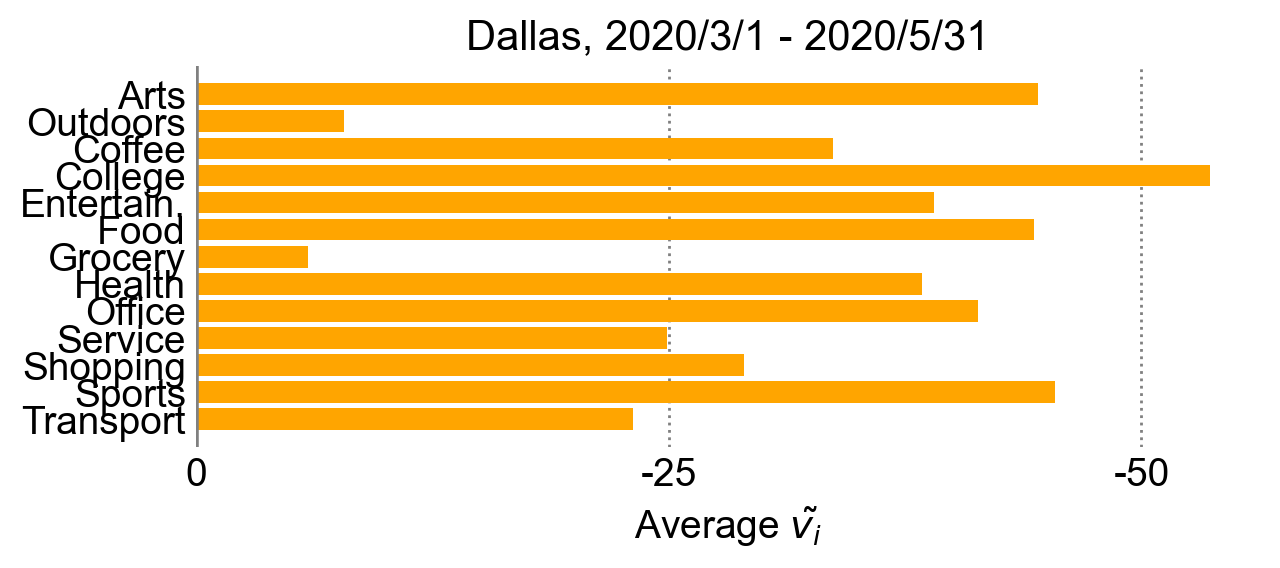}} \\
\caption[Change in visits to POIs in New York, Boston, Seattle, Los Angeles, and Dallas during the pandemic period.]{\textbf{Change in visits to POIs in New York, Boston, Seattle, Los Angeles, and Dallas during the pandemic period.} The left panels show the distribution of the change in visits, and the right panels show the average effects per POI category.}
\label{fig:s5vis1}
\end{figure}

To investigate the usefulness of the dependency network for understanding the resilience of businesses, we construct a regression modeling framework that predicts the change in visitation patterns to a POI using information about the change in visitation patterns to its alters and the dependency network. 
The observed change in visits to different places is computed by:
\begin{equation}
    \Tilde v_i = \big( \frac{v_i^{after}}{v_i^{before}} - 1 \big) * 100 (\%)
\end{equation}
where $v_i^{before}$ and $v_i^{after}$ denote the number of visits to place $i$ before the pandemic (September - December 2019) and during the pandemic period (March - June 2020), respectively.  
The left side panels in Figure \ref{fig:s5vis1} plots the distribution of the change in visits $\Tilde v_i$ in the five metropolitan areas. 
Overall, as expected, we observe a peak at around $\Tilde v_i = -75$ for all cities, suggesting the overall substantial decrease in visits during the initial stages of the pandemic. 
The right side panels in Figure \ref{fig:s5vis1} plots the average change in visits across different POI categories. 
We observe heterogeneity across POI categories, where Arts, Colleges, Food, and Sports POIs experienced the largest losses during the initial stages of the pandemic. On the other hand, Outdoor and Grocery places had a minor impact compared to the other POI types. 
Moreover, prior studies have found that POIs located in urban areas were affected more compared to rural areas (e.g., \cite{yabe2023behavioral}). 
Therefore, we take into consideration the effects of the category of the places and the area in which the POIs are located to predict the visitation losses to POIs.   

\begin{figure}
\centering
\subfloat[New York]{\includegraphics[width=0.4\linewidth]{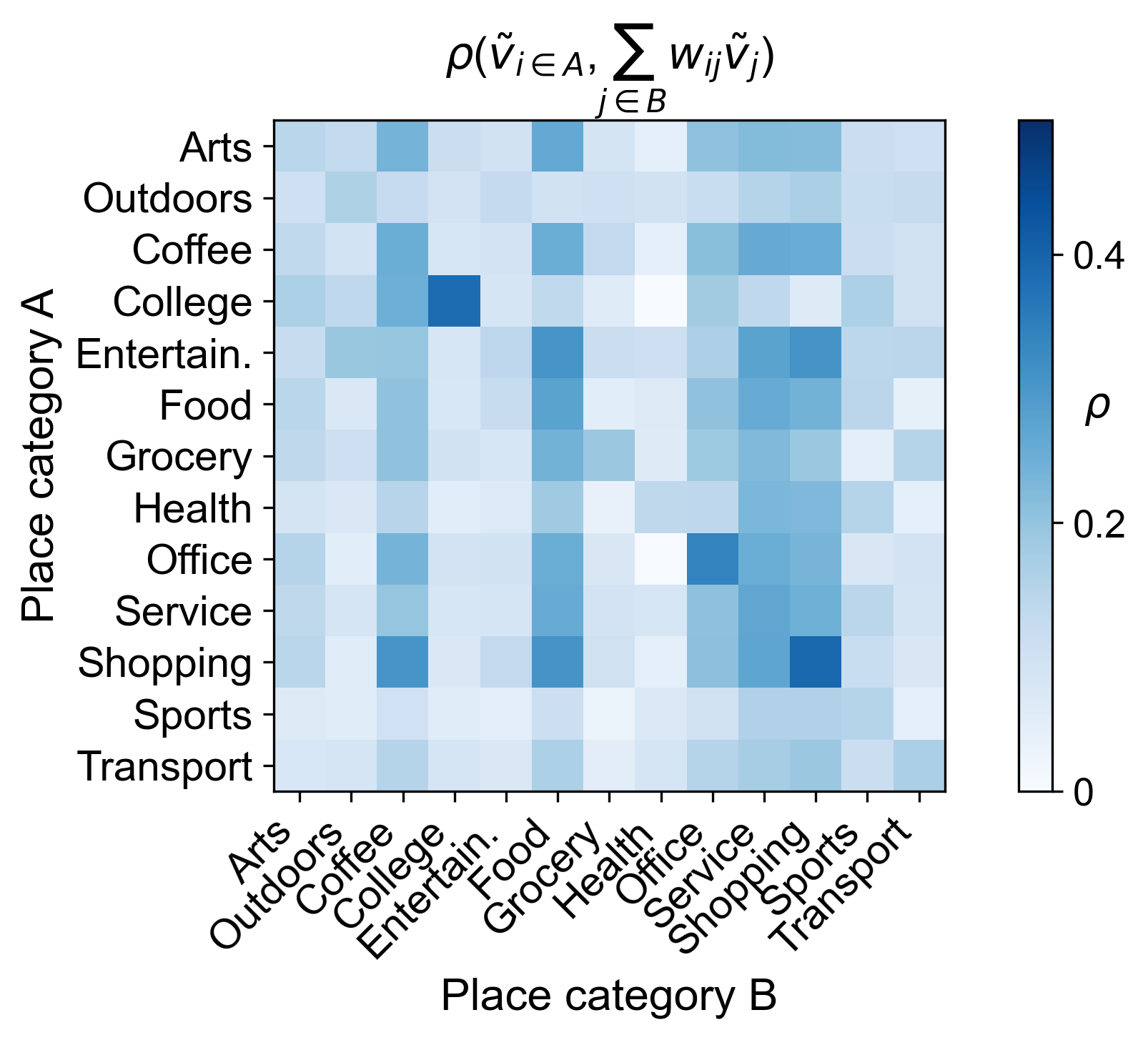}} 
\hspace{0.1\textwidth}
\subfloat[Boston]{\includegraphics[width=0.4\linewidth]{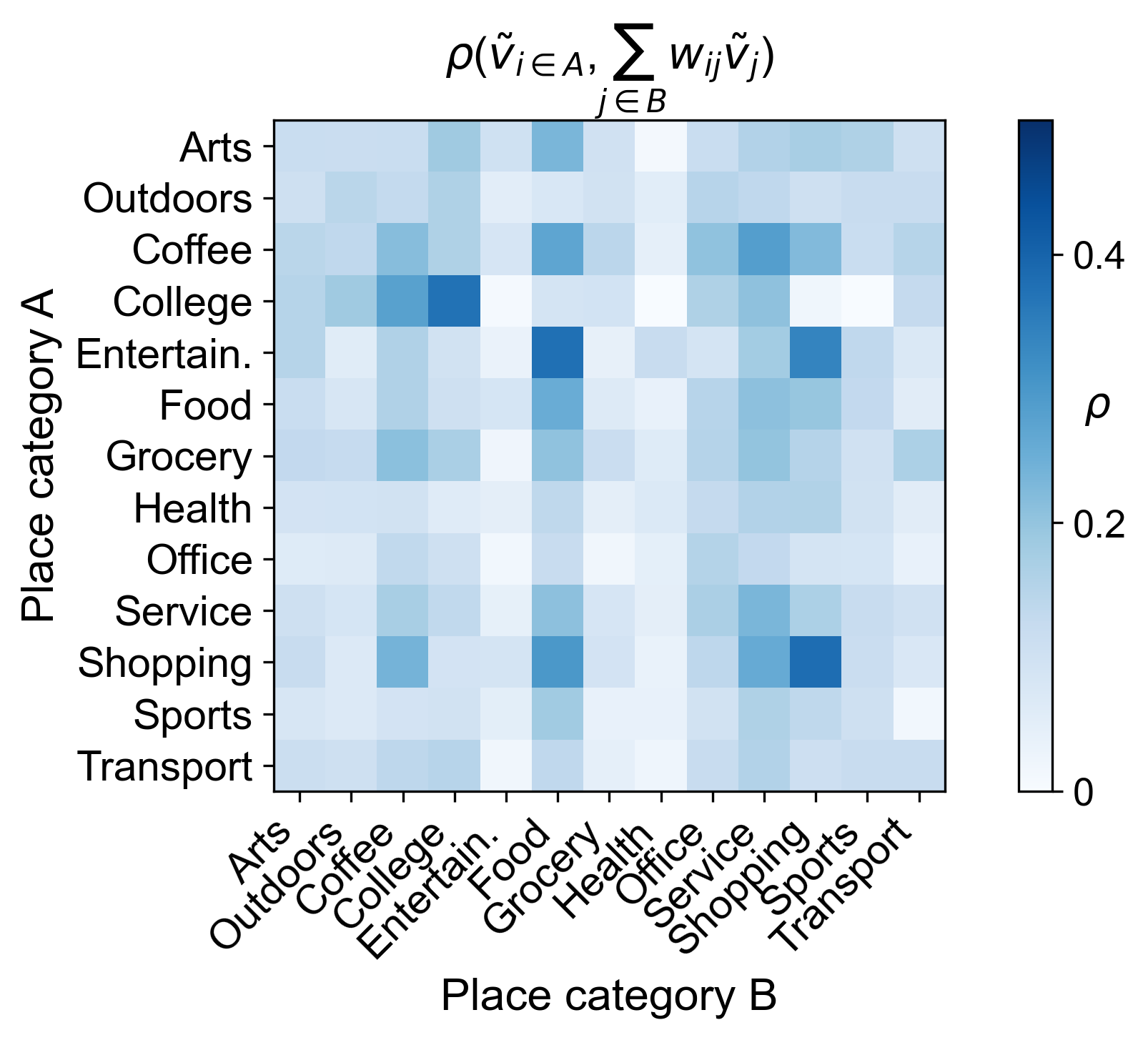}} \\
\subfloat[Seattle]{\includegraphics[width=0.4\linewidth]{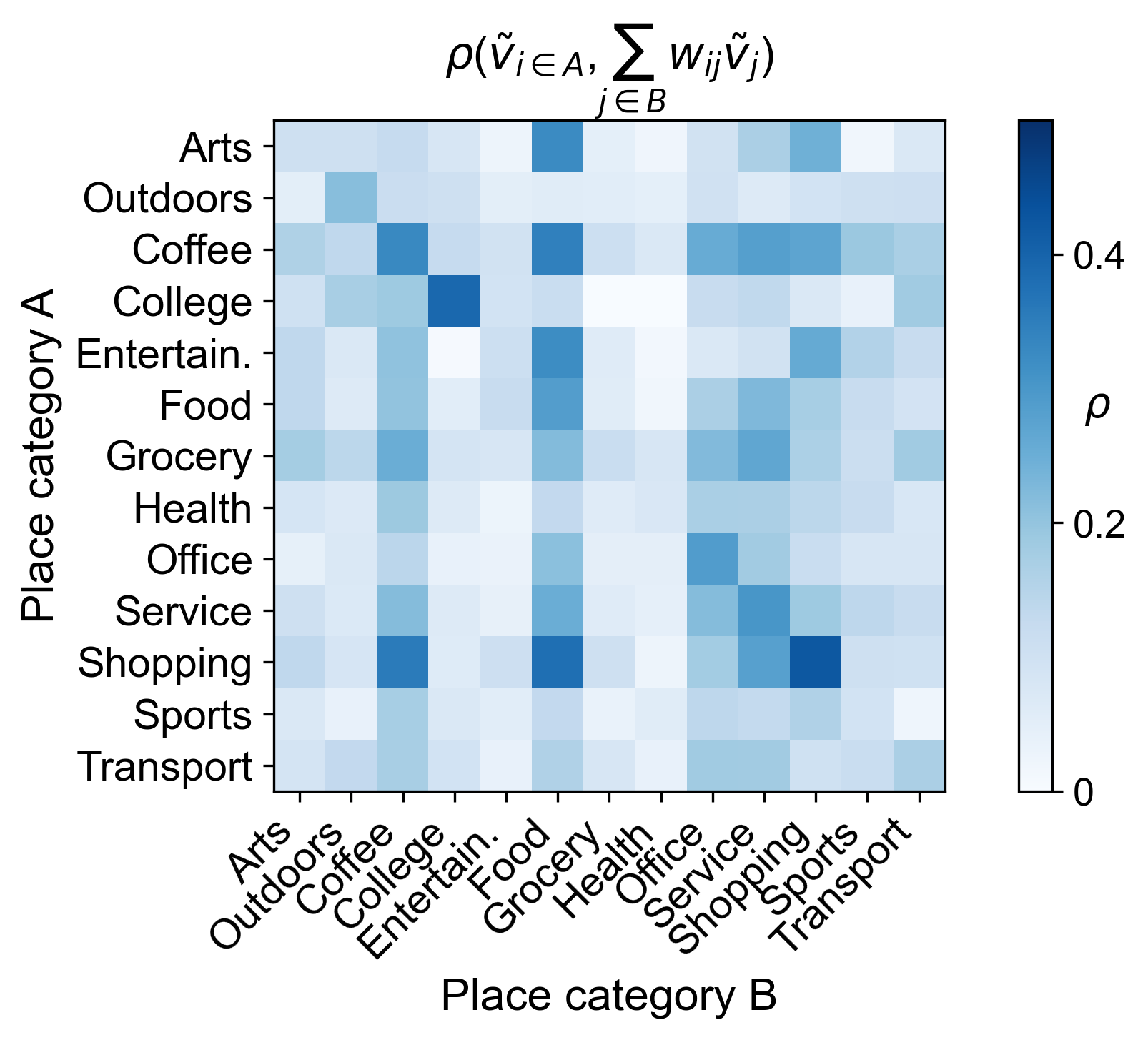}}
\hspace{0.1\textwidth}
\subfloat[Los Angeles]{\includegraphics[width=0.4\linewidth]{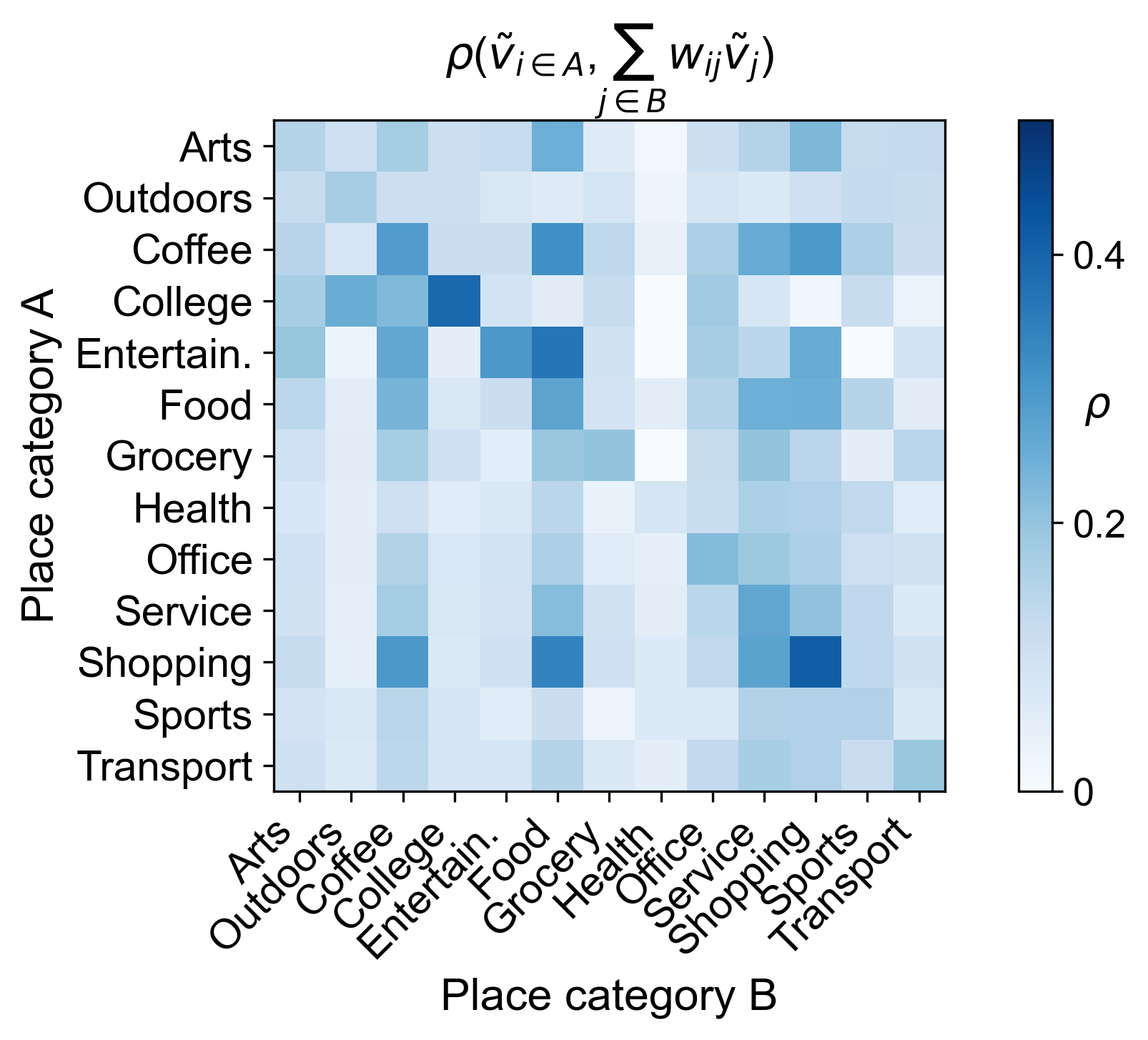}} \\
\subfloat[Dallas]{\includegraphics[width=0.4\linewidth]{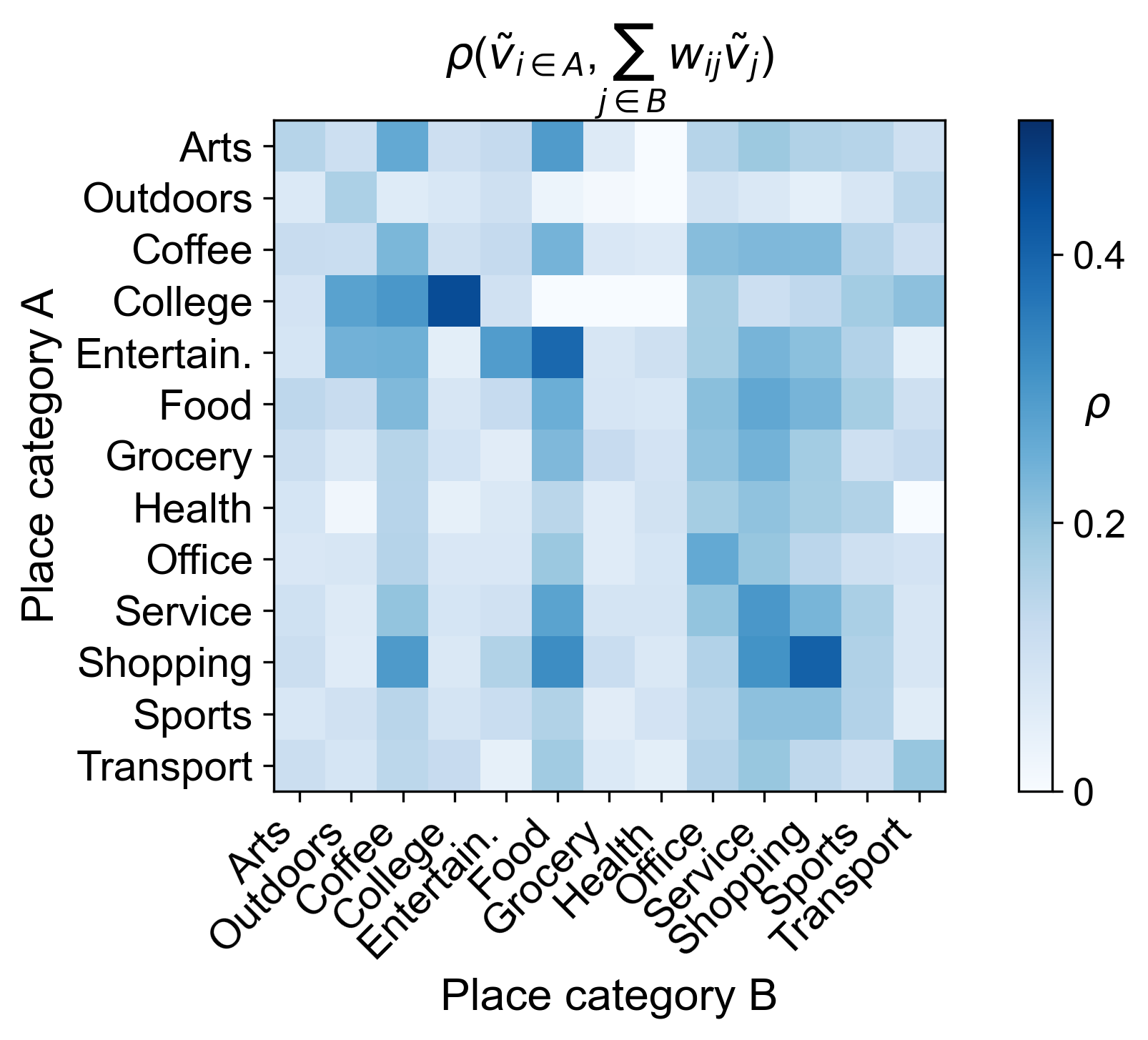}}
\caption[Correlation between change in visits of ego and the weighted change in visits of alters by category pairs]{\textbf{Correlation between change in visits of ego and the weighted change in visits of alters by category pairs.} Significant correlation between several category pairs suggest predictability of $\Tilde v_i$ using information of alters.}
\label{fig:vis_corr}
\end{figure}

% \subsection{Correlational Analysis}
To test the hypothesis that the change in visitation patterns to a POI can be modeled using information about the change in visitation patterns to its alters and the dependency network, we first investigate the simple  correlation between the two metrics. 
Figure \ref{fig:vis_corr} shows the partial correlation $\rho(\Tilde v_i, \sum_j w_{ij}\Tilde v_i)$, where POI $i$ and $j$s' categories are $A$ (vertical) and $B$ (horizontal), respectively.
Overall, several category pairs exhibit significant and strong correlation. For example, coffee POIs' changes in visits are strongly correlated with the weighted changes in visits to connected food, service, and shopping POIs across the five metropolitan areas. 
This significant and strong correlation suggests that the changes in the visitation of the ego may be predicted using the information of the alter POIs connected via the dependency network. 

\subsection{Model specification and estimation results}
Using the correlations revealed in the previous section, we build a simple linear regression model of the form:
\begin{equation}
    \Tilde v_i \sim \sum_j w_{ij} \Tilde v_j + \sum_j \delta_{ij} \Tilde v_j + \eta_i + \theta_i 
\end{equation}
where $\Tilde v_i$ denotes the change in visitations to POI $i$ during the initial stages of the pandemic (March 1st - May 21st, 2020), and:
\begin{itemize}
    \item $\sum_j w_{ij} \Tilde v_j$ is the sum of the neighbors' (POIs $j$) change in visitations ($\Tilde v_j$) weighted by the dependency network weights $w_{ij}$. 
    \item $\sum_j \delta_{ij} \Tilde v_j$ is the sum of the neighbors' (POIs $j$) change in visitations ($\Tilde v_j$) weighted by the distance based network weights $\delta_{ij}$. The distance based network weights mimic the patterns observed between physical distance and dependency weights, which are constant until 100 meters, but decays as the inverse squared of the actual distance beyond 100 meters. 
    \item $\eta_i$ is the fixed effect (FE) for POI $i$'s subcategory. There are 97 subcategories in the POI dataset, and examples are shown in Figure \ref{s2categoryrank}. 
    \item $\theta_i$ is the fixed effect for POI $i$'s located Public Use Microdata Area (PUMA). 
\end{itemize}

\renewcommand\labelenumi{(\theenumi)}

Tables \ref{covidregboston} to \ref{covidregdallas} show the regression results for New York, Boston, Seattle, Los Angeles, and Dallas, under the following four models:
\begin{enumerate}
    \item using only the subcategory and PUMA fixed effects, 
    \item using the distance-based network effects $\sum_j \delta_{ij} \Tilde v_j$ in addition to the subcategory and PUMA fixed effects, 
    \item using the behavior-based dependency network effects $\sum_j w_{ij} \Tilde v_j$ in addition to the subcategory and PUMA fixed effects, and
    \item using all of the factors introduced in the full model. 
\end{enumerate}
Both the distance-based network effects and behavior-based dependency network effects were standardized (by subtracting the mean and dividing by the standard deviation) before analysis, therefore the magnitude of the coefficients can be compared. 

In all cities, the model performance significantly increases when we use the behavior-based dependency network effects in model (3), compared to the baseline model (1) and distance-based network effects model (2).
Including both distance-based network effects and the behavior-based network effects slightly increases the $R^2$, however, the estimated coefficients for $\sum_j w_{ij} \Tilde v_j$ are substantially larger (5 to 15 fold) than the coefficients of $\sum_j \delta_{ij} \Tilde v_j$, indicating the significance of the dependency network effects. 

\begin{table}[!htbp] \centering \caption[Linear regression models predicting the change in visits to POIs during the pandemic (2019 March - May) in New York] {Linear regression models predicting the change in visits to POIs during the pandemic (2019 March - May) in New York.} \begin{tabular}{@{\extracolsep{5pt}}lcccc} \\[-1.8ex]\hline \hline \\[-1.8ex] & \multicolumn{4}{c}{\textit{Dependent variable: $\Tilde v_i $ (Change in visits during the pandemic)}} \ \cr \cline{2-5} \\[-1.8ex] & (1) & (2) & (3) & (4) \\ \hline \\[-1.8ex]  Constant & -40.9$^{***}$ & -41.6$^{***}$ & -41.7$^{***}$ & -42.2$^{***}$ \\               & (5.345) & (5.110)     & (5.110)     & (5.089) \\  Distance effect $\sum_j \delta_{ij} \Tilde v_j $ & & 4.278$^{***}$ & & 3.410$^{***}$ \\   & & (0.085) & & (0.082) \\  Dependency effect $\sum_j w_{ij} \Tilde v_j $ && & 14.59$^{***}$ & 14.26$^{***}$ \\   & & & (0.103) & (0.103) \\ \hline \\[-1.8ex] Subcategory FE  & Y & Y & Y & Y  \\  PUMA FE  & Y & Y & Y & Y \\  \hline \\[-1.8ex]  Observations & 213,140 & 213,140 & 213,140 & 213,140 \\  $R^2$ & 0.158 & 0.168 & 0.230 & 0.237 \\  Adjusted $R^2$ & 0.157 & 0.167 & 0.230 & 0.236 \\ \hline \hline \\[-1.8ex] \textit{Note:} & \multicolumn{4}{r}{$^{*}$p$<$0.1; $^{**}$p$<$0.05; $^{***}$p$<$0.01} \\ \end{tabular} \label{covidregny} \end{table}

\begin{table}[!htbp] \centering \caption[Linear regression models predicting the change in visits to POIs during the pandemic (2019 March - May) in Boston] {Linear regression models predicting the change in visits to POIs during the pandemic (2019 March - May) in Boston.} \begin{tabular}{@{\extracolsep{5pt}}lcccc} \\[-1.8ex]\hline \hline \\[-1.8ex] & \multicolumn{4}{c}{\textit{Dependent variable: $\Tilde v_i $ (Change in visits during the pandemic)}} \ \cr \cline{2-5} \\[-1.8ex] & (1) & (2) & (3) & (4) \\ \hline \\[-1.8ex]  Constant & -20.1$^{***}$ & -22.1$^{***}$ & -33.7$^{***}$ & -34.9$^{***}$ \\               & (6.512) & (6.253)     & (6.253)     & (6.221) \\  Distance effect $\sum_j \delta_{ij} \Tilde v_j $ & & 4.919$^{***}$ & & 3.946$^{***}$ \\   & & (0.182) & & (0.176) \\  Dependency effect $\sum_j w_{ij} \Tilde v_j $ && & 13.00$^{***}$ & 12.60$^{***}$ \\   & & & (0.201) & (0.200) \\ \hline \\[-1.8ex] Subcategory FE  & Y & Y & Y & Y  \\  PUMA FE  & Y & Y & Y & Y \\  \hline \\[-1.8ex]  Observations & 48,848 & 48,848 & 48,848 & 48,848 \\  $R^2$ & 0.145 & 0.158 & 0.213 & 0.221 \\  Adjusted $R^2$ & 0.143 & 0.155 & 0.210 & 0.218 \\ \hline \hline \\[-1.8ex] \textit{Note:} & \multicolumn{4}{r}{$^{*}$p$<$0.1; $^{**}$p$<$0.05; $^{***}$p$<$0.01} \\ \end{tabular} \label{covidregboston} \end{table}

\begin{table}[!htbp] \centering \caption[Linear regression models predicting the change in visits to POIs during the pandemic (2019 March - May) in Seattle] {Linear regression models predicting the change in visits to POIs during the pandemic (2019 March - May) in Seattle.} \begin{tabular}{@{\extracolsep{5pt}}lcccc} \\[-1.8ex]\hline \hline \\[-1.8ex] & \multicolumn{4}{c}{\textit{Dependent variable: $\Tilde v_i $ (Change in visits during the pandemic)}} \ \cr \cline{2-5} \\[-1.8ex] & (1) & (2) & (3) & (4) \\ \hline \\[-1.8ex]  Constant & -21.5$^{**}$ & -22.2$^{**}$ & -26.3$^{***}$ & -26.6$^{***}$ \\               & (10.58) & (10.06)     & (10.06)     & (10.00) \\  Distance effect $\sum_j \delta_{ij} \Tilde v_j $ & & 5.021$^{***}$ & & 3.993$^{***}$ \\   & & (0.187) & & (0.179) \\  Dependency effect $\sum_j w_{ij} \Tilde v_j $ && & 13.72$^{***}$ & 13.32$^{***}$ \\   & & & (0.205) & (0.205) \\ \hline \\[-1.8ex] Subcategory FE  & Y & Y & Y & Y  \\  PUMA FE  & Y & Y & Y & Y \\  \hline \\[-1.8ex]  Observations & 41,451 & 41,451 & 41,451 & 41,451 \\  $R^2$ & 0.168 & 0.182 & 0.249 & 0.258 \\  Adjusted $R^2$ & 0.166 & 0.180 & 0.247 & 0.256 \\ \hline \hline \\[-1.8ex] \textit{Note:} & \multicolumn{4}{r}{$^{*}$p$<$0.1; $^{**}$p$<$0.05; $^{***}$p$<$0.01} \\ \end{tabular} \label{covidregseattle} \end{table}

\begin{table}[!htbp] \centering \caption[Linear regression models predicting the change in visits to POIs during the pandemic (2019 March - May) in Los Angeles] {Linear regression models predicting the change in visits to POIs during the pandemic (2019 March - May) in Los Angeles.} \begin{tabular}{@{\extracolsep{5pt}}lcccc} \\[-1.8ex]\hline \hline \\[-1.8ex] & \multicolumn{4}{c}{\textit{Dependent variable: $\Tilde v_i $ (Change in visits during the pandemic)}} \ \cr \cline{2-5} \\[-1.8ex] & (1) & (2) & (3) & (4) \\ \hline \\[-1.8ex]  Constant & -17.3$^{***}$ & -17.8$^{***}$ & -21.2$^{***}$ & -21.4$^{***}$ \\               & (4.820) & (4.554)     & (4.554)     & (4.534) \\  Distance effect $\sum_j \delta_{ij} \Tilde v_j $ & & 4.860$^{***}$ & & 3.531$^{***}$ \\   & & (0.095) & & (0.091) \\  Dependency effect $\sum_j w_{ij} \Tilde v_j $ && & 14.30$^{***}$ & 13.89$^{***}$ \\   & & & (0.100) & (0.100) \\ \hline \\[-1.8ex] Subcategory FE  & Y & Y & Y & Y  \\  PUMA FE  & Y & Y & Y & Y \\  \hline \\[-1.8ex]  Observations & 168,385 & 168,385 & 168,385 & 168,385 \\  $R^2$ & 0.137 & 0.150 & 0.229 & 0.236 \\  Adjusted $R^2$ & 0.136 & 0.149 & 0.228 & 0.235 \\ \hline \hline \\[-1.8ex] \textit{Note:} & \multicolumn{4}{r}{$^{*}$p$<$0.1; $^{**}$p$<$0.05; $^{***}$p$<$0.01} \\ \end{tabular} \label{covidregla} \end{table}

\begin{table}[!htbp] \centering \caption[Linear regression models predicting the change in visits to POIs during the pandemic (2019 March - May) in Dallas] {Linear regression models predicting the change in visits to POIs during the pandemic (2019 March - May) in Dallas.} \begin{tabular}{@{\extracolsep{5pt}}lcccc} \\[-1.8ex]\hline \hline \\[-1.8ex] & \multicolumn{4}{c}{\textit{Dependent variable: $\Tilde v_i $ (Change in visits during the pandemic)}} \ \cr \cline{2-5} \\[-1.8ex] & (1) & (2) & (3) & (4) \\ \hline \\[-1.8ex]  Constant & 3.944$^{}$ & 1.362$^{}$ & -11.3$^{}$ & -12.6$^{}$ \\               & (10.25) & (9.730)     & (9.730)     & (9.671) \\  Distance effect $\sum_j \delta_{ij} \Tilde v_j $ & & 5.985$^{***}$ & & 4.427$^{***}$ \\   & & (0.146) & & (0.140) \\  Dependency effect $\sum_j w_{ij} \Tilde v_j $ && & 15.28$^{***}$ & 14.66$^{***}$ \\   & & & (0.160) & (0.160) \\ \hline \\[-1.8ex] Subcategory FE  & Y & Y & Y & Y  \\  PUMA FE  & Y & Y & Y & Y \\  \hline \\[-1.8ex]  Observations & 81,457 & 81,457 & 81,457 & 81,457 \\  $R^2$ & 0.140 & 0.158 & 0.227 & 0.236 \\  Adjusted $R^2$ & 0.139 & 0.156 & 0.225 & 0.235 \\ \hline \hline \\[-1.8ex] \textit{Note:} & \multicolumn{4}{r}{$^{*}$p$<$0.1; $^{**}$p$<$0.05; $^{***}$p$<$0.01} \\ \end{tabular} \label{covidregdallas} \end{table}

\subsection{Robustness against co-visit detection parameters}

\begin{figure}
\centering
\subfloat[Step difference parameter]{\includegraphics[width=\linewidth]{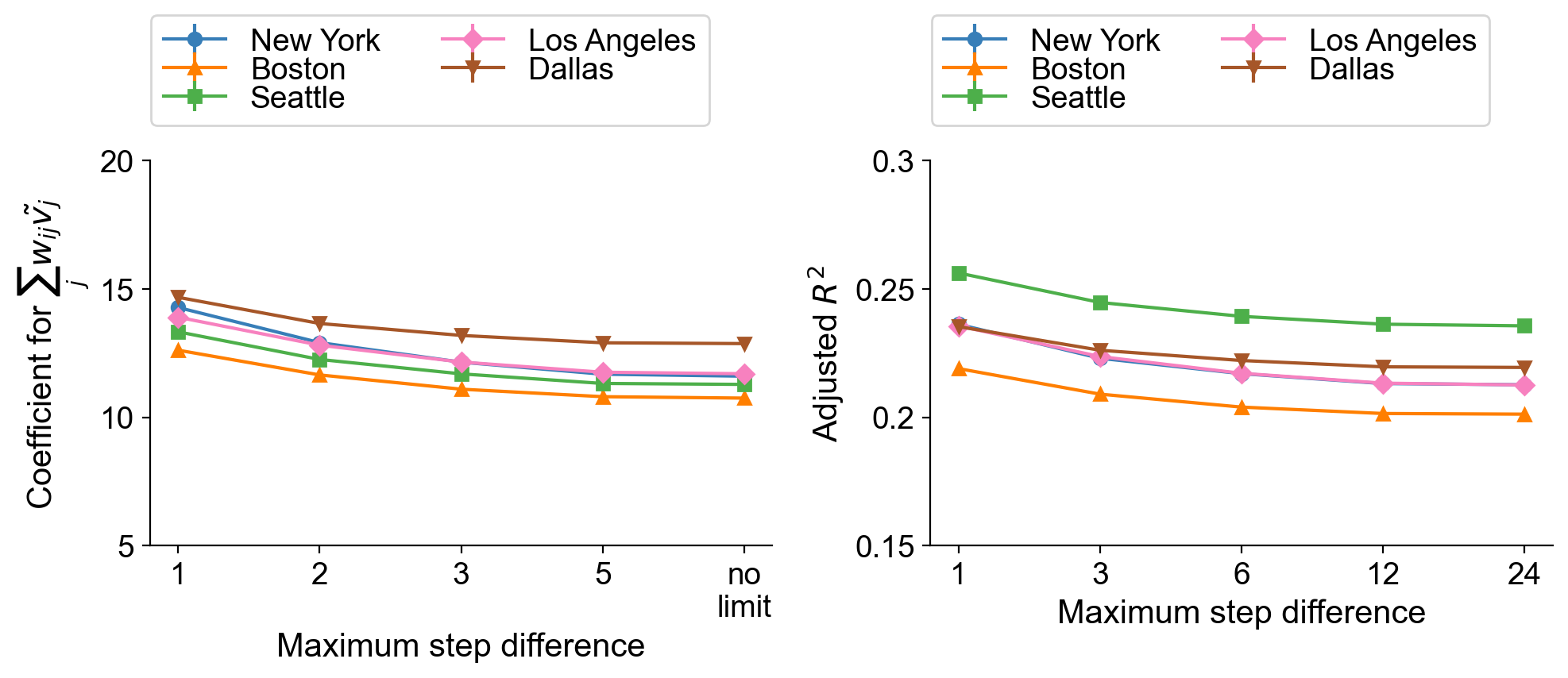}} \\
\subfloat[Time difference parameter]{\includegraphics[width=\linewidth]{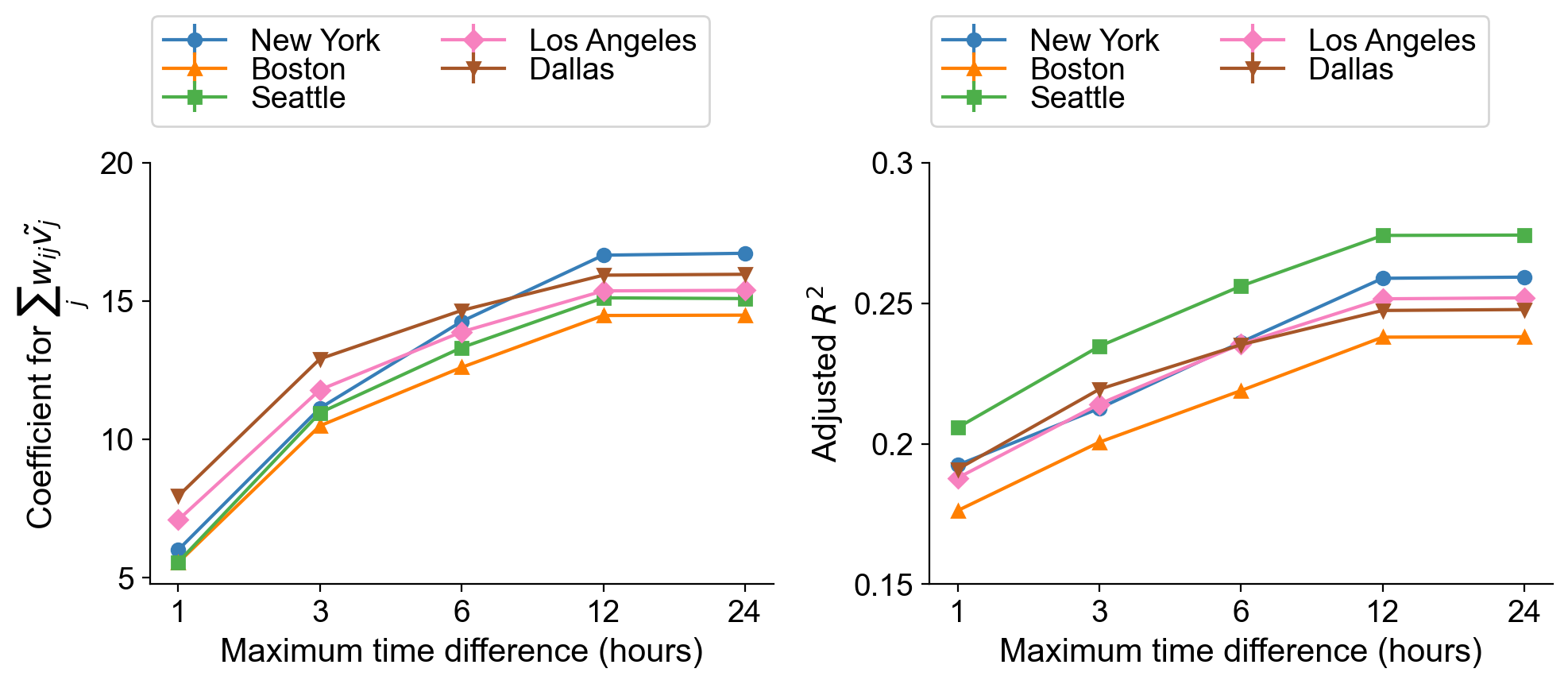}} \\
\caption[Regression results for dependency weights when using different co-visit detection parameters, contd.]{\textbf{Regression results for dependency weights when using different co-visit detection parameters, continued for Los Angeles and Dallas.} Except for the networks generated using maximum time difference of 1 hour, the results (adjusted $R^2$ and coefficient for Haversine distance) are relatively similar as the results in Table \ref{table:wijreg1}.}
\label{fig:s5covisitparams}
\end{figure}

To test the robustness of the prediction power of the models using the dependency-based network effects, we test the adjusted $R^2$ of the models using dependency networks generated using different parameters for the co-visit detection algorithm. 
Similar to Supplementary Note 4.3, we test the regression models under 25 pairs of parameters ($T_s = [1,2,3,5,unlimited]$ and $T_c = [1,3,6,12,24]$ hours). 
Figure \ref{fig:s5covisitparams} shows the adjusted $R^2$ values of the models using dependency networks generated with the corresponding co-visit detection parameters.
Across all five metropolitan areas, we observe the general trend of the performance of the models increasing with longer maximum time differences and with smaller number of between-visit steps. 
This suggests that, for detecting the impact of the COVID-19 pandemic on businesses, we need to observe 1-step (thus direct) visitation patterns between POIs, regardless of the interval time between the entry and exit.

% \begin{figure}
% \centering
% \subfloat[New York]{\includegraphics[width=0.4\linewidth]{ny_r2_regression_params.png}} 
% \hspace{0.1\textwidth}
% \subfloat[Boston]{\includegraphics[width=0.4\linewidth]{boston_r2_regression_params.png}} \\
% \subfloat[Seattle]{\includegraphics[width=0.4\linewidth]{seattle_r2_regression_params.png}}
% \hspace{0.1\textwidth}
% \subfloat[Los Angeles]{\includegraphics[width=0.4\linewidth]{la_r2_regression_params.png}} \\
% \subfloat[Dallas]{\includegraphics[width=0.4\linewidth]{dallas_r2_regression_params.png}}
% \caption[Regression model performance under different co-visit detection parameters]{\textbf{Regression model performance under different co-visit detection parameters.} The performance of the models increases with longer maximum time difference and with smaller number of between-visit steps. In all cities, $T_s=1$ steps and $T_c=24$ hours yields the best performance.}
% \label{fig:s5covisitparams}
% \end{figure}

\subsection{Results using changes in visits during different time periods}

To test whether the behavior-based dependency network effects significantly improves the prediction performance of visitation losses to POIs regardless of the time period, we test the same regression models with dependent variables calculated from different periods during the pandemic. 
Tables \ref{covidreg2ny} to \ref{covidreg2dallas} show the regression models for predicting the change of visits during 2020/6/1 and 2020/8/31, compared to the pre-pandemic period (2019/9/1 to 2019/11/30). Thus, the dependent variable is computed as:
\begin{equation}
    \Tilde v_i  = \big( \frac{v_i^{2020 June \sim 2020 August}}{v_i^{2019 September \sim 2019 November}} - 1 \big) * 100 (\%)
\end{equation}

In all metropolitan areas, the adjusted $R^2$ using the behavior-based dependency network effects (model 3) outperforms models 1 and 2 by 50\% to 100\%, showing substantial improvement in predictability. The regression coefficients for the behavior-based dependency network effects is significantly larger (30 fold) than the distance based effects, similar to the findings from the experiments in Tables \ref{covidregboston} to \ref{covidregdallas}. 

Similarly, Tables \ref{covidreg3ny} to \ref{covidreg3dallas} show the regression models for predicting the change of visits during 2020/9/1 and 2020/11/30, compared to the pre-pandemic period (2019/9/1 to 2019/11/30). 
In all metropolitan areas, the adjusted $R^2$ using the behavior-based dependency network effects (model 3) outperforms models 1 and 2 by 50\% to 100\%, showing substantial improvement in predictability. The regression coefficients for the behavior-based dependency network effects is significantly larger (30 fold) than the distance based effects, similar to the findings from the previous experiments.
Overall the magnitude of the $R^2$ across all 4 models are smaller than those in the previous experiments. This suggests that the dependency network observed during one year prior to the dependent variable loses predictive power due to structural changes in the urban form. 
Nevertheless, experiments using change in visits during different time periods during the pandemic showed that the behavior-based dependency network effects substantially improve predictability compared to using the distance-based network effects.

\begin{table}[!htbp] \centering \caption[Linear regression models predicting the change in visits to POIs during the pandemic (2019 June - August) in New York] {Linear regression models predicting the change in visits to POIs during the pandemic (2019 June - August) in New York.} \begin{tabular}{@{\extracolsep{5pt}}lcccc} \\[-1.8ex]\hline \hline \\[-1.8ex] & \multicolumn{4}{c}{\textit{Dependent variable: $\Tilde v_i $ (Change in visits during the pandemic)}} \ \cr \cline{2-5} \\[-1.8ex] & (1) & (2) & (3) & (4) \\ \hline \\[-1.8ex]  Constant & -20.6$^{}$ & -21.1$^{}$ & -15.8$^{}$ & -16.3$^{}$ \\               & (17.00) & (15.66)     & (15.66)     & (15.62) \\  Distance effect $\sum_j \delta_{ij} \Tilde v_j $ & & 26.41$^{***}$ & & 10.12$^{***}$ \\   & & (0.275) & & (0.275) \\  Dependency effect $\sum_j w_{ij} \Tilde v_j $ && & 57.97$^{***}$ & 53.96$^{***}$ \\   & & & (0.297) & (0.316) \\ \hline \\[-1.8ex] Subcategory FE  & Y & Y & Y & Y  \\  PUMA FE  & Y & Y & Y & Y \\  \hline \\[-1.8ex]  Observations & 213,140 & 213,140 & 213,140 & 213,140 \\  $R^2$ & 0.072 & 0.111 & 0.213 & 0.218 \\  Adjusted $R^2$ & 0.071 & 0.110 & 0.212 & 0.217 \\ \hline \hline \\[-1.8ex] \textit{Note:} & \multicolumn{4}{r}{$^{*}$p$<$0.1; $^{**}$p$<$0.05; $^{***}$p$<$0.01} \\ \end{tabular} \label{covidreg2ny} \end{table}

\begin{table}[!htbp] \centering \caption[Linear regression models predicting the change in visits to POIs during the pandemic (2019 June - August) in Boston] {Linear regression models predicting the change in visits to POIs during the pandemic (2019 June - August) in Boston.} \begin{tabular}{@{\extracolsep{5pt}}lcccc} \\[-1.8ex]\hline \hline \\[-1.8ex] & \multicolumn{4}{c}{\textit{Dependent variable: $\Tilde v_i $ (Change in visits during the pandemic)}} \ \cr \cline{2-5} \\[-1.8ex] & (1) & (2) & (3) & (4) \\ \hline \\[-1.8ex]  Constant & -19.6$^{}$ & -24.9$^{}$ & -33.7$^{*}$ & -35.2$^{**}$ \\               & (18.67) & (17.52)     & (17.52)     & (17.38) \\  Distance effect $\sum_j \delta_{ij} \Tilde v_j $ & & 25.83$^{***}$ & & 14.70$^{***}$ \\   & & (0.519) & & (0.521) \\  Dependency effect $\sum_j w_{ij} \Tilde v_j $ && & 44.40$^{***}$ & 39.43$^{***}$ \\   & & & (0.546) & (0.570) \\ \hline \\[-1.8ex] Subcategory FE  & Y & Y & Y & Y  \\  PUMA FE  & Y & Y & Y & Y \\  \hline \\[-1.8ex]  Observations & 48,848 & 48,848 & 48,848 & 48,848 \\  $R^2$ & 0.062 & 0.107 & 0.173 & 0.187 \\  Adjusted $R^2$ & 0.059 & 0.104 & 0.171 & 0.185 \\ \hline \hline \\[-1.8ex] \textit{Note:} & \multicolumn{4}{r}{$^{*}$p$<$0.1; $^{**}$p$<$0.05; $^{***}$p$<$0.01} \\ \end{tabular} \label{covidreg2boston} \end{table}

\begin{table}[!htbp] \centering \caption[Linear regression models predicting the change in visits to POIs during the pandemic (2019 June - August) in Seattle] {Linear regression models predicting the change in visits to POIs during the pandemic (2019 June - August) in Seattle.} \begin{tabular}{@{\extracolsep{5pt}}lcccc} \\[-1.8ex]\hline \hline \\[-1.8ex] & \multicolumn{4}{c}{\textit{Dependent variable: $\Tilde v_i $ (Change in visits during the pandemic)}} \ \cr \cline{2-5} \\[-1.8ex] & (1) & (2) & (3) & (4) \\ \hline \\[-1.8ex]  Constant & 3.507$^{}$ & 2.605$^{}$ & -1.16$^{}$ & -1.55$^{}$ \\               & (16.98) & (16.36)     & (16.36)     & (16.27) \\  Distance effect $\sum_j \delta_{ij} \Tilde v_j $ & & 8.777$^{***}$ & & 6.400$^{***}$ \\   & & (0.303) & & (0.296) \\  Dependency effect $\sum_j w_{ij} \Tilde v_j $ && & 19.09$^{***}$ & 17.98$^{***}$ \\   & & & (0.338) & (0.340) \\ \hline \\[-1.8ex] Subcategory FE  & Y & Y & Y & Y  \\  PUMA FE  & Y & Y & Y & Y \\  \hline \\[-1.8ex]  Observations & 41,451 & 41,451 & 41,451 & 41,451 \\  $R^2$ & 0.106 & 0.124 & 0.170 & 0.180 \\  Adjusted $R^2$ & 0.104 & 0.121 & 0.168 & 0.177 \\ \hline \hline \\[-1.8ex] \textit{Note:} & \multicolumn{4}{r}{$^{*}$p$<$0.1; $^{**}$p$<$0.05; $^{***}$p$<$0.01} \\ \end{tabular} \label{covidreg2seattle} \end{table}

\begin{table}[!htbp] \centering \caption[Linear regression models predicting the change in visits to POIs during the pandemic (2019 June - August) in Los Angeles] {Linear regression models predicting the change in visits to POIs during the pandemic (2019 June - August) in Los Angeles.} \begin{tabular}{@{\extracolsep{5pt}}lcccc} \\[-1.8ex]\hline \hline \\[-1.8ex] & \multicolumn{4}{c}{\textit{Dependent variable: $\Tilde v_i $ (Change in visits during the pandemic)}} \ \cr \cline{2-5} \\[-1.8ex] & (1) & (2) & (3) & (4) \\ \hline \\[-1.8ex]  Constant & 8.252$^{}$ & 7.685$^{}$ & 3.501$^{}$ & 3.390$^{}$ \\               & (7.296) & (6.885)     & (6.885)     & (6.857) \\  Distance effect $\sum_j \delta_{ij} \Tilde v_j $ & & 8.702$^{***}$ & & 5.212$^{***}$ \\   & & (0.145) & & (0.140) \\  Dependency effect $\sum_j w_{ij} \Tilde v_j $ && & 21.46$^{***}$ & 20.42$^{***}$ \\   & & & (0.149) & (0.151) \\ \hline \\[-1.8ex] Subcategory FE  & Y & Y & Y & Y  \\  PUMA FE  & Y & Y & Y & Y \\  \hline \\[-1.8ex]  Observations & 168,385 & 168,385 & 168,385 & 168,385 \\  $R^2$ & 0.080 & 0.099 & 0.180 & 0.187 \\  Adjusted $R^2$ & 0.079 & 0.098 & 0.179 & 0.186 \\ \hline \hline \\[-1.8ex] \textit{Note:} & \multicolumn{4}{r}{$^{*}$p$<$0.1; $^{**}$p$<$0.05; $^{***}$p$<$0.01} \\ \end{tabular} \label{covidreg2la} \end{table}

\begin{table}[!htbp] \centering \caption[Linear regression models predicting the change in visits to POIs during the pandemic (2019 June - August) in Dallas] {Linear regression models predicting the change in visits to POIs during the pandemic (2019 June - August) in Dallas.} \begin{tabular}{@{\extracolsep{5pt}}lcccc} \\[-1.8ex]\hline \hline \\[-1.8ex] & \multicolumn{4}{c}{\textit{Dependent variable: $\Tilde v_i $ (Change in visits during the pandemic)}} \ \cr \cline{2-5} \\[-1.8ex] & (1) & (2) & (3) & (4) \\ \hline \\[-1.8ex]  Constant & 38.63$^{**}$ & 33.88$^{*}$ & 12.24$^{}$ & 10.28$^{}$ \\               & (18.73) & (18.16)     & (18.16)     & (18.11) \\  Distance effect $\sum_j \delta_{ij} \Tilde v_j $ & & 8.472$^{***}$ & & 5.673$^{***}$ \\   & & (0.269) & & (0.265) \\  Dependency effect $\sum_j w_{ij} \Tilde v_j $ && & 20.41$^{***}$ & 19.47$^{***}$ \\   & & & (0.283) & (0.285) \\ \hline \\[-1.8ex] Subcategory FE  & Y & Y & Y & Y  \\  PUMA FE  & Y & Y & Y & Y \\  \hline \\[-1.8ex]  Observations & 81,457 & 81,457 & 81,457 & 81,457 \\  $R^2$ & 0.062 & 0.073 & 0.118 & 0.123 \\  Adjusted $R^2$ & 0.060 & 0.071 & 0.116 & 0.121 \\ \hline \hline \\[-1.8ex] \textit{Note:} & \multicolumn{4}{r}{$^{*}$p$<$0.1; $^{**}$p$<$0.05; $^{***}$p$<$0.01} \\ \end{tabular} \label{covidreg2dallas} \end{table}

\begin{table}[!htbp] \centering \caption[Linear regression models predicting the change in visits to POIs during the pandemic (2019 September - November) in New York] {Linear regression models predicting the change in visits to POIs during the pandemic (2019 September - November) in New York.} \begin{tabular}{@{\extracolsep{5pt}}lcccc} \\[-1.8ex]\hline \hline \\[-1.8ex] & \multicolumn{4}{c}{\textit{Dependent variable: $\Tilde v_i $ (Change in visits during the pandemic)}} \ \cr \cline{2-5} \\[-1.8ex] & (1) & (2) & (3) & (4) \\ \hline \\[-1.8ex]  Constant & -11.7$^{}$ & -12.4$^{}$ & -11.4$^{}$ & -11.9$^{}$ \\               & (9.700) & (9.413)     & (9.413)     & (9.382) \\  Distance effect $\sum_j \delta_{ij} \Tilde v_j $ & & 9.426$^{***}$ & & 6.109$^{***}$ \\   & & (0.163) & & (0.162) \\  Dependency effect $\sum_j w_{ij} \Tilde v_j $ && & 22.57$^{***}$ & 21.12$^{***}$ \\   & & & (0.196) & (0.199) \\ \hline \\[-1.8ex] Subcategory FE  & Y & Y & Y & Y  \\  PUMA FE  & Y & Y & Y & Y \\  \hline \\[-1.8ex]  Observations & 213,140 & 213,140 & 213,140 & 213,140 \\  $R^2$ & 0.081 & 0.095 & 0.135 & 0.140 \\  Adjusted $R^2$ & 0.080 & 0.094 & 0.134 & 0.139 \\ \hline \hline \\[-1.8ex] \textit{Note:} & \multicolumn{4}{r}{$^{*}$p$<$0.1; $^{**}$p$<$0.05; $^{***}$p$<$0.01} \\ \end{tabular} \label{covidreg3ny} \end{table}

\begin{table}[!htbp] \centering \caption[Linear regression models predicting the change in visits to POIs during the pandemic (2019 September - November) in Boston] {Linear regression models predicting the change in visits to POIs during the pandemic (2019 September - November) in Boston.} \begin{tabular}{@{\extracolsep{5pt}}lcccc} \\[-1.8ex]\hline \hline \\[-1.8ex] & \multicolumn{4}{c}{\textit{Dependent variable: $\Tilde v_i $ (Change in visits during the pandemic)}} \ \cr \cline{2-5} \\[-1.8ex] & (1) & (2) & (3) & (4) \\ \hline \\[-1.8ex]  Constant & -13.4$^{}$ & -15.8$^{}$ & -19.8$^{}$ & -21.2$^{}$ \\               & (15.04) & (14.94)     & (14.94)     & (14.90) \\  Distance effect $\sum_j \delta_{ij} \Tilde v_j $ & & 8.396$^{***}$ & & 6.962$^{***}$ \\   & & (0.427) & & (0.429) \\  Dependency effect $\sum_j w_{ij} \Tilde v_j $ && & 11.95$^{***}$ & 10.84$^{***}$ \\   & & & (0.469) & (0.472) \\ \hline \\[-1.8ex] Subcategory FE  & Y & Y & Y & Y  \\  PUMA FE  & Y & Y & Y & Y \\  \hline \\[-1.8ex]  Observations & 48,848 & 48,848 & 48,848 & 48,848 \\  $R^2$ & 0.081 & 0.088 & 0.093 & 0.097 \\  Adjusted $R^2$ & 0.078 & 0.085 & 0.090 & 0.095 \\ \hline \hline \\[-1.8ex] \textit{Note:} & \multicolumn{4}{r}{$^{*}$p$<$0.1; $^{**}$p$<$0.05; $^{***}$p$<$0.01} \\ \end{tabular} \label{covidreg3boston} \end{table}

\begin{table}[!htbp] \centering \caption[Linear regression models predicting the change in visits to POIs during the pandemic (2019 September - November) in Seattle] {Linear regression models predicting the change in visits to POIs during the pandemic (2019 September - November) in Seattle.} \begin{tabular}{@{\extracolsep{5pt}}lcccc} \\[-1.8ex]\hline \hline \\[-1.8ex] & \multicolumn{4}{c}{\textit{Dependent variable: $\Tilde v_i $ (Change in visits during the pandemic)}} \ \cr \cline{2-5} \\[-1.8ex] & (1) & (2) & (3) & (4) \\ \hline \\[-1.8ex]  Constant & 8.718$^{}$ & 7.900$^{}$ & 4.898$^{}$ & 4.468$^{}$ \\               & (16.05) & (15.56)     & (15.56)     & (15.48) \\  Distance effect $\sum_j \delta_{ij} \Tilde v_j $ & & 7.389$^{***}$ & & 5.628$^{***}$ \\   & & (0.286) & & (0.281) \\  Dependency effect $\sum_j w_{ij} \Tilde v_j $ && & 16.64$^{***}$ & 15.80$^{***}$ \\   & & & (0.323) & (0.324) \\ \hline \\[-1.8ex] Subcategory FE  & Y & Y & Y & Y  \\  PUMA FE  & Y & Y & Y & Y \\  \hline \\[-1.8ex]  Observations & 41,451 & 41,451 & 41,451 & 41,451 \\  $R^2$ & 0.071 & 0.086 & 0.127 & 0.136 \\  Adjusted $R^2$ & 0.069 & 0.083 & 0.125 & 0.133 \\ \hline \hline \\[-1.8ex] \textit{Note:} & \multicolumn{4}{r}{$^{*}$p$<$0.1; $^{**}$p$<$0.05; $^{***}$p$<$0.01} \\ \end{tabular} \label{covidreg3seattle} \end{table}

\begin{table}[!htbp] \centering \caption[Linear regression models predicting the change in visits to POIs during the pandemic (2019 September - November) in Los Angeles] {Linear regression models predicting the change in visits to POIs during the pandemic (2019 September - November) in Los Angeles.} \begin{tabular}{@{\extracolsep{5pt}}lcccc} \\[-1.8ex]\hline \hline \\[-1.8ex] & \multicolumn{4}{c}{\textit{Dependent variable: $\Tilde v_i $ (Change in visits during the pandemic)}} \ \cr \cline{2-5} \\[-1.8ex] & (1) & (2) & (3) & (4) \\ \hline \\[-1.8ex]  Constant & 9.110$^{}$ & 8.471$^{}$ & 4.076$^{}$ & 3.920$^{}$ \\               & (7.274) & (7.055)     & (7.055)     & (7.024) \\  Distance effect $\sum_j \delta_{ij} \Tilde v_j $ & & 7.655$^{***}$ & & 5.468$^{***}$ \\   & & (0.144) & & (0.142) \\  Dependency effect $\sum_j w_{ij} \Tilde v_j $ && & 15.62$^{***}$ & 14.69$^{***}$ \\   & & & (0.151) & (0.152) \\ \hline \\[-1.8ex] Subcategory FE  & Y & Y & Y & Y  \\  PUMA FE  & Y & Y & Y & Y \\  \hline \\[-1.8ex]  Observations & 168,385 & 168,385 & 168,385 & 168,385 \\  $R^2$ & 0.053 & 0.069 & 0.109 & 0.117 \\  Adjusted $R^2$ & 0.052 & 0.068 & 0.108 & 0.116 \\ \hline \hline \\[-1.8ex] \textit{Note:} & \multicolumn{4}{r}{$^{*}$p$<$0.1; $^{**}$p$<$0.05; $^{***}$p$<$0.01} \\ \end{tabular} \label{covidreg3la} \end{table}

\begin{table}[!htbp] \centering \caption[Linear regression models predicting the change in visits to POIs during the pandemic (2019 September - November) in Dallas] {Linear regression models predicting the change in visits to POIs during the pandemic (2019 September - November) in Dallas.} \begin{tabular}{@{\extracolsep{5pt}}lcccc} \\[-1.8ex]\hline \hline \\[-1.8ex] & \multicolumn{4}{c}{\textit{Dependent variable: $\Tilde v_i $ (Change in visits during the pandemic)}} \ \cr \cline{2-5} \\[-1.8ex] & (1) & (2) & (3) & (4) \\ \hline \\[-1.8ex]  Constant & 17.86$^{}$ & 14.28$^{}$ & -4.65$^{}$ & -6.44$^{}$ \\               & (19.67) & (19.18)     & (19.18)     & (19.14) \\  Distance effect $\sum_j \delta_{ij} \Tilde v_j $ & & 7.407$^{***}$ & & 5.340$^{***}$ \\   & & (0.282) & & (0.278) \\  Dependency effect $\sum_j w_{ij} \Tilde v_j $ && & 18.95$^{***}$ & 18.28$^{***}$ \\   & & & (0.292) & (0.293) \\ \hline \\[-1.8ex] Subcategory FE  & Y & Y & Y & Y  \\  PUMA FE  & Y & Y & Y & Y \\  \hline \\[-1.8ex]  Observations & 81,457 & 81,457 & 81,457 & 81,457 \\  $R^2$ & 0.037 & 0.045 & 0.084 & 0.089 \\  Adjusted $R^2$ & 0.035 & 0.043 & 0.083 & 0.087 \\ \hline \hline \\[-1.8ex] \textit{Note:} & \multicolumn{4}{r}{$^{*}$p$<$0.1; $^{**}$p$<$0.05; $^{***}$p$<$0.01} \\ \end{tabular} \label{covidreg3dallas} \end{table}

\subsection{Robustness of regression results when using different time periods to generate dependency networks}

The robustness checks conducted in Supplementary Notes 3.1 and 4.4 showed high correlation in edge weights between different time periods. 
Nevertheless, in this section, we tested whether the time periods used to generate the behavior-based dependency networks affects the performance of the regression models. 
Tables \ref{covidreg4ny} to \ref{covidreg4dallas} show the regression models for predicting the change of visits during 2020/3/1 and 2020/5/31, using the dependency network computed from data between 2019/1/1 and 2019/4/30. 

In all metropolitan areas, the adjusted $R^2$ using the behavior-based dependency network effects (model 3) outperforms models 1 and 2 by around 50\%, showing substantial improvement in predictability. The regression coefficients for the behavior-based dependency network effects is significantly larger (around 5 fold) than the distance based effects, similar to the findings from the previous experiments. 

Similarly, Tables \ref{covidreg5ny} to \ref{covidreg5dallas} show the regression models for predicting the change of visits during 2020/3/1 and 2020/5/31, using the dependency network computed from data between 2019/5/1 and 2019/8/31. 
Again, in all metropolitan areas, the adjusted $R^2$ using the behavior-based dependency network effects (model 3) outperforms models 1 and 2 by around 50\%, showing the robustness against the choice of time period to generated the dependency networks. 
In summary, the results on the predictability of shocks using the behavior-based dependency networks are robust against the choice of the time periods used to generate the dependency networks.

\begin{table}[!htbp] \centering \caption[Linear regression models predicting the change in visits to POIs during the pandemic (2019 September - November) in New York when using dependency network generated from data between 2019 January to April] {Linear regression models predicting the change in visits to POIs during the pandemic (2019 September - November) in New York when using dependency network generated from data between 2019 January to April.} \begin{tabular}{@{\extracolsep{5pt}}lcccc} \\[-1.8ex]\hline \hline \\[-1.8ex] & \multicolumn{4}{c}{\textit{Dependent variable: $\Tilde v_i $ (Change in visits during the pandemic)}} \ \cr \cline{2-5} \\[-1.8ex] & (1) & (2) & (3) & (4) \\ \hline \\[-1.8ex]  Constant & -38.3$^{***}$ & -38.9$^{***}$ & -40.3$^{***}$ & -40.8$^{***}$ \\               & (5.251) & (5.033)     & (5.033)     & (5.009) \\  Distance effect $\sum_j \delta_{ij} \Tilde v_j $ & & 4.089$^{***}$ & & 3.614$^{***}$ \\   & & (0.084) & & (0.081) \\  Dependency effect $\sum_j w_{ij} \Tilde v_j $ && & 13.63$^{***}$ & 13.44$^{***}$ \\   & & & (0.100) & (0.100) \\ \hline \\[-1.8ex] Subcategory FE  & Y & Y & Y & Y  \\  PUMA FE  & Y & Y & Y & Y \\  \hline \\[-1.8ex]  Observations & 206,108 & 206,108 & 206,108 & 206,108 \\  $R^2$ & 0.166 & 0.175 & 0.234 & 0.241 \\  Adjusted $R^2$ & 0.165 & 0.174 & 0.233 & 0.240 \\ \hline \hline \\[-1.8ex] \textit{Note:} & \multicolumn{4}{r}{$^{*}$p$<$0.1; $^{**}$p$<$0.05; $^{***}$p$<$0.01} \\ \end{tabular} \label{covidreg4ny} \end{table}

\begin{table}[!htbp] \centering \caption[Linear regression models predicting the change in visits to POIs during the pandemic (2019 September - November) in Boston when using dependency network generated from data between 2019 January to April] {Linear regression models predicting the change in visits to POIs during the pandemic (2019 September - November) in Boston when using dependency network generated from data between 2019 January to April.} \begin{tabular}{@{\extracolsep{5pt}}lcccc} \\[-1.8ex]\hline \hline \\[-1.8ex] & \multicolumn{4}{c}{\textit{Dependent variable: $\Tilde v_i $ (Change in visits during the pandemic)}} \ \cr \cline{2-5} \\[-1.8ex] & (1) & (2) & (3) & (4) \\ \hline \\[-1.8ex]  Constant & -22.0$^{***}$ & -24.0$^{***}$ & -33.1$^{***}$ & -34.6$^{***}$ \\               & (6.518) & (6.277)     & (6.277)     & (6.242) \\  Distance effect $\sum_j \delta_{ij} \Tilde v_j $ & & 4.779$^{***}$ & & 4.102$^{***}$ \\   & & (0.185) & & (0.178) \\  Dependency effect $\sum_j w_{ij} \Tilde v_j $ && & 12.30$^{***}$ & 12.00$^{***}$ \\   & & & (0.202) & (0.201) \\ \hline \\[-1.8ex] Subcategory FE  & Y & Y & Y & Y  \\  PUMA FE  & Y & Y & Y & Y \\  \hline \\[-1.8ex]  Observations & 46,643 & 46,643 & 46,643 & 46,643 \\  $R^2$ & 0.147 & 0.159 & 0.210 & 0.219 \\  Adjusted $R^2$ & 0.145 & 0.157 & 0.208 & 0.217 \\ \hline \hline \\[-1.8ex] \textit{Note:} & \multicolumn{4}{r}{$^{*}$p$<$0.1; $^{**}$p$<$0.05; $^{***}$p$<$0.01} \\ \end{tabular} \label{covidreg4boston} \end{table}

\begin{table}[!htbp] \centering \caption[Linear regression models predicting the change in visits to POIs during the pandemic (2019 September - November) in Seattle when using dependency network generated from data between 2019 January to April] {Linear regression models predicting the change in visits to POIs during the pandemic (2019 September - November) in Seattle when using dependency network generated from data between 2019 January to April.} \begin{tabular}{@{\extracolsep{5pt}}lcccc} \\[-1.8ex]\hline \hline \\[-1.8ex] & \multicolumn{4}{c}{\textit{Dependent variable: $\Tilde v_i $ (Change in visits during the pandemic)}} \ \cr \cline{2-5} \\[-1.8ex] & (1) & (2) & (3) & (4) \\ \hline \\[-1.8ex]  Constant & -27.2$^{**}$ & -27.5$^{**}$ & -30.1$^{***}$ & -30.3$^{***}$ \\               & (11.03) & (10.52)     & (10.52)     & (10.46) \\  Distance effect $\sum_j \delta_{ij} \Tilde v_j $ & & 4.856$^{***}$ & & 3.937$^{***}$ \\   & & (0.191) & & (0.183) \\  Dependency effect $\sum_j w_{ij} \Tilde v_j $ && & 13.12$^{***}$ & 12.75$^{***}$ \\   & & & (0.211) & (0.210) \\ \hline \\[-1.8ex] Subcategory FE  & Y & Y & Y & Y  \\  PUMA FE  & Y & Y & Y & Y \\  \hline \\[-1.8ex]  Observations & 39,582 & 39,582 & 39,582 & 39,582 \\  $R^2$ & 0.173 & 0.186 & 0.247 & 0.256 \\  Adjusted $R^2$ & 0.171 & 0.184 & 0.245 & 0.253 \\ \hline \hline \\[-1.8ex] \textit{Note:} & \multicolumn{4}{r}{$^{*}$p$<$0.1; $^{**}$p$<$0.05; $^{***}$p$<$0.01} \\ \end{tabular} \label{covidreg4seattle} \end{table}

\begin{table}[!htbp] \centering \caption[Linear regression models predicting the change in visits to POIs during the pandemic (2019 September - November) in Los Angeles when using dependency network generated from data between 2019 January to April] {Linear regression models predicting the change in visits to POIs during the pandemic (2019 September - November) in Los Angeles when using dependency network generated from data between 2019 January to April.} \begin{tabular}{@{\extracolsep{5pt}}lcccc} \\[-1.8ex]\hline \hline \\[-1.8ex] & \multicolumn{4}{c}{\textit{Dependent variable: $\Tilde v_i $ (Change in visits during the pandemic)}} \ \cr \cline{2-5} \\[-1.8ex] & (1) & (2) & (3) & (4) \\ \hline \\[-1.8ex]  Constant & -18.6$^{***}$ & -19.3$^{***}$ & -20.9$^{***}$ & -21.3$^{***}$ \\               & (4.828) & (4.563)     & (4.563)     & (4.542) \\  Distance effect $\sum_j \delta_{ij} \Tilde v_j $ & & 4.651$^{***}$ & & 3.561$^{***}$ \\   & & (0.097) & & (0.092) \\  Dependency effect $\sum_j w_{ij} \Tilde v_j $ && & 14.26$^{***}$ & 13.91$^{***}$ \\   & & & (0.102) & (0.102) \\ \hline \\[-1.8ex] Subcategory FE  & Y & Y & Y & Y  \\  PUMA FE  & Y & Y & Y & Y \\  \hline \\[-1.8ex]  Observations & 162,861 & 162,861 & 162,861 & 162,861 \\  $R^2$ & 0.138 & 0.150 & 0.230 & 0.237 \\  Adjusted $R^2$ & 0.137 & 0.149 & 0.229 & 0.236 \\ \hline \hline \\[-1.8ex] \textit{Note:} & \multicolumn{4}{r}{$^{*}$p$<$0.1; $^{**}$p$<$0.05; $^{***}$p$<$0.01} \\ \end{tabular} \label{covidreg4la} \end{table}

\begin{table}[!htbp] \centering \caption[Linear regression models predicting the change in visits to POIs during the pandemic (2019 September - November) in Dallas when using dependency network generated from data between 2019 January to April] {Linear regression models predicting the change in visits to POIs during the pandemic (2019 September - November) in Dallas when using dependency network generated from data between 2019 January to April.} \begin{tabular}{@{\extracolsep{5pt}}lcccc} \\[-1.8ex]\hline \hline \\[-1.8ex] & \multicolumn{4}{c}{\textit{Dependent variable: $\Tilde v_i $ (Change in visits during the pandemic)}} \ \cr \cline{2-5} \\[-1.8ex] & (1) & (2) & (3) & (4) \\ \hline \\[-1.8ex]  Constant & 7.027$^{}$ & 4.521$^{}$ & -8.57$^{}$ & -9.93$^{}$ \\               & (10.15) & (9.647)     & (9.647)     & (9.585) \\  Distance effect $\sum_j \delta_{ij} \Tilde v_j $ & & 5.827$^{***}$ & & 4.493$^{***}$ \\   & & (0.147) & & (0.141) \\  Dependency effect $\sum_j w_{ij} \Tilde v_j $ && & 14.91$^{***}$ & 14.37$^{***}$ \\   & & & (0.162) & (0.161) \\ \hline \\[-1.8ex] Subcategory FE  & Y & Y & Y & Y  \\  PUMA FE  & Y & Y & Y & Y \\  \hline \\[-1.8ex]  Observations & 78,222 & 78,222 & 78,222 & 78,222 \\  $R^2$ & 0.145 & 0.161 & 0.228 & 0.238 \\  Adjusted $R^2$ & 0.143 & 0.160 & 0.227 & 0.237 \\ \hline \hline \\[-1.8ex] \textit{Note:} & \multicolumn{4}{r}{$^{*}$p$<$0.1; $^{**}$p$<$0.05; $^{***}$p$<$0.01} \\ \end{tabular} \label{covidreg4dallas} \end{table}

% ---- 

\begin{table}[!htbp] \centering \caption[Linear regression models predicting the change in visits to POIs during the pandemic (2019 September - November) in New York when using dependency network generated from data between 2019 May to August] {Linear regression models predicting the change in visits to POIs during the pandemic (2019 September - November) in New York when using dependency network generated from data between 2019 May to August.} \begin{tabular}{@{\extracolsep{5pt}}lcccc} \\[-1.8ex]\hline \hline \\[-1.8ex] & \multicolumn{4}{c}{\textit{Dependent variable: $\Tilde v_i $ (Change in visits during the pandemic)}} \ \cr \cline{2-5} \\[-1.8ex] & (1) & (2) & (3) & (4) \\ \hline \\[-1.8ex]  Constant & -38.2$^{***}$ & -38.7$^{***}$ & -40.5$^{***}$ & -40.9$^{***}$ \\               & (5.249) & (5.016)     & (5.016)     & (5.002) \\  Distance effect $\sum_j \delta_{ij} \Tilde v_j $ & & 3.398$^{***}$ & & 2.733$^{***}$ \\   & & (0.085) & & (0.081) \\  Dependency effect $\sum_j w_{ij} \Tilde v_j $ && & 14.70$^{***}$ & 14.50$^{***}$ \\   & & & (0.104) & (0.103) \\ \hline \\[-1.8ex] Subcategory FE  & Y & Y & Y & Y  \\  PUMA FE  & Y & Y & Y & Y \\  \hline \\[-1.8ex]  Observations & 210,540 & 210,540 & 210,540 & 210,540 \\  $R^2$ & 0.164 & 0.170 & 0.236 & 0.240 \\  Adjusted $R^2$ & 0.163 & 0.169 & 0.235 & 0.240 \\ \hline \hline \\[-1.8ex] \textit{Note:} & \multicolumn{4}{r}{$^{*}$p$<$0.1; $^{**}$p$<$0.05; $^{***}$p$<$0.01} \\ \end{tabular} \label{covidreg5ny} \end{table}

\begin{table}[!htbp] \centering \caption[Linear regression models predicting the change in visits to POIs during the pandemic (2019 September - November) in Boston when using dependency network generated from data between 2019 May to August] {Linear regression models predicting the change in visits to POIs during the pandemic (2019 September - November) in Boston when using dependency network generated from data between 2019 May to August.} \begin{tabular}{@{\extracolsep{5pt}}lcccc} \\[-1.8ex]\hline \hline \\[-1.8ex] & \multicolumn{4}{c}{\textit{Dependent variable: $\Tilde v_i $ (Change in visits during the pandemic)}} \ \cr \cline{2-5} \\[-1.8ex] & (1) & (2) & (3) & (4) \\ \hline \\[-1.8ex]  Constant & -19.8$^{***}$ & -21.4$^{***}$ & -33.8$^{***}$ & -34.9$^{***}$ \\               & (6.580) & (6.324)     & (6.324)     & (6.302) \\  Distance effect $\sum_j \delta_{ij} \Tilde v_j $ & & 3.816$^{***}$ & & 3.244$^{***}$ \\   & & (0.185) & & (0.178) \\  Dependency effect $\sum_j w_{ij} \Tilde v_j $ && & 13.20$^{***}$ & 13.00$^{***}$ \\   & & & (0.208) & (0.207) \\ \hline \\[-1.8ex] Subcategory FE  & Y & Y & Y & Y  \\  PUMA FE  & Y & Y & Y & Y \\  \hline \\[-1.8ex]  Observations & 47,995 & 47,995 & 47,995 & 47,995 \\  $R^2$ & 0.146 & 0.153 & 0.212 & 0.217 \\  Adjusted $R^2$ & 0.144 & 0.151 & 0.210 & 0.215 \\ \hline \hline \\[-1.8ex] \textit{Note:} & \multicolumn{4}{r}{$^{*}$p$<$0.1; $^{**}$p$<$0.05; $^{***}$p$<$0.01} \\ \end{tabular} \label{covidreg5boston} \end{table}

\begin{table}[!htbp] \centering \caption[Linear regression models predicting the change in visits to POIs during the pandemic (2019 September - November) in Seattle when using dependency network generated from data between 2019 May to August] {Linear regression models predicting the change in visits to POIs during the pandemic (2019 September - November) in Seattle when using dependency network generated from data between 2019 May to August.} \begin{tabular}{@{\extracolsep{5pt}}lcccc} \\[-1.8ex]\hline \hline \\[-1.8ex] & \multicolumn{4}{c}{\textit{Dependent variable: $\Tilde v_i $ (Change in visits during the pandemic)}} \ \cr \cline{2-5} \\[-1.8ex] & (1) & (2) & (3) & (4) \\ \hline \\[-1.8ex]  Constant & -21.9$^{**}$ & -22.5$^{**}$ & -26.3$^{***}$ & -26.6$^{***}$ \\               & (10.61) & (10.07)     & (10.07)     & (10.03) \\  Distance effect $\sum_j \delta_{ij} \Tilde v_j $ & & 4.667$^{***}$ & & 3.622$^{***}$ \\   & & (0.190) & & (0.181) \\  Dependency effect $\sum_j w_{ij} \Tilde v_j $ && & 14.01$^{***}$ & 13.64$^{***}$ \\   & & & (0.210) & (0.210) \\ \hline \\[-1.8ex] Subcategory FE  & Y & Y & Y & Y  \\  PUMA FE  & Y & Y & Y & Y \\  \hline \\[-1.8ex]  Observations & 40,800 & 40,800 & 40,800 & 40,800 \\  $R^2$ & 0.172 & 0.184 & 0.253 & 0.260 \\  Adjusted $R^2$ & 0.169 & 0.181 & 0.251 & 0.258 \\ \hline \hline \\[-1.8ex] \textit{Note:} & \multicolumn{4}{r}{$^{*}$p$<$0.1; $^{**}$p$<$0.05; $^{***}$p$<$0.01} \\ \end{tabular} \label{covidreg5seattle} \end{table}

\begin{table}[!htbp] \centering \caption[Linear regression models predicting the change in visits to POIs during the pandemic (2019 September - November) in Los Angeles when using dependency network generated from data between 2019 May to August] {Linear regression models predicting the change in visits to POIs during the pandemic (2019 September - November) in Los Angeles when using dependency network generated from data between 2019 May to August.} \begin{tabular}{@{\extracolsep{5pt}}lcccc} \\[-1.8ex]\hline \hline \\[-1.8ex] & \multicolumn{4}{c}{\textit{Dependent variable: $\Tilde v_i $ (Change in visits during the pandemic)}} \ \cr \cline{2-5} \\[-1.8ex] & (1) & (2) & (3) & (4) \\ \hline \\[-1.8ex]  Constant & -17.6$^{***}$ & -17.8$^{***}$ & -20.5$^{***}$ & -20.6$^{***}$ \\               & (4.839) & (4.561)     & (4.561)     & (4.543) \\  Distance effect $\sum_j \delta_{ij} \Tilde v_j $ & & 4.597$^{***}$ & & 3.372$^{***}$ \\   & & (0.096) & & (0.091) \\  Dependency effect $\sum_j w_{ij} \Tilde v_j $ && & 14.72$^{***}$ & 14.37$^{***}$ \\   & & & (0.102) & (0.102) \\ \hline \\[-1.8ex] Subcategory FE  & Y & Y & Y & Y  \\  PUMA FE  & Y & Y & Y & Y \\  \hline \\[-1.8ex]  Observations & 166,292 & 166,292 & 166,292 & 166,292 \\  $R^2$ & 0.136 & 0.148 & 0.233 & 0.239 \\  Adjusted $R^2$ & 0.135 & 0.147 & 0.232 & 0.238 \\ \hline \hline \\[-1.8ex] \textit{Note:} & \multicolumn{4}{r}{$^{*}$p$<$0.1; $^{**}$p$<$0.05; $^{***}$p$<$0.01} \\ \end{tabular} \label{covidreg5la} \end{table}

\begin{table}[!htbp] \centering \caption[Linear regression models predicting the change in visits to POIs during the pandemic (2019 September - November) in Dallas when using dependency network generated from data between 2019 May to August] {Linear regression models predicting the change in visits to POIs during the pandemic (2019 September - November) in Dallas when using dependency network generated from data between 2019 May to August.} \begin{tabular}{@{\extracolsep{5pt}}lcccc} \\[-1.8ex]\hline \hline \\[-1.8ex] & \multicolumn{4}{c}{\textit{Dependent variable: $\Tilde v_i $ (Change in visits during the pandemic)}} \ \cr \cline{2-5} \\[-1.8ex] & (1) & (2) & (3) & (4) \\ \hline \\[-1.8ex]  Constant & 8.334$^{}$ & 5.849$^{}$ & -8.60$^{}$ & -9.90$^{}$ \\               & (10.10) & (9.581)     & (9.581)     & (9.530) \\  Distance effect $\sum_j \delta_{ij} \Tilde v_j $ & & 5.621$^{***}$ & & 4.216$^{***}$ \\   & & (0.149) & & (0.143) \\  Dependency effect $\sum_j w_{ij} \Tilde v_j $ && & 15.68$^{***}$ & 15.16$^{***}$ \\   & & & (0.164) & (0.165) \\ \hline \\[-1.8ex] Subcategory FE  & Y & Y & Y & Y  \\  PUMA FE  & Y & Y & Y & Y \\  \hline \\[-1.8ex]  Observations & 80,431 & 80,431 & 80,431 & 80,431 \\  $R^2$ & 0.141 & 0.156 & 0.228 & 0.236 \\  Adjusted $R^2$ & 0.140 & 0.154 & 0.227 & 0.235 \\ \hline \hline \\[-1.8ex] \textit{Note:} & \multicolumn{4}{r}{$^{*}$p$<$0.1; $^{**}$p$<$0.05; $^{***}$p$<$0.01} \\ \end{tabular} \label{covidreg5dallas} \end{table}

\subsection{Robustness of regression results when using only short non-work stays}

In this section, we test whether the behavior-based dependency network generated using only short (i.e., non-work) stays yields similar performance metrics in the regression models. 
Tables \ref{covidreg6ny} to \ref{covidreg6dallas} show the regression models for predicting the change of visits during 2020/3/1 and 2020/5/31, using the dependency network computed with stays shorter than 4 hours. 
In all metropolitan areas, the adjusted $R^2$ using the behavior-based dependency network effects (model 3) outperforms models 1 and 2, showing improvement in predictability. The regression coefficients for the behavior-based dependency network effects are significantly larger (around 5 fold) than the distance based effects, similar to the findings from the previous experiments. 
This results suggest that our estimation results are robust against the limiting the analysis to the use of short non-work trips.

\begin{table}[!htbp] \centering \caption[Linear regression models predicting the change in visits to POIs during the pandemic (2019 September - November) in New York when using dependency network generated from stays shorter than 4 hours] {Linear regression models predicting the change in visits to POIs during the pandemic (2019 September - November) in New York when using dependency network generated from stays shorter than 4 hours.} \begin{tabular}{@{\extracolsep{5pt}}lcccc} \\[-1.8ex]\hline \hline \\[-1.8ex] & \multicolumn{4}{c}{\textit{Dependent variable: $\Tilde v_i $ (Change in visits during the pandemic)}} \ \cr \cline{2-5} \\[-1.8ex] & (1) & (2) & (3) & (4) \\ \hline \\[-1.8ex]  Constant & -41.1$^{***}$ & -41.8$^{***}$ & -42.7$^{***}$ & -43.2$^{***}$ \\               & (5.230) & (4.943)     & (4.943)     & (4.928) \\  Distance effect $\sum_j \delta_{ij} \Tilde v_j $ & & 4.217$^{***}$ & & 2.896$^{***}$ \\   & & (0.084) & & (0.080) \\  Dependency effect $\sum_j w_{ij} \Tilde v_j $ && & 16.66$^{***}$ & 16.26$^{***}$ \\   & & & (0.105) & (0.106) \\ \hline \\[-1.8ex] Subcategory FE  & Y & Y & Y & Y  \\  PUMA FE  & Y & Y & Y & Y \\  \hline \\[-1.8ex]  Observations & 208,329 & 208,329 & 208,329 & 208,329 \\  $R^2$ & 0.165 & 0.174 & 0.254 & 0.258 \\  Adjusted $R^2$ & 0.164 & 0.173 & 0.253 & 0.257 \\ \hline \hline \\[-1.8ex] \textit{Note:} & \multicolumn{4}{r}{$^{*}$p$<$0.1; $^{**}$p$<$0.05; $^{***}$p$<$0.01} \\ \end{tabular} \label{covidreg6ny} \end{table}

\begin{table}[!htbp] \centering \caption[Linear regression models predicting the change in visits to POIs during the pandemic (2019 September - November) in Boston when using dependency network generated from stays shorter than 4 hours] {Linear regression models predicting the change in visits to POIs during the pandemic (2019 September - November) in Boston when using dependency network generated from stays shorter than 4 hours.} \begin{tabular}{@{\extracolsep{5pt}}lcccc} \\[-1.8ex]\hline \hline \\[-1.8ex] & \multicolumn{4}{c}{\textit{Dependent variable: $\Tilde v_i $ (Change in visits during the pandemic)}} \ \cr \cline{2-5} \\[-1.8ex] & (1) & (2) & (3) & (4) \\ \hline \\[-1.8ex]  Constant & -20.6$^{***}$ & -22.6$^{***}$ & -36.6$^{***}$ & -37.5$^{***}$ \\               & (6.562) & (6.261)     & (6.261)     & (6.236) \\  Distance effect $\sum_j \delta_{ij} \Tilde v_j $ & & 4.844$^{***}$ & & 3.456$^{***}$ \\   & & (0.183) & & (0.176) \\  Dependency effect $\sum_j w_{ij} \Tilde v_j $ && & 14.16$^{***}$ & 13.68$^{***}$ \\   & & & (0.205) & (0.206) \\ \hline \\[-1.8ex] Subcategory FE  & Y & Y & Y & Y  \\  PUMA FE  & Y & Y & Y & Y \\  \hline \\[-1.8ex]  Observations & 47,466 & 47,466 & 47,466 & 47,466 \\  $R^2$ & 0.148 & 0.160 & 0.225 & 0.231 \\  Adjusted $R^2$ & 0.145 & 0.158 & 0.223 & 0.229 \\ \hline \hline \\[-1.8ex] \textit{Note:} & \multicolumn{4}{r}{$^{*}$p$<$0.1; $^{**}$p$<$0.05; $^{***}$p$<$0.01} \\ \end{tabular} \label{covidreg6boston} \end{table}

\begin{table}[!htbp] \centering \caption[Linear regression models predicting the change in visits to POIs during the pandemic (2019 September - November) in Seattle when using dependency network generated from stays shorter than 4 hours] {Linear regression models predicting the change in visits to POIs during the pandemic (2019 September - November) in Seattle when using dependency network generated from stays shorter than 4 hours.} \begin{tabular}{@{\extracolsep{5pt}}lcccc} \\[-1.8ex]\hline \hline \\[-1.8ex] & \multicolumn{4}{c}{\textit{Dependent variable: $\Tilde v_i $ (Change in visits during the pandemic)}} \ \cr \cline{2-5} \\[-1.8ex] & (1) & (2) & (3) & (4) \\ \hline \\[-1.8ex]  Constant & -21.7$^{**}$ & -22.4$^{**}$ & -28.8$^{***}$ & -29.0$^{***}$ \\               & (10.50) & (9.905)     & (9.905)     & (9.860) \\  Distance effect $\sum_j \delta_{ij} \Tilde v_j $ & & 4.933$^{***}$ & & 3.439$^{***}$ \\   & & (0.188) & & (0.179) \\  Dependency effect $\sum_j w_{ij} \Tilde v_j $ && & 14.94$^{***}$ & 14.44$^{***}$ \\   & & & (0.211) & (0.212) \\ \hline \\[-1.8ex] Subcategory FE  & Y & Y & Y & Y  \\  PUMA FE  & Y & Y & Y & Y \\  \hline \\[-1.8ex]  Observations & 40,135 & 40,135 & 40,135 & 40,135 \\  $R^2$ & 0.171 & 0.185 & 0.263 & 0.270 \\  Adjusted $R^2$ & 0.169 & 0.183 & 0.261 & 0.268 \\ \hline \hline \\[-1.8ex] \textit{Note:} & \multicolumn{4}{r}{$^{*}$p$<$0.1; $^{**}$p$<$0.05; $^{***}$p$<$0.01} \\ \end{tabular} \label{covidreg6seattle} \end{table}

\begin{table}[!htbp] \centering \caption[Linear regression models predicting the change in visits to POIs during the pandemic (2019 September - November) in Los Angeles when using dependency network generated from stays shorter than 4 hours] {Linear regression models predicting the change in visits to POIs during the pandemic (2019 September - November) in Los Angeles when using dependency network generated from stays shorter than 4 hours.} \begin{tabular}{@{\extracolsep{5pt}}lcccc} \\[-1.8ex]\hline \hline \\[-1.8ex] & \multicolumn{4}{c}{\textit{Dependent variable: $\Tilde v_i $ (Change in visits during the pandemic)}} \ \cr \cline{2-5} \\[-1.8ex] & (1) & (2) & (3) & (4) \\ \hline \\[-1.8ex]  Constant & -18.5$^{***}$ & -19.0$^{***}$ & -22.9$^{***}$ & -23.1$^{***}$ \\               & (4.834) & (4.544)     & (4.544)     & (4.527) \\  Distance effect $\sum_j \delta_{ij} \Tilde v_j $ & & 4.905$^{***}$ & & 3.241$^{***}$ \\   & & (0.096) & & (0.091) \\  Dependency effect $\sum_j w_{ij} \Tilde v_j $ && & 15.07$^{***}$ & 14.61$^{***}$ \\   & & & (0.102) & (0.103) \\ \hline \\[-1.8ex] Subcategory FE  & Y & Y & Y & Y  \\  PUMA FE  & Y & Y & Y & Y \\  \hline \\[-1.8ex]  Observations & 164,195 & 164,195 & 164,195 & 164,195 \\  $R^2$ & 0.139 & 0.152 & 0.239 & 0.245 \\  Adjusted $R^2$ & 0.138 & 0.151 & 0.238 & 0.244 \\ \hline \hline \\[-1.8ex] \textit{Note:} & \multicolumn{4}{r}{$^{*}$p$<$0.1; $^{**}$p$<$0.05; $^{***}$p$<$0.01} \\ \end{tabular} \label{covidreg6la} \end{table}

\begin{table}[!htbp] \centering \caption[Linear regression models predicting the change in visits to POIs during the pandemic (2019 September - November) in Dallas when using dependency network generated from stays shorter than 4 hours] {Linear regression models predicting the change in visits to POIs during the pandemic (2019 September - November) in Dallas when using dependency network generated from stays shorter than 4 hours.} \begin{tabular}{@{\extracolsep{5pt}}lcccc} \\[-1.8ex]\hline \hline \\[-1.8ex] & \multicolumn{4}{c}{\textit{Dependent variable: $\Tilde v_i $ (Change in visits during the pandemic)}} \ \cr \cline{2-5} \\[-1.8ex] & (1) & (2) & (3) & (4) \\ \hline \\[-1.8ex]  Constant & 3.572$^{}$ & 0.977$^{}$ & -13.2$^{}$ & -14.2$^{}$ \\               & (10.11) & (9.521)     & (9.521)     & (9.476) \\  Distance effect $\sum_j \delta_{ij} \Tilde v_j $ & & 5.910$^{***}$ & & 3.843$^{***}$ \\   & & (0.146) & & (0.140) \\  Dependency effect $\sum_j w_{ij} \Tilde v_j $ && & 16.34$^{***}$ & 15.65$^{***}$ \\   & & & (0.161) & (0.163) \\ \hline \\[-1.8ex] Subcategory FE  & Y & Y & Y & Y  \\  PUMA FE  & Y & Y & Y & Y \\  \hline \\[-1.8ex]  Observations & 78,929 & 78,929 & 78,929 & 78,929 \\  $R^2$ & 0.144 & 0.162 & 0.242 & 0.250 \\  Adjusted $R^2$ & 0.143 & 0.160 & 0.241 & 0.248 \\ \hline \hline \\[-1.8ex] \textit{Note:} & \multicolumn{4}{r}{$^{*}$p$<$0.1; $^{**}$p$<$0.05; $^{***}$p$<$0.01} \\ \end{tabular} \label{covidreg6dallas} \end{table}

\subsection{Case study on school closure period}

\begin{figure}
\centering
\subfloat{\includegraphics[width=0.3\linewidth]{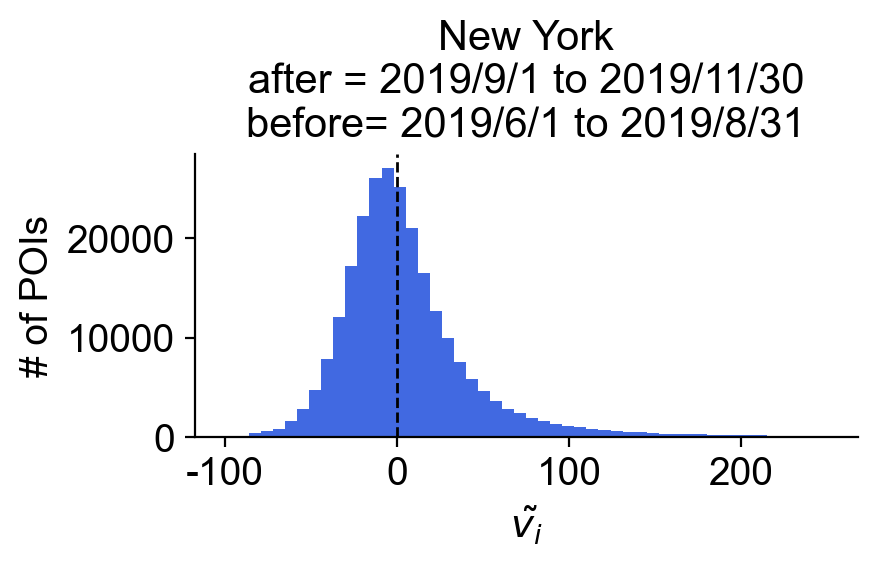}} 
\subfloat{\includegraphics[width=0.4\linewidth]{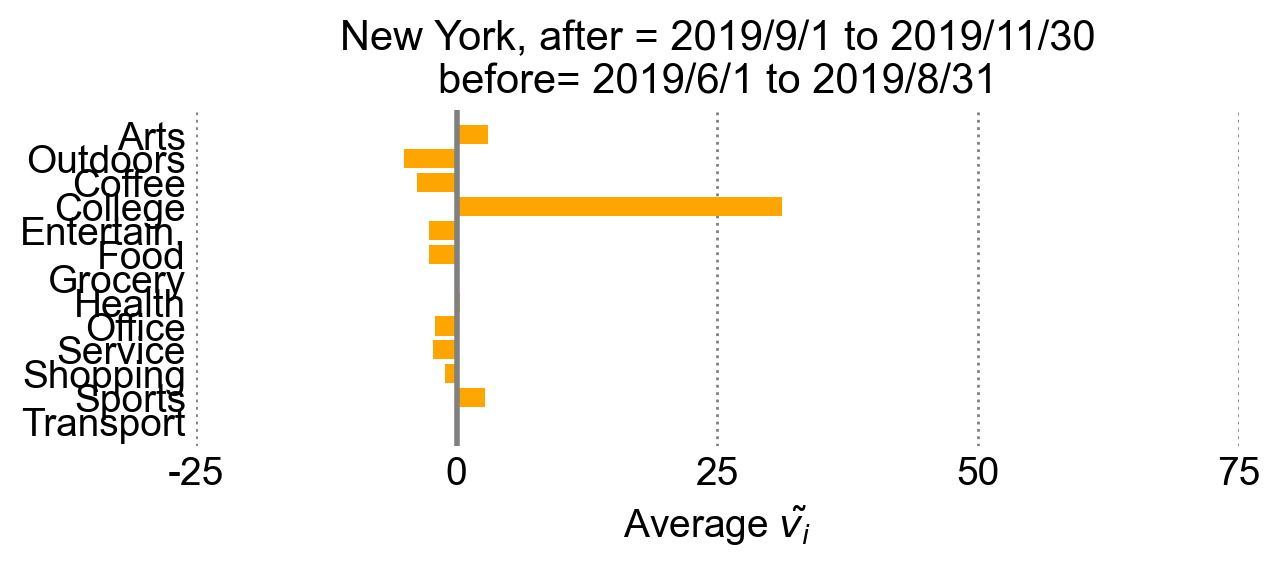}} \\
\subfloat{\includegraphics[width=0.3\linewidth]{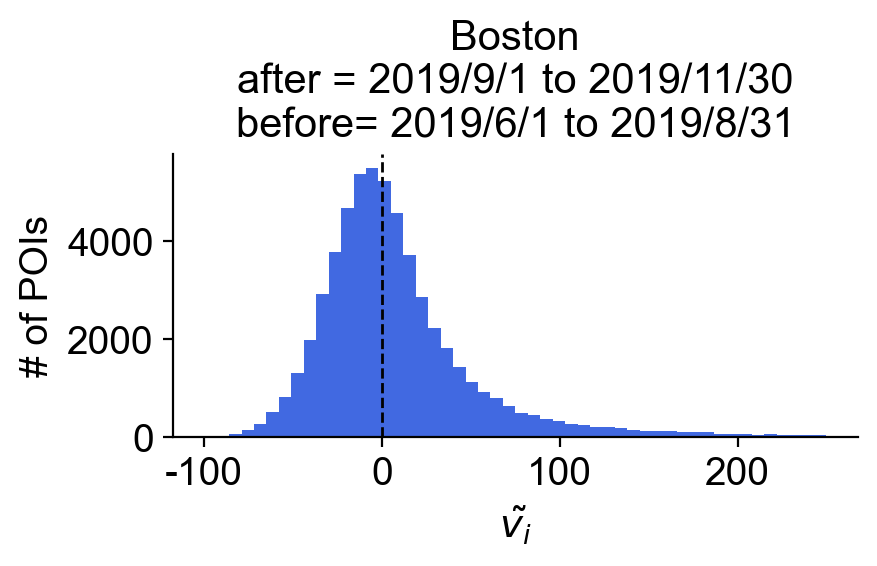}} 
\subfloat{\includegraphics[width=0.4\linewidth]{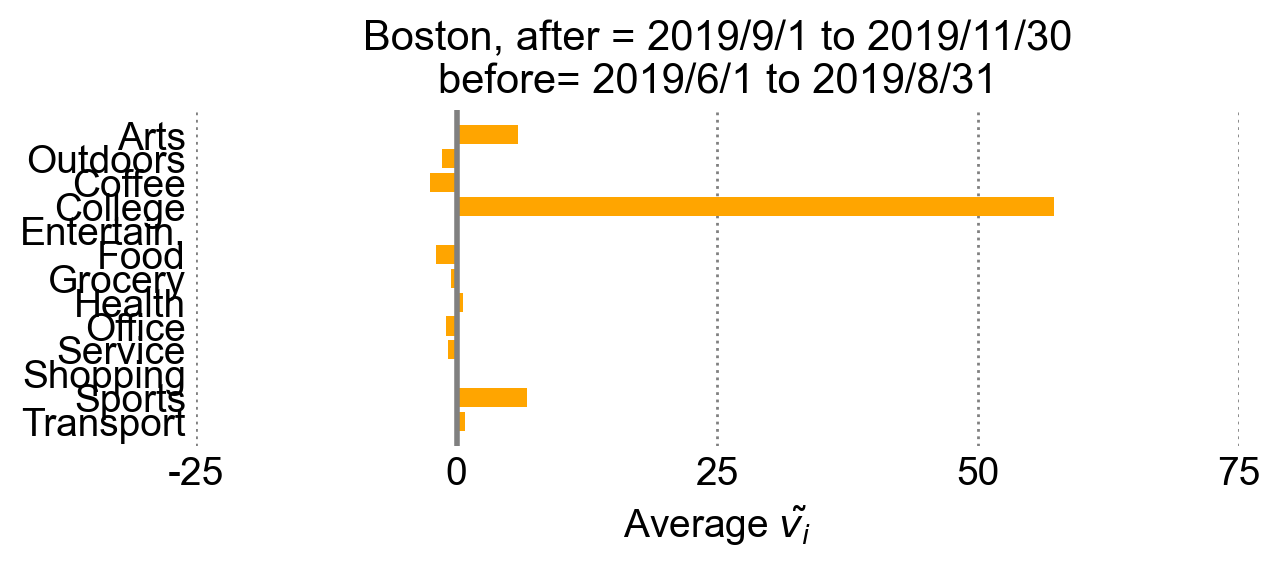}} \\
\subfloat{\includegraphics[width=0.3\linewidth]{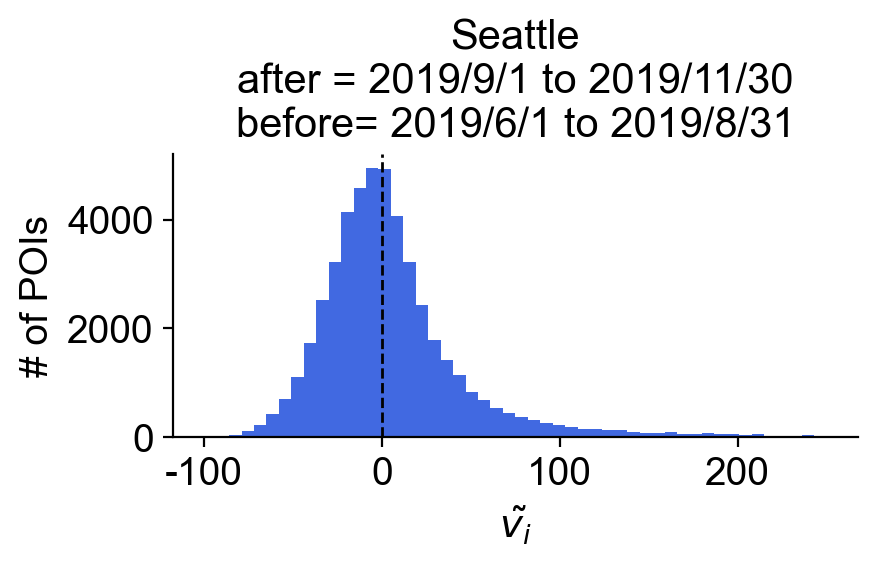}} 
\subfloat{\includegraphics[width=0.4\linewidth]{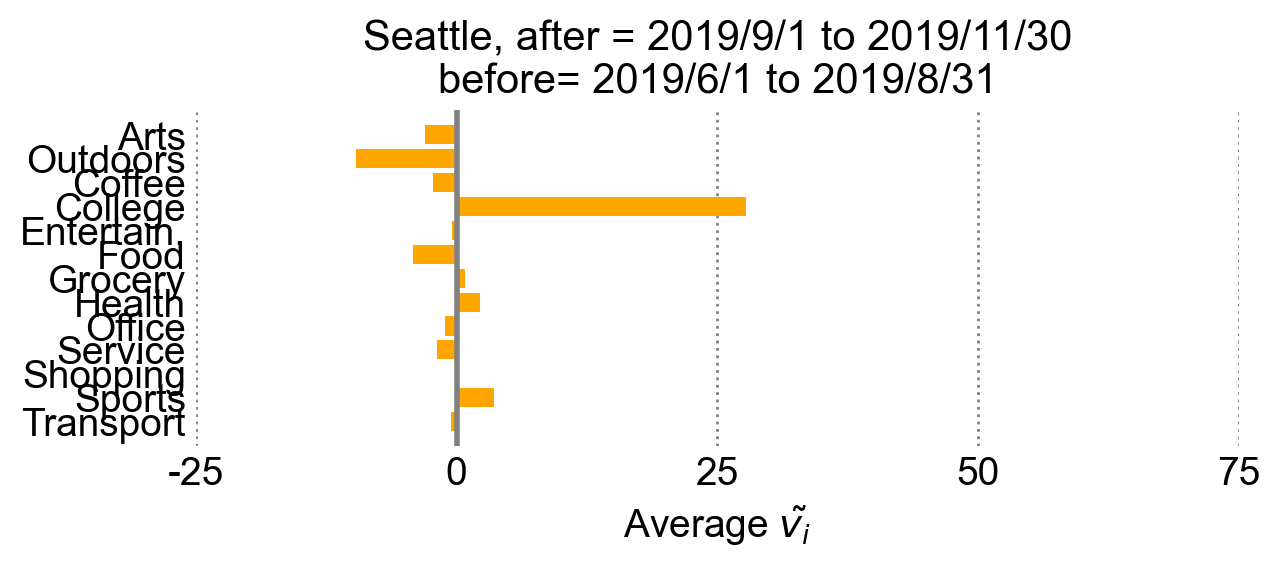}} \\
\subfloat{\includegraphics[width=0.3\linewidth]{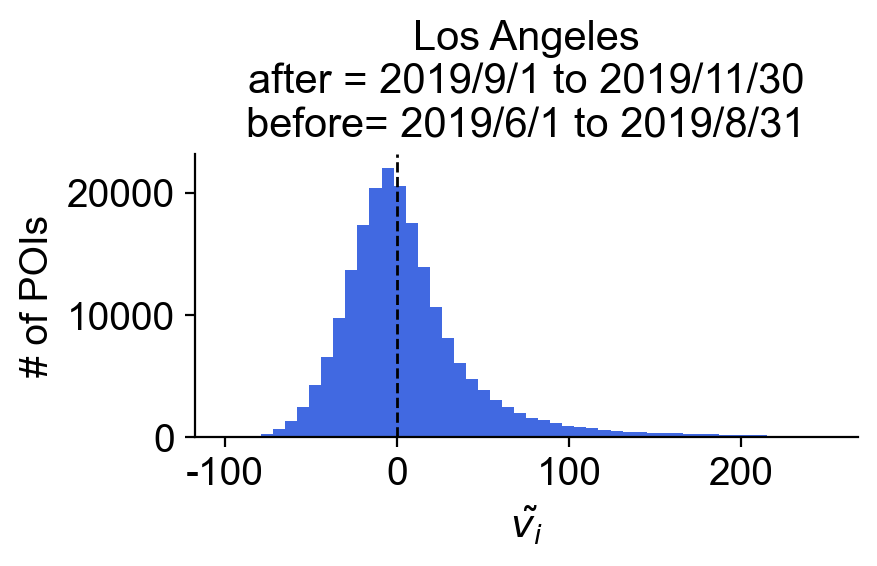}} 
\subfloat{\includegraphics[width=0.4\linewidth]{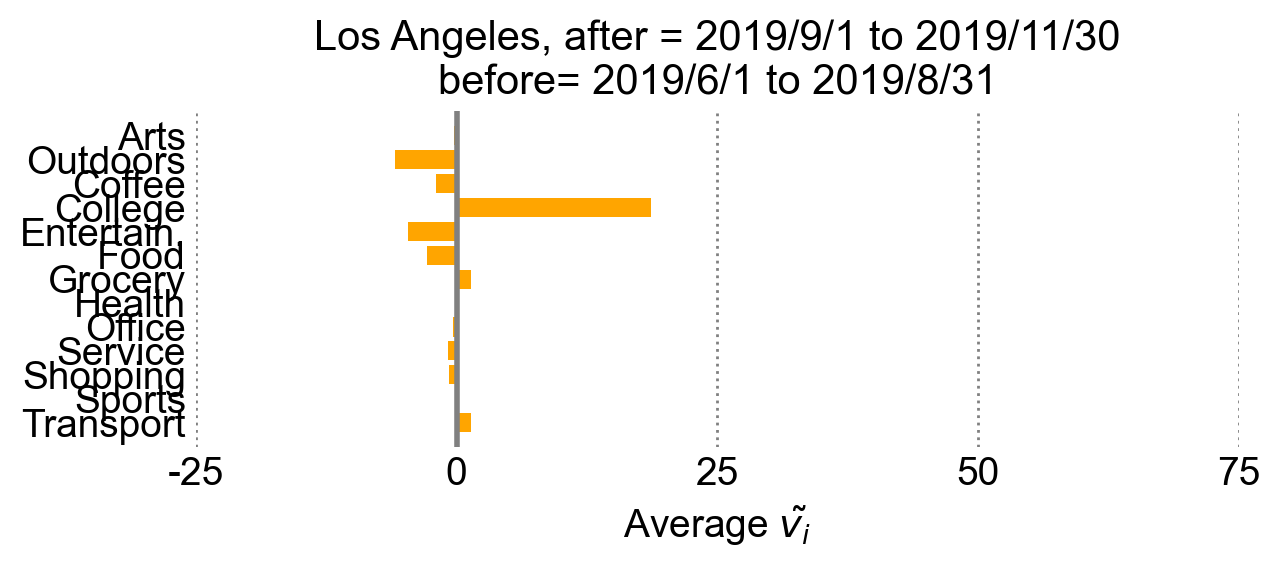}} \\
\subfloat{\includegraphics[width=0.3\linewidth]{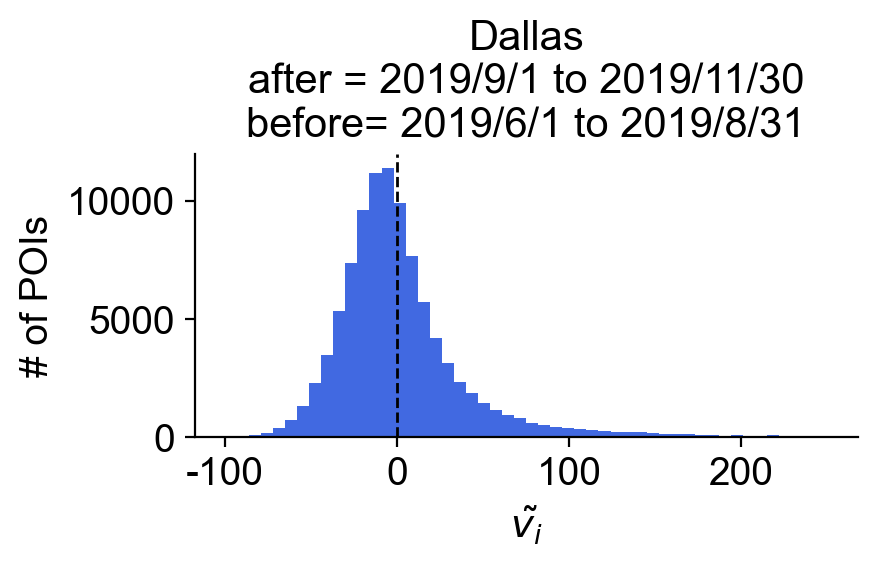}} 
\subfloat{\includegraphics[width=0.4\linewidth]{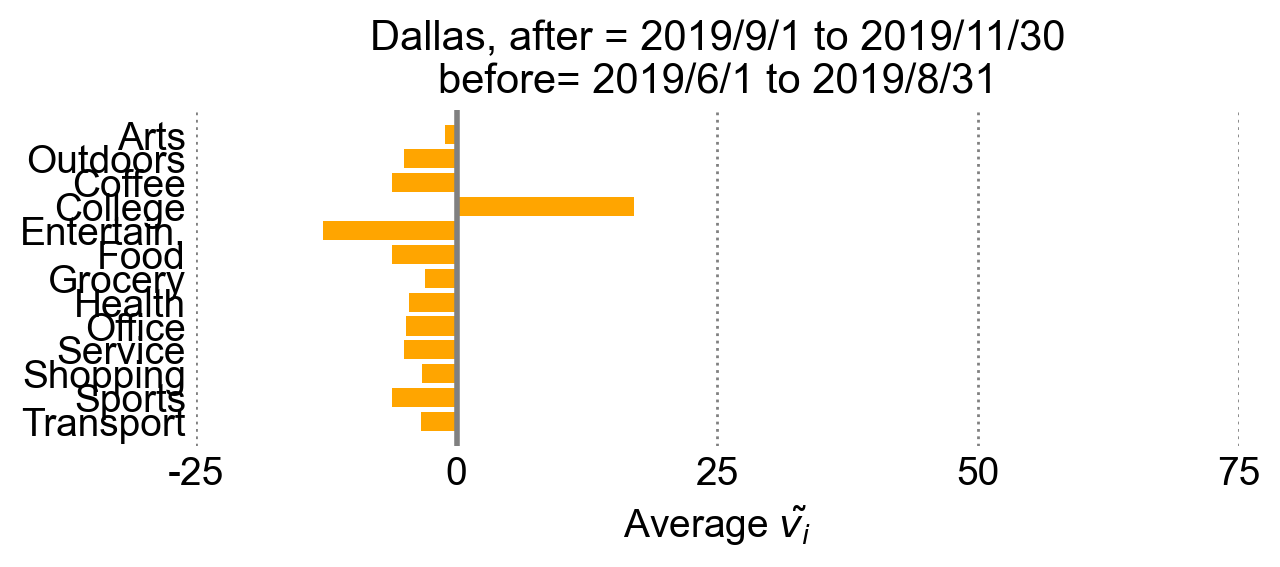}} \\
\caption[Change in visits to POIs in New York, Boston, Seattle, Los Angeles, and Dallas during the fall semester period.]{\textbf{Change in visits to POIs in New York, Boston, Seattle, Los Angeles, and Dallas during the fall semester period.} The left panels show the distribution of the change in visits, and the right panels show the average effects per POI category.}
\label{fig:s5vis1semester}
\end{figure}

To further test the effectiveness of using the dependency network for predicting the cascade of shocks, we used school holidays as a separate case study. The assumption is that, places that are connected to schools (in our case, college places) tend to have an additional decrease in visits than other places that have less dependency on college POIs.
To compute the change in visits, we used the school semester period (September to November 2019) as the post period and the summer break period (June to August 2019) as the pre- period.
\begin{equation}
    \Tilde v_i  = \big( \frac{v_i^{2019 September \sim 2019 November}}{v_i^{2019 June \sim 2019 August}} - 1 \big) * 100 (\%)
\end{equation}
Figure \ref{fig:s5vis1semester} shows the changes in visits to POIs in the five cities during the fall semester period compared to the summer break period. 
As the right panels clearly show, College POIs have a significant (25\% to 60\%) increase in the number of visits during the semester compared to the breaks. 
The same model specification was used to test the effectiveness of using the dependency network. Tables \ref{schoolreg1ny} to \ref{schoolreg1dallas} show the regression results. 
Although the predictability of the change in visits is generally lower than the pandemic results (due to the smaller impacts of summer holidays compared to the pandemic), in all five cities, the model performance is highest when we use the behavior-based dependency network compared to when we use the distance-based null network. 

\begin{table}[!htbp] \centering \caption[Linear regression models predicting the change in visits to POIs during the school semester (2019 September - November) compared to summer break (2019 June - August) in New York] {Linear regression models predicting the change in visits to POIs during the school semester (2019 September - November) compared to summer break (2019 June - August) in New York.} \begin{tabular}{@{\extracolsep{5pt}}lcccc} \\[-1.8ex]\hline \hline \\[-1.8ex] & \multicolumn{4}{c}{\textit{Dependent variable: $\Tilde v_i $ (Change in visits during school semester)}} \ \cr \cline{2-5} \\[-1.8ex] & (1) & (2) & (3) & (4) \\ \hline \\[-1.8ex]  Constant & -5.92$^{}$ & -2.48$^{}$ & -4.70$^{}$ & -3.06$^{}$ \\               & (14.83) & (14.42)     & (14.42)     & (14.39) \\  Distance effect $\sum_j \delta_{ij} \Tilde v_j $ & & 15.11$^{***}$ & & 7.714$^{***}$ \\   & & (0.240) & & (0.247) \\  Dependency effect $\sum_j w_{ij} \Tilde v_j $ && & 26.13$^{***}$ & 23.74$^{***}$ \\   & & & (0.235) & (0.247) \\ \hline \\[-1.8ex] Subcategory FE  & Y & Y & Y & Y  \\  PUMA FE  & Y & Y & Y & Y \\  \hline \\[-1.8ex]  Observations & 212,008 & 212,008 & 212,008 & 212,008 \\  $R^2$ & 0.015 & 0.033 & 0.069 & 0.073 \\  Adjusted $R^2$ & 0.014 & 0.032 & 0.068 & 0.072 \\ \hline \hline \\[-1.8ex] \textit{Note:} & \multicolumn{4}{r}{$^{*}$p$<$0.1; $^{**}$p$<$0.05; $^{***}$p$<$0.01} \\ \end{tabular} \label{schoolreg1ny} \end{table}

\begin{table}[!htbp] \centering \caption[Linear regression models predicting the change in visits to POIs during the school semester (2019 September - November) compared to summer break (2019 June - August) in Boston] {Linear regression models predicting the change in visits to POIs during the school semester (2019 September - November) compared to summer break (2019 June - August) in Boston.} \begin{tabular}{@{\extracolsep{5pt}}lcccc} \\[-1.8ex]\hline \hline \\[-1.8ex] & \multicolumn{4}{c}{\textit{Dependent variable: $\Tilde v_i $ (Change in visits during school semester)}} \ \cr \cline{2-5} \\[-1.8ex] & (1) & (2) & (3) & (4) \\ \hline \\[-1.8ex]  Constant & 12.09$^{}$ & 17.42$^{}$ & 18.28$^{}$ & 20.11$^{}$ \\               & (16.29) & (15.45)     & (15.45)     & (15.39) \\  Distance effect $\sum_j \delta_{ij} \Tilde v_j $ & & 20.47$^{***}$ & & 9.685$^{***}$ \\   & & (0.456) & & (0.475) \\  Dependency effect $\sum_j w_{ij} \Tilde v_j $ && & 33.63$^{***}$ & 29.87$^{***}$ \\   & & & (0.458) & (0.492) \\ \hline \\[-1.8ex] Subcategory FE  & Y & Y & Y & Y  \\  PUMA FE  & Y & Y & Y & Y \\  \hline \\[-1.8ex]  Observations & 48,652 & 48,652 & 48,652 & 48,652 \\  $R^2$ & 0.049 & 0.087 & 0.144 & 0.151 \\  Adjusted $R^2$ & 0.046 & 0.084 & 0.141 & 0.148 \\ \hline \hline \\[-1.8ex] \textit{Note:} & \multicolumn{4}{r}{$^{*}$p$<$0.1; $^{**}$p$<$0.05; $^{***}$p$<$0.01} \\ \end{tabular} \label{schoolreg1boston} \end{table}

\begin{table}[!htbp] \centering \caption[Linear regression models predicting the change in visits to POIs during the school semester (2019 September - November) compared to summer break (2019 June - August) in Seattle] {Linear regression models predicting the change in visits to POIs during the school semester (2019 September - November) compared to summer break (2019 June - August) in Seattle.} \begin{tabular}{@{\extracolsep{5pt}}lcccc} \\[-1.8ex]\hline \hline \\[-1.8ex] & \multicolumn{4}{c}{\textit{Dependent variable: $\Tilde v_i $ (Change in visits during school semester)}} \ \cr \cline{2-5} \\[-1.8ex] & (1) & (2) & (3) & (4) \\ \hline \\[-1.8ex]  Constant & -7.76$^{}$ & -8.32$^{}$ & -8.75$^{}$ & -9.05$^{}$ \\               & (56.04) & (55.94)     & (55.94)     & (55.92) \\  Distance effect $\sum_j \delta_{ij} \Tilde v_j $ & & 6.630$^{***}$ & & 4.304$^{***}$ \\   & & (0.991) & & (1.011) \\  Dependency effect $\sum_j w_{ij} \Tilde v_j $ && & 12.02$^{***}$ & 11.15$^{***}$ \\   & & & (0.974) & (0.995) \\ \hline \\[-1.8ex] Subcategory FE  & Y & Y & Y & Y  \\  PUMA FE  & Y & Y & Y & Y \\  \hline \\[-1.8ex]  Observations & 41,294 & 41,294 & 41,294 & 41,294 \\  $R^2$ & 0.004 & 0.005 & 0.008 & 0.008 \\  Adjusted $R^2$ & 0.001 & 0.002 & 0.005 & 0.005 \\ \hline \hline \\[-1.8ex] \textit{Note:} & \multicolumn{4}{r}{$^{*}$p$<$0.1; $^{**}$p$<$0.05; $^{***}$p$<$0.01} \\ \end{tabular} \label{schoolreg1seattle} \end{table}

\begin{table}[!htbp] \centering \caption[Linear regression models predicting the change in visits to POIs during the school semester (2019 September - November) compared to summer break (2019 June - August) in Los Angeles] {Linear regression models predicting the change in visits to POIs during the school semester (2019 September - November) compared to summer break (2019 June - August) in Los Angeles.} \begin{tabular}{@{\extracolsep{5pt}}lcccc} \\[-1.8ex]\hline \hline \\[-1.8ex] & \multicolumn{4}{c}{\textit{Dependent variable: $\Tilde v_i $ (Change in visits during school semester)}} \ \cr \cline{2-5} \\[-1.8ex] & (1) & (2) & (3) & (4) \\ \hline \\[-1.8ex]  Constant & 4.979$^{}$ & 5.578$^{}$ & 6.292$^{}$ & 6.486$^{}$ \\               & (12.90) & (12.63)     & (12.63)     & (12.61) \\  Distance effect $\sum_j \delta_{ij} \Tilde v_j $ & & 11.14$^{***}$ & & 5.416$^{***}$ \\   & & (0.253) & & (0.260) \\  Dependency effect $\sum_j w_{ij} \Tilde v_j $ && & 21.40$^{***}$ & 19.81$^{***}$ \\   & & & (0.251) & (0.262) \\ \hline \\[-1.8ex] Subcategory FE  & Y & Y & Y & Y  \\  PUMA FE  & Y & Y & Y & Y \\  \hline \\[-1.8ex]  Observations & 166,784 & 166,784 & 166,784 & 166,784 \\  $R^2$ & 0.009 & 0.020 & 0.050 & 0.052 \\  Adjusted $R^2$ & 0.008 & 0.019 & 0.049 & 0.051 \\ \hline \hline \\[-1.8ex] \textit{Note:} & \multicolumn{4}{r}{$^{*}$p$<$0.1; $^{**}$p$<$0.05; $^{***}$p$<$0.01} \\ \end{tabular} \label{schoolreg1la} \end{table}

\begin{table}[!htbp] \centering \caption[Linear regression models predicting the change in visits to POIs during the school semester (2019 September - November) compared to summer break (2019 June - August) in Dallas] {Linear regression models predicting the change in visits to POIs during the school semester (2019 September - November) compared to summer break (2019 June - August) in Dallas.} \begin{tabular}{@{\extracolsep{5pt}}lcccc} \\[-1.8ex]\hline \hline \\[-1.8ex] & \multicolumn{4}{c}{\textit{Dependent variable: $\Tilde v_i $ (Change in visits during school semester)}} \ \cr \cline{2-5} \\[-1.8ex] & (1) & (2) & (3) & (4) \\ \hline \\[-1.8ex]  Constant & 15.96$^{}$ & 32.60$^{}$ & 25.89$^{}$ & 33.77$^{}$ \\               & (45.82) & (41.70)     & (41.70)     & (40.70) \\  Distance effect $\sum_j \delta_{ij} \Tilde v_j $ & & 65.77$^{***}$ & & 40.32$^{***}$ \\   & & (0.598) & & (0.634) \\  Dependency effect $\sum_j w_{ij} \Tilde v_j $ && & 75.00$^{***}$ & 57.42$^{***}$ \\   & & & (0.579) & (0.629) \\ \hline \\[-1.8ex] Subcategory FE  & Y & Y & Y & Y  \\  PUMA FE  & Y & Y & Y & Y \\  \hline \\[-1.8ex]  Observations & 81,103 & 81,103 & 81,103 & 81,103 \\  $R^2$ & 0.006 & 0.135 & 0.177 & 0.216 \\  Adjusted $R^2$ & 0.004 & 0.134 & 0.175 & 0.214 \\ \hline \hline \\[-1.8ex] \textit{Note:} & \multicolumn{4}{r}{$^{*}$p$<$0.1; $^{**}$p$<$0.05; $^{***}$p$<$0.01} \\ \end{tabular} \label{schoolreg1dallas} \end{table}

\clearpage
\section{Cascading impacts of hypothetical shocks}

\subsection{Leontief open model}

% \textcolor{red}{Figure: (map, category impact heatmap, and distance decay) for different beta parameters, initial impact, also using bootstrap network? [can compute locally]}

To explore what the behavior-based dependency network can tell us about other types of future shocks, we apply the network effects regression model to simulate the spatial cascades of such shocks in different cities. 
More specifically, re-writing and reorganizing the regression model in matrix form, we obtain
\begin{equation}
    \Vec{v} = W \Vec{v} + \Vec{f}   
\end{equation}
where $\Vec{v}$ is a vector of $\Tilde v_i$ for all $N$ places, $W$ is an $N \times N$ matrix where each element is $\Tilde w_{ij} = \hat \beta_W w_{ij}$, and vector $\Vec{f}$ is an aggregation of all fixed effects $\beta_0$, $\eta_i$, and $\theta_i$. This model specification is known as the Leontief Open Model, which is a simplified and linear economic model for an economy in which input equals output \cite{leontief1986input}.
To predict the propagation of shocks throughout places in the city, the shocks are modeled in the fixed effect vector $f$ (e.g., all colleges experience an external shock of $-50\%$ visits reduction due to uptake of online education), and the production vector $\Vec{v}$ is computed by solving the linear system $\hat{\Vec{v}} = (I-W)^{-1} \Vec{f}$ via the generalized minimal residual iteration method.

% In the following sections, we simulate the effects of three urban shocks: online education (thus reduction of mobility behavior to university campuses), remote health services (thus reduction of mobility behavior to hospitals), and electric vehicle adoption (thus reduction of mobility behavior to gas stations). 
% Different levels of shocks ($f$) as well as different levels of impact parameters $\beta_W$ are tested and compared with the distance-based dependency network counterpart to assess the effects of behavior-based dependency on the spatial cascade of shocks. 

The shift to online education which occurred during the pandemic is reported to have a continuing effect, with roughly 20 percent of school systems planning to or have already started online school programs \cite{RR-A956-1}. 
Previous studies \cite{bonner1968economic} as well as analysis in Figure 2 in the main manuscript have pointed out that college campuses have a substantial impact on the local economy. 
If online learning and remote education were permanent and increased with the help of advanced technology (e.g., augmented reality), what impacts would it have on other businesses and amenities?

\begin{figure}
\centering
\subfloat[Map of effects using dependency network]{\includegraphics[width=0.58\linewidth]{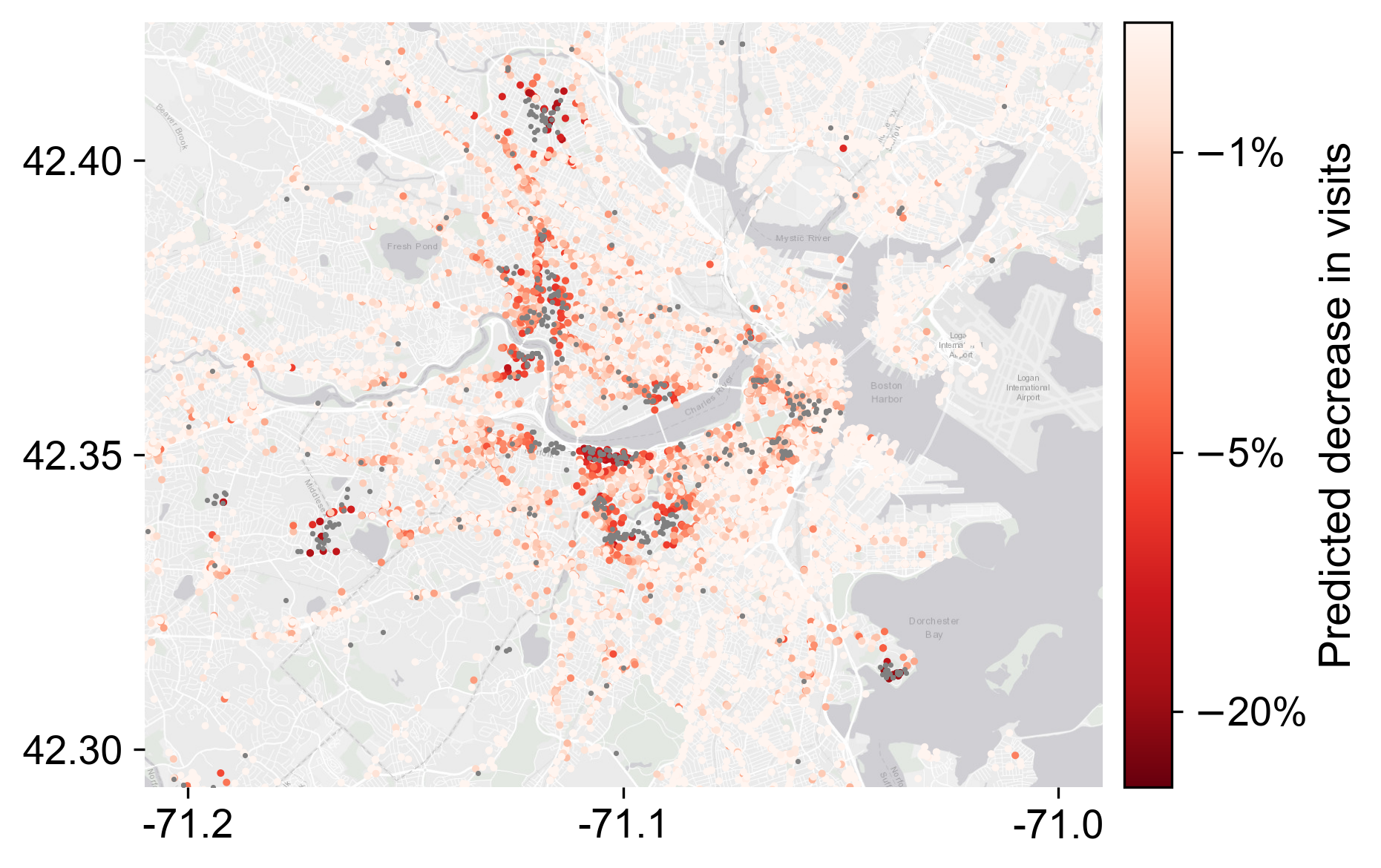}} 
\subfloat[Category and distance effects using dependency network]{\includegraphics[width=0.38\linewidth]{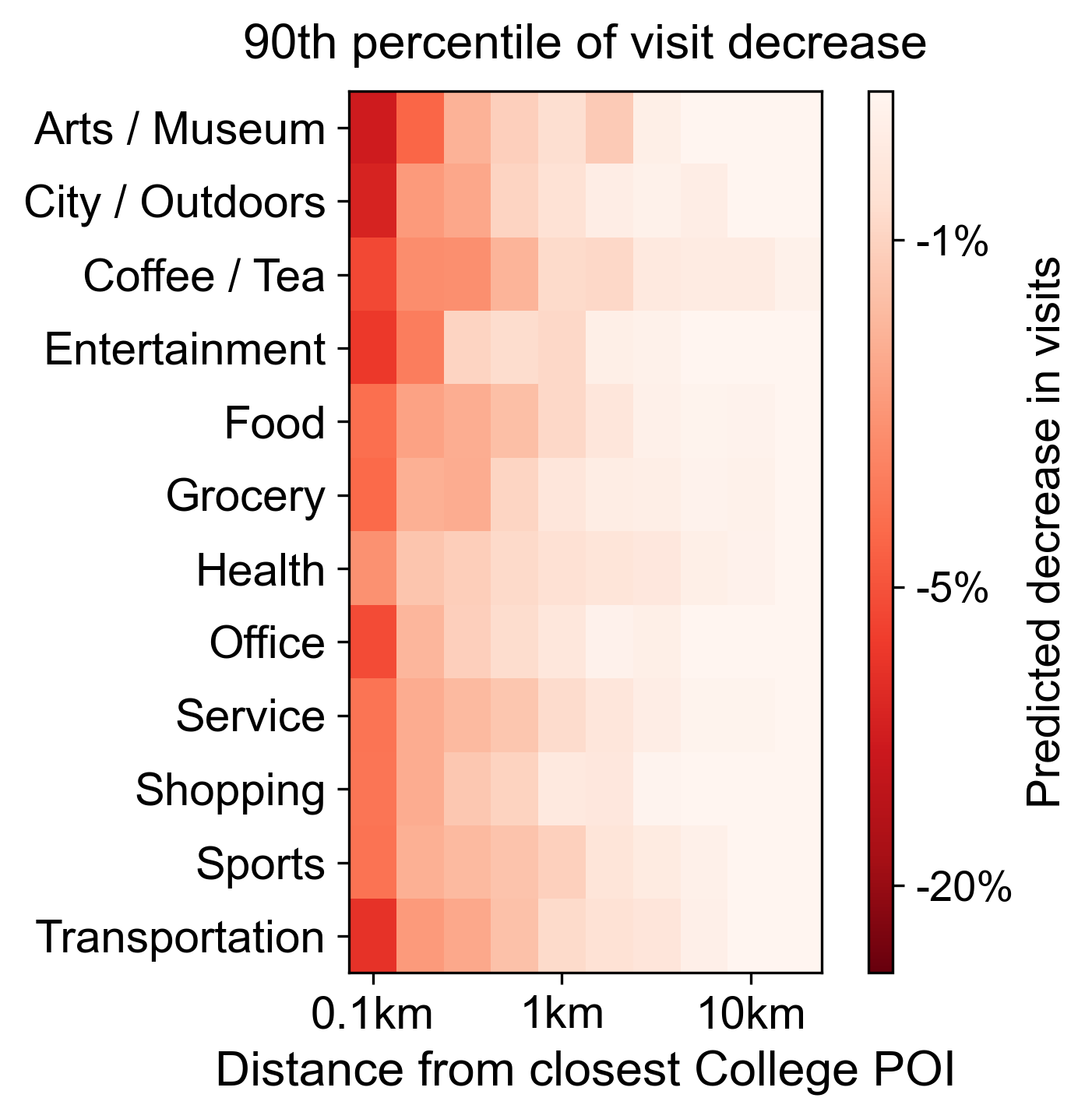}} \\
\subfloat[Map of effects using null network]{\includegraphics[width=0.58\linewidth]{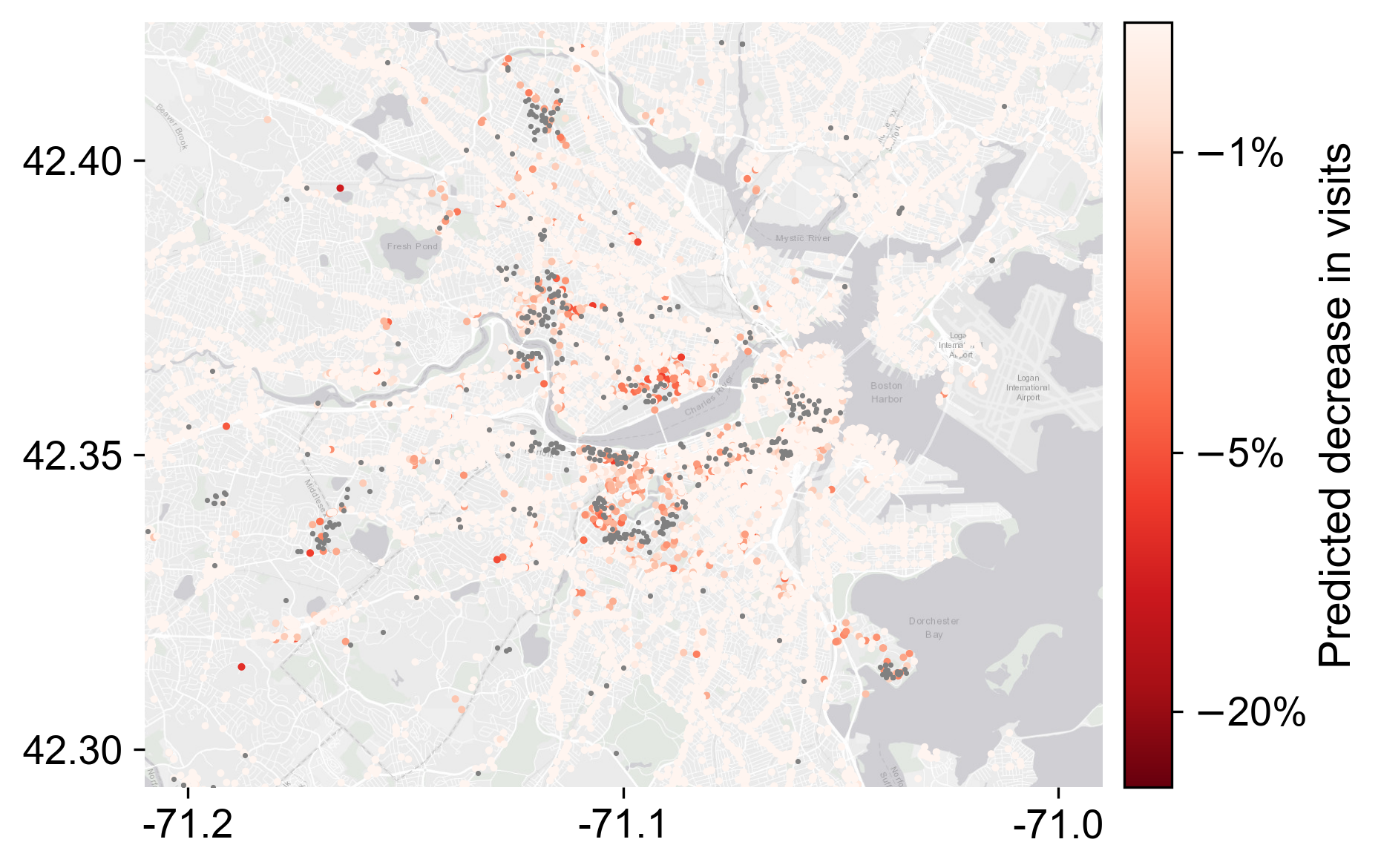}} 
\subfloat[Category and distance effects using null network]{\includegraphics[width=0.38\linewidth]{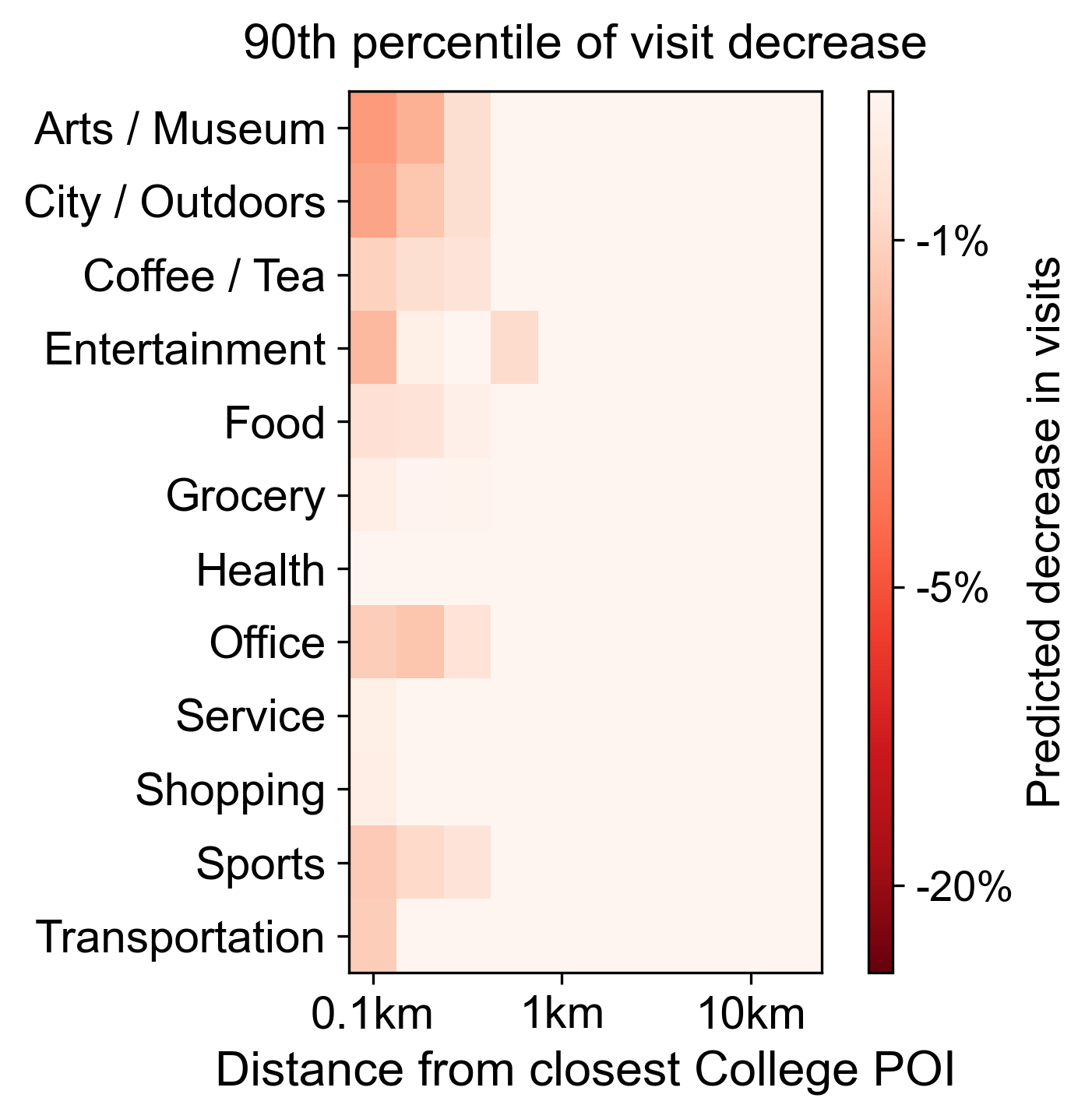}} 
\caption[Cascading impacts of a 50\% visit reduction to colleges.]{\textbf{Cascading impacts of a 50\% visit reduction to colleges.} 
a. Simulated effects of a 50\% reduction in visits to college POIs (gray points) on nearby non-college POIs (red points; darker red indicates larger negative impacts), using the fitted Leontief Open Model. Impacted POIs are not limited to those in proximity to college POIs. 
b. Impacts of the 50\% visit reduction to colleges on places by category and distance (90th percentile decrease in visits are shown, log scaled). 
c. and d. show results using the null networks.}
\label{fig:s6collegesim}
\end{figure}

Figure \ref{fig:s6collegesim}a shows the simulated effects of a 50\% reduction in visits to college POIs (gray points) on nearby non-college POIs (red points; darker red indicates larger negative impacts). Impacted POIs are not limited to those in proximity to college POIs, but also in locations that are popular with college students, for example, Massachusetts Avenue which connects MIT and Harvard University. 
For comparison, we simulated the shocks to non-college POIs using the physical distance network $\hat{W}$, where $\hat w_{ij}$ is used as the matrix elements instead of behavior-based dependency $w_{ij}$ in Figure \ref{fig:s6collegesim}c. 
Comparing the simulation results using the dependency network and the null network shows that neglecting the behavior-based dependencies results in a substantial underestimation of the effects on POIs that are located further away from colleges. 
Model parameters, $\beta_W = 0.70$ and $\beta_{null}=0.15$ were derived from prior regressions on estimation of changes in visits during pre-pandemic periods. 

\begin{figure}
\centering
\subfloat[Map of effects using dependency network]{\includegraphics[width=0.58\linewidth]{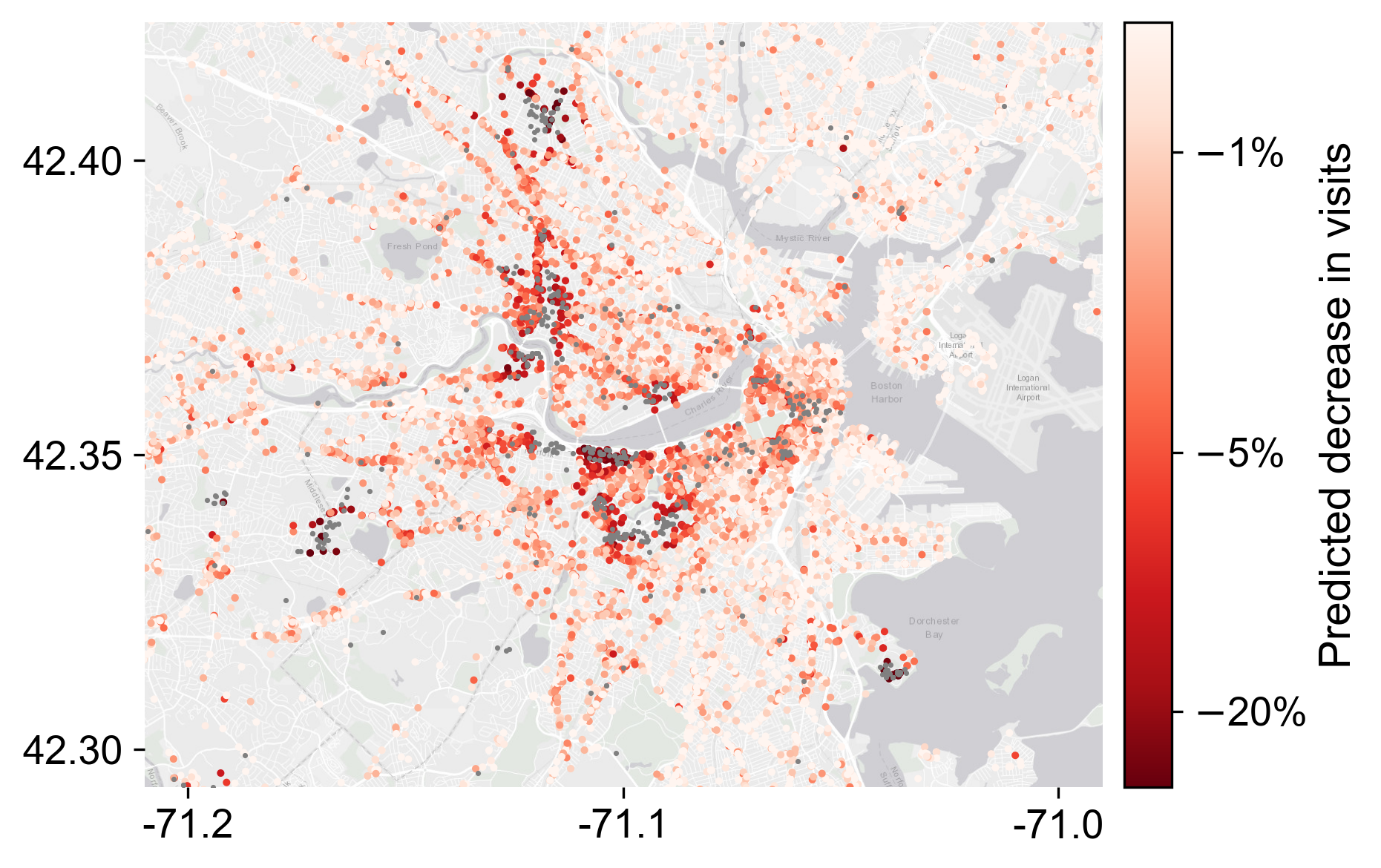}} 
\subfloat[Category and distance effects using dependency network]{\includegraphics[width=0.38\linewidth]{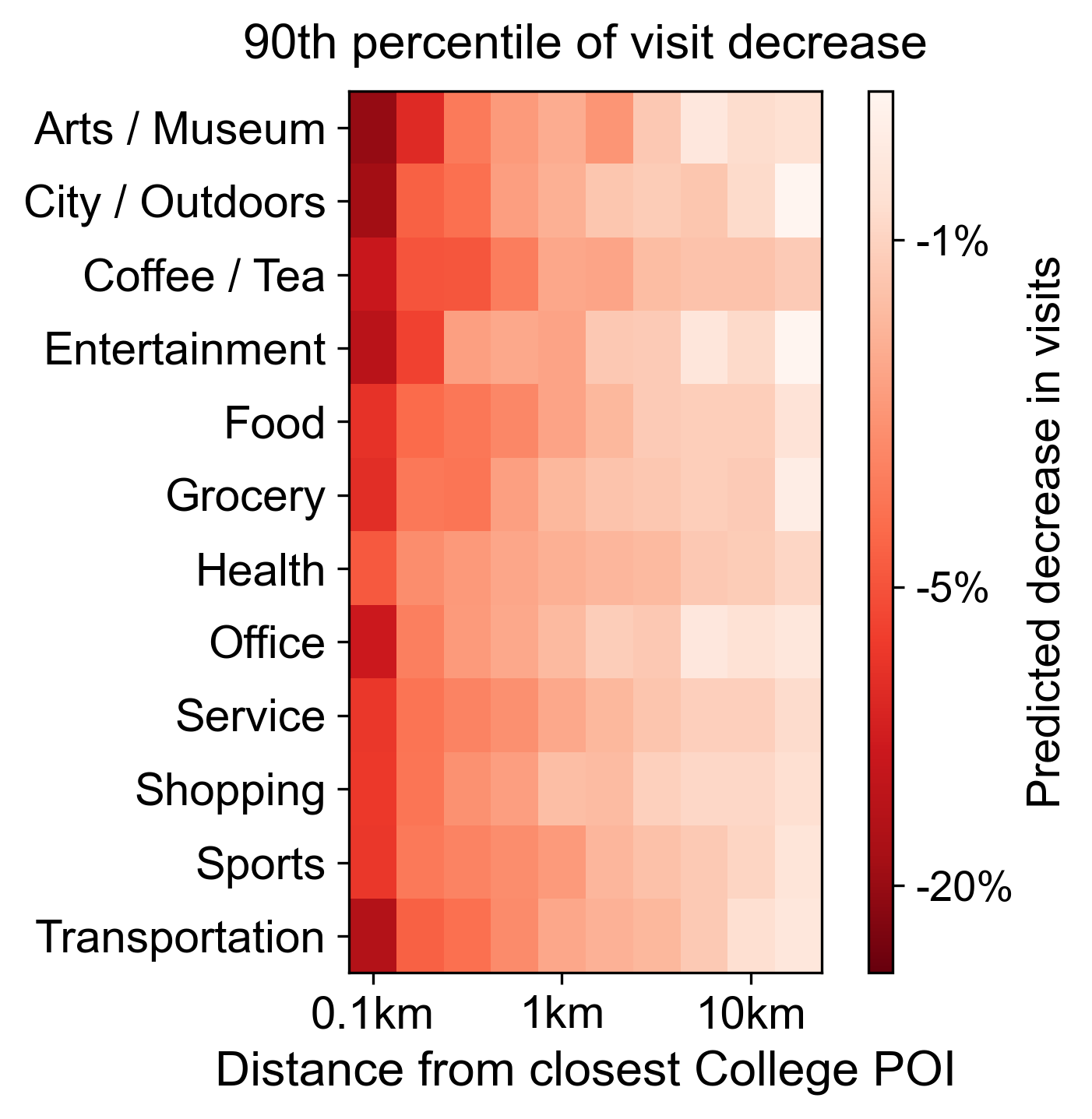}} \\
\subfloat[Map of effects using null network]{\includegraphics[width=0.58\linewidth]{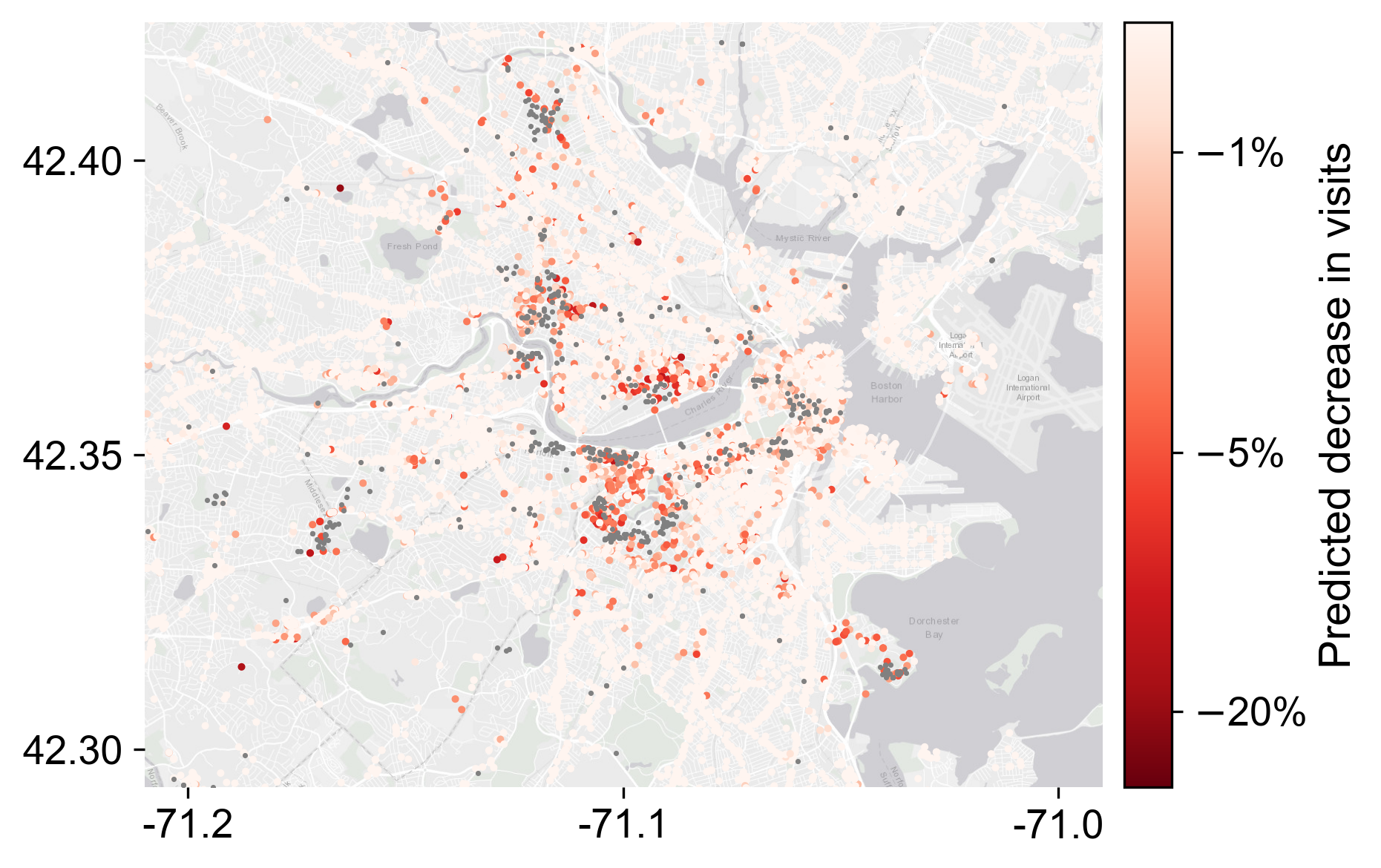}} 
\subfloat[Category and distance effects using null network]{\includegraphics[width=0.38\linewidth]{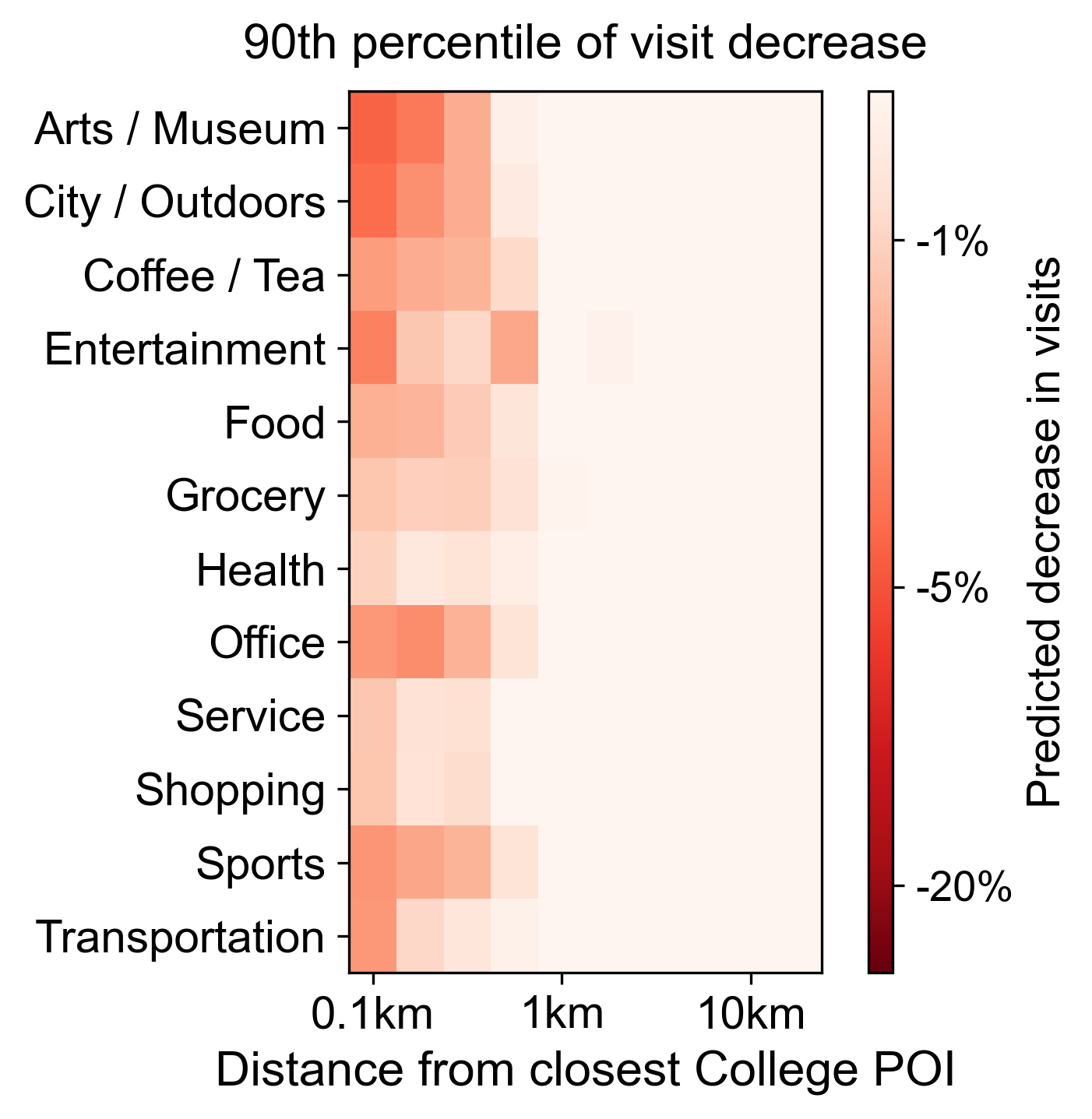}} 
\caption[Cascading impacts of a 100\% visit reduction to colleges.]{\textbf{Cascading impacts of a 100\% visit reduction to colleges.} 
a. Simulated effects of a 50\% reduction in visits to college POIs (gray points) on nearby non-college POIs (red points; darker red indicates larger negative impacts), using the fitted Leontief Open Model. Impacted POIs are not limited to those in proximity to college POIs. 
b. Impacts of the 100\% visit reduction to colleges on places by category and distance (90th percentile decrease in visits are shown, log scaled). 
c. and d. show results using the null networks.}
\label{fig:s6collegesim2}
\end{figure}

\begin{figure}
\centering
\subfloat[Map of effects using dependency network]{\includegraphics[width=0.58\linewidth]{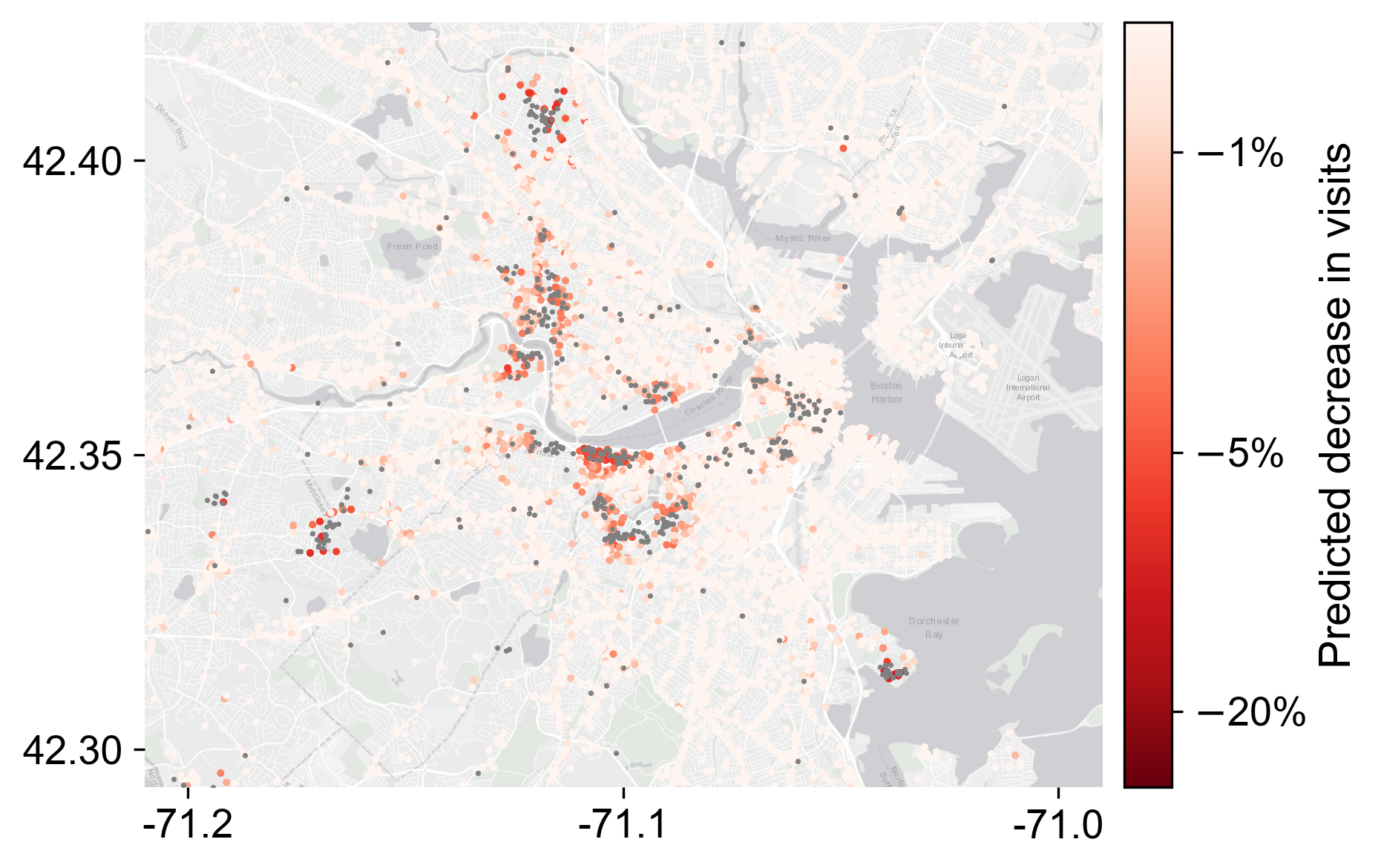}} 
\subfloat[Category and distance effects using dependency network]{\includegraphics[width=0.38\linewidth]{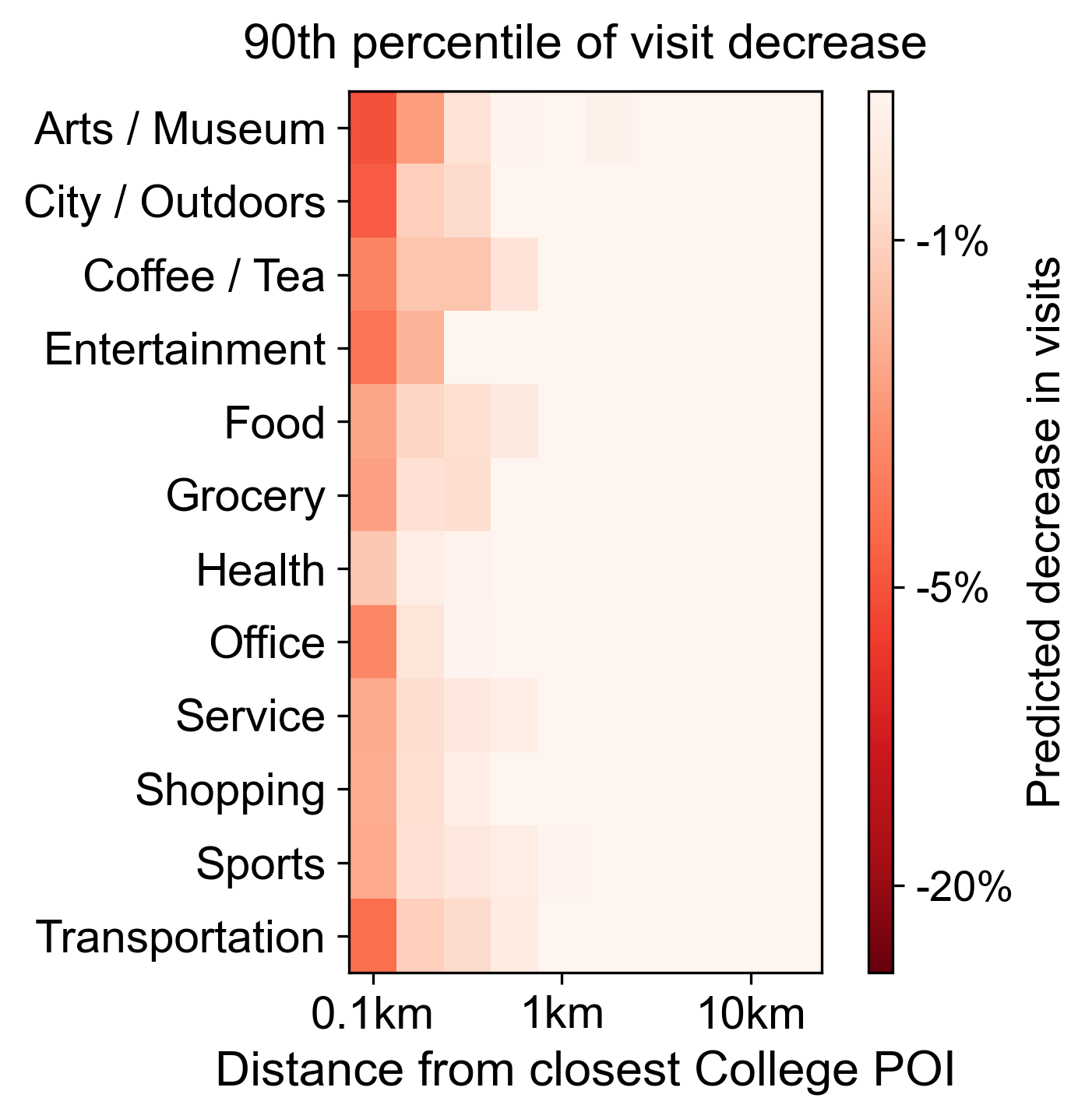}} \\
\subfloat[Map of effects using null network]{\includegraphics[width=0.58\linewidth]{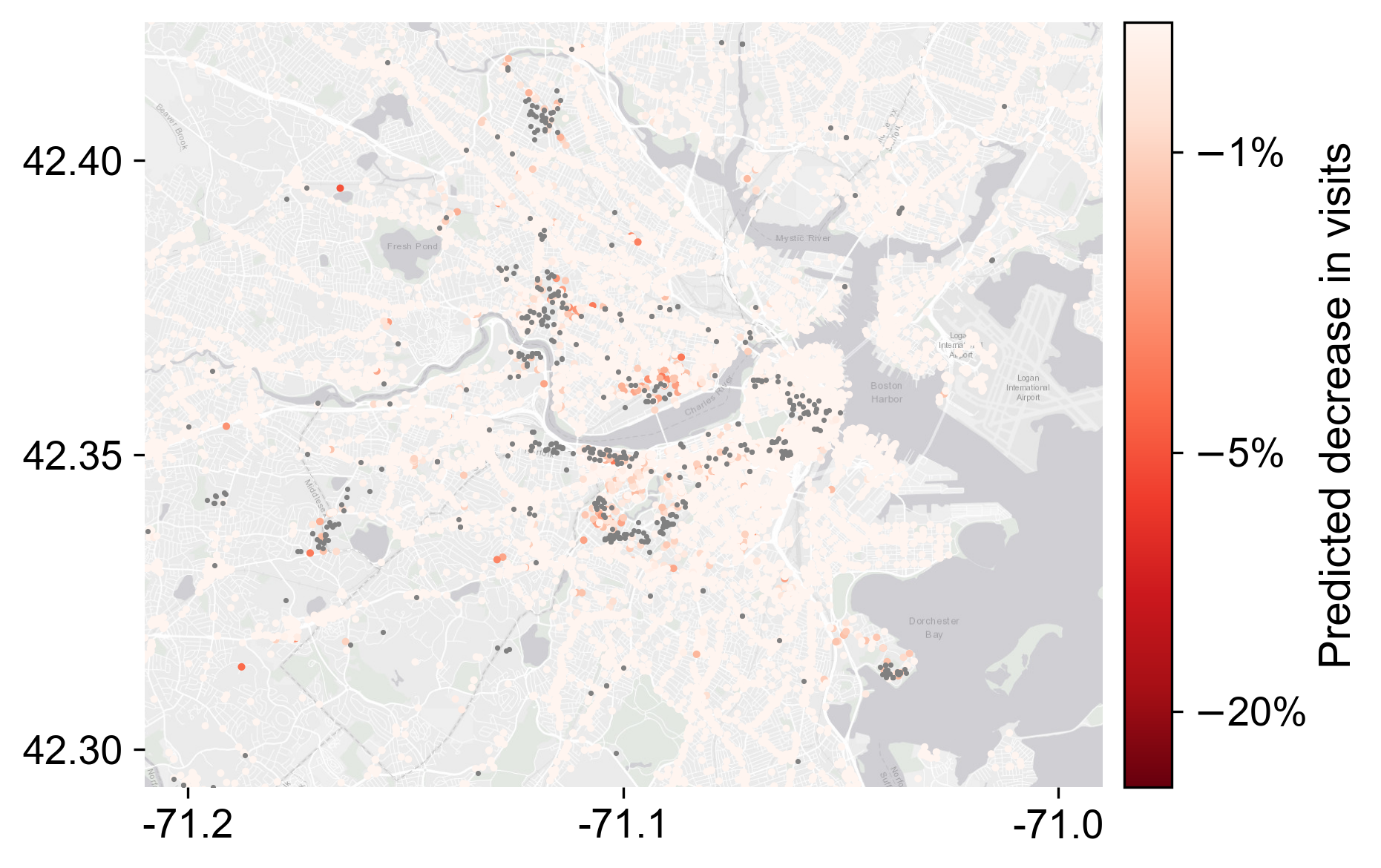}} 
\subfloat[Category and distance effects using null network]{\includegraphics[width=0.38\linewidth]{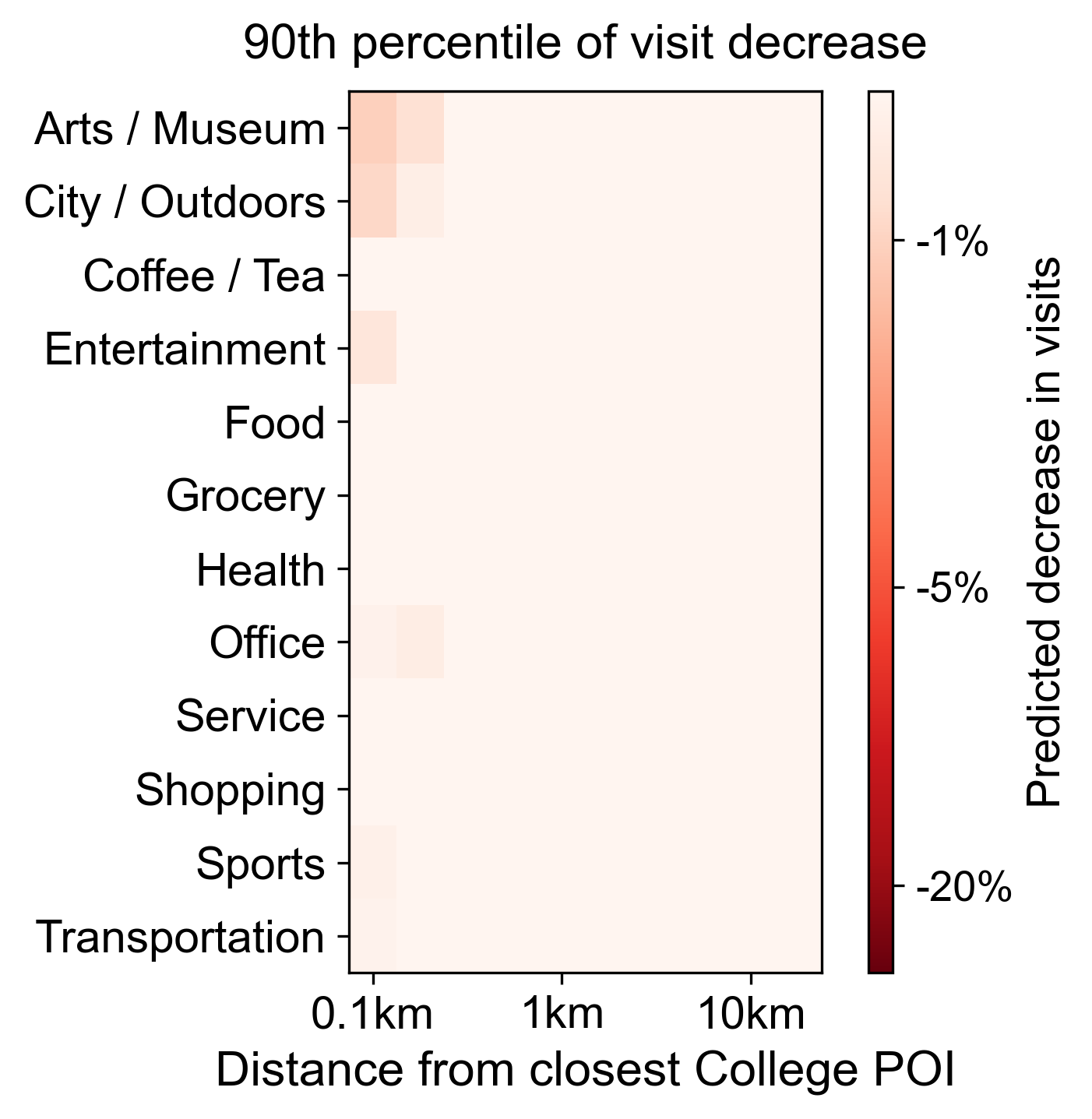}} 
\caption[Cascading impacts of a 25\% visit reduction to colleges.]{\textbf{Cascading impacts of a 25\% visit reduction to colleges.} 
a. Simulated effects of a 25\% reduction in visits to college POIs (gray points) on nearby non-college POIs (red points; darker red indicates larger negative impacts), using the fitted Leontief Open Model. Impacted POIs are not limited to those in proximity to college POIs. 
b. Impacts of the 50\% visit reduction to colleges on places by category and distance (90th percentile decrease in visits are shown, log scaled). 
c. and d. show results using the null networks.}
\label{fig:s6collegesim3}
\end{figure}

The effects of online education were heterogeneous for different place categories located at different distances from colleges. Figure \ref{fig:s6collegesim}b shows the 90th percentile of impacts on POIs by category and distance (log scaled). 
While most significant impacts occur within 0.5 km, places such as arts and museums, food, and service places experience substantial long-distance impacts. 
% To quantitatively compare the decay of impact with respect to the distance from the closest college between the two models, Figure \ref{fig:fig4}d plots the normalized (divided by the average impact to POIs within 100 meters) visit decrease at different distance intervals. The figure highlights the long-distance cascade of shocks from colleges mediated by the complex interdependencies created by human mobility behavior. 
Furthermore, simulations assuming different levels of visit decrease to colleges (e.g., -100\%, -25\%) show a similar long-distance cascade of shocks (Figures \ref{fig:s6collegesim2} and \ref{fig:s6collegesim3}).
These persistent spatial cascades emphasize the importance of considering behavior-based dependency relationships between places to grasp the holistic impact of such urban shocks for resilient urban planning.

% \noindent
% \textbf{Impact of remote health services (New York)}

% \noindent
% \textbf{Impact of EV adoption (Los Angeles)}

\subsection{Cascading effects of individual place closures}

\begin{figure}
\centering
\subfloat[New York]{\includegraphics[width=\linewidth]{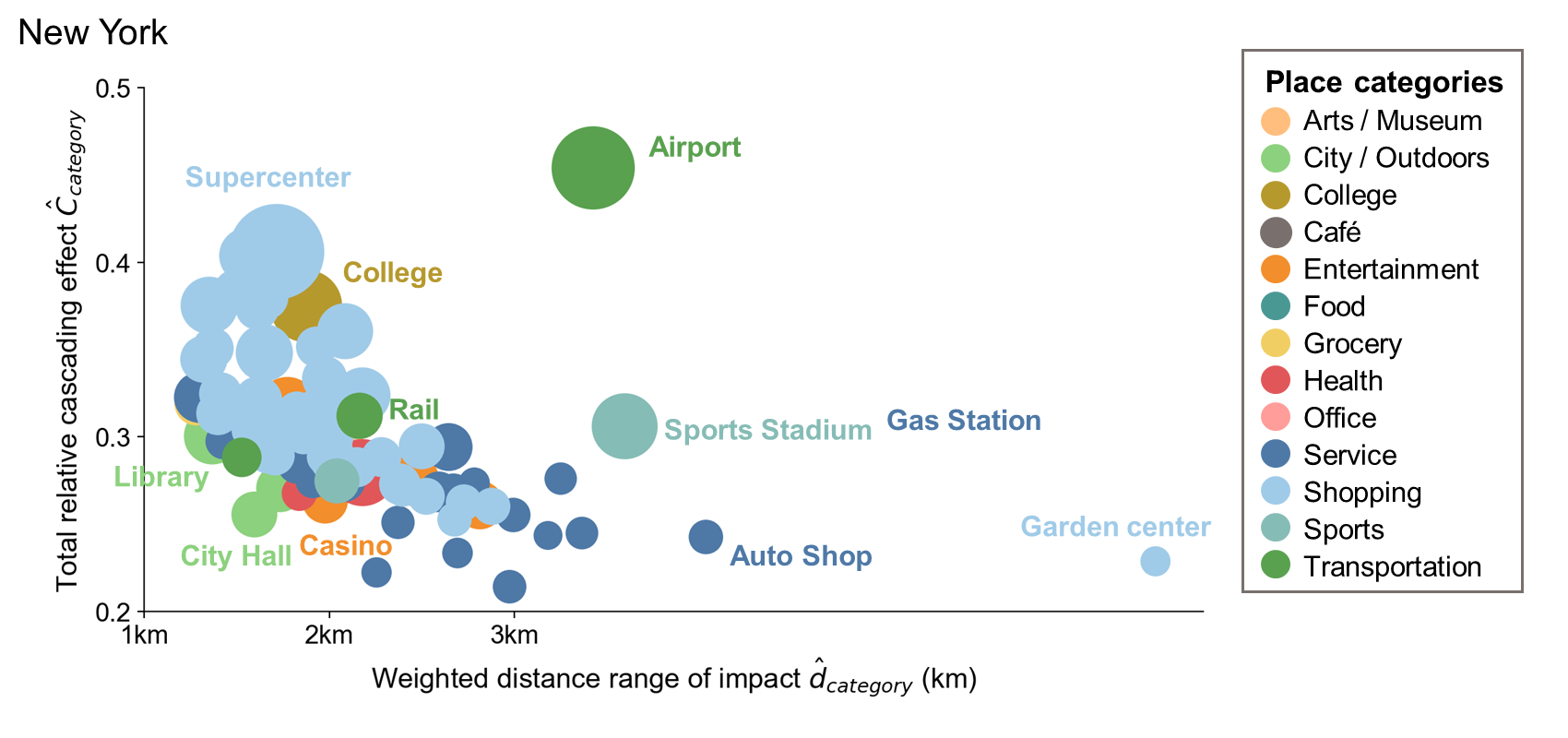}} \\
\subfloat[Boston]{\includegraphics[width=\linewidth]{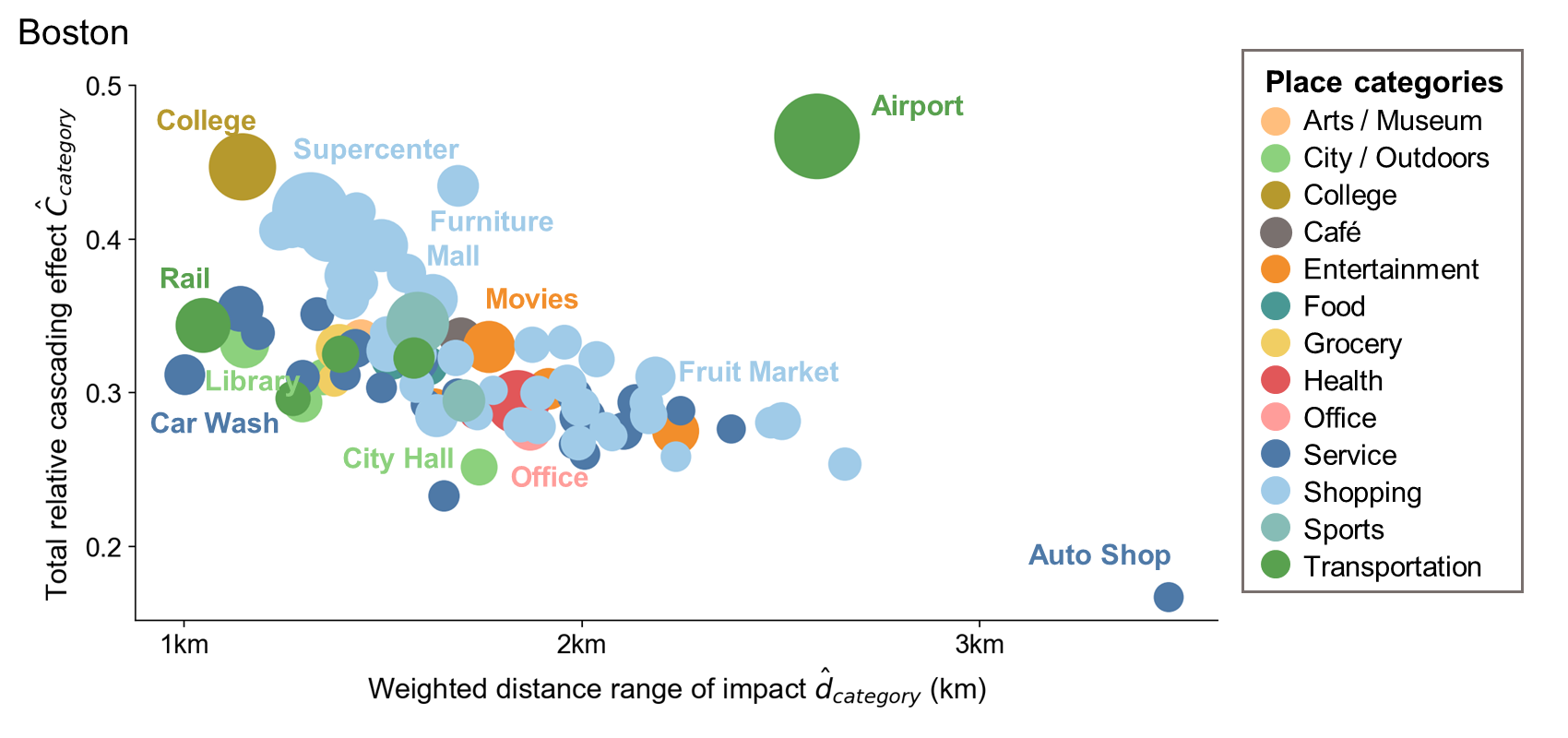}} \\
\caption[Total cascading impact of closing places on other locations in New York and Boston]{\textbf{Total cascading impact of closing places on other locations.} Total cascading impact of closing places on other locations, relative to its own size (x-axis) and the weighted distance range of the impact (y-axis) for different POI subcategories in New York and Boston.}
\label{fig:nodeclosure1}
\end{figure}

\begin{figure}
\centering
\subfloat[Seattle]{\includegraphics[width=\linewidth]{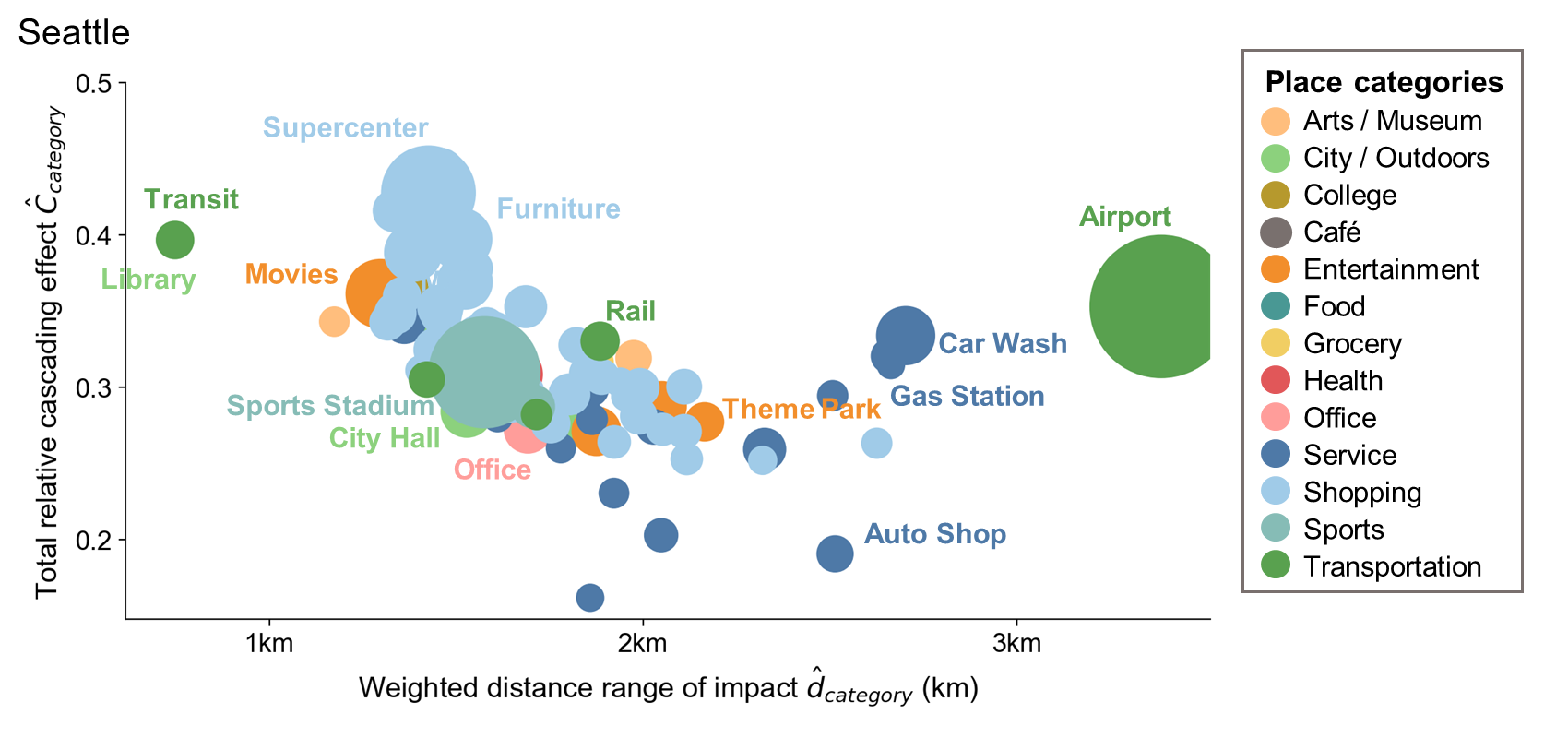}} \\
\subfloat[Los Angeles]{\includegraphics[width=\linewidth]{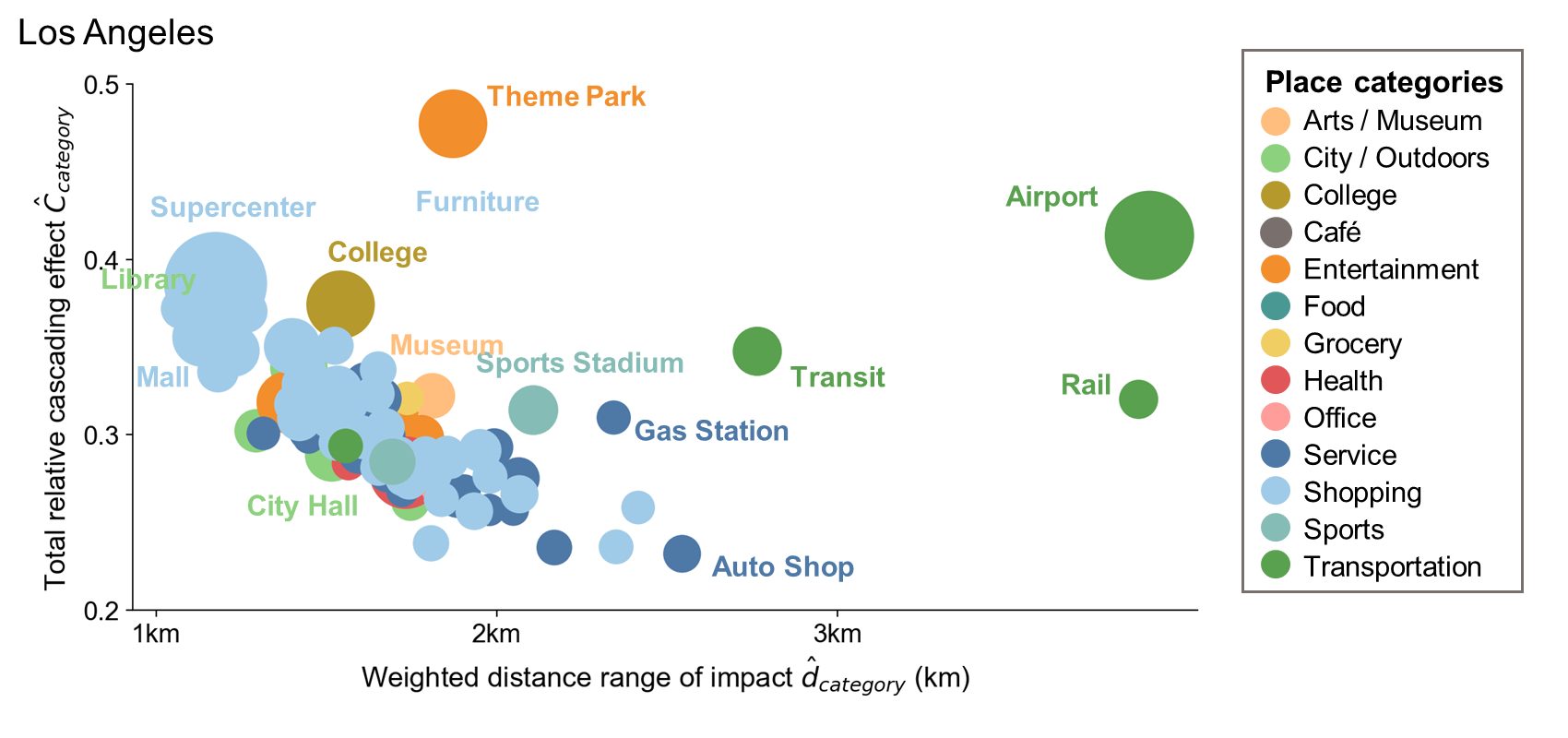}} \\
\caption[Total cascading impact of closing places on other locations in Seattle and Los Angeles]{\textbf{Total cascading impact of closing places on other locations.} Total cascading impact of closing places on other locations, relative to its own size (x-axis) and the weighted distance range of the impact (y-axis) for different POI subcategories in Seattle and Los Angeles.}
\label{fig:nodeclosure2}
\end{figure}

\begin{figure}
\centering
\subfloat[Dallas]{\includegraphics[width=\linewidth]{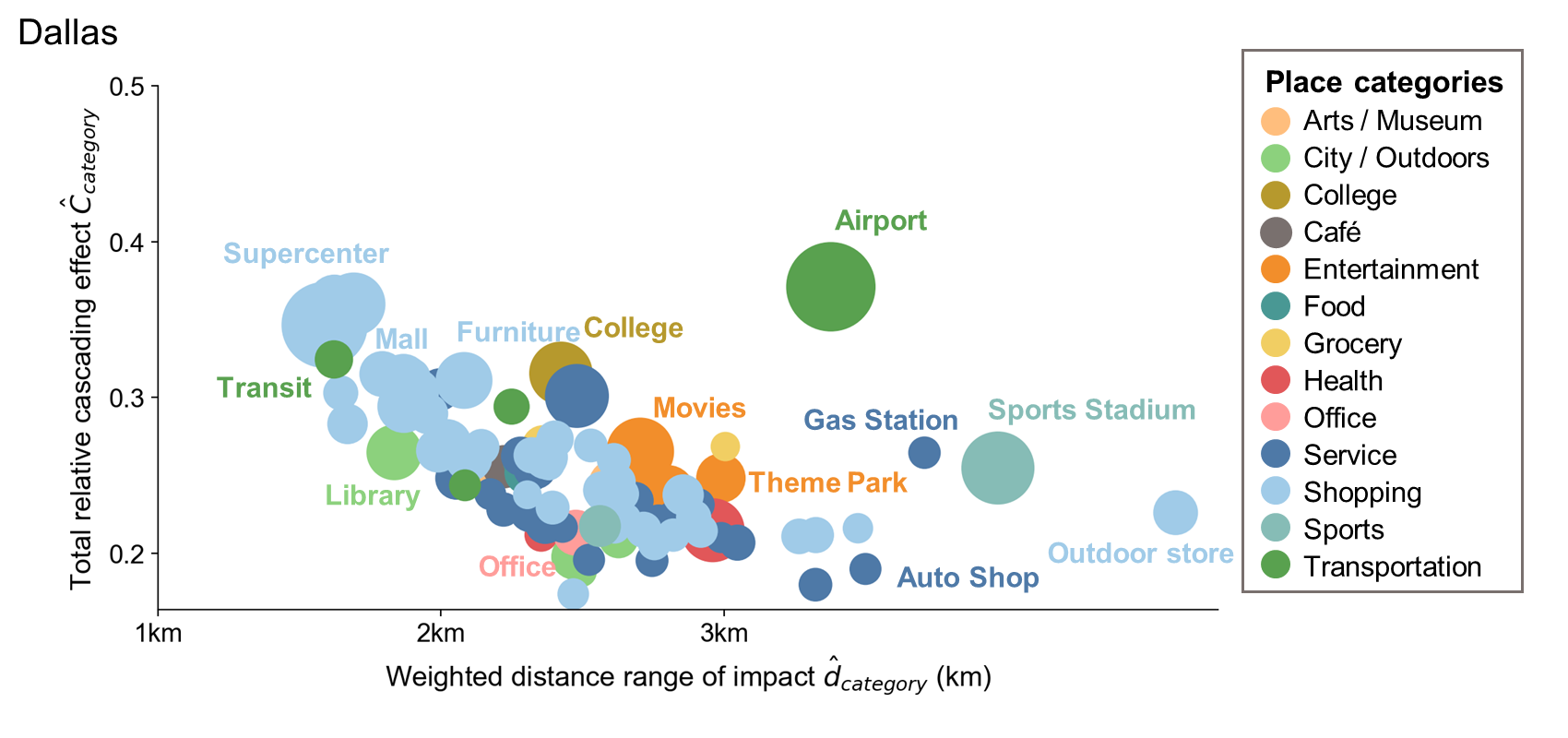}} \\
\caption[Total cascading impact of closing places on other locations in Dallas]{\textbf{Total cascading impact of closing places on other locations.} Total cascading impact of closing places on other locations, relative to its own size (x-axis) and the weighted distance range of the impact (y-axis) for different POI subcategories in Dallas.}
\label{fig:nodeclosure3}
\end{figure}

Further leveraging the network model, we are able to simulate the impacts of POI closure scenarios and identify the seed nodes (POIs) that have the largest cascading effects on other POIs if inflicted by other urban shocks. 
For each node, we simulate the cascading impacts of a 100\% visit change to a single node $i$, by computing $\hat{\vec{\nu}}^{(i)} = (I-W)^{-1} \vec e^{(i)}$, where $\vec e^{(i)}$ is a one-hot encoding vector of the initial shock that assigns a change in visits of $+1$ to node $i$ and $0$ otherwise, and $\hat{\vec{\nu}}^{(i)}$ is the resulting vector of the cascading impacts, where each element measures the impacts of the initial shock to all nodes.
The total impacts of changes in the number of visits to all nodes can be computed by multiplying $\hat{\vec{\nu}}^{(i)} = (\hat \nu_1^{(i)},\ldots,\hat \nu_N^{(i)})$ with the vector of total visits to each POI, $\vec n = (n_1, \cdots, n_N)$. Thus, the total impacts of the initial shock to node $i$ can be computed by $C_i = \sum_{j;j\neq i} \hat{\nu}^{(i)}_j  n_j$. 
By further scaling the impact to its own size $n_i$, we obtain the total relative cascading effect as $\hat{C_i} = C_i/n_i$. 
$\hat{C_i}=0.3$ indicates that increasing the number of visits to node $i$ by 100\% ($=n_i$) results in a total of 30\%$*n_i$ increase in visits across all other nodes. 
The mean relative cascading impacts of each POI category, $\hat{C}_{category}$ are shown in the y-axis of Figures \ref{fig:nodeclosure1} to \ref{fig:nodeclosure3}. 
POI categories such as airports, supercenters, colleges, furniture stores, theme parks, railway stations, and sports stadiums have a high impact on other POIs in urban areas propagated through behavior-based dependency networks. 

When implementing policies to close down certain POIs for emergency response (e.g., lockdowns during pandemics), it is important to understand the spatial extent of the cascade. 
To quantify this, we defined the distance range of the cascade by computing the average distance to impacted nodes, weighted by the magnitude of the impacts. More specifically, we compute the weighted distance range of POI $i$ by $\hat{d}_{i} = \sum_{j; j\neq i} \hat{\nu}^{(i)}_j d_{ij} / \sum_{j; j\neq i} \hat{\nu}^{(i)}_j$.

The weighted distance range of impact for each POI category, $\hat{d}_{category}$ are shown in the x-axis of Figures \ref{fig:nodeclosure1} to \ref{fig:nodeclosure3}. 
Supercenters and colleges have high cascading effects but are focused locally ($\sim$ 1.5km around the POI). On the other hand, the impacts of airports, stadiums, theme parks, and gas stations are both large and far-reaching ($\sim$ 2.5km to 3.5km). Understanding the magnitude and spatial extent of the cascading effects could be applied to design emergency management policies to effectively close places while minimizing economic losses. 
% the total cascading impact under four different seed node selection criteria: we select 100 nodes with the 1) largest total in-weights, 2) highest PageRank scores, 3) largest in-degrees, and 4) largest elements of the leading eigenvector of matrix $(I-W)^{-1}$. 
% Simulations show that closing POIs with the largest in-weights have an 11-fold impact on other POIs. 
% Figure \ref{fig:fig4}d shows the same results but when closing POIs with the largest in-weights across categories. 
% Results show that closing shopping POIs with the largest in-weights (i.e., supercenters and department stores) has the largest impact on other POIs, followed by service and college POIs, which agrees with the POI subcategory dependency structure shown in Figure \ref{fig:fig1}b. 
The large magnitude of the spatial cascades that occur due to behavior-based dependency networks calls for new urban policy-making approaches that balance the benefits of mobility restriction measures (e.g., preventing the spread of diseases) while minimizing the total cascading economic impacts to urban places and amenities. 

\clearpage
\section{Software}
Analysis was conducted using Python, Jupyter Lab, and the following libraries and software:
\begin{itemize}
    \item \texttt{NumPy} \cite{harris2020array} for general computation on Python.
    \item \texttt{Pandas} \cite{mckinney2011pandas} for loading, transforming, and analyzing data tables.
    \item \texttt{Matplotlib} \cite{hunter2007matplotlib} for creating plots and figures. 
    \item \texttt{GeoPandas} \cite{jordahl2014geopandas} for spatial analysis and plotting map figures. 
    \item \texttt{networkx} \cite{SciPyProceedings_11} for network analysis.
    \item \texttt{Statsmodels} \cite{seabold2010statsmodels} for statistical modeling and econometric analysis. 
    \item A Python implementation of the R \texttt{Stargazer} multiple regression model creation tool\footnote{\url{https://github.com/mwburke/stargazer}} was used to create the regression tables.
\end{itemize}

% \clearpage
% \bibliographystyle{plain}
% \bibliography{references}